\documentclass[12pt,a4paper,english]{book}
\usepackage[T1]{fontenc}
\usepackage[latin1]{inputenc}
\usepackage{float}
\usepackage{graphicx}
\usepackage{amsmath}

\usepackage{amsfonts}
\usepackage{latexsym}

\usepackage{babel}

\makeatletter

\newcommand{\beq}{\begin{equation}}
\newcommand{\eeq}{\end{equation}}

\newcommand{\CC}{\mathbb{C}}
\newcommand{\ZZ}{\mathbb{Z}}
\newcommand{\RR}{\mathbb{R}}

\newcommand{\mH}{\mathcal{H}}
\newcommand{\mA}{\mathcal{A}}
\newcommand{\CPT}{\tilde{C}\tilde{P}\tilde{T}}
\newcommand{\hlf}{\frac{1}{2}}
\newcommand{\bra}[1]{\langle #1 |}
\newcommand{\ket}[1]{|#1 \rangle}
\newcommand{\braket}[2]{\langle #1 | #2 \rangle}

\newcommand{\brakettt}[3]{\langle #1 | #2 |#3 \rangle}

\makeatother

\begin{document}

\thispagestyle{empty}

\vspace*{1cm}

\begin{center}
\LARGE
\bf 
Investigations in Two-Dimensional\\[2mm]
Quantum Field Theory\\[2mm]
by the Bootstrap and TCSA Methods
\vspace{1.8cm}

\large
\rm
Ph.D.\ thesis\\[2cm]

\LARGE
G\'abor Zsolt T\'oth
\vspace{2cm}
\end{center}

{
\large
\noindent
\hspace{1cm}
Supervisor: Professor L\'aszl\'o Palla, D.Sc.\\[0.8cm]
\hspace*{1cm}
E\"otv\"os  University, Budapest\\
\noindent
\hspace*{1cm}
Physics Doctoral School\\
\hspace*{1cm}
Particle Physics and Astronomy program\\
\hspace*{1cm}
Doctoral School leader: Zal\'an Horv\'ath\\
\hspace*{1cm}
Program leader: Ferenc Csikor
\vspace{1cm}

\begin{center}
Theoretical Physics Research Group of the\\[0cm]
Hungarian Academy of Sciences,\\[0cm] 
Theoretical Physics Department\\[0cm]
E\"otv\"os University, Budapest\\
\vspace{0.5cm}
2006
\end{center}
}

\newpage
\
\thispagestyle{empty}

\newpage

\renewcommand{\baselinestretch}{1.24}
\large\normalsize

\setcounter{page}{1}

\tableofcontents
\markboth{}{}

\chapter{Introduction}
\label{sec.introduction}
\markboth{INTRODUCTION}{}

Relativistic quantum field theory in two space-time dimensions has important
role in physics, for example in the branches of  string theory and statistical
mechanics. It has also played  important role in the development of a
non-perturbative understanding of quantum field theory in general.
In the latter respect massive integrable models and conformal
field theory attract most interest,
as they usually allow the exact determination of several physical quantities.
Since  W.\ Thirring proposed the first exactly solvable quantum field
theoretical model in 1958 \cite{WT}  and J.\ Schwinger presented the exact
solution of Quantum Electrodynamics in 1+1 dimensions \cite{JSch1,JSch2}, a  remarkable complexity and richness of
the non-perturbative structure of relativistic quantum field theories has been
revealed. 
In the field of the
two-dimensional  models of statistical physics, which are
  closely related to two-dimensional quantum field theory, H.\ Bethe's results \cite{Bethe} in 1931 and L.\ Onsager's solution of the Ising
model in 1943 \cite{LO} can be regarded as the beginning of the study of
integrable models.

In massive integrable field theories the spectrum,  
S-matrix and the form factors of local operators can often be determined exactly,
mainly by means of the bootstrap method. This is a  feature that
deserves high appreciation in itself, regarding the difficulty of 
making non-perturbative statements about these quantities in general. 

The bootstrap programme for the spectrum and S-matrix 
 proves to be  manageable in integrable quantum field theory because  the
 existence of higher spin conserved quantities, which is the criterion of integrability, severely constrains the
 scattering theory. The special kind of scattering theory related to
 integrable models is called factorized scattering theory. The bootstrap
 programme for form factors in integrable quantum field theory formulated 
in \cite{BKW,KW,W} is also based on factorized scattering.

Conformal field theory in two dimensions is a distinct branch of quantum field theory
which has important role in string theory and in the description of
physical systems at critical points. A remarkable feature of the conformal symmetry
algebra is that it is infinite dimensional in two space-time dimensions, therefore it imposes
very severe constraints on the spectrum and correlation functions of
conformally symmetric theories. 

A link between massive integrable models and conformal field theory also
exists: two dimensional quantum field theories can generally be regarded as conformal
field theories perturbed by  suitable operators \cite{Z1,Z2,Z3}. Certain
perturbations preserve a part of the conformal symmetry and render the perturbed
theory integrable.  This perturbed conformal field theory framework also
serves as a basis for useful approximation methods like the conformal perturbation
theory or the TCSA (truncated conformal space approach).   
\vspace{0.5cm}

In this thesis we study problems in three areas of two-dimensional
quantum field theory, especially integrable and conformal field
theory. In accordance with this, the thesis is divided into three
chapters, apart from the introduction. The investigations in the three
chapters are largely independent, although certain connections between them 
exist. 

The three areas, the particular problems and the contents of the chapters are
introduced in the next sections.

The results of Chapter \ref{sec.chap2},  \ref{sec.chap3} and \ref{sec.chap4} have been
published in 

\begin{itemize}
\item
G.Zs.\ T\'oth, N=1 boundary supersymmetric bootstrap,
\emph{Nucl.\ Phys.} \textbf{B676}, 2003, 497-536, hep-th/0308146

\item
G.Zs.\ T\'oth,  A Study of truncation effects in boundary flows of the Ising
model on a strip, \emph{J.\ Stat.\ Mech.} P04005, 2007, hep-th/0612256

\item
G.Zs.\ T\'oth, A non-perturbative study of phase transitions
in the multi-frequency sine-Gordon model, \emph{J.\ Phys.} \textbf{A37}, 2004,
9631-9650, hep-th/0406139.

\end{itemize}

\section{N=1 supersymmetric boundary bootstrap}

The description of certain physical phenomena demands two-dimensional  boundary quantum field
theories, i.e. quantum field theories defined on manifolds with boundaries. 
 Examples for such phenomena are impurity effects like the Kondo effect
 \cite{AL3,AL}, junctions in quantum wires, 
 absorption of polymers on a surface 
and transport properties of
 Luttinger liquids \cite{qhall}. (See also the introduction of \cite{C2} and \cite{KLeM}.  A
 review can also be found in \cite{Sal}.) Boundary quantum field theory is 
 important in open string theory as well. 

In Chapter \ref{sec.chap2} we consider massive  quantum
field theories  with one boundary, i.e.\ boundary quantum field theory  
defined on the space-time $(-\infty,0]\times \RR$.
In these boundary quantum field theories the role of the S-matrix is taken over by the
reflection matrix (describing the bouncing back of the particles from the
boundary) and in addition to the particles the spectrum contains states
---called boundary bound
states--- which are localized at the boundary. The interior of the space $(-\infty,0)$
is referred to as the bulk. Boundary field theories can usually be derived,
especially at the classical level, from
ordinary field theories by imposing a boundary condition. 
For integrable theories (without boundary) it is often possible to find boundary conditions which
preserve the integrability in a sense described in \cite{GZ}. 
In this case the scattering theory of the boundary model will be a factorized
boundary 
scattering theory \cite{GZ}. It should be noted that a factorized
boundary 
scattering theory incorporates a (bulk) factorized scattering theory, the factorized scattering theory of the bulk
part of the model under consideration. 
The complete procedure of the calculation of the S-matrix and reflection
matrix and the full particle and boundary state
spectrum of an integrable model (in the presence of a boundary) generally consists of
several subsequent steps. Usually the bulk part is completed
first, then the ground state reflection matrix is determined, and finally the
spectrum of the higher level boundary states and the reflection matrix blocks
on these boundary states are obtained. 

The full reflection matrix and the
spectrum of boundary states are known only for a few integrable models so
far. The list of these models include the sine-Gordon model \cite{sajat1,Mattsson,MattssonDorey},
$a_2^{(1)}$ and $a_4^{(1)}$ affine Toda field theories \cite{DeliusGand}, the free boson on
the half-line and the sinh-Gordon model \cite{CorTaor,CorDel}.   

It is a further step to calculate the form factors of the local operators.
The generalization of the form factor program to the boundary case
has  been proposed \cite{BPT2} (see also \cite{JKKMW,Q,HSWY}) only recently.
In \cite{BPT2} the minimal form factors of the boundary operators of the
 free boson, free fermion, Lee-Yang and sinh-Gordon models with certain boundary conditions have been
investigated.

A formalism for constructing supersymmetric factorized scattering theories from
non-supersymmetric factorized scattering theories is
developed in \cite{Sch,BL,ABL,Ahn,HolMav}. This consists mainly of replacing
the particles by
supermultiplets and multiplying the S-matrix blocks by suitable supersymmetric
factors. These supersymmetric S-matrix factors satisfy the axioms of
factorized scattering theory in themselves  with certain modifications.
 
An essential step in the construction is the choice of the supersymmetry
representations in which the new particle multiplets will transform, this
choice must be compatible with the fusion rules of the non-supersymmetric
theory. If the possible representations in which the particles may transform
are fixed, then by solving the axioms  one can derive necessary and sufficient
conditions that have to be satisfied by the particle spectrum and fusion rules
of a non-supersymmetric theory to which one wants to apply the
construction. Such conditions have been obtained in the case when the
possible representations are the kink and the boson-fermion representations \cite{HolMav},
and several factorized scattering theories ---the $a_{n-1}^{(1)}$,
$d_{n}^{(1)},$ $(c_{n}^{(1)},d_{n+1}^{(2)})$ and $(b_{n}^{(1)},a_{2n-1}^{(2)})$
affine Toda theories and the sine-Gordon
model--- have been found to satisfy these
conditions  \cite{HolMav}. However, the Lagrangian field theories 
underlying the corresponding
supersymmetric scattering theories are not known in every cases.
The supersymmetric $SU(2)$ principal chiral model,
 the supersymmetric $O(2n)$ sigma model \cite{HolMav} and
the multicomponent supersymmetric Yang-Lee  minimal models (or supersymmetric
FKM models) \cite{SM} have also been found to fit in the framework described
in \cite{HolMav}.

In Chapter \ref{sec.chap2} we study the
construction above in the presence of a boundary. For this  a concept of supersymmetry in the
presence of a boundary is needed, the description of which is an important part of
Chapter \ref{sec.chap2}.    
Assuming that the supersymmetrization of the bulk part is already done,  the first
step of the construction is the choice of a representation of the boundary supersymmetry algebra
for the ground state.   
Next the supersymmetric factors for the
ground state one-particle reflection matrix should be determined
using the boundary Yang-Baxter, unitarity and crossing symmetry equations,
analyticity requirements  and
the supersymmetry condition for these factors.  Finally the boundary
bootstrap and fusion equations for supersymmetric factors can be used
to obtain the representations in which the excited boundary bound
states transform together with the supersymmetric factors of the one-particle
reflection matrix on these states. 
The first and especially the second steps have been considered in
the literature \cite{BPT,SM,Nep2,AK1,AK2,Chim}, whereas
the last step has been completed only in the case of the sine-Gordon model
\cite{BPT}. Our main purpose, motivated by  \cite{HolMav} and \cite{BPT}, is
to generalize the result of \cite{BPT} and formulate 
rules that can be applied to any particular model. We assume that the
particles in the bulk transform
either in the kink or in the boson-fermion representation, mainly because this
is the simplest and most natural choice, and this is the  case for which the 
necessary results (concerning the bulk part and the first two steps) are
sufficiently developed in the literature.

For the ground state we take the singlet representations with RSOS
label $\hlf$, this being the simplest case (see also Section \ref{sec.repbsa}).
The general maximally analytic supersymmetric one-particle
ground state reflection factors have been determined for this case
in \cite{AK1,AK2,Chim}, but without imposing the supersymmetry
condition. 
We rederive these reflection
factors  imposing the supersymmetry condition at the beginning, which simplifies the
calculation considerably.

As the main result of Chapter \ref{sec.chap2}  we present rules for the  determination of the representations and
supersymmetric one-particle reflection factors for excited boundary
bound states. 
It should  be considered that in general there are several
ways to generate a higher level boundary state by fusion, and it is not
obvious that our rules give the same representations and supersymmetric factors for each way. We
verify this statement in specific models and present an 
argument that it can be expected to hold generally. In
particular, we complete the verification of the statement for the boundary sine-Gordon
model started in \cite{BPT}. The other examples to which we apply our rules are the
boundary
$a_{2}^{(1)}$ and $a_{4}^{(1)}$ affine Toda field theories \cite{DeliusGand},
the free boson on the half-line and  the boundary sinh-Gordon
model \cite{CorTaor,CorDel}.

In Section \ref{sec.fst} and \ref{sec.fstb} we review factorized scattering theory in
the bulk and in the presence of a boundary. The main characteristics are
presented as axioms, we refer the reader to other
reviews \cite{Dorey,M,AAR,Mattsson,Alvaredo,Riva,CH} and the articles
\cite{ZZ,GZ}  for their derivation or explanation.  
In our review we emphasize the linear algebraic structure of the axioms of factorized scattering theory. The main reason for this approach
is that we found this linear algebraic form much more suitable
for dealing with our problem than the component based form. 
Our review also includes a discussion of
symmetries and the construction of representations on multi-particle states,
i.e.\ the multiplication of representations. In the boundary case it is not entirely obvious what the proper
algebraic formulation is. We adopt the structure described in
\cite{DeliusMacKay, DeliusGeorge}.
We remark that we   describe  the
correspondence between the Coleman-Thun diagrams and the singularities of the
S-matrix only briefly.

In Section \ref{sec.bootstrap} we outline  the bootstrap procedure usually followed to find solutions
to the axioms of factorized scattering theory.   

In Section \ref{sec.sa} we describe the supersymmetry algebra in 1+1 dimensions, the
multiplication of representations and the construction of multi-particle representations from one-particle
representations, and the one-particle representations which we use. We also
discuss the vacuum representation and the decomposition of products of
representations.

In Section \ref{sec.sab} we describe the supersymmetry algebra  in 1+1
dimensions in the presence of a
boundary, the construction of multi-particle representations and the
one-dimensional  representations for the ground state. Our formulation differs from the formulations that can
be found in the literature in that we apply the algebraic structure proposed
in \cite{DeliusMacKay, DeliusGeorge}.

In Section \ref{sec.ansatz} we describe the ansatz for constructing
supersymmetric factorized scattering theory.

In Section \ref{sec.ansatzb} we describe the extension of the construction
 to the case when a
boundary is also present.

In Section \ref{sec.susybootstrap} we describe the bootstrap procedure for the
supersymmetric factors briefly.

In Section \ref{sec.sssf} we discuss the particular supersymmetric S-matrix
factors for kinks and boson-fermion states, their
important linear algebraic properties, supersymmetry properties, the bootstrap
structure and fusion rules. The discussion of the bootstrap structure and
the fusion rules are based on the results of \cite{HolMav}, which we have
brought to a new form.

In Section \ref{sec.ssrmf}  the supersymmetric ground state reflection matrix factors
and their most important linear algebraic, singularity, supersymmetry and
bootstrap properties are described. 

In Section \ref{sec.hlssbs} we  describe the  boundary supersymmetric bootstrap
structure, i.e.\ the supersymmetric boundary fusion
rules and the supersymmetric reflection factors on higher level supersymmetric
boundary states. These are the main results of Chapter \ref{sec.chap2}.

In Section \ref{sec.ex..} we present examples for the application of the fusion rules
described in Section \ref{sec.hlssbs}.

The Appendix (Section \ref{sec.app..}) contains the normalization factors for the S-matrix and
reflection matrix factors.

Chapter \ref{sec.chap2}, especially the sections \ref{sec.sa}-\ref{sec.app..},
is largely based on the paper
\cite{sajat}, nevertheless several parts have been rewritten.

\section{Truncation effects in the boundary flows of the Ising model on a strip}

Chapter \ref{sec.chap3} is devoted to an investigation of the method called TCSA
(truncated conformal space approach), which is a numerical method for the
calculation of the spectra and eigenvectors of Hamiltonian operators of the
form $H=H_0+hH_I$, where $H_0$ has a known discrete spectrum, $h$ is a coupling
constant. This method is applied mainly to two-dimensional quantum field
theories in finite volume formulated as perturbed conformal field
theories.  An advantage of the TCSA is that  integrability is not necessary
for its applicability. Application to theories in higher space-time dimensions is also
possible in principle.
The uses of the data obtained by TCSA include the  verification of results obtained by
other methods, example in 
\cite{FRT4},
the extraction of resonance widths \cite{PT},  the mapping of the phase structure of
certain quantum field theories as in \cite{BPTW} and in Chapter \ref{sec.chap4}, and the finding of renormalization group
flow fixed points as in \cite{Kormos}. These last two uses are similar. Our
investigation in this chapter is related to the  use of TCSA
for the study of  renormalization group flows between minimal boundary conformal
field theories.

Boundary
conformal field theory is defined on surfaces with
boundaries, e.g.\ on the strip $[0,L]\times
\RR$, which is the case that we consider.  
A brief
review of the areas where boundary
conformal field theory plays important role can be found in \cite{RRS}: it provides the framework for a world sheet analysis of
D-branes in string theory, it also has applications to various systems of
condensed matter physics such as the three-dimensional Kondo effect \cite{AL},
fractional quantum Hall fluids (see e.g.\ \cite{qhall}) and other quantum impurity problems.   

A boundary flow is a one-parameter family of models, the parameter being the
width (or volume) $L$ of the strip.  In the simplest case the parameter
can  be taken to be  a coupling constant $h$ instead of $L$ and the models have the Hamiltonian operators
$H=H_0+hH_I$. $H_0$ is  the Hamiltonian operator of a boundary conformal field theory, $h$ is allowed to vary from $0$ to $\infty$ or from $0$ to
$-\infty$. $H_I$ is a relevant boundary field taken at a certain initial time. 
The study of such flows or deformations  away from the critical point should provide some
insight into the structure of the space of boundary theories. Such deformations are also important in string
theory (see the introduction of \cite{RRS}) and they 
may have applications in
condensed matter physics. 
A particular problem of interest is that of finding values of $h$ other than $0$,
called fixed points,  where the model
corresponding to  $H_0+hH_I$ is a conformal field theory, and identifying these
conformal field theories.  The TCSA can be used for this purpose if the Hilbert space of the  conformal field theory
corresponding to $h=0$ consists of finitely many (or countably many)
irreducible representations of the
Virasoro algebra. Boundary conformal minimal models 
satisfy
this condition. These boundary minimal models can be obtained from the usual 
(bulk) minimal models by imposing suitable conformal boundary conditions. 
Regarding the spectra at nonzero values of $h$,
conformal symmetry can be recognized by  the equal distances between neighbouring energy levels; 
the representation content can be identified by the degeneracy of
the energy levels.  It should be
noted that perturbation
theory and other methods can also be used \cite{RRS, GW, Fr, LSS, CAZ} to explore flows. An important 
problem regarding the
TCSA is that it gives  approximate data, and our knowledge of the precise relation
between this data and the exact spectrum is still limited (see \cite{CLM} for
already existing results).
A good understanding of the
effect of the truncation could be used to improve the TCSA data and to explain
the qualitative features of TCSA pictures of flows, and generally it would make the
results obtained using TCSA data better founded. 

An idea proposed by G.M.T.\ Watts and K.\ Graham \cite{prc,talk}
is that the
effect of the truncation on the spectrum can be taken into consideration by a
suitable change of the coefficients of the terms in $H$, i.e.\ the spectrum of
$H^{TCSA}$ ($H^{TCSA}$ is the finite approximating Hamiltonian operator used
in the TCSA method) is equal, in certain approximation at least, to the spectrum of
\beq
\label{eq.i}
H^r=s_0(h,n_c)H_0+s_1(h,n_c)H_I+s_2(h,n_c)H_{I,2}+\dots\ ,
\eeq
which we shall call
renormalized Hamiltonian operator.
$n_c$ is the truncation parameter, 
$s_0,s_1,\dots$ are suitable functions and $H_{I,2},\dots$ are suitable operators. $H_{I,2},\dots$
should be primary or descendant bulk or boundary operators. Our main purpose
in Chapter \ref{sec.chap3} is to investigate
the validity of this picture.

We  consider the  perturbed boundary conformal field theory on the strip
$[0,L]\times \RR$
with Hamiltonian operator  
\beq
\label{eq.yyy}
H=\frac{\pi}{L}L_0+hL^{-1/2}\phi_{1/2}(x=L,t=0)\ .
\eeq
The unperturbed model is the $c=1/2$ unitary conformal minimal model, i.e.\ the
continuum limit of the critical Ising model, 
with the Cardy boundary condition
$0$ on the left and $1/16$ on the right. $\frac{\pi}{L}L_0$ is the Hamiltonian
operator of this model. $L_0$ is the `zero index' Virasoro generator. $L$ will be kept fixed at the value $L=1$.  
The Hilbert space of the unperturbed model is the single $c=1/2$, $h=1/16$
irreducible highest weight representation of the  Virasoro
algebra. (Here and in Chapter \ref{sec.chap3} the coupling constant and the highest
weight are both denoted by
$h$, but it should be clear from the context which one is meant.) 
It should be noted that a whole series of similar boundary conformal minimal models exist, they can be obtained
from other unitary minimal models by imposing boundary conditions.  Imposing the $0$ boundary condition on
one side and another Cardy boundary condition on the other side always results
in that the Hilbert space consists of a single
irreducible representation. 
The field $\phi_{1/2}(L,t)$ is the
weight $1/2$  boundary primary field on the right boundary, which is
also known in the literature as the  boundary spin operator \cite{Cardy1,CL,GZ}. The
normalization of $\phi_{1/2}$ is given by
$\brakettt{1/16}{\phi_{1/2}(L,0)}{1/16}=1$, where $\ket{1/16}$ is the highest
weight state, $\braket{1/16}{1/16}=1$.
 The coupling constant $h$ can also be regarded as a constant external
boundary magnetic field, which is coupled to the boundary spin operator. The
model (\ref{eq.yyy}) is also referred to as the critical Ising model on a
strip with boundary magnetic field. 
The perturbation $h\phi_{1/2}(L,t=0)$ violates the conformal symmetry, which is nevertheless
restored in the $h\to \pm\infty$
limit. It is known that in the $h \to \infty$ limit the $c=1/2, h=1/2$
representation is realized; in the $h \to -\infty$ limit the $c=1/2, h=0$
representation is realized (see e.g.\ \cite{AL2,GZ,Fr,GW}). 
This can be written in a shorthand form as 
\begin{align}
\label{eq.flow1}
(1/2,1/16)+\phi_{1/2} &\  \to \  (1/2,1/2)\\
\label{eq.flow2}
(1/2,1/16)-\phi_{1/2} &\  \to \  (1/2,0)\ .
\end{align}
We have chosen the model (\ref{eq.yyy})  because it is integrable, furthermore
it is relatively easy to handle;
in particular the spectrum can be calculated analytically in terms of simple
functions and the application of
Rayleigh-Schr\"odinger perturbation theory  is also relatively easy. 
Investigations in other perturbed conformal minimal models, especially
in the case of the tricritical Ising model and generally in the case of a
perturbation by the field $\phi_{(13)}$, are carried
out in \cite{FGPW} (see also \cite{talk}). 

It should be noted that the Ising model with boundary magnetic field has been
studied much, especially on the lattice and on the half-line.
We refer the reader to \cite{C1,C2,CZ,GZ} and the references in them for further
information. 

To recognize conformal symmetry and to identify the representation content it is sufficient to look at the ratios of the energy gaps; therefore
one often considers normalized spectra  which are obtained by subtracting 
the ground state energy and dividing by the lowest energy gap. 
The normalized exact and  TCSA spectra for the flows (\ref{eq.flow1}), (\ref{eq.flow2}) as a function of the logarithm of $h$ can
be seen in Figure \ref{fig.tcsa2}. 
An interesting feature of these TCSA spectra is that
they appear to correspond to the  flows $(1/2,1/16) \to (1/2,1/2) \to (1/2,
1/16)$ and   $(1/2,1/16) \to (1/2,0) \to (1/2,
1/16)$, i.e.\ second flows appear to be present after the normal flows. 
Flows in models similar to (\ref{eq.yyy}) mentioned above also show this behaviour.   
One application of the picture (\ref{eq.i}) could be the explanation of this phenomenon.

Following G.M.T.\ Watts' proposal  based on the look of the TCSA spectra shown
in
Figure \ref{fig.tcsa2}  we
assume that only the first two terms are nonzero in (\ref{eq.i}): 
\beq
\label{eq.i2}
H^r=s_0(h,n_c)H_0+s_1(h,n_c)H_I\ .
\eeq

In summary, in Chapter \ref{sec.chap3} we look for answers for the following
questions: 1.\ Does the spectrum of (\ref{eq.i2}) agree with the TCSA spectrum in some
approximation with a suitable choice of
the functions  $s_0$ and $s_1$? 2.\ How can we explain the 'second' flows in
the TCSA spectra?

The main difficulty we encounter is that it is hard to handle the TCSA spectra analytically
even if the nontruncated model is exactly solvable. Therefore,
hoping that we can gain some insight by looking at a similar but exactly solvable
truncation method, we tried another
truncation method which we call mode truncation. The mode truncated model can
be solved exactly, but it turns out, rather unexpectedly, that the behaviour
of the spectrum for large values of $h$ is different from the behaviour of the
TCSA spectrum, namely the  qualitative behaviour of the mode truncated
spectrum is very similar to that of the exact spectrum; the second flows are
not present. This, besides leaving the second question open, raises the
problem of finding the possible   behaviours  for large values of $h$ and their dependence on the truncation
method, and whether the mode truncation method can be generalized to other models.

We  also applied the Rayleigh-Schr\"odinger perturbation theory  to verify
the validity of (\ref{eq.i2}) for both the TCSA and mode truncation methods. 
In the mode truncation scheme, using the exact analytic expressions for the
eigenvalues,
we also obtained a result that is non-perturbative in $h$ and
perturbative in $1/n_c$. 

The third calculation that we did is
a  numerical comparison of the exact and TCSA and the exact
and mode truncated spectra. The TCSA calculation was done by a program written
entirely by ourselves.

We determined the  exact spectrum using an essentially  known (see e.g.\
\cite{GZ,CZ,C1,LMSS,Kon,KLeM})  quantum field theoretic representation of
the operator (\ref{eq.yyy}). In this representation the operator
(\ref{eq.yyy}) is a quadratic expression of fermionic fields. We extracted
the spectrum from the field equations which are linear.    
Besides the spectrum we also  considered the
interacting fermion 
fields and their matrix elements, and certain other issues.  The  field theoretic approach also raises the problem of
defining distributions (or similar objects) on a closed interval. We do not know of a systematic
exposition of this subject (neither for a closed interval nor for the half-line), although it would be needed for  boundary
field theory.  
In this thesis we use distributions on  closed intervals, nevertheless we
restrict to the most necessary formulae only and do not work out a complete
theory. 

The
 field theoretic model mentioned above  was studied in \cite{LMSS} and especially  in
\cite{C1} (at finite temperature).  Our approach and aim are  different, however, and the
overlap between the results of \cite{C1} and our results is only
partial. 
Our quantum field theoretic calculations are
partially independent of the problem of the TCSA approximation. Some of these
 calculations are included only because we think that they are generally
 interesting from the point of view of quantum field theory.

We propose a description of (\ref{eq.yyy}) as a perturbation of the $h\to
\pm \infty$ limiting case, which cannot be found in the literature. An
interesting feature of this description, which we call reverse description, is that
the  perturbing operator is non-relevant. 
We
calculate the exact spectrum in a similar way as in the case of the standard
description mentioned above.

We also present the description of the spectrum using the Bethe-Yang equations,
which give the exact result in this case.

We treat the mode truncated model along the same lines as the
nontruncated model, we restrict to the spectrum in this case, however.  

We remark that our TCSA program relies on the conformal transformation
properties only and does   
not make use of the 
representation mentioned above, therefore it can be used with  other  values of
the central charge,  highest weight  and  weight of the perturbing field.

The contents of Chapter \ref{sec.chap3} are the following:

In Sections \ref{sec.def}-\ref{sec.tcsades} definitions and other introductory 
information are collected, which is followed by the presentation of the results
in Sections
\ref{sec.eel}-\ref{sec.numres}.

In Section \ref{sec.def} the definition of  quantum field theory on
a strip is discussed briefly.  

In Section \ref{sec.repvir} we describe certain well known results in conformal field
theory which are important for our work. 

In Section \ref{sec.CFTstrip} we give a basic definition of conformal field theory on
a strip. We restrict to those elements  which are essential for our work. We
refer the reader to \cite{IR,DRTW,FMS} for  more advanced
exposition. In Subsection \ref{sec.bflows} we give a definition of flows.

In Section \ref{sec.tcsades} we describe TCSA in general and its application to the type
of models that we consider. 

Section \ref{sec.eel} contains the field theoretical description of the model
(\ref{eq.yyy}), in particular the calculation of the exact spectrum.
Results concerning the boundary conditions, the normalization of interacting
creation and annihilation operators, 
the relation between free and
interacting creation and annihilation operators, matrix elements of the
interacting fields,
nontrivial identities for the  Dirac-delta and the expression of eigenstates in
terms of the unperturbed eigenstates are obtained.   
The section also
contains the reverse description of the model and the description of the spectrum using
the Bethe-Yang equations. It is found that the latter gives exact result in
this case. 

Section \ref{sec.modetr} contains the description of the mode truncated model and the
calculation of its spectrum. 

Section \ref{sec.RS} contains the power series for the exact, TCSA and mode
truncated energy levels up to third
order in $h$ obtained by the Rayleigh-Schr\"odinger perturbation theory.    

Section \ref{sec.pertres} contains the perturbative results for the $s_0$ and $s_1$
functions.

Section \ref{sec.numres} contains the results of the numerical test of the approximation by
(\ref{eq.i2}) for the TCSA and the mode truncation schemes. 

Section \ref{sec.scaling} contains a description of the scaling properties,
i.e.\ the truncation level dependence of the
$s_0$, $s_1$, $s_1/s_0$ functions.

The results of Chapter \ref{sec.chap3} have been published in \cite{isingtcsa}.

\section{A nonperturbative study of phase transitions in the multi-frequency
  sine-Gordon model}

In Chapter \ref{sec.chap4} we present an application of the TCSA to the mapping of the
phase structure of the multi-frequency sine-Gordon model.   

The sine-Gordon model has attracted interest
long time ago for the reason that it appears in several areas of physics,
and it is an integrable field theory.  The areas of
application include statistical mechanics of one-dimensional quantum
spin chains and nonlinear optics among many others ---see the introduction
of \cite{DM} for a 
list with references.

The multi-frequency sine-Gordon model is a 
non-integrable extension of
the sine-Gordon model in
which the scalar potential consists of several cosine terms with different
frequencies. It is suggested in \cite{DM} that this model can be
used to give more refined approximation to some of the physical situations
where the ordinary sine-Gordon model can be used. A feature of the
multi-frequency model that is new compared to the usual sine-Gordon
model is ---apart form non-integrability--- that phase transition
can occur as the coupling constants are tuned. We concentrate our
attention to this property. Such a phase transition
is related to the evolution of the spectrum of the theory
as the coupling constants vary. In accordance with this we shall use the massgap and other
characteristics of the energy spectrum to identify the different phases.

Our investigation can be regarded as a continuation of the work done on the double-frequency
case in \cite{BPTW}. We use the truncated conformal space approach
(TCSA);  the applicability and reliability
of this method was thoroughly investigated in \cite{BPTW} and it
was shown that the existence, nature and location of the phase transition
can be established by this method, although rather large  truncated space is needed for satisfactory precision. In particular, the
existence and location of an Ising type transition was established
in the double-frequency model (DSG) for the ratio $1/2$ of the frequencies,
verifying a prediction by \cite{DM} based on perturbation theory
and classical field theoretic arguments. We extend these investigations to the ratio
$1/3$ and to the three-frequency model (at the ratio $1/2/3$ of
the frequencies), in which a tricritical point and first order transition
are expected to be found. The numerical nature of the TCSA makes it
necessary to choose specific values for the frequencies.

We remark that the calculations of Chapter \ref{sec.chap4} were completed before the
work presented in Chapter \ref{sec.chap3} was done.

The contents of Chapter \ref{sec.chap4} are the following:

In Section \ref{sec: ketto} we introduce the multi-frequency sine-Gordon model and describe
basic properties of it.

In Section \ref{sec: harom} we briefly review
the formulation of the model in the perturbed conformal field theory framework (for the two-frequency case this can be found
in  \cite{BPTW}, the extension to the multi-frequency model is
straightforward), which is necessary for the application of the TCSA.

In Section \ref{sec: negy} we give a description of the phase structure
of the classical two- and three-frequency model, which serves as a
reference for the investigations in quantum theory. The $n$-frequency
case is also considered briefly. Exact and elementary analytic methods
can be applied to the classical case, and the results are more general
than in the quantum case. 

Section \ref{sec: ot} is devoted to theoretical
considerations on the signatures of 1st and 2nd order phase transitions
in the framework of perturbed conformal field theory in finite volume.
Most of these considerations, which are necessary for the evaluation
of the TCSA data, can also be found in \cite{BPTW}. 

In Section
\ref{sec: hat} and \ref{sec: het} we present the results we obtained
by TCSA on the phase structure of the two- and three-frequency model,
which are the main results of Chapter \ref{sec.chap4}. The calculations were done by
a program written by L.\ Palla, Z.\ Bajnok, G.\ Tak\'acs and F.\ W\'agner on
which we performed certain modifications.

Chapter \ref{sec.chap4} is based on the article \cite{sajat2}.
\vspace{0.2cm}

\begin{center}
-----------------------
\end{center}
\vspace{0.5cm}

Finally, we list a selection of books, articles and PhD theses which can be
used as references:

\begin{itemize}
\item Two-dimensional quantum field theory: \cite{AAR}
\item Factorized scattering theory:  \cite{Dorey,M,AAR,Mattsson,Alvaredo,Riva,CH,ZZ,GZ} 
\item Factorized scattering theory in the presence of a boundary:
  \cite{GZ,Mattsson} 
\item Form factor bootstrap programme: \cite{BKW,BFKZ,Smirnov,Alvaredo}
\item Form factor bootstrap programme in the presence of a boundary: \cite{BPT2}
\item Conformal field theory: \cite{FMS,CH,Ginsparg}
\item Boundary conformal field theory: \cite{IR,FMS,CH,Cardy1,CL,Lew,Runkel1,Runkel2,PetkZub,BPPZ1,BPPZ2,BPZ,Quella}
\item Flows: \cite{Watts2,Watts4,Watts1,Fr,M,Ravanini}
\end{itemize}

\chapter{On N=1 supersymmetric boundary bootstrap}
\label{sec.chap2}
\markboth{CHAPTER \thechapter.\ \ SUPERSYMMETRIC BOUNDARY BOOTSTRAP}{}

\section{Factorized scattering theory}
\label{sec.fst}
\markright{\thesection.\ \ FACTORIZED SCATTERING THEORY}

Factorized scattering theory describes  collisions of quantum particles or
particle-like quantum objects (e.g.\ solitons)  which travel in  1+1 dimensional 
space-time. We shall consider relativistic scattering theory of finitely many massive particles.
Specific factorized scattering theories are  usually associated to  integrable
relativistic quantum field theories which are characterized by the existence
of higher spin conserved quantities. 

The main constituents of  scattering theory  in general is a Hilbert space of
asymptotic states of the particles, an operator $S$ on the Hilbert space
describing the scattering of particles, a representation of the Poincare group
on the Hilbert space with respect to which $S$ is equivariant, and 
representations of further possible symmetry algebras.

The characteristics of factorized scattering  are listed
and explicitly described below. Some of these properties are special to
factorized scattering theory, others (like unitarity) are general properties
of scattering theory applied to factorized scattering theory.

Factorized scattering theory is characterized by the following properties:\\

1) The particle number is conserved.\\

2) The sets of incoming and outgoing momenta are equal.\\

3) Factorization and Yang-Baxter equation: An arbitrary N-particle scattering process can be described as a sequence
   of 2-particle collisions, all possible descriptions (of which there are
   $N(N-1)/2$) are equivalent. The most interesting quantity in factorized
   scattering theory is therefore the   two-particle S-matrix block that
   describes the  2-particle collisions. \\

The momentum $(p_0,p_1)$ of a free particle of mass $m$ satisfies
$p_0^2-p_1^2=m^2$, $p_0>0$. Instead of $p_0$ and $p_1$ one often uses $m$ and the
rapidity parameter $\Theta$:
$$p_0=m\cosh(\Theta)\ ,\qquad p_1=m\sinh(\Theta)\ .$$

The physical values $\Theta$ are the real numbers, however it is useful to allow
complex values (i.e.\ 
$\Theta\in \CC$) unless there is a reason to restrict to real values. 

The asymptotic in and out states are denoted in the following way:
\begin{equation}
\label{eq.1}
\ket{a_1(\Theta_1)a_2(\Theta_2)\dots a_N(\Theta_N)}_{in/out}
\end{equation}
where 
$\Theta_1\ge\Theta_2\ge\dots \ge\Theta_N$
for in states and 
$\Theta_1\le\Theta_2\le\dots\le\Theta_N$
for out states.  

The Hilbert space can be written as a sum of $N$-particle subspaces:
$$\mathcal{H}=\oplus_{N=0}^\infty \mathcal{H}_N$$
where $ \mathcal{H}_0$ is the vacuum subspace, which is a finite dimensional
space in general.

The $N$-particle Hilbert space can be written as
$$\mathcal{H}_N=\underbrace{\mathcal{H}_1\otimes \mathcal{H}_1 \otimes \dots \otimes
\mathcal{H}_1}_N\ .$$
$\mathcal{H}_N$ can be decomposed into a (not direct) sum of spaces of in, out and intermediate states:
$$\mathcal{H}_N=\mathcal{H}_{N,in} + \mathcal{H}_{N,out} + \mathcal{H}_{N,intermed}$$
for 
$N\ge 3$, for $N=0$ and $N=1$ we have
$$\mathcal{H}_{0,in}=\mathcal{H}_{0,out}\ ,\qquad
\mathcal{H}_{1,in}=\mathcal{H}_{1,out}$$
and there are
no intermediate states for $N=0,1,2$.

The space of intermediate states is spanned by those elements of the form (\ref{eq.1})
which do not satisfy the ordering prescription for in and out states. 

The notation in (\ref{eq.1}) means 
\begin{equation}
\label{eq.2}
a_1(\Theta_1)\otimes a_2(\Theta_2)\otimes \dots \otimes a_N(\Theta_N)
\end{equation}
where $a_i(\Theta_i)\in \mathcal{H}_1$ are one-particle states.

$\mathcal{H}_1$ can be written as
$$\mathcal{H}_1=\int_{\Theta\in \RR} d\Theta\ V(\Theta)$$
where $V(\Theta)$ is a finite dimensional vector space for any fixed value of
$\Theta$. $\Theta \in \CC$ can also  be allowed, but we do not introduce
further notation for this case.

The action of a boost $B(\Theta)$ of rapidity $\Theta$ is 
$$B(\Theta)a(\Theta_1)=a(\Theta_1+\Theta)\ .$$
We introduce a finite dimensional vector space $V$ with a basis labeled by $a$
which  also labels  the particles, and the linear map $i_{\Theta}:
V(\Theta)\to V,\ a(\Theta)\mapsto a$ for all $\Theta$. $V$ will be called
internal space. The notation $a(\Theta)$ is used both for elements of
$\mathcal{H}_1$ and for elements of $V(\Theta)$. 

$\mathcal{H}_1$, $V(\Theta)$ and $V$ can also be decomposed into a sum of mass
eigenspaces:
\begin{equation}
\label{eq.decm}
\mathcal{H}_1=\oplus_m (\mathcal{H}_1)_m\ ,\qquad V(\Theta)=\oplus_m
V(\Theta)_m\ ,\qquad V=\oplus_m V_m\ .
\end{equation}

The normalization of the states is
$$\braket{\Omega_i}{\Omega_j}=\delta_{ij}$$
and 
$$\braket{a_1(\Theta_1)}{a_2(\Theta_2)}=\delta(\Theta_1^*-\Theta_2)\delta_{a_1, a_2}$$
for  $\mathcal{H}_0$ and  $\mathcal{H}_1$. The ${}^*$ denotes complex conjugation. For multi-particle states the
canonical normalization for tensor products 
is used. For $V$, $V(\Theta)$ and $V(\Theta^*)$ we have 
\begin{equation}
\braket{a_1}{a_2}=\delta_{a_1,a_2}\ ,\qquad
\braket{a_1(\Theta^*)}{a_2(\Theta)}=\delta_{a_1,a_2}\ ,\qquad
V(\Theta)^\dagger=V(\Theta^*)\ .
\end{equation}

The particles usually correspond to basis vectors in a distinguished
orthonormal basis of $V$ denoted by $\mathcal{B}$. $V$ has a
corresponding decomposition into one-dimensional subspaces:
\begin{equation}
\label{eq.resz}
V=\oplus_l V_l\ .
\end{equation}
These basis vectors are mass eigenstates, i.e.\ the decomposition is a
refinement of (\ref{eq.decm}).

If some of the particles are soliton-like, then, assuming that we have chosen
the appropriate basis, only  part of the  multi-particle states of the form (\ref{eq.2}) are
physical states, and physical subspaces are spanned by
these physical states. The physical states are usually selected by adjacency
conditions:
in a physical state of the form (\ref{eq.2}) only certain ordered pairs $(a,b)$ may
occur, i.e.\ if a pair $(c,d)$ is not allowed, then a state of the form
$\dots \otimes c(\Theta_i)\otimes d(\Theta_{i+1})\otimes \dots$ is not
physical. Physical and non-physical states of the form (\ref{eq.2}) are orthogonal, and
one can introduce the orthogonal projector $P: V\otimes V \to V\otimes V$ that
projects onto the physical subspace of $V\otimes V$. The physical
subspace of $V\otimes V \otimes V$ can be obtained as 
$[P\otimes I][I\otimes P](V\otimes V \otimes V)$ where $P\otimes I$ and
$I\otimes P$ commute, and similar formulae can be written for general
$N$-particle spaces. More general adjacency conditions are also possible,
as we shall see in Section \ref{subsec.rsa}.\\

Instead of the physical S-operator one usually considers an auxiliary
S-operator $S_{aux}$.  The relation between $S$ and $S_{aux}$  is that the
latter is defined for any complex value of the rapidities (although it can have
singularities).  
The $aux$ subscript is often omitted.   

The $\mathcal{H}_N$ are invariant subspaces of $S_{aux}$, the
$N$-particle auxiliary
S-operator $S_N$ is obtained by restricting the full auxiliary S-operator to $\mathcal{H}_N$: 
$S_N:\ \mathcal{H}_N\to \mathcal{H}_N$.  
In particular
$$S_N(\mathcal{H}_{N,in})= \mathcal{H}_{N,out}$$
and
$$S_0=Id_{\mathcal{H}_0}\ ,\qquad
S_1=Id_{\mathcal{H}_1}\ .$$

For $N\ge 2$ the operators 
$$\hat{S}_N(\Theta_{1},\Theta_{2},\dots ,\Theta_{N}): V(\Theta_1)\otimes V(\Theta_2)
\otimes \dots  \otimes V(\Theta_N) \to  V(\Theta_N)\otimes V(\Theta_{N-1})
\otimes \dots  \otimes V(\Theta_1)$$
can be introduced so that
\begin{multline}
\notag
\brakettt{b_N(\Theta_N')b_{N-1}(\Theta_{N-1}')\dots b_1(\Theta_1')}{S_N}{a_1(\Theta_1)a_{2}(\Theta_{2})\dots a_N(\Theta_N)}=\\[3mm]
{\qquad=\delta(\Theta_1-\Theta_1'^*)\delta(\Theta_2-\Theta_2'^*)\dots \delta(\Theta_N-\Theta_N'^*)\times}\\
\times \brakettt{b_N(\Theta_N^*)b_{N-1}(\Theta_{N-1}^*)\dots b_1(\Theta_1^*)}
{ \hat{S}_N(
  \Theta_1,\Theta_{2},\dots ,\Theta_N)}{a_1(\Theta_1)a_{2}(\Theta_{2})\dots a_N(\Theta_N)}
\end{multline}

For the matrix elements of the $\hat{S}_N$-s the notation
\begin{multline}
\notag
(\hat{S}_N)^{b_{N}b_{N-1}\dots b_1}_{a_1 a_2\dots  a_N}(\Theta_1,\Theta_2,\dots ,\Theta_N)=\\[2mm]
=\brakettt{b_N(\Theta_N^*)b_{N-1}(\Theta_{N-1}^*)\dots b_1(\Theta_1^*)}
{ \hat{S}_N(
  \Theta_1,\Theta_{2},\dots ,\Theta_N)}{a_1(\Theta_1)a_{2}(\Theta_{2})\dots a_N(\Theta_N)}
\end{multline}
is used. One can also introduce the operators
$$\tilde{S}_N(\Theta_{12},\Theta_{13},\dots ,\Theta_{1N}): \underbrace{V\otimes V\otimes
\dots  \otimes V}_N \to \underbrace{V\otimes V\otimes \dots  \otimes V}_N$$
$$\tilde{S}_N(\Theta_{12},\Theta_{13},\dots ,\Theta_{1N})=[i_{\Theta_N}\otimes
i_{\Theta_{N-1}}\otimes \dots  \otimes i_{\Theta_1}] \hat{S}_N
(\Theta_1,\Theta_2,\dots ,\Theta_N) [i^{-1}_{\Theta_1}\otimes
i^{-1}_{\Theta_{2}}\otimes \dots  \otimes i^{-1}_{\Theta_N}]$$
where $\Theta_{ij}=\Theta_i-\Theta_j$. The matrix elements of the $\tilde{S}_N$-s are
denoted as 
$$(\tilde{S}_N)^{b_{N}b_{N-1}\dots b_1}_{a_1
  a_2\dots  a_N}(\Theta_{12},\Theta_{13},\dots ,\Theta_{1N})\ .$$
For the matrix elements of $\hat{S}_N$ and $\tilde{S}_N$ the equation
$$(\tilde{S}_N)^{b_{N}b_{N-1}\dots b_1}_{a_1 a_2\dots  a_N}(\Theta_{12},\Theta_{13},\dots ,\Theta_{1N})=
(\hat{S}_N)^{b_{N}b_{N-1}\dots b_1}_{a_1 a_2\dots  a_N}(\Theta_1,\Theta_2,\dots ,\Theta_N)$$
holds.

One can also consider restrictions of  $\hat{S}_N$ and $\tilde{S}_N$
to subspaces of states with definite mass of the particles:    
$$(\tilde{S}_N)_{m_1 m_2 \dots m_N}(\Theta_{12},\Theta_{13},\dots ,\Theta_{1N}): V_{m_1}\otimes V_{m_2}\otimes
\dots \otimes V_{m_N} \to V_{m_N}\otimes V_{m_{N-1}}\otimes \dots \otimes V_{m_1}$$
and similarly for  $\hat{S}_N$. $\hat{S}_N$ and $\tilde{S}_N$ are composed of such
blocks. 

If 
$$
\tilde{S}_2(\Theta)(V_1\otimes V_2)\subseteq V_2\otimes V_1
$$
for some subspaces $V_1\subseteq V_{m_1}$,  $V_2\subseteq V_{m_2}$ with some
masses $m_1,m_2$, then we can use the notation $\tilde{S}_{V_1V_2}(\Theta)$
for $\tilde{S}_2(\Theta)|_{V_1\otimes V_2}$ and say that the multiplets $V_1$
and $V_2$ scatter on each other and the two-particle S matrix of the
multiplets $V_1$ and $V_2$ is  $\tilde{S}_{V_1V_2}(\Theta)$.

If the particles satisfy adjacency conditions, they have to be observed in the
formulae above, however we
do not introduce explicit notation for this case.

If $\hat{S}_2(\Theta_1,\Theta_2)$ or $\tilde{S}_2(\Theta)$ is written in matrix form, i.e.\ as a table of entries, then
the upper indices specify the rows and the lower indices specify the columns. 

\begin{figure}
\begin{center}
\includegraphics[scale=0.7]{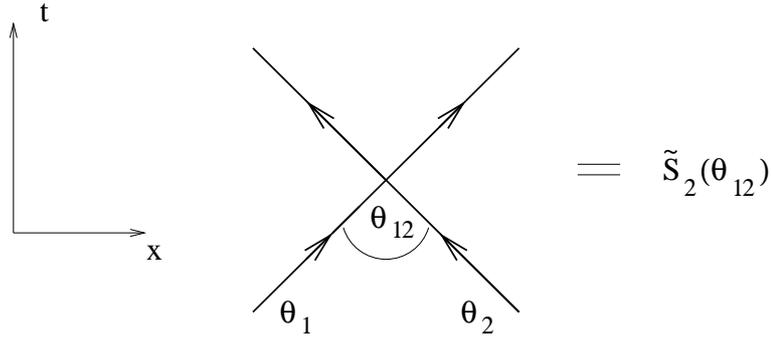}
\caption{\label{fig.s}Two-particle S-matrix $\tilde{S}_2(\Theta_{12})$}
\end{center}
\end{figure}

The graphical representation of $\tilde{S}_2(\Theta)$ is shown in Figure \ref{fig.s}. \\

In the $N=3$ case the
   factorization property is expressed by the equation
$$\tilde{S}_3(\Theta_{12}, \Theta_{13})= [\tilde{S}_2(\Theta_{23})\otimes I][I\otimes
\tilde{S}_2(\Theta_{13})][\tilde{S}_2(\Theta_{12})\otimes I]$$
or 
$$\tilde{S}_3(\Theta_{12}, \Theta_{13})=
[I\otimes \tilde{S}_2(\Theta_{12})][\tilde{S}_2(\Theta_{13})\otimes
I][I\otimes \tilde{S}_2(\Theta_{23})]\ .$$
The two equations correspond to the two possible decompositions of the
3-particle scattering into 2-particle scatterings.

The equality of the two expressions on the right-hand side 
$$[\tilde{S}_2(\Theta_{23})\otimes I][I\otimes
\tilde{S}_2(\Theta_{13})][\tilde{S}_2(\Theta_{12})\otimes I]=
[I\otimes \tilde{S}_2(\Theta_{12})][\tilde{S}_2(\Theta_{13})\otimes I][I\otimes \tilde{S}_2(\Theta_{23})]$$
is called the 
Yang-Baxter equation. If the Yang-Baxter equations are satisfied, then the
analogous equations expressing the equality of the possible factorizations of
the $N$-particle scattering ($N>3$)  into 2-particle scatterings are also satisfied.

The above equations (with the obvious modifications) are also satisfied by $S_{2}$,
$S_{3}$ and
$\hat{S}_2$, $\hat{S}_3$ and by the appropriate blocks of them corresponding
to definite masses. Similar statement applies to several formulae below. We
shall usually write down the tilde  versions only.    \\

Most of the equations  of factorized scattering theory for transition 
amplitudes like the Yang-Baxter
equation admit a
graphical representation which is very useful for 
grasping and handling these equations. The graphs representing these equations are similar to the
Feynman graphs, their vertices correspond to tensors, especially  to the two-particle S matrix or its
blocks and to the fusion and decay tensors described below.  Outgoing and
incoming lines (distinguished by arrows) at a vertex correspond to the two
type of indices (normal or upper and dual
or lower), an outgoing line can be joined to an incoming line which
corresponds to the contraction of the corresponding indices. We usually do not
introduce a vertex for the identity tensor, although in the equations it can
be useful to insert identity tensors explicitly. Momenta or
rapidities are 
also associated to the lines. In several cases the graphs  reflect
the kinematical situation geometrically faithfully, i.e.\ the graphs can be
regarded as geometrical pictures of scattering, fusion and decay
processes. In the graphical formulation the Yang-Baxter and the bootstrap
equations (described below) express the parallel shiftability of the individual
lines in
the diagrams. This shiftability is a consequence of the existence of higher
spin conserved charges.

\begin{figure}
\begin{center}
\includegraphics[scale=0.6]{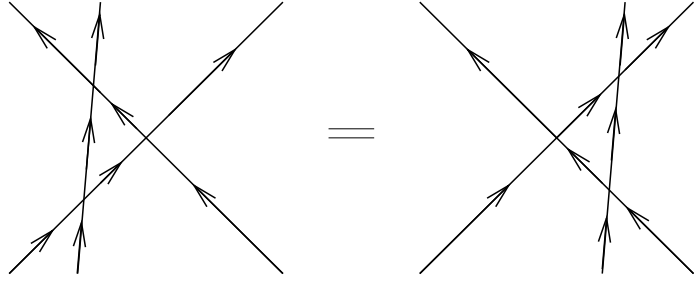}
\caption{\label{fig.yb}Yang-Baxter equation}
\end{center}
\end{figure}

A graphical representation of the Yang-Baxter equation is shown in
Figure \ref{fig.yb}.\\

4) Analytic properties: $\tilde{S}_2(\Theta)$ is an analytic function with
   possible pole singularities which are located in $i\RR$. The domain
   $0<Im(\Theta)<\pi$ is called the `physical strip'. The singular 
   part of $\tilde{S}_2(\Theta)$ at a pole in the physical strip 
   is generally a sum of various contributions related to bound states and
   anomalous thresholds. \\

5) Real analyticity 
\begin{equation}
(\tilde{S}_2(\Theta^*))^\dagger=\tilde{S}_2(-\Theta)\ .
\end{equation}
For the blocks corresponding to definite masses, for example, this reads as
\begin{equation}
\notag
((\tilde{S}_2)_{m_1m_2}(\Theta^*))^\dagger=(\tilde{S}_2)_{m_2m_1}(-\Theta)\ .
\end{equation}

6) Unitarity
\begin{equation}
\label{eq.un}
(\tilde{S}_2(\Theta^*))^\dagger\tilde{S}_2(\Theta)=I\ .
\end{equation}
This condition
   and the real analyticity condition imply
\begin{equation}
\tilde{S}_2(-\Theta)\tilde{S}_2(\Theta)=I\ .
\end{equation}
It is usual in factorized scattering theory  to refer to the latter condition rather than (\ref{eq.un}) as unitarity.\\

7) Bound states: at some poles of $(\tilde{S}_2)_{m_1m_2}(\Theta)$ in the
physical strip the
singular part has a contribution 
corresponding to
direct-channel (or s-channel)  bound
states: 
\begin{equation}
\label{eq.fusion}
(\tilde{S}_2)_{m_1 m_2}(\Theta)=\frac{\tilde{d}_{m_3}^{m_2
    m_1}\tilde{f}_{m_1m_2}^{m_3}}{\Theta-iu}+s(\Theta)+ reg(\Theta)   
\end{equation}
where $iu$, $0<u<\pi$, is the location of the pole, $reg(\Theta)$ is regular
at $iu$, $s(\Theta)$ contains further possible singular terms. $u$ is called
    fusion angle.
$$\tilde{d}_{m_3}^{m_2 m_1}: V_{m_3} \to V_{m_2}\otimes V_{m_1} $$
is the decay tensor,
$$\tilde{f}_{m_1m_2}^{m_3}: V_{m_1}\otimes V_{m_2} \to V_{m_3}$$
is the fusion tensor. 
We also have the hatted versions
$$\hat{d}_{m_3}^{m_2 m_1}: V_{m_3}(\Theta)\to V_{m_2}(\Theta+iu_2)\otimes V_{m_1}(\Theta-iu_1)$$
and
$$\hat{f}_{m_1m_2}^{m_3}: V_{m_1}(\Theta-iu_1)\otimes V_{m_2}(\Theta+iu_2) \to
V_{m_3}(\Theta)\ ,$$
where
$0<u_1$ and $0<u_2$, $u_1+u_2=u$. $u_1$ and $u_2$ are determined in terms of $m_1$, $m_2$, $m_3$ by momentum
conservation and by the mass shell condition. The fusion map projects onto
the bound states in question.

\begin{figure}
\begin{center}
\includegraphics[scale=1]{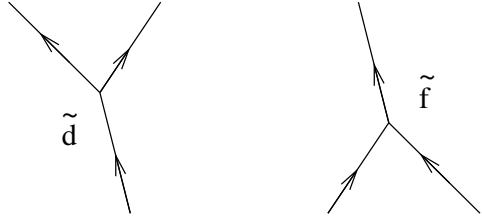}
\caption{\label{fig.fd}Fusion and decay tensors}
\end{center}
\end{figure}

\begin{figure}[!h]
\begin{center}
\includegraphics[scale=0.3]{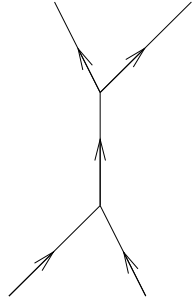}
\caption{\label{fig.bstate}Contribution of bound states}
\end{center}
\end{figure}

Graphical representation of $\tilde{f}_{m_1m_2}^{m_3}$ and
$\tilde{d}_{m_3}^{m_2 m_1}$ can be seen in Figure \ref{fig.fd}. The graphical
representation of the residue $\tilde{d}_{m_3}^{m_2
    m_1}\tilde{f}_{m_1m_2}^{m_3}$ is shown in Figure \ref{fig.bstate}.

The image space of $\tilde{f}_{m_1m_2}^{m_3}$ has zero intersection with the kernel
of  $\tilde{d}_{m_3}^{m_2 m_1}$.

The image space of  $\tilde{f}_{m_1m_2}^{m_3}$ can usually be decomposed into
a sum of some  one-particle subspaces occuring in (\ref{eq.resz}). 

Generally if 
$$V_1= \oplus_{l_1} V_{l_1}\ ,\qquad V_2=\oplus_{l_2} V_{l_2}\ ,\qquad
V_3=\oplus_{l_3} V_{l_3}\ ,$$
where $l_1$, $l_2$ and $l_3$ take certain values from all possible values of
$l$ defined in (\ref{eq.resz}) and the states in $V_1$, $V_2$, $V_3$ have mass $m_1$, $m_2$, $m_3$ and 
there exists a fusion tensor $f_{m_1m_2}^{m_3}$ and
$$f_{m_1m_2}^{m_3} (V_1\otimes V_2) =V_3\ ,$$
then we say that the particles  labeled by $l_3$ (i.e.\ the particle multiplet
$V_3$) are bound states of the
particles labeled by $l_1$ and $l_2$ (i.e.\ of the particle multiplets $V_1$
and $V_2$). We can write
\begin{equation}
V_1 +V_2\ \to \ V_3\qquad (u)\ ,
\end{equation}
where $u$ is the fusion angle, and call this expression a fusion rule.

A rule denoted as
\begin{equation}
a + b\  \to\  c\qquad (u)
\end{equation}
can be written down and also called a fusion rule if there exists a fusion tensor
$\tilde{f}$ with fusion angle $u$ with  nonzero matrix
element between the one-particle states $\ket{a}$, $\ket{b}$, $\ket{c}$: 
$\braket{c}{\tilde{f}(a\otimes b)}\ne 0$.\\

8) Crossing symmetry
\begin{equation}
\label{eq.cross1}
\braket{c\otimes d}{\tilde{S}_2(\Theta)(a\otimes b)}=\braket{d\otimes
  (\tilde{C}\tilde{P}\tilde{T}b)}{\tilde{S}_2(i\pi-\Theta)((\tilde{C}\tilde{P}\tilde{T}c)\otimes a)}
\end{equation}
and
\begin{equation}
\braket{c}{\tilde{f}_{m_1m_2}^{m_3} (a\otimes b)}=\braket{\CPT
  a}{\tilde{f}_{m_2m_3}^{m_1} (b\otimes (\CPT c))}
\end{equation}
\begin{equation}
\braket{b\otimes c}{\tilde{d}_{m_1}^{m_2m_3} a}=\braket{(\CPT
  a)\otimes b}{\tilde{d}_{m_3}^{m_1m_2}\CPT c}
\end{equation}
\begin{equation}
\label{eq.cross4}
\braket{b\otimes c}{\tilde{d}_{m_1}^{m_2 m_3} a}=\braket{\CPT
  a}{\tilde{f}_{m_3m_2}^{m_1} ((\CPT c)\otimes (\CPT b))}\ .
\end{equation}

$\tilde{C}\tilde{P}\tilde{T}$ will be discussed in 12).\\

9) Bootstrap equation\\

Let $iu$ be the location of the pole of $(\tilde{S}_2)_{m_1m_2}(\Theta)$ for some
masses $m_1$ and $m_2$ that corresponds to a direct-channel bound state. The
following equation called the bootstrap equation is satisfied:  
\begin{equation}
\label{eq.boot}
(\tilde{S_2})_{mm_3}(\Theta_{12})[I\otimes \tilde{f}_{m_1m_2}^{m_3}]=[\tilde{f}_{m_1m_2}^{m_3}\otimes I][I\otimes
  (\tilde{S_2})_{mm_2}(\Theta_{12}+iu_2)][(\tilde{S}_2)_{mm_1}(\Theta_{12}-iu_1)  \otimes I]
\end{equation}
This equation is analogous to the Yang-Baxter equation and can be regarded as
an equation expressing the parallel shiftability of lines in the diagrams
corresponding to factorized scattering processes. 

\begin{figure}
\begin{center}
\includegraphics[scale=0.4]{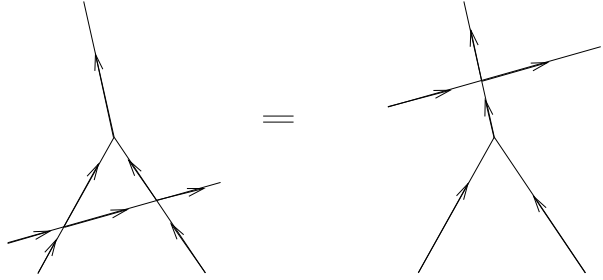}
\caption{\label{fig.bootstrap}The bootstrap equation}
\end{center}
\end{figure}

A graphical representation of (\ref{eq.boot})  is shown in Figure \ref{fig.bootstrap}.\\

10) Coleman-Thun diagrams\\

The  poles of $(\tilde{S}_2)_{m_1m_2}(\Theta)$ lying in the physical strip 
can be associated to bound
states, as already mentioned in 4),  and to anomalous
thresholds \cite{CT}.  The latter can also be related to
on-shell diagrams, which are called Coleman-Thun diagrams in factorized
scattering theory.  
The graphs corresponding to bound states can also be
regarded as Coleman-Thun diagrams.

 If $(\tilde{S}_2)_{m_1m_2}(\Theta)$ has a pole at $iu$ in the physical strip,
 then the singular part of the Laurent series of
 $(\tilde{S}_2)_{m_1m_2}(\Theta)$ around $iu$ is the sum of the contributions of
 all possible Coleman-Thun diagrams. 

 The Coleman-Thun diagrams are planar geometrical graphs with oriented lines and can be drawn in the Euclidean
  space-time plane.
 Each diagram can be interpreted as a non-physical
  two-particle scattering process (and so has 4 external lines).
   The
lines of the diagrams correspond to particles (or particle multiplets) which
can be thought to be freely
flying between pointlike events represented by the vertices.    A mass and a
  momentum are associated with each line
  of the diagram. The momenta have  real time components and
   imaginary space components. The `length' of the momenta associated with
  the lines equals the masses assigned to the lines (i.e.\ the usual
   $p_0^2-p_1^2=m^2$ mass-shell condition is satisfied).  Momentum is
   conserved at each vertex (i.e.\ $p_{in}=p_{out}$, where $p_{in}$ is the sum
   of the incoming momenta, $p_{out}$ is the sum of the outgoing
   momenta. Incoming and outgoing momenta are distinguished  by the
   orientation of the lines with respect to the vertex.)  The momenta are parallel to the lines with which
   they are associated.
Either four or three lines can join in a vertex, the possible
   vertices are the two-particle scattering vertex as shown in Figure \ref{fig.s} and the
   fusion and decay vertices of Figure \ref{fig.fd}.  

The possible angles in a diagram are completely determined by the mass
 spectrum, so if the mass spectrum is known, then it is a geometrical and
 combinatorial problem to find all  possible Coleman-Thun diagrams. 

In the simplest case the order of the pole of the term corresponding to a Coleman-Thun diagram 
is equal to the number of `degrees of freedom', i.e.\ the number of internal
lengths in the diagram which can be adjusted independently \cite{Mattsson}. In the bound state
diagram (Figure \ref{fig.bstate}), for example, there is one internal line,
the length of which can be set freely. 

In general only the  bound state diagram and its crossed version give first
order pole, all other Coleman-Thun diagrams give higher order poles. This general rule is often modified, however.\\

11) Symmetries \\

Let $\mathcal{A}$ be an associative algebra  over $\RR$.
Representations of  $\mathcal{A}$ on multi-particle spaces are 
defined in the following way: as the
multi-particle states are elements of tensor products of one-particle
spaces, it is sufficient to  construct representations on these tensor product
spaces from the one-particle representations. The in,  out and intermediate
subspaces  must be invariant, as well as the physical subspaces, if adjacency
conditions are satisfied. 
On the tensor product of two one-particle spaces the construction is done by taking the
tensor product of the two one-particle representations, which is a
representation of $\mathcal{A}\otimes \mathcal{A}$, and then composing it with
an algebra homomorphism $\Delta:\ \mathcal{A}\to \mathcal{A}\otimes
\mathcal{A}$: 
$$D_1\times D_2=(D_1\otimes D_2)\circ \Delta$$
where $D_1, D_2$ are one-particle representations of $\mathcal{A}$ on the
spaces $H_1$ and $H_2$, respectively:
$$D_1:\ \mathcal{A}\to End(H_1)\ ,\qquad   D_2:\ \mathcal{A}\to End(H_2)\ .$$
$D_1 \times D_2$ is called the product of the representations $D_1$ and
$D_2$. This definition can be applied generally to any two representations, so it is also
suitable to define products of several representations recursively. 
For the product of any three representations the following two definitions can
be used: 
\begin{gather*}
D_1\times D_2\times D_3=D_1\times (D_2\times D_3)=[D_1\otimes ((D_2\otimes
D_3)\circ \Delta)]\circ \Delta=\\
=(D_1\otimes D_2\otimes D_3)\circ
(id_{\mathcal{A}}\otimes \Delta)\circ \Delta
\end{gather*}
and
\begin{gather*}
D_1\times D_2\times D_3=(D_1 \times D_2)\times D_3=
[((D_1\otimes D_2)\circ \Delta) \otimes D_3]\circ \Delta=\\
=(D_1\otimes D_2\otimes D_3)\circ
(\Delta \otimes id_{\mathcal{A}})\circ \Delta\ .
\end{gather*}
However, it is usually required that $\Delta$ be co-associative, i.e.\
$$(id_{\mathcal{A}}\otimes \Delta)\circ \Delta = (\Delta \otimes
id_{\mathcal{A}})\circ \Delta\ ,$$
which implies that the two definitions above give identical results. Moreover,
the analogous statement holds for products of more than three
representations without further restriction on $\Delta$, i.e.\  the multiplication
of representations is associative if $\Delta$ is co-associative.    

One expects that a representation $D_\Omega$ on the vacuum subspace has
the property 
$$D \times D_\Omega \simeq D_\Omega \times D \simeq D$$
for any one-particle representation $D$. $\simeq$ denotes the equivalence of
representations.

An element $A\in \mathcal{A}$ is a symmetry of the $S$ operator if $S_2$ has the
intertwining property
\begin{equation}
\label{eq.intert}
D(A)S_2=S_2D(A)
\end{equation}
where $D(A)$ is the representation of $A$ on $\mathcal{H}_2$. (\ref{eq.intert}) implies that
$S_N$ has the intertwining property for the $N$-particle representation for
any value of $N$. A symmetry of a fusion or a decay tensor is defined
similarly. A symmetry of the factorized scattering theory is 
a symmetry of the S operator and all fusion and decay tensors.

In 1+1 dimensions the universal enveloping algebra of the Poincare algebra (supplemented with a
unit element) is generated by the boost generator $N$, the time translation
generator $H$, the space translation generator $P$, and the unit element $I$.   
They are subject to the relations
\begin{equation}
[N,H+P]=H+P\ ,\qquad
[N,H-P]=-(H-P)\ ,\qquad
[H,P]=0\ .
\end{equation}
The co-product takes the standard form: 
\begin{align}
& \Delta(H)=  H\otimes I+I\otimes H\ , && \Delta(N)=  N\otimes I+I\otimes N\ ,
\notag \\
&  \Delta(P)=  P\otimes I+I\otimes P\ , && \Delta(I)=  I\otimes I\ .
\end{align} 

A  local conserved quantity $A$ of spin $s$ has the properties
\begin{equation}
\label{eq.hs1}
[A,H]=0,\qquad [A,P]=0,\qquad A(V(\Theta))=V(\Theta),
\end{equation}
\begin{equation}
\label{eq.hs2}
A(a(\Theta))=e^{s\Theta}q_Aa(\Theta)\ ,
\end{equation}
where $q_A$ is a $\Theta$-independent linear mapping, and
\begin{multline}
\label{eq.hs3}
A[a_1(\Theta_1)\otimes a_2(\Theta_2)\otimes \dots \otimes a_N(\Theta_N)]=
e^{s\Theta_1}(q_Aa_1(\Theta_1))\otimes a_2(\Theta_2)\otimes \dots \otimes
a_N(\Theta_N)+\\
+e^{s\Theta_2}a_1(\Theta_1)\otimes (q_Aa_2(\Theta_2))\otimes \dots \otimes
a_N(\Theta_N)+ \dots + \\+
e^{s\Theta_N}a_1(\Theta_1)\otimes a_2(\Theta_2)\otimes \dots \otimes
(q_Aa_N(\Theta_N))\ .
\end{multline}
$A$ commutes with the S operator as well. 

In integrable models there exist  commuting local higher spin
(i.e.\ not spin 1) conserved quantities, their number is usually
infinite. From this property of integrable models it is possible to derive the
main factorization properties of factorized scattering theory. \\

12) Charge conjugation and reflections\\

The representation of some transformations do not always fit in the scheme described
above in 11). For instance, the charge conjugation and space reflection are such transformations. 

Charge conjugation $C$ acts as a linear transformation and has the property
$C(V(\Theta))=V(\Theta)$ (the representation map is suppressed here and
further on).

The space reflection $P$ also acts as a unitary transformation, it has the
property $P(V(\Theta))=V(-\Theta)$ and it has a tilde-d version $\tilde{P}:
V\to V$. $P(a(\Theta))=(\tilde{P}a)(-\Theta)$.   On multi-particle states it acts in the following way:
$$P(a_1(\Theta_1)\otimes a_2(\Theta_2)\otimes \dots  \otimes a_N(\Theta_N))=
P(a_N(\Theta_N))\otimes P(a_{N-1}(\Theta_{N-1}))\otimes \dots  \otimes P(a_1(\Theta_1))$$

The time reflection $T$ is represented by an antiunitary operator with the
property $T(V(\Theta))=V(-\Theta)$. It has a tilde-d version $\tilde{T}: V\to
V$. On $N$-particle states it is given by
\begin{displaymath}
T_N=\underbrace{T_1\otimes T_1 \otimes \dots  \otimes T_1}_N\ ,
\end{displaymath}
$T_N$ denoting its restriction to $N$-particle states.

The product $CPT$ of $C$, $P$ and $T$ is  always represented (even if some of
$C$, $P$ and $T$ are not represented) and it is always a symmetry. It is
represented by an antiunitary operator.  It has
the property that $CPT(V(\Theta))=V(\Theta)$ and it has a tilde-d version
$\tilde{C}\tilde{P}\tilde{T}: V \to V$. On $N$-particle states it is given by 
\begin{multline}
\notag
CPT(a_1(\Theta_1)\otimes a_2(\Theta_2)\otimes \dots  \otimes a_N(\Theta_N))=\\
=CPT(a_N(\Theta_N))\otimes CPT(a_{N-1}(\Theta_{N-1}))\otimes \dots  \otimes
CPT(a_1(\Theta_1))\ .
\end{multline}

Energy and mass are invariant with respect to the $C$, $P$, $T$ and $CPT$  transformations.

The $\tilde{C}\tilde{P}\tilde{T}$ transformation
usually has the property that   $\tilde{C}\tilde{P}\tilde{T}(V_l)=V_{\bar{l}}$
where the $V_l$-s are the one-dimensional subspaces appearing in
(\ref{eq.resz}) and
$\bar{\bar{l}}=l$. The mapping $l\mapsto \bar{l}$ on the label of the
particles is the particle-antiparticle correspondence. We also say that the
particle $\bar{l}$ is the conjugate of $l$.

Invariance of the $S$ operator  with respect to CPT transformation means 
\begin{equation}
\label{eq.cpts}
\CPT \tilde{S}_2(\Theta)=(\tilde{S}_2(\Theta))^\dagger \CPT
\end{equation}
or equivalently
\begin{equation}
\braket{a\otimes b}{\tilde{S}_2(\Theta) (c\otimes d)}=\braket{\CPT
  (c\otimes d)}{\tilde{S}_2(\Theta)\CPT (a\otimes b)}\ ,
\end{equation}
which is the same as (\ref{eq.cross1}) applied twice. (\ref{eq.cpts}) implies that
$S_N$ also has the symmetry  property for
any value of $N$. 

CPT-invariance of fusion and decay tensors means 
$$\CPT \tilde{d}_{m_1}^{m_2m_3}=(\tilde{f}_{m_3m_2}^{m_1})^\dagger \CPT\ ,$$
which is the same as (\ref{eq.cross4}).

CPT-invariance of the factorized scattering theory means 
the CPT-invariance of the S operator and all fusion and decay tensors.

Space reflection invariance of the $S$-operator means
$$\tilde{P}\tilde{S}_2(\Theta)=\tilde{S}_2(\Theta) \tilde{P}\ .$$

Space reflection invariance of fusion and decay tensors means 
$$\tilde{P} \tilde{d}_{m_1}^{m_2m_3}=\tilde{d}^{m_3m_2}_{m_1}\tilde{P}
\qquad \textrm{and}\qquad
\tilde{P} \tilde{f}^{m_3}_{m_1m_2}=\tilde{f}_{m_2m_1}^{m_3}\tilde{P}\ .$$

Time reflection invariance of the $S$-operator means
$$\tilde{T} \tilde{S}_2(\Theta)=(\tilde{S}_2(\Theta))^\dagger \tilde{T}\ .$$

Time reflection invariance of fusion and decay tensors means 
$$\tilde{T} \tilde{d}_{m_1}^{m_2m_3}=(\tilde{f}_{m_2m_3}^{m_1})^\dagger
\tilde{T}\ .$$

Charge conjugation invariance of the $S$-operator means 
$$\tilde{C}\tilde{S}_2(\Theta)=\tilde{S}_2(\Theta) \tilde{C}\ .$$

Charge conjugation invariance of fusion and decay tensors means 
$$\tilde{C} \tilde{d}_{m_1}^{m_2m_3}=\tilde{d}^{m_2m_3}_{m_1}\tilde{C}
\qquad\textrm{and}\qquad
\tilde{C} \tilde{f}^{m_3}_{m_1m_2}=\tilde{f}_{m_1m_2}^{m_3}\tilde{C}\ .$$
\vspace{0.8cm}

We remark, finally, that if a factorized  scattering theory is given, then it
is possible, in a straightforward way, to introduce particle groups or multiplets of
particle groups and their scattering. A particle group is a multi-particle state with fixed
(possible imaginary) relative rapidities of the constituting particles. These
particle groups behave in the same way as single particles. They can be
regarded as compound states. 
The S-matrix elements for these particle groups can be built from the two-particle
S-matrix  in straightforward way, they are
 certain multi-particle S matrix elements, in fact.  The Yang-Baxter equation
 and the bootstrap equation can be regarded as equations which express the
 isomorphism or homomorphism between 
 certain particle groups or single particles, regarding their scattering and
 symmetry properties. The homomorphism maps are two-particle S-matrix blocks
 or fusion and decay tensors.

We also remark that the particle statistics has been left completely
unspecified in this section, and the space of $in$ and $out$ and $intermediate$ states were defined to be
different. This, however, is only a minor  deviation from the more usual
formalism, in which  both the $in$ and the  $out$ states span the whole
Hilbert space and the S-matrix elements are j conversion coefficients
between the $in$ and $out$ bases.  The more usual formalism corresponds to quotienting out the  Hilbert space 
by the subspace defined by the relations $\ket{v}=\dots \otimes I\otimes \dots \otimes S_2\otimes
\dots \otimes I \otimes  \dots
\ket{v}$,  where $\ket{v}$ is any multi-particle state with rapidities so that $S_2$ is nonsingular and has nonzero determinant.  The
statistics of the particles is then specified by the value of the two-particle
S-matrix elements at zero relative rapidity. 

It is possible to draw other figures which are similar to Figure \ref{fig.bootstrap}, and
to write down the corresponding bootstrap equations. For example, one can
consider the mirror image of Figure \ref{fig.bootstrap}. However, it can be shown that the new bootstrap
equations obtained in this way can be derived from the axioms that have been
written down.

\section{Factorized scattering theory with a boundary}
\label{sec.fstb}
\markright{\thesection.\ \ FACTORIZED SCATTERING THEORY WITH A BOUNDARY}

If the space is a half-line, then scattering theory describes interactions of
particle-like objects and boundary states. The particle-like objects  travel
in  the half space, whereas the boundary states are localized at
the boundary of the (half) space. The ground states are boundary
states. The scattering takes place in the following way: at the beginning 
particles far form each other and  the boundary travel towards the boundary,
then the particles scatter on each other and reflect from the boundary,
and finally particles travel away from the boundary and each other. The
boundary state participates in the reflection and generally changes. The
scattering operator is called reflection operator and denoted by $R$.

The characteristics of factorized boundary scattering  are listed
and explicitly described below. Some of these properties are special to
factorized boundary scattering theory, others (like unitarity) are general properties
of boundary scattering theory applied to factorized boundary scattering theory.

Factorized scattering theory in the presence of a boundary is characterized by the following properties:\\

1) A factorized boundary scattering theory incorporates a  standard (i.e.\
   bulk) factorized
   scattering theory.\\

2) The particle number is conserved.\\

3) The  outgoing rapidities equal $-1$ times the incoming rapidities. \\

4) Factorization and boundary Yang-Baxter equation: An arbitrary N-particle scattering process can be described as a sequence
   of 2-particle collisions and one-particle reflections, all possible
   descriptions  are equivalent. The 2-particle collisions are described by the
   two-particle S-matrix block, the one-particle reflections are described by
   the one-particle reflection matrix block. These are the most interesting
   quantities in factorized boundary scattering
   theory.  \\

The Hilbert space $\mathcal{H}^b$ can be decomposed into a sum of $N$-particle subspaces:
\begin{equation}
\mathcal{H}^b=\oplus_{N=0}^\infty \mathcal{H}_N^b\ ,
\end{equation}
where
$$\mathcal{H}_N^b=\underbrace{\mathcal{H}_1\otimes \dots  \otimes
  \mathcal{H}_1}_N \otimes
\mathcal{H}_B$$
for $N\ge1$, and
$$\mathcal{H}_0^b=\mathcal{H}_B\ .$$
$\mathcal{H}_B$ is the space of boundary states, the space $\mathcal{H}_{B0}$
of the ground states is a subspace of it. $\mathcal{H}_{B0}$ is usually finite
dimensional, most frequently one-dimensional.

$\mathcal{H}_N^b$ can be decomposed into a (not direct) sum of spaces of in, out and intermediate states:
$$\mathcal{H}_N^b=\mathcal{H}_{N,in}^b + \mathcal{H}_{N,out}^b + \mathcal{H}_{N,intermed}^b$$
for 
$N\ge 1$, for $N=0$ we have
$$\mathcal{H}_{0,in}^b=\mathcal{H}_{0,out}^b$$
and there are no intermediate states for $N=0,1$.

A multi-particle state of the form 
$$a_1(\Theta_1)\otimes a_2(\Theta_2)\otimes \dots  \otimes a_N(\Theta_N)\otimes B$$
is an in state if $\Theta_1\ge\Theta_2\ge\dots \ge\Theta_N>0$, an out state if 
$\Theta_1\le\Theta_2\le\dots \le\Theta_N<0$, and an intermediate state
otherwise. We shall also use the ket notation for the multi-particle states.

$\mathcal{H}_B$ can be decomposed into energy eigenspaces:
\begin{equation}
\label{eq.decmb}
\mathcal{H}_B=\oplus_E \mathcal{H}_{B,E}\ .
\end{equation}

The boundary states are normalized as 
$$\braket{B_i}{B_j}=\delta_{ij}\ .$$

It is often useful to consider a distinguished orthonormal basis of $\mH_B$
and the corresponding decomposition into a sum of one-dimensional subspaces:
\begin{equation}
\label{eq.decbpar}
\mH_B=\oplus_{l^b} \mH_{B,l^b}\ ,
\end{equation} 
which should be a refinement of the decomposition (\ref{eq.decmb}).

Boundary states can also be involved in adjacency conditions.\\

Instead of the physical R-operator one usually uses an auxiliary R-operator
$R_{aux}$, which is related to the physical R-operator in the same way as
$S_{aux}$ is related to $S$. The $aux$ subscript will be suppressed.  
The $\mathcal{H}_N^b$ are invariant subspaces of $R_{aux}$, the
$N$-particle auxiliary
R-operator $R_N$ is obtained by restricting the full auxiliary R-operator to $\mathcal{H}_N^b$: 
$R_N:\ \mathcal{H}_N^b\to \mathcal{H}_N^b$.  
In particular
$$R_N^b(\mathcal{H}_{N,in}^b)= \mathcal{H}_{N,out}^b$$
and
$$R_0=Id_{\mathcal{H}_0^b}\ .$$

For $N\ge 1$ the operators
\begin{multline}
\notag 
\hat{R}_N(\Theta_{1},\Theta_{2},\dots ,\Theta_{N}):\\[2mm]
V(\Theta_1)\otimes V(\Theta_2)
\otimes \dots  \otimes V(\Theta_N)\otimes \mathcal{H}_B \to  V(-\Theta_1)\otimes V(-\Theta_{2})
\otimes \dots  \otimes V(-\Theta_N) \otimes \mathcal{H}_B
\end{multline}
can be introduced so that
\begin{multline}
\notag
\brakettt{b_1(\Theta_1')b_{2}(\Theta_{2}')\dots b_N(\Theta_N')B_j}{R_N}{a_1(\Theta_1)a_{2}(\Theta_{2})\dots a_N(\Theta_N)
B_i}=\\[3mm]
=\delta(\Theta_1+\Theta_1'^*)\delta(\Theta_2+\Theta_2'^*)\dots \delta(\Theta_N+\Theta_N'^*)\times\\
\times \brakettt{b_1(-\Theta_1^*)b_{2}(-\Theta_{2}^*)\dots b_N(-\Theta_N^*)B_j}
{ \hat{R}_N(
  \Theta_1,\Theta_{2},\dots ,\Theta_N)}{a_1(\Theta_1)a_{2}(\Theta_{2})\dots a_N(\Theta_N) B_i }
\end{multline}

For the matrix elements of the $\hat{R}_N$-s the notation
\begin{multline}
\notag
(\hat{R}_N)^{b_{1}b_{2}\dots b_N,j}_{a_1 a_2\dots  a_N,i}(\Theta_1,\Theta_2,\dots ,\Theta_N)=\\[2mm]
=\brakettt{b_1(-\Theta_1^*)b_{2}(-\Theta_{2}^*)\dots b_N(-\Theta_N^*)B_j}
{ \hat{R}_N(
  \Theta_1,\Theta_{2},\dots ,\Theta_N)}{a_1(\Theta_1)a_{2}(\Theta_{2})\dots a_N(\Theta_N)B_i}
\end{multline}
is used. One can also introduce the operators
$$\tilde{R}_N(\Theta_{1},\Theta_{2},\dots ,\Theta_{N}): \underbrace{V\otimes V\otimes
\dots  \otimes V}_N \otimes \mathcal{H}_B \to \underbrace{V\otimes V\otimes \dots  \otimes V}_N
\otimes \mathcal{H}_B $$
$$\tilde{R}_N(\Theta_{1},\Theta_{2},\dots ,\Theta_{N})=[i_{-\Theta_1}\otimes
i_{-\Theta_{2}}\otimes \dots  \otimes i_{-\Theta_N} \otimes I  ] \hat{R}_N
(\Theta_1,\Theta_2,\dots ,\Theta_N) [i^{-1}_{\Theta_1}\otimes
i^{-1}_{\Theta_{2}}\otimes \dots  \otimes i^{-1}_{\Theta_N} \otimes I  ]\ .$$
The matrix elements of the $\tilde{R}_N$-s are
denoted as 
$$(\tilde{R}_N)^{b_{1}b_{2}\dots b_N,j}_{a_1
  a_2\dots  a_N,i}(\Theta_{1},\Theta_{2},\dots ,\Theta_{N})\ .$$
For the matrix elements of $\hat{R}_N$ and $\tilde{R}_N$ the equation
$$(\tilde{R}_N)^{b_{1}b_{2}\dots b_N,j}_{a_1 a_2\dots  a_N,i}(\Theta_{1},\Theta_{2},\dots ,\Theta_{N})=
(\hat{R}_N)^{b_{1}b_{2}\dots b_N,j}_{a_1 a_2\dots  a_N,i}(\Theta_1,\Theta_2,\dots ,\Theta_N)$$
holds.\\
One can introduce $(\tilde{R}_N)_{m_1 m_2 \dots  m_N,E}(\Theta_{1},\Theta_{2},\dots ,\Theta_{N})$
and
$(\hat{R}_N)_{m_1 m_2 \dots  m_N,E}(\Theta_{1},\Theta_{2},\dots ,\Theta_{N})$ as
in Section \ref{sec.fst}. $\tilde{R}_N$ and $\hat{R}_N$ are composed of
these blocks.

\begin{figure}
\begin{center}
\includegraphics[scale=0.7]{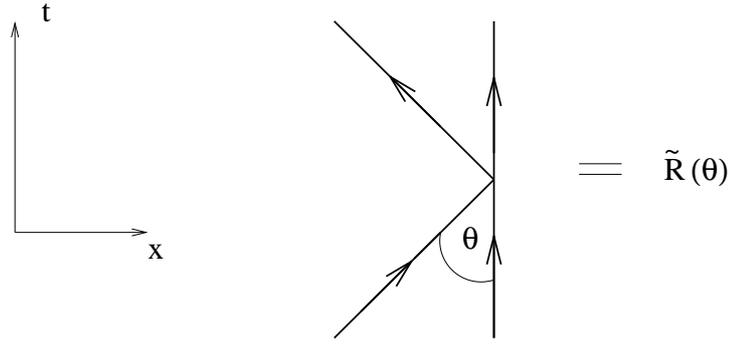}
\caption{\label{fig.r}One-particle reflection matrix $\tilde{R}_1(\Theta)$}
\end{center}
\end{figure}

Graphical representation of $\tilde{R}_1(\Theta)$ is shown in Figure \ref{fig.r}. 

If 
$$
\tilde{R}_1(\Theta)(V_1\otimes \mH_{B,2}) \subseteq  V_1\otimes \mH_{B,2}
$$ 
for some subspaces $V_1\subseteq V_{m_1}$ and $\mH_{B,2}\subseteq \mH_{B,E_2}$
with some mass $m_1$ and energy $E_2$, then we can introduce the notation 
$\tilde{R}_{V_1 \mH_{B,2}}(\Theta)=\tilde{R}_1(\Theta)|_{V_1\otimes \mH_{B,2}}$ 
and say that $\tilde{R}_{V_1 \mH_{B,2}}(\Theta)$ is the reflection matrix of
the multiplet $V_1$ on the boundary multiplet $\mH_{B,2}$. The case when
$\mH_{B,2}$ is $\mH_{B0}$ is considered frequently, the reflection matrix blocks with
$\mH_{B,2}=\mH_{B0}$ are called ground state reflection matrices.\\

In the $N=2$ case the factorization property is expressed by the equation
$$\tilde{R}_2(\Theta_1,\Theta_2)=[I\otimes
\tilde{R}_1(\Theta_2)][\tilde{S}_2(\Theta_2+\Theta_1)\otimes I][I\otimes
\tilde{R}_1(\Theta_1)][\tilde{S}_2(\Theta_1-\Theta_2)\otimes I]$$
or
$$\tilde{R}_2(\Theta_1,\Theta_2)=[\tilde{S}_2(\Theta_1-\Theta_2)  \otimes I][I\otimes
 \tilde{R}_1(\Theta_1)][\tilde{S}_2(\Theta_1+\Theta_2)\otimes I][I\otimes
 \tilde{R}_1(\Theta_2)]\ .$$
The two equations correspond to the two possible decompositions of the
 2-particle reflection into 2-particle scatterings and one-particle
 reflections. 

\begin{figure}
\begin{center}
\includegraphics[scale=0.5]{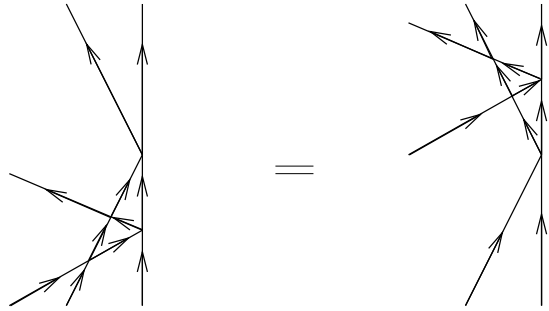}
\caption{\label{fig.byb}Boundary Yang-Baxter equation}
\end{center}
\end{figure}

The equality of the two expressions on the right-hand side 
\begin{multline}
\label{eq.byb}
[I\otimes
\tilde{R}_1(\Theta_2)][\tilde{S}_2(\Theta_2+\Theta_1)\otimes I][I\otimes
\tilde{R}_1(\Theta_1)][\tilde{S}_2(\Theta_1-\Theta_2)\otimes I]=\\[2mm]
=[\tilde{S}_2(\Theta_1-\Theta_2)  \otimes I][I\otimes
 \tilde{R}_1(\Theta_1)][\tilde{S}_2(\Theta_1+\Theta_2)\otimes I][I\otimes
 \tilde{R}_1(\Theta_2)]
\end{multline}
is called the boundary Yang-Baxter equation.  A graphical representation of this equation is shown in
Figure \ref{fig.byb}. If the boundary Yang-Baxter is satisfied, then the
analogous equations expressing the equivalence of the possible factorizations
of the $N$-particle reflection (for $N>2$) into two-particle scatterings and one-particle
reflections are also satisfied.

The above equations (with the obvious modifications) are also satisfied by
$R_2$, $R_1$, $S_2$ and $\hat{R}_2$, $\hat{R}_1$, $\hat{S}_2$ and by the
appropriate blocks of them corresponding to definite masses and energies.\\

5) Analytic properties: $\tilde{R}_1(\Theta)$ is an analytic function with
   possible pole singularities which are located in $i\RR$. The `physical
   strip' for $\tilde{R}_1(\Theta)$ is the domain $0<Im(\Theta)<\pi/2$. The
   singular part of $\tilde{R}_1(\Theta)$ at a pole lying in the physical
   strip  is generally a sum of
   various contributions related to bound states and anomalous thresholds.\\

6) Real analyticity
$$(\tilde{R}_1(\Theta^*))^\dagger =\tilde{R}_1(-\Theta)\ .$$

7) Unitarity
$$(\tilde{R}_1(\Theta^*))^\dagger \tilde{R}_1(\Theta)=I\ .$$
This condition and the real analyticity condition imply
$$\tilde{R}_1(-\Theta)\tilde{R}_1(\Theta)=I\ .$$
In factorized scattering theory  usually the latter condition is referred to
as unitarity.\\   

8) Crossing symmetry
\begin{multline}
\label{eq.bcu}
\braket{c\otimes B_j}{\tilde{R}_1(\Theta)(a\otimes B_i)}=\\[2mm]
=\sum_{b,d \in \mathcal{B}}
\braket{c\otimes d}{\tilde{S}_2(2\Theta)(a\otimes b)}
\braket{(\tilde{C}\tilde{P}\tilde{T}d)\otimes
  B_j}{\tilde{R}_1(i\pi-\Theta)((\tilde{C}\tilde{P}\tilde{T}b)\otimes B_i)}
\end{multline}
(\ref{eq.bcu}) is called `boundary cross-unitarity' condition. Any
orthonormal basis could be used instead of $\mathcal{B}$.\\

9) Boundary bound states: at some poles of  $(\tilde{R}_1)_{m,E}(\Theta)$ in
the physical strip the
singular part has a contribution 
corresponding to boundary bound states:
\begin{equation}
\label{eq.bfusion}
(\tilde{R}_1)_{m,E}(\Theta)=\frac{\tilde{h}_{E_1}^{m,E} \tilde{g}_{m,E}^{E_1}}{\Theta-iu}+s(\Theta)+
reg(\Theta)
\end{equation}
where $iu$, $0<u<\pi/2$, is the location of the pole, $reg(\Theta)$ is regular
at $iu$, $s(\Theta)$ contains further possible singular terms. $u$ is called
fusion angle. It will also be denoted by $\nu$. 
$$\tilde{h}_{E_1}^{m,E}: \mathcal{H}_{B,E_1} \to V_m\otimes \mathcal{H}_{B,E}$$
is the boundary decay tensor, 
$$\tilde{g}_{m,E}^{E_1}: V_m\otimes \mathcal{H}_{B,E} \to \mathcal{H}_{B,E_1}$$
is the boundary fusion tensor. We also have the hatted versions
$$\hat{h}_{E_1}^{m,E}: \mathcal{H}_{B,E_1} \to V_m(-iu)\otimes \mathcal{H}_{B,E}$$
and
$$\hat{g}_{m,E}^{E_1}: V_m(iu)\otimes \mathcal{H}_{B,E} \to \mathcal{H}_{B,E_1}.$$
$E$, $E_1$, $m$ and $u$ are related by energy conservation: $E_1=E+m\cos(u)$.

\begin{figure}
\begin{center}
\includegraphics[scale=0.5]{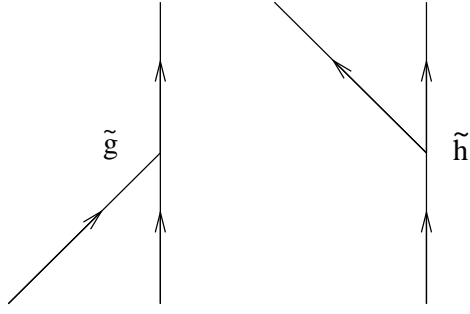}
\caption{\label{fig.gh}Boundary fusion and decay tensors}
\end{center}
\end{figure}

\begin{figure}[!h]
\begin{center}
\includegraphics[scale=0.5]{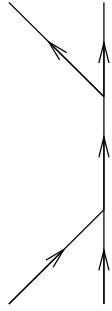}
\caption{\label{fig.bbstate}Contribution of boundary bound states}
\end{center}
\end{figure}

Graphical representation of $\tilde{g}_{m,E}^{E_1}$  and
$\hat{h}_{E_1}^{m,E}$ can be seen in Figure \ref{fig.gh}. The graphical
representation of the residue $\tilde{h}_{E_1}^{m,E}\tilde{g}_{m,E}^{E_1}$
 is shown in Figure \ref{fig.bbstate}.

The image space of $\tilde{g}_{m,E}^{E_1}$ has zero intersection with the kernel
of  $\tilde{h}_{E_1}^{m,E}$. 

Generally if 
$$V_1= \oplus_{l_1} V_{l_1}\ ,\qquad \mH_{B,1}=\oplus_{l^b_1} \mH_{B,l^b_1}\ ,\qquad  \mH_{B,2}=\oplus_{l^b_2} \mH_{B,l^b_2}$$
where $l_1$, $l^b_1$ and $l^b_2$ take certain values from all possible values of
$l$, $l^b$  defined in (\ref{eq.resz}) and (\ref{eq.decmb})  and the states in $V_1$, $\mH_{B,1}$, $\mH_{B,2}$
have mass $m$ and energy  $E$, $E_1$ and 
there exists a fusion tensor $\tilde{g}_{m,E}^{E_1}$ and
$$\tilde{g}_{m,E}^{E_1} (V_1\otimes \mH_{B,1}) =\mH_{B,2}\ ,$$
then we say that the boundary states labeled by $l^b_2$ (i.e.\ the multiplet $\mH_{B,2}$) are bound states of the
particles and boundary states labeled by $l_1$ and $l^b_1$ (i.e.\ of the
multiplet $V_1$ and $\mH_{B,1}$). We can write 
\begin{equation}
V_1 + \mH_{B,1}\  \to\  \mH_{B,2}\qquad (u)\ ,
\end{equation}
where $u$ is the boundary fusion angle,
and call this expression a boundary fusion rule.

A rule denoted as
\begin{equation}
a + B_1\  \to\  B_2\qquad (u)
\end{equation}
can be written down and also called a boundary fusion rule if
there exists a boundary fusion tensor
$\tilde{g}$ with fusion angle $u$ with  nonzero matrix
element between the one-particle states $\ket{a}$ and the boundary states  $\ket{B_1}$, $\ket{B_2}$: 
$\braket{B_2}{\tilde{g}(a\otimes B_1)}\ne 0$.\\

\begin{figure}[t]
\begin{center}
\includegraphics[scale=0.4]{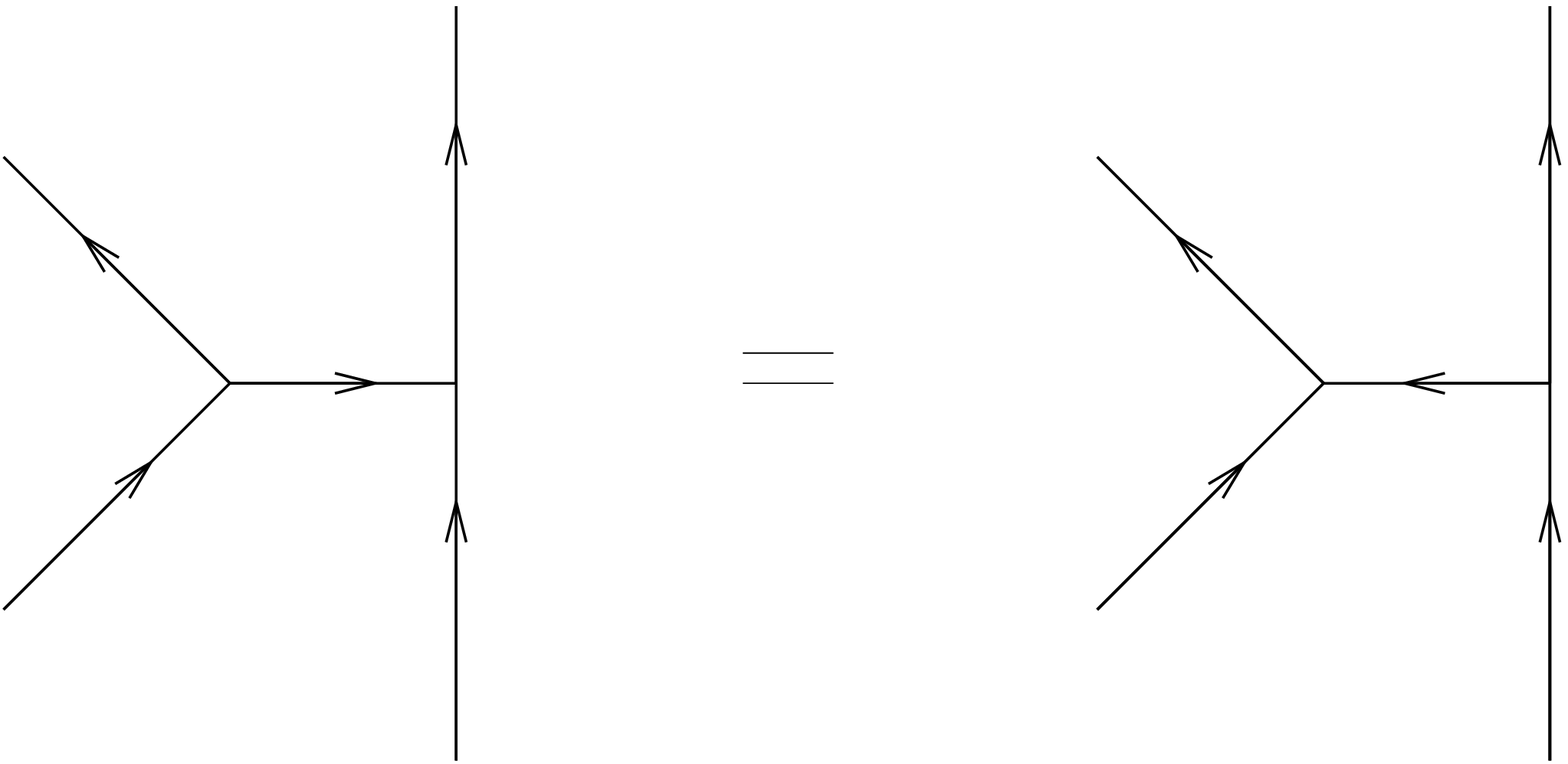}
\caption{\label{fig.GHcont}Graphical representation for (\ref{eq.ss1})}
\end{center}
\end{figure}

10) At some poles the singular part of $(\tilde{R}_1)_{m,E}(\Theta)$ may have contributions related to the existence of a bulk fusion
    tensor $\tilde{f}_{mm}^{m'}$. If such a fusion tensor exists, then 
    $(\tilde{R}_1)_{m,E}(\Theta)$ can be written as  
\begin{equation}
\label{eq.contrib1}
(\tilde{R}_1)_{m,E}(\Theta)=\frac{1}{2}\frac{[I\otimes
    \tilde{G}_{m',E}^{E}][\tilde{d}_{m}^{mm'}\otimes
    I]
}
{\Theta-iu}+s(\Theta)+reg(\Theta)\ ,
\end{equation}
where $u$ is determined by the kinematical condition that
   $$m(\cosh(iu),\sinh(iu))=m(\cosh(-iu),\sinh(-iu))+m'(\cosh(i\pi/2),\sinh(i\pi/2))\ .$$ 
The residue can be written in another form as well:
\begin{equation}
\label{eq.ss1}
[I\otimes
    \tilde{G}_{m',E}^{E}][\tilde{d}_{m}^{mm'}\otimes I]=[\tilde{f}^{m}_{mm'}\otimes I][I\otimes
    \tilde{H}^{m',E}_{E}]\ .
\end{equation}
A diagrammatic representation 
 is shown in Figure \ref{fig.GHcont}.

\begin{figure}[t]
\begin{center}
\includegraphics[scale=0.4]{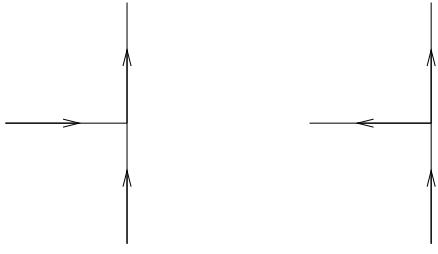}
\caption{\label{fig.GH}$\tilde{G}_{m,E}^{E}$ and $\tilde{H}_{E}^{m,E}$}
\end{center}
\end{figure}

$$\tilde{H}_{E}^{m,E}: \mathcal{H}_{B,E} \to V_m\otimes \mathcal{H}_{B,E}$$
can be regarded as a decay tensor, 
$$\tilde{G}_{m,E}^{E}: V_m\otimes \mathcal{H}_{B,E} \to \mathcal{H}_{B,E}$$
as a fusion tensor. We also have the hatted versions
$$\hat{H}_{E}^{m,E}: \mathcal{H}_{B,E} \to V_m(-i\pi/2)\otimes \mathcal{H}_{B,E}$$
and
$$\hat{G}_{m,E}^{E}: V_m(i\pi/2)\otimes \mathcal{H}_{B,E} \to \mathcal{H}_{B,E}\ .$$
The graphical representation of $\tilde{G}_{m,E}^{E}$ and $\tilde{H}_{E}^{m,E}$
is shown in Figure \ref{fig.GH}. 

The existence of $\tilde{f}_{mm}^{m'}$ does not necessarily imply that a contribution
(\ref{eq.contrib1}) really exists, as $\tilde{G}_{m',E}^{E}$
and $\tilde{H}_{E}^{m',E}$ may be zero.

\begin{figure}
\begin{center}
\includegraphics[scale=0.5]{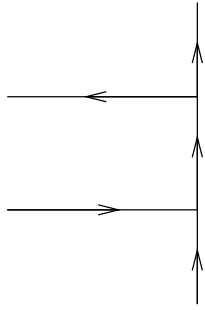}
\caption{\label{fig.GHcont2} Graphical representation for $\tilde{H}_E^{m,E}\tilde{G}^E_{m,E}$    }
\end{center}
\end{figure}

If $\tilde{G}_{m',E}^{E}$ and $\tilde{H}_{E}^{m',E}$  are nonzero, then $(\tilde{R}_1)_{m',E}(\Theta)$ can
be written as
\begin{equation}
\label{eq.contrib2}
(\tilde{R}_1)_{m,E}(\Theta)=\frac{1}{2}\frac{\tilde{H}_E^{m,E}\tilde{G}^E_{m,E}}{\Theta-i\pi/2}+s(\Theta)+reg(\Theta)\
\end{equation}
and so it usually has a pole at  $i\pi/2$.
The graphical representation of $\tilde{H}_E^{m,E}\tilde{G}^E_{m,E}$ is shown in
Figure \ref{fig.GHcont2}. 

We remark that in some cases $\tilde{g}_{m',E}^{E}$ and $\tilde{h}^{m',E}_{E}$
exist and $s(\Theta)$ contains a contribution of the form (\ref{eq.bfusion}). 

The image space of  $\tilde{G}_{m',E}^{E}$ has zero intersection with the kernel of  $\tilde{H}^{m',E}_{E}$.

\begin{figure}
\begin{center}
\includegraphics[scale=0.5]{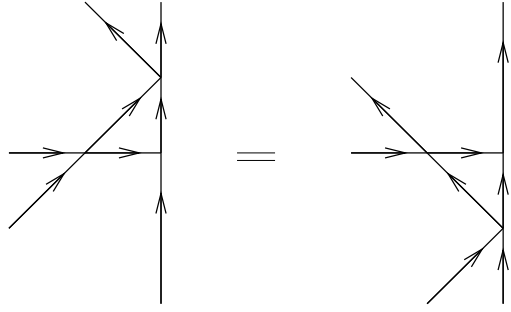}
\caption{\label{fig.GSR} Graphical representation for (\ref{eq.GSR}) }
\end{center}
\end{figure}

We also have the relation
\begin{multline}
\label{eq.GSR}
(\tilde{R}_1)_{m',E}^{m'}(\Theta)[I\otimes
  \tilde{G}_{m,E}^E][(\tilde{S}_2)_{mm'}(\frac{i\pi}{2}-\Theta)\otimes I]=\\[0mm]
=[I\otimes \tilde{G}_{m,E}^E][(\tilde{S}_2)_{mm'}(\frac{i\pi}{2}+\Theta)\otimes
  I][I\otimes (\tilde{R}_1)_{m',E}(\Theta)]\ ,
\end{multline}
which can be obtained from (\ref{eq.byb}) and (\ref{eq.boot}). Its graphical
representation is shown in Figure \ref{fig.GSR}.\\

11) Bootstrap equations\\

There are two kinds of bootstrap equations in the presence of a boundary. The
first kind is related to reflections on boundary bound states:
let $iu$ be the location of the pole of $(\tilde{R}_1)_{m,E}(\Theta)$ for some
mass $m$ and energy $E$ that corresponds to a boundary bound state. The
following equation called the boundary bootstrap equation is satisfied:
\begin{multline}
\label{eq.bboot1}
[(\tilde{R}_1)_{m_1,E_1}(\Theta)][I\otimes
  \tilde{g}_{m,E}^{E_1}]=\\[2mm]
=[I\otimes \tilde{g}_{m,E}^{E_1}][(\tilde{S}_2)_{m,m_1}(iu+\Theta)\otimes I][I\otimes (\tilde{R}_1)_{m_1,E}(\Theta)    ]
[(\tilde{S}_2)_{m_1,m}(\Theta-iu)\otimes I]\ .
\end{multline}
We shall also call it first boundary bootstrap equation.

\begin{figure}
\begin{center}
\includegraphics[scale=0.5]{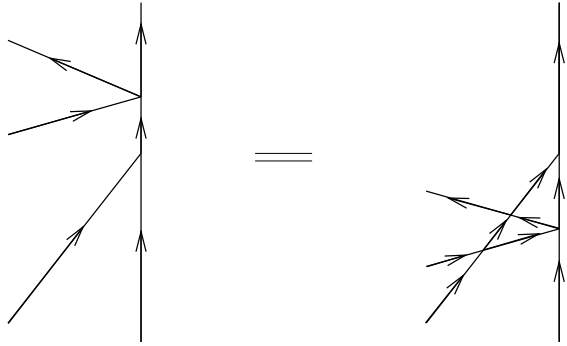}
\caption{\label{fig.bstr2}The first boundary bootstrap equation}
\end{center}
\end{figure}

A graphical representation of (\ref{eq.bboot1})  is shown in Figure \ref{fig.bstr2}.

\begin{figure}
\begin{center}
\includegraphics[scale=0.5]{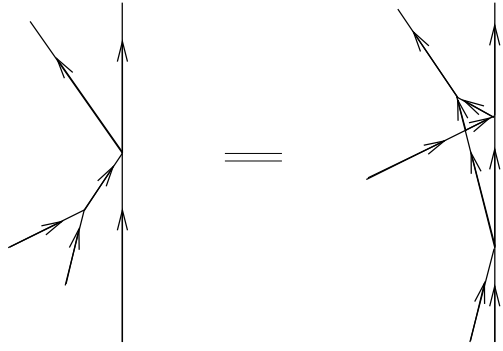}
\caption{\label{fig.bstr1}The second boundary bootstrap equation}
\end{center}
\end{figure}

The second kind is related to reflections of bound states of particles: 
let us assume that the underlying bulk factorized scattering theory has the
property that
if $(\tilde{S}_2)_{m_1m_2}(\Theta)$ has a pole at $iu$ that corresponds to bound
states, then $(\tilde{S}_2)_{m_2m_1}(\Theta)$ also has a pole at $iu$
corresponding to bound states. This property is ensured if, for example, the scattering is
invariant with respect to space reflection.
If  $(\tilde{S}_2)_{m_1m_2}(\Theta)$ has a pole that corresponds to a
direct channel bound state, then the following equation is satisfied: 
\begin{multline}
\label{eq.bboot2}
[(\tilde{R}_1)_{m_3,E}(\Theta)][\tilde{f}_{m_1m_2}^{m_3}\otimes I]=\\[2mm]
=[\tilde{f}_{m_2m_1}^{m_3}\otimes I][I\otimes (\tilde{R}_1)_{m_1,E}(\Theta+iu_1)]
[(\tilde{S}_2)_{m_1m_2}(2\Theta+iu_1-iu_2)\otimes I][I\otimes
(\tilde{R})_{m_2,E}(\Theta-iu_2)]\ .
\end{multline}
We shall call it second boundary bootstrap equation.
A graphical representation of (\ref{eq.bboot2})  is shown in Figure \ref{fig.bstr1}.\\

12) Boundary Coleman-Thun diagrams\\

The statements made in 10) in Section \ref{sec.fst}  can be generalized
 for the singularities of $\tilde{R}_1(\Theta)$ in a straightforward way 
\cite{DTW}.
In general, first order poles correspond only to the diagrams shown in Figures
\ref{fig.bbstate}, \ref{fig.GHcont}, \ref{fig.GHcont2}.  This general rule is
 often modified, however.

A more detailed description of boundary Coleman-Thun diagrams and the
singularities of  $\tilde{R}_1(\Theta)$   can be found in \cite{Mattsson}, for
example. Applications can be found in
\cite{Mattsson,MattssonDorey,sajat1,DeliusGand}.
We remark that significant steps were made in \cite{bbt1} and \cite{bbt2}
to develop perturbative quantum field theory in the presence of a boundary and
to justify the generalization of the description of singularities to the
boundary case. \\

13) Symmetries\\

Let $\mathcal{A}^b$ be an associative algebra over $\RR$. 
The representation of $\mathcal{A}^b$
has the following structure usually: $\mathcal{A}^b$ is usually a remnant of a
symmetry algebra $\mathcal{A}$ of the underlying bulk scattering, so it is
sufficient to have a representation $D_B: \mathcal{A}_B\to End(\mH_B)$  of $\mathcal{A}^b$ on
$\mathcal{H}_B$ and a co-product (an algebra-homomorphism)  $\Delta_B: \mathcal{A}^b\to
\mathcal{A}\otimes \mathcal{A}^b$ to define representations on the
$\mathcal{H}_N^b$ spaces. It is assumed that the representations of
$\mathcal{A}$ in the bulk scattering theory have the structure that is
described in Section \ref{sec.fst}.

Generally the  co-product $\Delta_B$ allows
the definition of the product $D_1\times D_2\times \dots  \times D_N\times D_B$
of arbitrary representations $D_1$, $D_2$, \dots  , $D_N$ of $\mathcal{A}$ and of
an arbitrary representation $D_B$ of $\mathcal{A}^b$, which will be a
representation of $\mathcal{A}^b$. The definition of the product is analogous
to that described in Section \ref{sec.fst}.  For two representations 
$$D_1\times
D_B=(D_1\otimes D_B)\circ \Delta_B\ .$$
For $D_1$, $D_2$, $D_B$ one can choose either
\begin{equation}
\notag
D_1\times D_2\times D_B=D_1\times (D_2\times D_B)=[D_1\otimes ((D_2\otimes
D_B)\circ \Delta_B)]\circ \Delta_B=
\end{equation}
\begin{equation}
\label{eq.z1}
=(D_1\otimes D_2\otimes
D_B)\circ(Id_{\mathcal{A}}\otimes \Delta_B)\circ \Delta_B
\end{equation}
or
\begin{equation}
\notag
D_1\times D_2\times D_B=(D_1\times D_2)\times D_B=[((D_1\otimes D_2)\circ
\Delta)\otimes D_B]\circ \Delta_B=
\end{equation}
\begin{equation}
\label{eq.z2}
=(D_1\otimes D_2\otimes D_B)\circ
(\Delta\otimes Id_{\mathcal{A}^b})\circ \Delta_B\ .
\end{equation}
$\Delta_B$ is usually required to have the co-associativity
property
$$(Id_\mathcal{A} \otimes \Delta_B)\circ \Delta_B=(\Delta\otimes
Id_{\mathcal{A}^b})\circ  \Delta_B\ ,$$
where $\Delta$ is the co-product of $\mathcal{A}$,
which implies that the two definitions (\ref{eq.z1}) and (\ref{eq.z2}) are
equivalent. If $\Delta$ and $\Delta_B$ are both co-associative, then the  multiplication of
representations is associative. 

Usually $\mathcal{A}^b$ can be realized as a subalgebra of $\mathcal{A}$,
i.e.\ 
there exists a monomorphism $i: \mathcal{A}^b \to \mathcal{A}$. $i$ is not
necessarily unique. 

If $(\Delta \circ i) (\mathcal{A}^b)\subseteq \mathcal{A}\otimes
\mathcal{A}^b$, then a co-product $\Delta_B$ can be defined in terms of
$\Delta$ and $i$ as follows: 
\begin{equation}
\label{eq.bcp}
\Delta_B=  (Id_{\mathcal{A}}\otimes i)^{-1}\circ
\Delta\circ i\ .
\end{equation}
This co-product is co-associative if $\Delta$ is
co-associative. The co-product $\Delta_B$ that  actually appears in the
definition of representations on multi-particle spaces often admits the form (\ref{eq.bcp}).

In the definition of a symmetry transformation we assume that $\mathcal{A}$ is
a symmetry algebra of the underlying
bulk factorized scattering theory. 
An element $A_B\in \mathcal{A}^b$ is a symmetry of the $R$ operator if $R_1$ has the
intertwining property
\begin{equation}
\label{eq.intert2}
D(A_B)R_1=R_1D(A_B)\ ,
\end{equation}
where $D(A_B)$ is the representation of $A_B$ on
$\mathcal{H}_1^b$. (\ref{eq.intert2}) and the assumed symmetry property of
$\mathcal{A}$ imply that
$R_N$ has the intertwining property for the representation on $\mH_N^b$ for
any value of $N$. A symmetry of a fusion or a decay tensor is defined
similarly. A symmetry of the factorized scattering theory with boundary is 
a symmetry of the R operator and all fusion and decay tensors.

The remnant of the Poincare algebra is generated by the time translation
generator $H_B$ and the unit element $I_B$, the co-multiplication is
\begin{equation}
\Delta_B(H_B)=H\otimes I_B+I\otimes H_B\ , \qquad
\Delta_B(I_B)=I\otimes I_B\ .
\end{equation}

We refer the reader to \cite{GZ} for the discussion of higher spin conserved quantities.
The special case of the boundary supersymmetry algebra will be discussed, however, in
later sections. \\

14) Charge conjugation and reflections\\

Space reflection is not defined. It could be introduced as a transformation
that relates distinct theories with boundaries on the left- and
right-hand side, respectively.

The boundary charge conjugation and time reflection $C_B$ and $T_B$, if their
representation is defined in a theory, have the properties that $C_B(\mH_{B,E})=\mH_{B,E}$ and  
$T_B(\mH_{B,E})=\mH_{B,E}$. We also have
$$T_N=\underbrace{T_1\otimes T_1 \otimes \dots  \otimes T_1}_N\otimes T_B$$
on $\mH^b_N$. (Where $T_B$ and $T_N$ denote the restriction of $T$ to
$\mH_{B}$ and $\mH_{N}^b$.)

Invariance of the $R$ operator with respect to charge conjugation means invariance of the
underlying bulk scattering theory as defined in 12) in Section \ref{sec.fst} and 
$$\tilde{C}\tilde{R}_1(\Theta)=\tilde{R}_1(\Theta)\tilde{C}\ ,$$
invariance of boundary fusion and decay tensors means 
$$\tilde{C}\tilde{g}_{m,E}^{E_1}=\tilde{g}_{m,E}^{E_1}\tilde{C}\qquad \textrm{and} \qquad 
\tilde{C}\tilde{h}_{E_1}^{m,E}=\tilde{h}_{E_1}^{m,E}\tilde{C}\ .$$

Invariance of the $R$ operator with respect to time reflection means invariance of the
underlying bulk scattering theory as defined in 12) in Section \ref{sec.fst} and 
$$\tilde{T}\tilde{R}_1(\Theta)=(\tilde{R}_1(\Theta))^\dagger \tilde{T}\ ,$$
invariance of boundary fusion and decay tensors means 
$$\tilde{T}\tilde{g}_{m,E}^{E_1}=(\tilde{h}^{m,E}_{E_1})^\dagger \tilde{T}\ .$$

Invariance of factorized scattering  theory with boundary means that the $R$
operator and the fusion and decay tensors are both invariant. 
\vspace{0.8cm}

We remark, finally, that  we assume that all boundary states are bound
states, which implies that all boundary states can be generated in a few steps
form the ground state by boundary fusion. 

It
is possible to introduce boundary particle groups, which are in fact
multi-particle states (containing a boundary state) with fixed rapidities of
the constituent particles. 
These `boundary
particle groups' behave in the same way as boundary states. They can also be
regarded as compound boundary states. The reflection
matrix elements on the boundary particle groups can be built from the two-particle
S-matrix and one-particle reflection matrix in a straightforward way. 
One can also allow bulk `particle groups' introduced in Section
\ref{sec.fst}, their reflection matrix elements are also defined in a
straightforward way.
The boundary Yang-Baxter and bootstrap equations can be given a similar interpretation
as in the bulk.

We also remark that the space of $in$ and $out$ and $intermediate$ states were defined to be
different, which is a minor  deviation from the more standard
formalism, in which both the $in$ and the  $out$ states span the whole
Hilbert space.   This formalism corresponds to quotienting out the  Hilbert space 
by the subspace defined by relations $\ket{v}=\dots \otimes I\otimes \dots \otimes S_2\otimes \dots \otimes I
\ket{v}$,
and  $\ket{v}=\dots \otimes I\otimes \dots R_1 
\ket{v}$,  where $\ket{v}$ is any multi-particle state with rapidities so that
$S_2$ and $R_1$  is nonsingular and has nonzero determinant. 

As in the bulk case, one can draw further figures which are similar to Figure
\ref{fig.bstr2} and Figure \ref{fig.bstr1} and write down the corresponding bootstrap equations.
 These new equations, however, can be derived from the axioms written down and do not
 impose new restrictions.

\section{The bootstrap method}
\label{sec.bootstrap}
\markright{\thesection.\ \ THE BOOTSTRAP METHOD}

The bootstrap method is a method for finding factorized scattering theories,
i.e.\ solutions to the conditions described in Section \ref{sec.fst}. The boundary bootstrap method
is a straightforward extension of the bootstrap method to the boundary case.

A description of the bootstrap method is the following: let us assume that we already know a part of the particle spectrum that is
closed under scattering and  the action of CPT, and this action is also known or can be
guessed.  The corresponding internal space is denoted by
$V^0$. In this case we solve the Yang-Baxter equation for
$\tilde{S}_2(\Theta)|_{V^0\otimes V^0}$. In some cases it is done by
converting it into a linear differential equation by differentiation. 
If we look for theories with a given
symmetry the action of which on $V^0$ is known, then imposing it on
$\tilde{S}_2(\Theta)|_{V^0\otimes V^0}$ may also simplify the Yang-Baxter
equations. 
 The solution is obviously undetermined up to
an overall scalar function multiplier $F(\Theta)$ and it can contain further
undetermined parameters. Only analytic $F(\Theta)$ and $\tilde{S}_2(\Theta)|_{V^0\otimes
  V^0}$ with possible pole singularities are considered.   
$F(\Theta)$ and the solution is further restricted by
the real analyticity, unitarity and crossing symmetry conditions. After
imposing these conditions $F(\Theta)$ is constrained to satisfy the equations
$F(\Theta)=F(i\pi-\Theta)$, $F(\Theta)F(-\Theta)=1$ and
$F(\Theta^*)^*=F(-\Theta)$. The functions satisfying these equations are
called CDD (Castillejo-Dalitz-Dyson,  \cite{CDD}) factors. $F(\Theta)$ is usually chosen so that  $\tilde{S}_2(\Theta)|_{V^0\otimes
  V^0}$ have the minimum or nearly minimum number of poles in the physical
strip. After having determined  $\tilde{S}_2(\Theta)|_{V^0\otimes
  V^0}$, we check whether its poles lying in the physical strip can be related
to bound states and Coleman-Thun diagrams as described in 4), 7), 10) in Section \ref{sec.fst}, and
whether the bootstrap equations are satisfied. One often finds that not all simple
poles can be explained in terms of bound states and Coleman-Thun diagrams, and
in this case one extends $V^0$ by including new particles with masses
prescribed by the location of the unexplained poles and kinematics. This extension of
$V^0$ is denoted by $V^1$. The bootstrap equations allow to calculate the
still unknown parts of $\tilde{S}_2(\Theta)|_{V^1\otimes V^1}$, they are
linear algebraic equations for these parts. The action of possible symmetries
is also determined on the extension of $V^0$ by demanding that the intertwining
properties hold. The next step is to check again if the simple poles of
$\tilde{S}_2(\Theta)|_{V^1\otimes V^1}$ can be explained in terms of    
bound states and Coleman-Thun diagrams.
If this is not the case, then one
can continue by extending $V^1$ further into an internal space $V^2$,
calculate $\tilde{S}_2(\Theta)|_{V^2\otimes V^2}$ and so
on until no further extension is necessary. Finally it should be checked
whether all poles can be explained in terms of bound states and Coleman-Thun
diagrams. If they can, the procedure is
finished (`the bootstrap is closed').

In case of the sine-Gordon model, for instance, $V^0$ can be chosen to be a two-dimensional
space spanned by the soliton and anti-soliton, and the bootstrap is closed in
one step. The space $V^1$ is
spanned by the breathers together with the soliton and anti-soliton.   

The choice of the CDD factor has
an influence on the singularity structure of the S-matrix, so
the consistency of the axioms of Section \ref{sec.fst} imposes a
constraint on the CDD factor, although in a complicated way. 

It should also be noted that there is a large class of models (including the
sinh-Gordon model, for example) in which the
scattering is diagonal, and in this case the Yang-Baxter equation is
automatically satisfied and cannot be used to constrain the
S-matrix. However, the second part of the above method remains unaltered.    

The boundary  bootstrap method is the straightforward
adaptation of the above described steps to the boundary factorized scattering
theory. One usually  assumes that the ground state is unique and one starts
with $\mH_B^0=\mH_{B0}$. (The boundary analogue of $V^0, V^1,\dots $ is denoted by $\mH_B^0,
\mH_B^1, \dots $.) Then one determines the ground state reflection matrix
$\tilde{R}_1(\Theta)|_{V\otimes \mH_B^0}$ by the boundary Yang-Baxter equations,
unitarity, crossing symmetry and real analyticity conditions and by the second
boundary bootstrap equations. 
The second boundary bootstrap equation is a linear algebraic equation for the
reflection matrix of bulk bound states on the the ground state boundary, so if
the ground state reflection matrices of certain particles is already known, then the
second boundary bootstrap equation generally allows one to compute the ground
state reflection matrix of their bound states relatively easily.

If we multiply a reflection matrix which satisfies the Yang-Baxter equation by  an overall scalar factor
$F(\Theta)$, then the result  also satisfies the
Yang-Baxter equation. If the unitarity, crossing symmetry and real
analyticity conditions are also taken into consideration, then $F(\Theta)$ is constrained exactly
in the
same way as in the bulk case described above.  This $F(\Theta)$ is also
referred to as a CDD factor.

Having obtained the ground state reflection matrix one considers the poles of this reflection
matrix and determines $\mH_B^1 < \mH_B^2< \dots $ and
$\tilde{R}_1(\Theta)|_{V\otimes \mH_B^0},
 \tilde{R}_1(\Theta)|_{V\otimes \mH_B^1}, \dots $ until the bootstrap is
 closed. In summary, in the boundary case all the excited boundary states are
 obtained by bootstrap form the ground state. 

We remark that  to perform the complete bootstrap procedure as described
above can be rather laborious. In some cases the
equations are best handled by a suitable computer algebra program. It is not uncommon in
the literature that only certain parts, e.g.\ the solution of the Yang-Baxter
equation or the verification of the pole structure and the actual bootstrap
are considered, especially in the boundary case. Often the
check of the correspondence between poles and Coleman-Thun diagrams is not
performed completely.
 
 The boundary bootstrap is usually harder than the
bulk bootstrap, partially because it is built on the bulk part, 
partially because the equations themselves are larger and more difficult to
handle, the structure of boundary states is more complex than the
structure of bulk states, and much more Coleman-Thun diagrams exist than in
the bulk. 

It should be noted that if a solution to the axioms of factorized scattering
theory has been found, then it is usually a further problem to link this
solution to a model defined by a Lagrangian function, for instance, or to make
sure that the solution is really right for the model that one investigates.
In
some cases the problem is only to find the mapping between the parameters of
the Lagrangian function and the parameters of the solution. 
This matching of  S matrices with field theoretical models is usually done by
various other methods of quantum field theory. 
\vspace{4mm}
\begin{center}
-----------------------
\end{center}

\vspace{4mm}

In the rest of this chapter we shall generally omit the tilde and subscripts 
2 and 1 from $\tilde{S}_2(\Theta)$ and
$\tilde{R}_1(\Theta)$ and write  ${S}(\Theta)$ and
${R}(\Theta)$,  and we shall not use the other (e.g. hatted) versions of the
two-particle S-matrix. Similar change in the notation applies to the fusion
and decay tensors.

\section{The supersymmetry algebra in 1+1 dimensions }
\label{sec.sa}
\markright{\thesection.\ \ THE SUPERSYMMETRY ALGEBRA IN $1+1$ DIMENSIONS}

The supersymmetry algebra $\mathcal{A}$ is an associative algebra over $\RR$, generated by $Q$, $\bar{Q}$, $\hat{Z}$, $H$, $P$, $N$,
$\Gamma$, $I$.   $Q$ and $\bar{Q}$ are called supercharges, $\hat{Z}$ is the
supersymmetric central charge, $\Gamma$ is the
fermionic parity operator, $H$ and $P$ are the time and space translation
generators, $N$ is the boost generator and $I$ is the unit element of the algebra. These
generators satisfy the following relations: 
\begin{align}
& \{ \Gamma,\Gamma \} =2I && \{ Q,\bar{Q} \}=2\hat{Z} \notag \\
& \{ \Gamma, Q \} =0 && \{ \Gamma, \bar{Q} \} =0 \notag \\
& \{ Q, Q \}=2(H+P) && \{ \bar{Q}, \bar{Q} \} =2(H-P) \notag \\
& [N,Q]=\frac{1}{2} Q && [N,\bar{Q} ]=-\frac{1}{2} \bar{Q} \notag \\
& [N,\Gamma]=0 && \notag \\
& [N,H+P]=H+P && [N,H-P]=-(H-P)\ .
\end{align}

The supersymmetric central charge commutes with all elements of the algebra. 
$\mathcal{A}$ admits a $\ZZ_2$-grading, generators of grade $0$ are $H$, $P$,
$N$, $\hat{Z}$, $I$, generators of grade $1$ are $Q$, $\bar{Q}$, $\Gamma$.  

One can also introduce the boosts: $B(\varphi)=e^{\varphi N}$.

The co-product $\Delta$ used to define the action of $\mathcal{A}$ on
multi-particle states is given by 
\begin{align}
& \Delta(Q)=Q\otimes I+\Gamma\otimes Q && \Delta(\bar{Q})=\bar{Q}\otimes
I+\Gamma \otimes \bar{Q} \notag \\
& \Delta(\Gamma)=\Gamma\otimes \Gamma && \Delta(\hat{Z})=\hat{Z}\otimes
I+I\otimes\hat{Z} \notag \\
& \Delta(H)=H\otimes I+I\otimes H && \Delta(P)=P\otimes I+I\otimes P \notag \\
& \Delta(N)=N\otimes I+I\otimes N\ . &&
\end{align}
$\Delta$ satisfies the co-associativity property.

We call the algebra obtained by the omission of $N$ the internal part of
$\mathcal{A}$. We denote this internal part by $\tilde{\mathcal{A}}$. 
It has the property that
$B(\varphi)\tilde{\mathcal{A}}B(-\varphi)=\tilde{\mathcal{A}}$ and
$\Delta(\tilde{\mathcal{A}}) \subset \tilde{\mathcal{A}} \otimes
\tilde{\mathcal{A}}$. The latter property implies that the restriction of
$\Delta$ to $\tilde{\mathcal{A}}$ will be a (co-associative) co-product for $\tilde{\mathcal{A}}$. 

The one-particle representations of $\mathcal{A}$ generally take the form of
induced representations. A representation $D$ of $\mathcal{A}$ that is induced
from a representation of $\tilde{\mathcal{A}}$ is characterized by the
following properties:
the representation space $H$ can be written as 
\begin{equation}
H=\int d\Theta\ W(\Theta)\ ,
\end{equation}
where the space $W(\Theta)$ at any fixed $\Theta$ is invariant with respect to
$\tilde{\mathcal{A}}$ and contains states with rapidity $\Theta$. The notation
$W(D(\Theta))$, $W_{D(\Theta)}$, $W_D(\Theta)$ are also used instead of
$W(\Theta)$. The  representation of $\tilde{\mathcal{A}}$ on $W(\Theta)$ is
denoted by $D(\Theta)$.
We have
\begin{equation}
D(B(\varphi))D(\Theta)(\tilde{A})D(B(-\varphi))=D(\Theta+\varphi)(B(\varphi)\tilde{A}B(-\varphi))
\end{equation}
for all $\tilde{A} \in \tilde{\mathcal{A}}$, this equation describes the relation
between the representations  $D(\Theta)$ at various values of $\Theta$.

We can introduce the linear isomorphisms $i_\Theta:W(\Theta)\to W,\
a(\Theta) \mapsto a$, where the space $W$ is called the internal space.
 The
notation $W_D$ or $W(D)$ is also used instead of $W$.
 The equation 
$D(B(\varphi))a(\Theta)=a(\Theta+\varphi)$ is satisfied. For any fixed
$\Theta$ a representation of 
$\tilde{\mathcal{A}}$ on $W$ can be defined in the following way: $\tilde{A}
\mapsto  i_{\Theta}D(\Theta)(\tilde{A})i_{\Theta}^{-1}$. We do not introduce
new notation for these representations, we denote them by $D(\Theta)$.

The product of two representations $D_1(\Theta_1)$ and $D_2(\Theta_2)$ of
$\tilde{\mathcal{A}}$ is obtained using the co-product
$\Delta|_{\tilde{\mathcal{A}}}$.  $D_1(\Theta_1)\times D_2(\Theta_2)$ is contained in $D_1 \times
D_2$.

\subsection{Representations of the  supersymmetry algebra}
\label{subsec.rsa}

We consider representations on Hilbert spaces.  The Hermitian adjoints of
the generators are
\begin{equation}
H^\dagger =H\quad P^\dagger =P\quad Q^\dagger =Q\quad \bar{Q}^\dagger
=\bar{Q}\quad \hat{Z}^\dagger =\hat{Z}\quad \Gamma^\dagger =\Gamma \quad
N^\dagger =-N\ .
\end{equation}

On a one-particle supersymmetric multiplet $\ket{a(\Theta)}$ the action of the
supersymmetry algebra takes the following general form:
\begin{align}
& Q\ket{a(\Theta)}=\sqrt{m}e^{\Theta/2}q\ket{a(\Theta)}\\
& \bar{Q}\ket{a(\Theta)}=\sqrt{m}e^{-\Theta/2}\bar{q}\ket{a(\Theta)}\\
& B(\varphi)\ket{a(\Theta)}=e^{\varphi
  N}\ket{a(\Theta)}=\ket{a(\varphi+\Theta)}\ ,
\end{align}
where $m$ is the mass of the multiplet and $q$ and $\bar{q}$
are $\Theta$-independent matrices which act on the states of the
supermultiplet and satisfy
$$
q^{2}=1,\qquad\bar{q}^{2}=1,\qquad\{ q,\bar{q}\}=2Z\ ,$$
where $Z=\frac{1}{m}\hat{Z}$ on the multiplet. The action of
$\Gamma$ is independent of $\Theta$ and 
$$
\{\Gamma,q\}=\{\Gamma,\bar{q}\}=0\ .
$$

The boson-fermion representation $P_m$ of mass $m$ is defined by 

\begin{equation}
\label{eq.bofer}
q=\left(\begin{array}{cc}
0 & \epsilon\\
\epsilon^{*} & 0\end{array}\right),\qquad\bar{q}=\left(\begin{array}{cc}
0 & \epsilon^{*}\\
\epsilon & 0\end{array}\right),\qquad\Gamma=\left(\begin{array}{cc}
1 & 0\\
0 & -1\end{array}\right)
\end{equation}
in the basis $\{ \phi(\Theta),\psi(\Theta) \}$,   where $\epsilon=\exp(i\pi/4)$. 
The basis vectors $\phi(\Theta)$ and $\psi(\Theta)$ correspond to bosons and
fermions, respectively. The central charge is zero in this
representation. The action of $CPT$ is $CPT\phi(\Theta)=\phi(\Theta)$, $CPT\psi(\Theta)=-\psi(\Theta)$. The boson-fermion
representation will also be called particle representation and the
boson-fermion states will also be called particle states. The word particle is
used in the general sense as well, not referring to any particular representation.

Another representation $\bar{P}_m$ is obtained if we multiply $\Gamma$ in (\ref{eq.bofer}) by $-1$. We
call it pseudo-boson-fermion representation. 

The kink representation $K_m$ of mass $m$ is given by 
\begin{alignat}{2}
\label{eq.qk}
& q=\left(\begin{array}{cccc}
0 & i & 0 & 0\\
-i & 0 & 0 & 0\\
0 & 0 & 1 & 0\\
0 & 0 & 0 & -1\end{array}\right),
&\qquad & \bar{q}=\left(\begin{array}{cccc}
0 & i & 0 & 0\\
-i & 0 & 0 & 0\\
0 & 0 & -1 & 0\\
0 & 0 & 0 & 1\end{array}\right), \notag \\
& \Gamma=\left(\begin{array}{cccc}
0 & 1 & 0 & 0\\
1 & 0 & 0 & 0\\
0 & 0 & 0 & 1\\
0 & 0 & 1 & 0\end{array}\right),
&&
Z=\left(\begin{array}{cccc}
1 & 0 & 0 & 0\\
0 & 1 & 0 & 0\\
0 & 0 & -1 & 0\\
0 & 0 & 0 & -1\end{array}\right)
\end{alignat}

in the basis $\{ K_{0\frac{1}{2}}(\Theta), K_{1\frac{1}{2}}(\Theta),
K_{\frac{1}{2} 0}(\Theta), K_{\frac{1}{2} 1}(\Theta) \} $.

There are further  representations $\bar{K}_m$ called pseudo-kink
representations, which are obtained by multiplying $q$ and $\bar{q}$
by $-1$ in (\ref{eq.qk}), or by interchanging the labels
$0\leftrightarrow 1$. 

It is also true that there exists a continuous family of (inequivalent) representations
similar to the kink representation and interpolating between the kink and pseudo-kink representations.

Multi-kink states have to respect an adjacency
condition: in the physical\\ 
 $\ket{\dots K_{ab}(\Theta_{1})K_{cd}(\Theta_{2})\dots }$ state $b=c$
must hold. Multi-kink states not satisfying this condition are set equal to
zero (so $(\Gamma\otimes I)\ket{K_{\hlf1}(\Theta_{1})K_{1\hlf}(\Theta_{2})}=0$,
for example). This adjacency condition gives kinks a character rather different
from that of usual particles. 

$CPT$ acts as follows: $K_{ab} \leftrightarrow K_{ba}$.

The following decomposition equations hold for 
two-particle states:
\begin{align}
\label{eq.dec}
& P_{m_1}\times P_{m_2}\simeq\sum_{m}(
P_m+\bar{P}_m)\ , &&  \quad\bar{P}_{m_1}\times\bar{P}_{m_2}\simeq\sum_{m}
(P_m+\bar{P}_m)\ ,\\
\label{eq.dec2}
& P_{m_1}\times\bar{P}_{m_2}\simeq\sum_{m} (P_m+\bar{P}_m )\ . &&
\end{align}
In (\ref{eq.dec}) the first equation means, for example, that a two-particle state
transforms in the sum of a boson-fermion and a pseudo-boson-fermion representation (of
appropriate mass). 
One can write down the decomposition equations for the $P_{m_1}(\Theta_1)$
etc.\ representations of $\tilde{\mathcal{A}}$ as well:   
\begin{align}
\label{eq.dect}
& P_{m_1}(\Theta_1)\times P_{m_2}(\Theta_2)\simeq 
P_m(\Theta)+\bar{P}_m(\Theta)\ , &&  \quad\bar{P}_{m_1}(\Theta_1)\times\bar{P}_{m_2}(\Theta_2)\simeq
P_m(\Theta)+\bar{P}_m(\Theta)\ ,\\
\label{eq.dec2t}
& P_{m_1}(\Theta_1)\times\bar{P}_{m_2}(\Theta_2)\simeq P_m(\Theta))+\bar{P}_m(\Theta) \ , &&
\end{align}
where $m$ and $\Theta$ are determined by $m_1$, $m_2$, $\Theta_1$, $\Theta_2$
kinematically.

Multi-kink states containing
even number of kinks can be arranged in two sectors: the first sector
contains the states which have left and right label $\hlf$,
the second sector contains the states which have left and right 
labels $0$ or $1$. These two sectors will be called $\hlf$ and
$01$ sector. 

For two-kink states we have the decomposition equation 
\begin{equation}
\label{eq.twokink}
K_m\times K_m\simeq\sum_{m'}(
[P_{m'}]_{\hlf}+[P_{m'}+\bar{P}_{m'}]_{01})\ .
\end{equation}
The subscripts refer to the sectors in which the subspaces lie. 
The values of $m'$ may have  multiplicities higher than 1, this is not denoted
explicitly. 
Similar
equations as (\ref{eq.dect}), (\ref{eq.dec2t}) can also be written down.

It will become clear that it is reasonable to give the (reducible)
representation $[P_{m'}]_{\hlf}+[P_{m'}+\bar{P}_{m'}]_{01}$ appearing on the right-hand side of (\ref{eq.twokink})  a name of its own,
which will be `two-kink-representation' of mass $m'$, denoted by
$K^{(2)}_{m'}$. The states transforming in this representation will be called
two-kink-states.

The following combinations of two-kink states span the invariant subspaces
(see also \cite{Ahn,AK2}) :\\[4mm]
$[P]_{\hlf}$\ :
\begin{align}
& \label{eq.pp1}
\ket{\phi_{1}(\Theta; u)}=\ket{K_{\hlf
    0}(\Theta+iu)K_{0\hlf}(\Theta-iu)}+\ket{K_{\hlf
    1}(\Theta+iu)K_{1\hlf}(\Theta-iu)}
\\
&
\label{eq.pp2}
\ket{\psi_{1}(\Theta; u)}=i\sqrt{\frac{\cos(\frac{\pi}{4}-\frac{u}{2})}{\cos(\frac{\pi}{4}+\frac{u}{2})}}   (\ket{K_{\hlf
    0}(\Theta+iu)K_{0\hlf}(\Theta-iu)}-\ket{K_{\hlf 1}(\Theta+iu)K_{1\hlf}(\Theta-iu)})
\end{align}
$[P]_{01}$\ :
\begin{align}
& \label{eq.pp3}
\ket{\phi_{2}(\Theta; u)}=\ket{K_{0\hlf}(\Theta+iu)K_{\hlf 0}(\Theta-iu)}+\ket{K_{1\hlf}(\Theta+iu)K_{\hlf
    1}(\Theta-iu)}
\\
&
\label{eq.pp4}
\ket{\psi_{2}(\Theta; u)}= \sqrt{\frac{\cos(\frac{\pi}{4}-\frac{u}{2})}{\cos(\frac{\pi}{4}+\frac{u}{2})}} 
(\ket{K_{1\hlf}(\Theta+iu)K_{\hlf 0}(\Theta-iu)}-\ket{K_{0\hlf}(\Theta+iu)K_{\hlf 1}(\Theta-iu)})
\end{align}
$[\bar{P}]_{01}$\ :
\begin{align}
& \label{eq.pp5}
\ket{\bar{\phi}(\Theta; u)}=\ket{K_{0\hlf}(\Theta+iu)K_{\hlf 0}(\Theta-iu)}-\ket{K_{1\hlf}(\Theta+iu)K_{\hlf
    1}(\Theta-iu)}
\\
&
\label{eq.pp6}
\ket{\bar{\psi}(\Theta; u)}= \sqrt{\frac{\cos(\frac{\pi}{4}-\frac{u}{2})}{\cos(\frac{\pi}{4}+\frac{u}{2})}} 
(\ket{K_{1\hlf}(\Theta+iu)K_{\hlf 0}(\Theta-iu)}+\ket{K_{0\hlf}(\Theta+iu)K_{\hlf
    1}(\Theta-iu)}) 
\end{align}
$\ket{\phi_1(\Theta; u)}$, $\ket{\phi_2(\Theta; u)}$ are boson states with $\Gamma=1$,
$\ket{\psi_1(\Theta; u)}$ and $\ket{\psi_2(\Theta; u)}$ are fermion states with $\Gamma=-1$.
The two states in (\ref{eq.pp5}) and (\ref{eq.pp6}) span the
pseudo-boson-fermion 
representation. The value of $\Gamma$ on the pseudo-boson state
$\ket{\bar{\phi}(\Theta; u)}$ is $-1$, on $\ket{\bar{\psi}(\Theta; u)}$ it is
$1$.  In the
basis (\ref{eq.pp1}), (\ref{eq.pp2});  (\ref{eq.pp3}), (\ref{eq.pp4});
(\ref{eq.pp5}), (\ref{eq.pp6}) 
the matrices of $q$ and $\bar{q}$ take the form written down above.

$CPT$ acts on these states as follows:
\begin{align}
\label{eq.cpt1}
& CPT\ket{\phi_1(\Theta; u)}=\ket{\phi_1(\Theta; u)} &&
CPT\ket{\psi_1(\Theta; u)}=-\ket{\psi_1(\Theta; u)}\\
\label{eq.cpt2}
& CPT\ket{\phi_2(\Theta; u)}=\ket{\phi_2(\Theta; u)} &&
CPT\ket{\psi_2(\Theta; u)}=-\ket{\psi_2(\Theta; u)}\\
\label{eq.cpt3}
& CPT\ket{\bar{\phi}(\Theta; u)}=\ket{\bar{\phi}(\Theta; u)} &&
CPT\ket{\bar{\psi}(\Theta; u)}=\ket{\bar{\psi}(\Theta; u)}\ .
\end{align}

It should be noted that $\ket{\phi_1(\Theta; u)}$, $\ket{\psi_1(\Theta; u)}$,
$\ket{\phi_2(\Theta; u)}$,   $\ket{\psi_2(\Theta; u)}$,
$\ket{\bar{\phi}(\Theta; u)}$ and $\ket{\bar{\psi}(\Theta; u)}$ are states
which have zero norm and are not orthogonal. States transforming in the boson-fermion and
pseudo-boson-fermion representations that are orthogonal and have  nonzero scalar product with
themselves  are the following:
\begin{align}
& \ket{\phi_1(\Theta; u)}+  \ket{\phi_1(\Theta; -u)} &&
 \ket{\psi_1(\Theta; u)}+  \ket{\psi_1(\Theta; -u)}\\
& \ket{\phi_2(\Theta; u)}+  \ket{\phi_2(\Theta; -u)} &&
 \ket{\psi_2(\Theta; u)}+  \ket{\psi_2(\Theta; -u)}\\
& \ket{\bar{\phi}(\Theta; u)}+\ket{\bar{\phi}(\Theta; -u)} &&
\ket{\bar{\psi}(\Theta; u)}+\ket{\bar{\psi}(\Theta; -u)}
\end{align}
and 
\begin{align}
& \ket{\phi_1(\Theta; u)}-  \ket{\phi_1(\Theta; -u)} &&
 \ket{\psi_1(\Theta; u)}-  \ket{\psi_1(\Theta; -u)}\\
& \ket{\phi_2(\Theta; u)}-  \ket{\phi_2(\Theta; -u)} &&
 \ket{\psi_2(\Theta; u)}-  \ket{\psi_2(\Theta; -u)}\\
& \ket{\bar{\phi}(\Theta; u)}-\ket{\bar{\phi}(\Theta; -u)} &&
\ket{\bar{\psi}(\Theta; u)}-\ket{\bar{\psi}(\Theta; -u)}\ .
\end{align}

Although $K^{(2)}$ can be decomposed into a sum of irreducible representations, the products
of elements of $[P]_{\hlf}$, $[P]_{01}$ and $[\bar{P}]_{01}$ satisfy
certain relations because of the kink adjacency conditions. For example,
$\ket{\bar{\phi}\bar{\phi}}=\ket{\phi_{2}\phi_{2}}$.

The eight particle-kink states $\ket{p(\Theta_{1})K(\Theta_{2})}$,
where $p$ stands for a boson or fermion and $K$ stands for a kink,
transform in the direct sum of a kink and a pseudo-kink representation
if and only if
\begin{equation}
\label{eq.kpf}
\Theta_{1}-\Theta_{2}=i(\pi-u)\quad{\textrm{and}}\quad
m=2M\cos(u)=2M\sin(\pi/2-u)\ ,
\end{equation}
where $m$ is the mass of the particle and $M$ is the mass of the
kink.
This is precisely the condition that the total mass of the particle-kink
state is also $M$. If this condition is not satisfied, then the decomposition
of the representation in which the particle-kink states transform
contains the general kink-like representations mentioned above but does not contain the kink, pseudo-kink, particle or pseudo-particle
representations. The same can be stated for the kink-particle states
$\ket{K(\Theta_{1})p(\Theta_{2})}$. Similar statements can also be made if we
replace the particle representation by the $K^{(2)}$ representations.

The simplest vacuum representations are  one-dimensional, spanned by a state $\ket{\Omega}$.
In a one-dimensional representation all operators act as a multiplication by a
number, so  $Q$, $\bar{Q}$, $\hat{Z}$, $H$, $P$ have to be represented by zero. $\Gamma$ can
be either $1$ or $-1$. It is reasonable to require that a ground state
representation $D_{\Omega}$ have the property
\begin{equation}
\label{eq.egyseg}
D_{\Omega}\times D\simeq D
\end{equation}
for any one-particle representation $D$. This requirement eliminates the case
$D_{\Omega}(\Gamma)=-1$. Different values for $N$ yield inequivalent one-dimensional
representations, all of which satisfy (\ref{eq.egyseg}), nevertheless the scattering data
does not depend on the eigenvalue of $N$ on $\ket{\Omega}$, so it is chosen to
be 0. The choice $D_{\Omega}(\Gamma)=-1$ would also leave the  scattering data unaltered. 
$CPT$ acts in the following way: $CPT\ket{\Omega}=\ket{\Omega}$.

Another natural vacuum representation $D_{01\hlf}$ is 3-dimensional, the
representation space is spanned by the vectors
$\ket{0}$, $\ket{1}$, $\ket{\frac{1}{2}}$. The  generators in this
representation are
$Q=\bar{Q}=\hat{Z}=H=P=N=0$,
$$\Gamma=
\left(\begin{array}{ccc}
0 & 1 & 0 \\
1 & 0 & 0 \\
0 & 0 & 1 \end{array}\right)\ .
$$
The states in this representation are subject to the the kink adjacency
conditions. $CPT$ acts in the following way: $CPT\ket{0}=\ket{0}$,
$CPT\ket{1}=\ket{1}$, $CPT\ket{\hlf}=\ket{\hlf}$.

Taking into consideration the  adjacency conditions we have 
$$D_{01\hlf}\times K_m\simeq K_m\times D_{01\hlf}\simeq K_m\ ,$$
and the same equation holds for  $[P_m]_\hlf$, $[P_m]_{01}$,
$[\bar{P}_m]_{01}$ and  $K^{(2)}_m$, and also if a rapidity is specified and
representations of $\tilde{\mathcal{A}}$ are considered. 
$D_{01\hlf}$ is not used together with the $P_m$ representations.

The representation $D_{01\hlf}$ appears in the supersymmetric sine-Gordon
model, for example (see  \cite{BDPTW}). It should be noted, however, that the  vacuum
 does not appear to play a role in scattering theory.

The states 
\begin{eqnarray}
V_0(\Theta) & = &\ket{K_{0\hlf}(\Theta+\frac{i\pi}{2})K_{\hlf 0}(\Theta-\frac{i\pi}{2})}\label{eq.null.1}\\
V_1(\Theta) & = &\ket{K_{1\hlf}(\Theta+\frac{i\pi}{2})K_{\hlf 1}(\Theta-\frac{i\pi}{2})}\label{eq.null.2}\\
V_\hlf (\Theta) & = &\ket{K_{\hlf 0}(\Theta+\frac{i\pi}{2})K_{0 \hlf}(\Theta-\frac{i\pi}{2})}+\ket{K_{\hlf
  1}(\Theta+\frac{i\pi}{2})K_{1 \hlf}(\Theta-\frac{i\pi}{2})}\label{eq.null.3}
\end{eqnarray} 
transform in a representation that differs from $D_{01\hlf}$ only in the
  action of the boosts. 
 $Q$ and $\bar{Q}$ are nilpotent on the states (\ref{eq.pp1})-(\ref{eq.pp6}) at
  $u=\pi/2$. The kernel and image spaces of $Q$
  and $\bar{Q}$ are both spanned by the boson and pseudo-boson states, or equivalently
  by $V_0$, $V_1$, $V_\hlf$. 

The adjacency condition for a multi-particle state containing both
boson-fermions and kinks is the following: the usual condition applies for
neighbouring kinks, and if in a state $\ket{\dots K_{ab}p\dots p\dots pK_{cd}\dots }$
where $p$ stands for boson-fermions there are only boson-fermions between
$K_{ab}$ and $K_{cd}$, then either $b=c$ or $|b-c|=1$.  

We remark that $P_m$ and $\bar{P}_m$  as well as  $K_m$ and $\bar{K}_m$ are
equivalent as ray representations, and so are all the one-dimensional (vacuum)
representations.

As we mentioned in the Introduction, we shall consider the boson-fermion ($P_m$)
and  kink ($K_m$) representations as representations in which bulk particle
multiplets of supersymmetric theories can transform.

\section{The boundary supersymmetry algebra in 1+1 dimensions}
\label{sec.sab}
\markright{\thesection.\ \ THE BOUNDARY SUPERSYMMETRY ALGEBRA IN $1+1$ D}

The boundary supersymmetry algebra $\mathcal{A}_B$ is an associative algebra
over $\RR$. It is generated by a
boundary supercharge $Q_B$, a boundary central charge $Z_B$, the time
translation generator $H_B$ and a unit element $I_B$. $\mathcal{A}_B$ is
commutative and the following relation is satisfied:
\begin{equation}
Q_B^2=2H_B+2Z_B\ .
\end{equation}

There are essentially two different co-products
$\Delta_B^{\pm}:\ \mathcal{A}_B \to \mathcal{A}\otimes \mathcal{A}_B$:
\begin{align}
\Delta_B^+(I_B) &=  I\otimes I_B \\
\Delta_B^+(H_B) &=  H\otimes I_B+I\otimes H_B\\
\Delta_B^+(Z_B) &=  \hat{Z}\otimes I_B + I\otimes Z_B\\
\Delta_B^+(Q_B) &=  (Q+\bar{Q})\otimes I_B +\Gamma \otimes Q_B
\end{align}
and
\begin{align}
\Delta_B^-(I_B) &=  I\otimes I_B\\ 
\Delta_B^-(H_B) &=  H\otimes I_B+I\otimes H_B \\
\Delta_B^-(Z_B) &=  -\hat{Z}\otimes I_B +I\otimes Z_B \\ 
\Delta_B^-(Q_B) &=  (Q-\bar{Q})\otimes I_B +\Gamma \otimes Q_B\ .
\end{align}
They both satisfy the co-associativity property.

$\Delta_B^+$ and $\Delta_B^-$  can be related by an automorphism $j$ of $\mathcal{A}$
that has the property $j^2=Id$:
\begin{align}
& j(\bar{Q})=-\bar{Q} && j(Q)=Q && j(N)=N \\
& j(\hat{Z})=-\hat{Z} && j(\Gamma)=\Gamma &&  &&\\[1mm]
&(j\otimes Id)\circ \Delta_B^+ =\Delta_B^-\ . && &&
\end{align}

$\Delta_B^+$ and $\Delta_B^-$ can be written in the form (\ref{eq.bcp}) with
the monomorphisms $i^+, i^-:\mathcal{A}_B \to \mathcal{A}$:
\begin{align}
& i^+(Q_B)=Q+\bar{Q} &&
i^+(H_B)=H\\
& i^+(Z_B)=\hat{Z} &&
i^+(I_B)=I \\[3mm]
& i^-(Q_B)=Q-\bar{Q} &&
i^-(H_B)=H\\
&i^-(Z_B)=-\hat{Z} &&
i^-(I_B)=I\ .
\end{align}
$i^+$ and $i^-$ are also related by $j$: $j\circ i^+=i^-$.

To describe situations when the fermionic parity is also conserved,
$\mathcal{A}_B$ can be supplemented with the boundary fermionic
parity generator $\Gamma_B$. It satisfies the following relations:
\begin{equation}
\{ \Gamma_B, \Gamma_B\} =2I_B\ ,\qquad[\Gamma_B,Z_B]=0\ ,\qquad[\Gamma_B,H_B]=0\ ,
\label{bgamma}
\end{equation}
and also 
\begin{equation}
\{\Gamma_B,Q_B\}=2gI_B\ ,\label{bgamma2}
\end{equation}
where $g$ is a parameter of the algebra. The co-product of $\Gamma_B$ is
\begin{equation}
\Delta_{B}(\Gamma_B)=\Gamma\otimes\Gamma_B\ .
\label{copr3}
\end{equation}
The co-associativity property remains valid.

Finally, as $\Delta_B^{\pm}(\tilde{\mathcal{A}}) \subset \tilde{\mathcal{A}}
\otimes \mathcal{A}_B$, $\Delta_B^{\pm}$ can be used to multiply
representations of  $ \tilde{\mathcal{A}}$ and one representation of 
$ \mathcal{A}_B$.

\subsection{Representations of the boundary supersymmetry algebra}
\label{sec.repbsa}

As in Section \ref{subsec.rsa}, we consider representations on Hilbert spaces.  The Hermitian adjoints of
the generators are
\begin{equation}
H_B^\dagger =H_B\quad  Q_B^\dagger =Q_B \quad Z_B^\dagger =Z_B\quad (\Gamma_B^\dagger =\Gamma_B)\ .
\end{equation}

There are several different one-dimensional representations of the boundary supersymmetry
algebra which can serve as representations in which the ground states
of various models transform. Moreover, higher dimensional representations
may also occur in some models. The adjacency condition between the ground state and the nearest kink
is required to be satisfied, and   we shall consider
one-dimensional representations only,  following \cite{BPT}, this being the simplest choice in the
absence of other guiding information.  In this case, as explained in \cite{BPT},  the supersymmetric kink
label for the boundary must in general be  $\frac{1}{2}$. We shall not
consider the cases when the ground state is singlet with label 0 or 1.

The possible ground state representations form  a one-parameter family. It is
convenient to write this parameter in slightly different forms in the cases
when $\Delta_B^+$ and $\Delta_B^-$ is used. The representations are denoted by
$D_{\gamma}$ 
in the $(+)$ case and $D_{e\gamma}$ in the $(-)$ case, $\gamma$ and $e\gamma$
being the two forms of the parameter.
We shall also use the notation $B_{\hlf}$ for the ground state representation,
in this notation the parameters are not written explicitly. The
representation space of $B_{\hlf}$ will be denoted by $W(B_{\hlf})$.

The action of the boundary
supersymmetry generators on $\ket{B_{\hlf}}$ in the $(-)$ case is 
\begin{equation}
Q_B\ket{B_{\hlf}}=e\gamma\ket{B_{\hlf}}\ ,\qquad Z_B\ket{B_{\hlf}}=0\ ,\qquad
 e=\pm 1\ ,
\label{gr st rep}
\end{equation}
where $\gamma\in\RR,\ \gamma<0$
and $\ket{B_{\hlf}}$ is the  basis vector for the representation space.

In the $(+)$ case
$$
Q_B\ket{B_{\hlf}}=\gamma\ket{B_{\hlf}}\ ,\qquad Z_B\ket{B_{\hlf}}=0\ ,
$$
where  $\gamma\in\RR$.
$e\gamma$ or $\gamma$ is a parameter of the model to be described
and is expected to be expressible in terms of the parameters of the
Lagrangian density. The reason for writing the parameter in the form
$e\gamma$ in the $(-)$ case will become clear in \ref{sec.ssrmf}
when the ground state kink reflection amplitudes are discussed.

It is not necessary to set the eigenvalue of $Z_B$ equal to zero, 
however the scattering data is completely independent of the eigenvalue
of $Z_B$.  

If $\Gamma_B$ is also included in $\mathcal{A}_B$, then 
$$\Gamma_B\ket{B_{\hlf}}=\epsilon \ket{B_{\hlf}}\ ,\qquad \epsilon=\pm 1$$
and $g=\epsilon e\gamma$ in the $(-)$ case and $g=\epsilon \gamma$ in the
$(+)$ case.

\section{Supersymmetric factorized scattering}
\label{sec.ansatz00}
\markright{\thesection.\ \ SUPERSYMMETRIC FACTORIZED SCATTERING}

\subsection{Ansatz for the supersymmetric scattering theory}
\label{sec.ansatz}
\markright{\thesubsection.\ \ ANSATZ FOR SUPERSYMMETRIC SCATTERING}

A formalism for constructing $N=1$ supersymmetric factorizable bulk scattering
theory is the following:\\

1) It is assumed that a known factorized scattering theory to be
   supersymmetrized is given as
   described in Section \ref{sec.fst}. \\

2) 
The internal space of the supersymmetrized theory will be
\begin{equation}
\label{eq.tot}
V_{tot}=\oplus_k (W_k\otimes V_k)\ , 
\end{equation}
where the $W_k$-s are internal spaces of  representations $D_k$ of the supersymmetry
algebra as described in Section \ref{sec.sa}, the $V_k$-s constitute a decomposition of $V$:
\begin{equation}
\label{eq.fack}
V=\oplus_k V_k\ ,
\end{equation}
and each $V_k$ can be  decomposed further
into a sum of one-particle spaces appearing in (\ref{eq.resz}), i.e.\ a
specific value of $k$ is assigned to each particle. The masses of the particles
appearing in this decomposition of $V_k$ are the same for specific values of $k$,
i.e.\ the decomposition (\ref{eq.fack}) is a refinement of (\ref{eq.decm}), if
two particles have the same value of $k$, then they also have the same mass.  
The $V_k$ spaces 
must be invariant subspaces of $\CPT$. The masses
of the representations $D_k$ are the same as the masses belonging to $k$ in
the non-supersymmetric theory. 

The scalar product on $V_{tot}$ is defined in the standard
way.  The representation of the supersymmetry algebra on the one-particle
Hilbert space is also defined in a straightforward way, i.e.\ the
generators, except for the boost generator,
act trivially on the non-supersymmetric part.

The $W_k$-s will be referred to as supersymmetric parts.

${S}(\Theta)$ must be block `diagonal' with respect to the decomposition
(\ref{eq.fack}), i.e.\ 
\begin{equation}
\label{eq.svvvv}
{S}(\Theta)(V_{k_1}\otimes V_{k_2}) \subseteq  V_{k_2}\otimes V_{k_1}\ .
\end{equation}
The fusion tensors must also have a similar property
$$f_{m_1m_2}^{m_3}(V_{k_1}\otimes V_{k_2}) \subseteq V_{k_3}$$
where $k_3$ is uniquely determined by $k_1$ and $k_2$ and the fusion angle, which is
denoted in the following way:
$$k_1+k_2 \to k_3 \qquad (u)\ .$$
Thus we can consider the blocks $S_{k_1k_2}(\Theta)$, $f_{k_1k_2}^{k_3}$ and
$d_{k_3}^{k_2k_1}$, which behave essentially in the same way as the blocks
corresponding to definite masses. 

We note that $V_{tot}(\Theta)=\oplus_k(W_k(\Theta)\otimes V_k(\Theta))$.\\

3) The full supersymmetric S-matrix will be `diagonal' with respect to (\ref{eq.tot}) and
   the blocks are given by 
\begin{equation}
(S_{tot})_{k_1k_2}(\Theta)=(S_{SUSY})_{k_1k_2}(\Theta)\otimes
   S_{k_1k_2}(\Theta)\ .  
\end{equation}
The $(S_{SUSY})_{k_1k_2}(\Theta): W_{k_1} \otimes W_{k_2} \to   W_{k_2} \otimes W_{k_1}  $ factors are called supersymmetric factors. 

The fusion tensors also take  similar form:
\begin{equation}
(f_{tot})_{k_1k_2}^{k_3}=(f_{SUSY})_{k_1k_2}^{k_3}\otimes f_{k_1k_2}^{k_3}
\end{equation}
\begin{equation}
(d_{tot})_{k_3}^{k_2k_1}=(d_{SUSY})_{k_3}^{k_2k_1}\otimes d_{k_3}^{k_2k_1}\ .
\end{equation}
$(f_{SUSY})_{k_1k_2}^{k_3}: W_{k_1}\otimes W_{k_2} \to W_{k_3}$ and
$(d_{SUSY})_{k_3}^{k_2k_1}: W_{k_3}\to W_{k_2}\otimes W_{k_1}$ are also referred to as
supersymmetric factors.  The image space of $(f_{SUSY})_{k_1k_2}^{k_3}$ should
have zero intersection with  the kernel of $(d_{SUSY})_{k_3}^{k_2k_1}$.

The collection of the supersymmetry factors $(f_{SUSY})_{k_1k_2}^{k_3}$ and
$(d_{SUSY})_{k_3}^{k_2k_1}$ and\\   $(S_{SUSY})_{k_1k_2}(\Theta)$ 
should satisfy every possible bootstrap equation of the form
(\ref{eq.boot}). This ensures that the full S-matrix and fusion and decay
tensors will satisfy the bootstrap equations.
The supersymmetry factors should also be invariant with respect to the
supersymmetry algebra.

If $S_{k_1k_2}(\Theta)$ has a pole at $iu$ with a contribution corresponding to a direct channel
bound state, then there will be such a contribution for
$(S_{tot})_{k_1k_2}(\Theta)$ with residue $(S_{SUSY})_{k_1k_2}(iu)\otimes
d_{k_3}^{k_1k_2}f_{k_1k_2}^{k_3}$ and the following equation is satisfied:
\begin{equation}
\label{eq.susyfusion}
(S_{SUSY})_{k_1k_2}(iu)=(d_{SUSY})_{k_3}^{k_1k_2}(f_{SUSY})_{k_1k_2}^{k_3}\ ,
\end{equation}
which is the version of (\ref{eq.fusion}) for the supersymmetric factor of the
S-matrix. (\ref{eq.susyfusion}) is called fusion equation.\\

4) The supersymmetric factors $(S_{SUSY})_{k_1k_2}(\Theta)$ are complex analytic
   functions. They satisfy the Yang-Baxter equation described in Section \ref{sec.fst},
   the unitarity, real analyticity and crossing symmetry conditions.
   They do not have poles in the
   physical strip.

\subsection{Ansatz for the supersymmetric scattering theory in the presence of a boundary}
\label{sec.ansatzb}
\markright{\thesubsection.\ \ ANSATZ FOR BOUNDARY SUPERSYMMETRIC SCATTERING}

{\hspace{1cm} }

1) It is assumed that a known factorized scattering theory with  boundary to be
   supersymmetrized is given as
   described in Section \ref{sec.fstb}, and its underlying bulk factorized
   scattering theory is supersymmetrized in the way described in Section \ref{sec.ansatz}.\\

2) 
The space of boundary states of the supersymmetrized theory will be
\begin{equation}
\label{eq.totb}
\mH_{B,tot}=\oplus_{k^b} (W_{B,k^b}\otimes \mH_{B,k^b})\ , 
\end{equation}
where the $W_{B,k^b}$-s are representation spaces of  representations $D_{k^b}$ of
the boundary supersymmetry
algebra. 
The $\mH_{B,k^b}$-s constitute a decomposition of $\mH_B$:
\begin{equation}
\label{eq.fackb}
\mH_B=\oplus_{k^b} \mH_{B,k^b}\ ,
\end{equation}
which is a refinement of (\ref{eq.decmb}), i.e.\ if two states belong to the
$\mH_{B,k^b}$ with a fixed value of $k^b$, then they also belong to a common
energy eigenspace $\mH_{B,E}$.  
The energy of the representations $D_{k^b}$ is determined by the energy of the
states belonging to $k^b$ in the non-supersymmetric model.
The scalar product on $\mH_{B,tot}$ is defined in the standard way. 
 The representation of the supersymmetry algebra on  $W_{B,k^b}\otimes
 \mH_{B,k^b}$ 
is also defined in a straightforward way, i.e.\ the
generators
act trivially on the non-supersymmetric part.

The $W_{B,k^b}$-s are referred to as supersymmetric parts.

${R}(\Theta)$ must be diagonal with respect to (\ref{eq.fackb}) and
(\ref{eq.fack}):
$${R}(\Theta)(V_k\otimes \mH_{B,k^b})\subseteq V_k\otimes \mH_{B,k^b}$$
and the boundary fusion and decay tensors must also have a similar property:
$$g_{m,E_1}^{E_2}(V_k\otimes \mH_{B,k^b_1}) \subseteq \mH_{B,k^b_2}\ ,$$
where $k^b_2$ is uniquely determined by $k$ and $k^b_1$ and the fusion angle, which is denoted in the
following way:
$$k+k^b_1 \to k^b_2\qquad (u)\ .$$
Thus one can consider the blocks ${R}_{k,k^b_1}(\Theta)$, $g_{k,k^b_1}^{k^b_1}$,
$h_{k^b_2}^{k,k^b_1}$, which behave essentially in the same way as the blocks
corresponding to definite masses  and energies. \\

3) The full supersymmetric R-matrix will be diagonal with respect to
$k$ and $k^b$ and the blocks are 
\begin{equation}
(R_{tot})_{k,k^b_1}(\Theta)=(R_{SUSY})_{k,k^b_1}(\Theta)\otimes
R_{k,k^b_1}(\Theta)\ .
\end{equation}
The boundary fusion and decay tensors also take  similar form:
\begin{equation}
(g_{tot})_{k,k^b_1}^{k^b_2}=(g_{SUSY})_{k,k^b_1}^{k^b_2}\otimes g_{k,k^b_1}^{k^b_2}
\end{equation} 
\begin{equation}
(h_{tot})^{k,k^b_1}_{k^b_2}=(h_{SUSY})^{k,k^b_1}_{k^b_2}\otimes h^{k,k^b_1}_{k^b_2}
\end{equation} 
and
\begin{equation}
(G_{tot})_{k,k^b_1}^{k^b_1}=(G_{SUSY})_{k,k^b_1}^{k^b_1}\otimes G_{k,k^b_1}^{k^b_1}
\end{equation} 
\begin{equation}
(H_{tot})^{k,k^b_1}_{k^b_1}=(H_{SUSY})^{k,k^b_1}_{k^b_1}\otimes
H^{k,k^b_1}_{k^b_1}\ .
\end{equation} 
The factors 
$$(R_{SUSY})_{k,k^b_1}(\Theta):   W_k
\otimes W_{B,k^b_1} \to  W_k
\otimes W_{B,k^b_1}\ ,$$
$$(g_{SUSY})_{k,k^b_1}^{k^b_2}: W_k
\otimes W_{B,k^b_1} \to W_{B,k_2^b}\ ,\quad 
(h_{SUSY})^{k,k^b_1}_{k^b_2}: W_{B,k_2^b}\to W_k \otimes W_{B,k_1^b}\ ,$$
$$(G_{SUSY})_{k,k^b_1}^{k^b_1} : W_k
\otimes W_{B,k^b_1} \to W_{B,k_1^b}\ ,\quad   
(H_{SUSY})^{k,k^b_1}_{k^b_1} : W_{B,k_1^b}\to W_k \otimes W_{B,k_1^b}$$
are called supersymmetry factors.
The image spaces of $(g_{SUSY})_{k,k^b_1}^{k^b_2}$ and
$(G_{SUSY})_{k,k^b_1}^{k^b_1} $ should have zero intersection with the kernel of $(h_{SUSY})^{k,k^b_1}_{k^b_2}$ and $(H_{SUSY})^{k,k^b_1}_{k^b_1}$.
  
The collection of all boundary supersymmetry factors together with the
bulk supersymmetry factors  
should satisfy every possible bootstrap equation of the form (\ref{eq.bboot1})
and (\ref{eq.bboot2}).  This ensures that the full R-matrix and fusion and decay
tensors will satisfy the bootstrap equations.
The supersymmetry factors should also be invariant with respect to the
boundary supersymmetry algebra.

If $R_{k,k_1^b}(\Theta)$ has a pole at $iu$ with a contribution corresponding to a boundary 
bound state, then there will be such a contribution for
$(R_{tot})_{k,k^b_1}(\Theta)$ with residue $(R_{SUSY})_{k,k_1^b}(iu)\otimes 
h_{k_2^b}^{k,k_1^b}g_{k,k_1^b}^{k_2^b}$ and the following equation will be  satisfied:
\begin{equation}
\label{eq.susybfusion}
(R_{SUSY})_{k,k_1^b}(iu)=(h_{SUSY})_{k_2^b}^{k,k_1^b}(g_{SUSY})_{k,k_1^b}^{k_2^b}\
,
\end{equation}
which is the version of (\ref{eq.bfusion}) for the supersymmetric factor of the
R-matrix. (\ref{eq.susybfusion}) is called boundary fusion equation.

If $R_{k,k_1^b}(\Theta)$ has a pole at $iu$ with a contribution corresponding to the existence of a
bulk fusion tensor $f_{kk'}^{k}$, then there will be such a contribution
for $(R_{tot})_{k,k^b_1}(\Theta)$ with residue  $(R_{SUSY})_{k,k_1^b}(iu)
\otimes ([I\otimes
    {G}_{k',k_1^b}^{k_1^b}][{d}_{k}^{kk'}\otimes
    I]
)$.
We require that 
\begin{equation}
\label{eq.susybvilla}
(R_{SUSY})_{k,k_1^b}(iu)=[I\otimes
    {(G_{SUSY})}_{k',k_1^b}^{k_1^b}][{(d_{SUSY})}_{k}^{kk'}\otimes
    I]\ , 
\end{equation}
which is the version of (\ref{eq.contrib1}) for the supersymmetric factors, 
and 
\begin{equation}
[I\otimes
    {(G_{SUSY})}_{k',k_1^b}^{k_1^b}][{(d_{SUSY})}_{k}^{kk'}\otimes I]=[{(f_{SUSY})}^{k}_{kk'}\otimes I][I\otimes
    {(H_{SUSY})}^{k',k_1^b}_{k_1^b}]\ ,
\end{equation}
which is the version of (\ref{eq.ss1}) for the supersymmetric factors.

Similar assumption is made about the pole at $i\pi/2$ and 
the version of (\ref{eq.contrib2}) for the supersymmetry factors will be
\begin{equation}
\label{eq.susycontrib2}
({R}_{SUSY})_{k,k_1^b}(i\pi/2)={(H_{SUSY})}_{k_1^b}^{k,k_1^b}{(G_{SUSY})}^{k_1^b}_{k,k_1^b}\
.
\end{equation}

4) The supersymmetric factors $(R_{SUSY})_{k,k^b_1}(\Theta)$ are complex analytic
   functions. They satisfy the boundary Yang-Baxter equation,
   the unitarity, real analyticity and crossing symmetry conditions described
   in Section \ref{sec.fstb}.
 The supersymmetric factors $(R_{SUSY})_{k,k^b_1}(\Theta)$ should not have poles in
   the physical strip.

\subsection{Supersymmetric bootstrap}
\label{sec.susybootstrap}

The supersymmetric factors of the ansatz described in Sections
\ref{sec.ansatz}  and \ref{sec.ansatzb} can be calculated in a way that is very similar to
that described in Section \ref{sec.bootstrap}. The main difference is that the fusion angles are not determined by
poles but are taken from the non-supersymmetric theory, which is assumed to be
known. 

Supersymmetric fusion rules, which are related strictly to the supersymmetric 
factors only,
can be defined in analogy with the
non-supersymmetric fusion rules. 

It is an interesting problem 
to find all the possible supersymmetric factors
and supersymmetric fusion rules for a given set of representations for the
bulk part, 
regardless of particular non-supersymmetric theories, 
and to find all the possible supersymmetric factors of the ground state
reflection matrix for the given set of representations for the bulk particles
and for the ground state, and finally to find the supersymetric factors
(including the reflection matrix factors) for
higher level boundary states  for arbitrary boundary fusion angles, and to
describe the possible supersymmetric boundary fusion rules.  
Such
results can then be  applied  to
particular non-supersymmetric theories.

% ********************************** supers18.tex vege ********************************************

\subsection{Supersymmetric S-matrix factors}
\label{sec.sssf}
\markright{\thesubsection.\ \ SUPERSYMMETRIC S-MATRIX FACTORS}

The general solution of the Yang-Baxter equations that describes the
(supersymmetric factor of the) scattering of two boson-fermion supermultiplets
is 
\begin{multline}
\notag
S_{PP}^{[i,j]}(\Theta,m_i,m_j,\alpha^{[i,j]})=\\[2mm]
=G^{[i,j]}(\Theta)\left[\frac{1}{2i}(q_{1}-q_{2})(\bar{q}_{1}-\bar{q}_{2})+\alpha^{[i,j]}
  F(\Theta)[1-t(\Theta,m_{i},m_{j})q_{1}q_{2}][1+t(\Theta,m_{j},m_{i})\bar{q}_{1}\bar{q}_{2}]\right],
\end{multline}
where 
$$
t(\Theta,m_{i},m_{j})=\tanh\left(\frac{\Theta+\log(m_{i}/m_{j})}{4}\right),\qquad
F(\Theta)=\frac{m_{i}+m_{j}+2\sqrt{m_{i}m_{j}}\cosh(\Theta/2)}{2i\sinh(\Theta)},
$$
$$q_{1}=q\otimes I\ ,\qquad q_{2}=\Gamma\otimes q\ ,\qquad
\bar{q}_{1}=\bar{q}\otimes I\ ,\qquad
\bar{q}_{2}=\Gamma\otimes\bar{q}\ .$$ 
$m_{i}$ and $m_{j}$ are the
masses of the multiplets, and $\alpha^{[i,j]}$ is a real constant, which is
interpreted as the 
measure of the strength of Bose-Fermi mixing. $\alpha^{[i,j]}=0$ corresponds to trivial scattering. $i$ and $j$ are indices of the
type introduced in (\ref{eq.fack}).     $G^{[i,j]}(\Theta)$ is a
scalar function.      $S_{PP}^{[i,j]}(\Theta)/G^{[i,j]}(\Theta)$
can depend on the conserved quantities $i,j$ through $m_{i}$, $m_{j}$
and $\alpha^{[i,j]}$ only. It can be shown that the Yang-Baxter equation
implies 
that the particles in a theory can be divided into disjoint sets with
the property that any two particles in a set have the same nonzero
$\alpha$, and $\alpha=0$ for two particles from different sets.
To each particle in
a theory we associate a value of $\alpha$, which is the value that
occurs in the scattering of the particle with itself. For simplicity
we consider only theories which have only one such set and thus $\alpha$
is the same for any two-particle scattering (and the upper indeces of $\alpha$
can be omitted). The scalar function $G^{[i,j]}(\Theta)$
is determined by unitarity and crossing symmetry up to CDD factors.
It is important here that $i,j$ are invariant under charge conjugation.
It is also required that $S_{PP}^{[i,j]}(\Theta)$ should have minimal number
of poles and overall zeroes in the physical strip,
what fixes $G^{[i,j]}(\Theta)$ completely. An explicit expression for
$G^{[i,j]}(\Theta)$ can
be found in the Appendix. $G^{[i,j]}(\Theta)$
contains the parameters $u_{i}$, $u_{j}$ for which 
\begin{equation}
\label{eq.u}
0<Re(u_{i}),Re(u_{j})\leq\pi/2\ ,\qquad m_{i}=2M\sin(u_{i})\ ,\quad
m_{j}=2M\sin(u_{j})\ ,
\end{equation}
where $M=|1/(2\alpha)|$. Consequently, we can assign an angle $u$ to each particle. We shall consider only real values
of $u_{i}$ and $u_{j}$.

The supersymmetric  factor that describes the scattering of two kinks of
equal mass is
$$
S_{KK}(\Theta)=K(\Theta)[\cosh(\gamma\Theta)-\sinh(\gamma\Theta)q_{1}\bar{q}_{1}][\cosh(\Theta/4)-\sinh(\Theta/4)q_{1}q_{2}]\
,
$$
where $\gamma=(\log2)/2\pi i.$ This factor does not depend on any parameters. The scalar function $K(\Theta)$
is determined by unitarity and crossing symmetry and the condition
that $S_{KK}(\Theta)$ should have a minimal number of poles and zeroes in the
physical strip. An explicit expression for $K(\Theta)$ can be found in the Appendix
or \cite{HolMav}.  
There is no solution of the Yang-Baxter equation for the  scattering of kinks of different
mass, so all the kinks in a theory have to have the
same mass. 

The kink--boson-fermion S-matrix factors $S_{PK}(\Theta,\alpha,u_i)$ and  $S_{KP}(\Theta,\alpha,u_i)$ will be considered
later. They depend on the $\alpha$ parameter and on the $u_i$ angle parameter
of the boson-fermion representation.

The important common feature of these minimal S-matrix factors, including
$S_{PK}$ and $S_{KP}$,  is that they have no poles and overall zeroes in the physical
strip (although they can be degenerate at particular values of $\Theta$).

In the light of (\ref{eq.dec}), if it is decided that some particles and their bound
states transform in the boson-fermion representation of the supersymmetry
algebra, then the fusion equation (\ref{eq.susyfusion}) can be satisfied  only if $S_{PP}(iu)$
is a projection onto the appropriate subspace carrying the boson-fermion
representation. This is a nontrivial condition on $S_{PP}(iu)$, because
$S_{PP}(\Theta)$ is bijective (of rank four) for general general values of $\Theta$. The other possible way to
assure that only boson-fermion states (and no pseudo-boson-fermion states) are produced in the fusion is to quotient
out the unwanted states by hand. We shall consider only the first, more
natural possibility.

$S_{PP}^{[i,j]}(\Theta)$ is of rank two if $\Theta=iu_{ij}^{k}$\ ,
where 
\begin{equation}
u_{ij}^{k}\in\{ u_{i}+u_{j}\ ,\ \  \pi-u_{i}+u_{j}\ ,\ \ u_{i}+\pi-u_{j}\}\ .
\end{equation}
Only two of these values can be in the physical strip, and $S_{PP}^{[i,j]}(\Theta)$
is nondegenerate at other values of $\Theta$ in the physical strip.
The image space of $S_{PP}^{[i,j]}(iu_{ij}^{k})$ carries the particle
representation if and only if $\alpha<0$, i.e.\ if $\alpha=-1/(2M)$.
We remark that if $\alpha=1/(2M)$, then the image space carries the
pseudo-particle representation. The value of $u_{k}$ (which is the angle
parameter of the particle representation carried by the image space) is the following:
\begin{alignat}{2}
& u_{k}= u_{i}+u_{j} &\qquad &\textrm{if}\quad u_{ij}^{k}=u_{i}+u_{j}<\pi/2\ ,\label{angle 1}\\
& u_{k}=\pi-(u_{i}+u_{j})&\qquad & \textrm{if}\quad
u_{ij}^{k}=u_{i}+u_{j}\geq\pi/2\ ,\label{angle 2}\\
& u_{k}=u_{i}-u_{j} &\qquad &\textrm{if}\quad u_{ij}^{k}=\pi-u_{i}+u_{j}\ ,\label{angle 3}\\
& u_{k}=u_{j}-u_{i} &\qquad &\textrm{if}\quad u_{ij}^{k}=u_{i}+\pi-u_{j}\ .\label{angle 4}
\end{alignat}
The conditions above are sufficient for the existence of unique fusion and decay tensors
$f_{PP}^P(u_i,u_j,u_k,M)$, $d_P^{PP}(u_i,u_j,u_k,M)$  which satisfy the fusion
equation and have the required symmetry properties. Explicit expressions for
them can be found in \cite{HolMav,Sch}.

We turn to the case of the fusion of two (supersymmetric) kinks of
equal mass now. $S_{KK}$ is bijective (of rank six)
everywhere in the physical strip, so there is no natural degeneracy condition
on $S_{KK}(iu)$ and no constraint arises on the fusion
angle in this way, and unique fusion and decay tensors satisfying the fusion
equation and having the required symmetry properties exist. Consequently, in the light of (\ref{eq.twokink}),
if one insists that no pseudo-particles should be formed in kink fusion,
then one has to quotient out the unwanted states from the Hilbert-space
by hand. 
Even if states of the form
(\ref{eq.pp5}) and (\ref{eq.pp6}) are quotiented out, kink fusion produces
two types of particles corresponding to (\ref{eq.pp1})-(\ref{eq.pp4}),
i.e.\ to the $\hlf$ and $01$ sectors. The two types of particles
will be referred to as type $\hlf$ and type $01$ particles.

There are adjacency conditions for particles produced in kink fusion,
which follow from the adjacency conditions for kinks: type $\hlf$
and type $01$ particles cannot be adjacent in a multi-particle state,
so they cannot scatter on each other. There are also appropriate adjacency
conditions for kinks and particles. Bootstrap gives the result \cite{Ahn,HolMav}
that the two types of particles have the same S-matrix factor which is equal to $S_{PP}$. Consequently,
the two types of particles can be identified (which is the same as
quotienting out certain combinations). If this identification is made,
then only the following adjacency condition applies: if in a state
$\ket{\dots K_{ab}p\dots p\dots pK_{cd}\dots }$ (where $p$ stands for a particle
and the rapidities are suppressed) there are only particles (at least one)  between
$K_{ab}$ and $K_{cd}$, then either $b=c$, or $|b-c|=1$. For adjacent kinks
$K_{ab}$ and $K_{cd}$ the condition $b=c$ applies as before. 

The elimination of the pseudo-particle states  and the identification of $01$ and
$\hlf$ states as described above is usually done in the literature (e.g.\ \cite{HolMav,Ahn}),
despite of its  unnatural character. 
It is more natural to accept that the fusion of two kinks produces states that
transform in the two-kink-representations $K_m^{(2)}$, and recent numerical
calculations \cite{BDPTW} in finite volume for the supersymmetric sine-Gordon model also
seem to support this version. These calculations  suggest that the breathers
of the supersymmetric sine-Gordon model transform in the $K_m^{(2)}$
representations. To conform to the literature we shall use
the boson-fermion representations (i.e.\ the $P_m$-s). Switching to  two-kink-representation,
however, is in most cases straightforward.

$S_{KK}(\Theta)$ is bijective (of rank six) everywhere in the physical strip. However, it is
degenerate at $\Theta=\pm i\pi$, and at this point it projects onto the
subspace spanned by the states (\ref{eq.null.1})-(\ref{eq.null.3}).

The supersymmetry factors $S_{PK}(\Theta)$ and  $S_{KP}(\Theta)$  for the scattering of a particle with
$\alpha<0$ and a kink of mass $M=-1/(2\alpha)$ can be obtained from
$S_{KK}$ by bootstrap \cite{Ahn,HolMav} applied to the $kink+kink\rightarrow particle$
vertex. It turns out that $S_{PK}(\Theta)$ and  $S_{KP}(\Theta)$ are also minimal and have
neither poles nor overall zeroes in the physical strip.

It is expected that a kink is produced in the kink-particle fusion.
The transformation properties of the kink-particle states discussed
earlier show that in this case it is necessary that (\ref{eq.kpf})
is satisfied. We checked that $S_{PK}$ (and $S_{KP}$) is bijective (of rank eight)
everywhere in the physical strip, except when (\ref{eq.kpf}) is satisfied.
In the latter case it is a projection onto the four dimensional kink
subspace. The $kink+particle\rightarrow kink$ fusion is thus possible, 
and there are no restrictions other than (\ref{eq.kpf}).
The $kink+particle\rightarrow kink$ fusion is a crossed version of
$kink+kink\rightarrow particle$ fusion. The produced kink is of the
same mass as the incoming one, so the fusion angle is in the domain
$[\pi/2,\pi].$ The fusion tensor (regarded as a linear mapping) is a projection.

Finally, there are bootstrap equations for  $S_{PP}$,
$S_{KK}$, $S_{PK}$, $S_{KP}$ and 
$f_{PP}^P$, $d_P^{PP}$, $f_{KK}^P$, $d_{P}^{KK}$, $f_{KP}^K$, $d_{K}^{PK}$,
$f_{PK}^K$, $d_{K}^{KP}$ which were found to be satisfied \cite{HolMav,Sch}. The fusion of
two particles with $\alpha<0$ produces a particle with the same value of
$\alpha$. The fusion of two kinks of mass $M$ produces a particle with
$\alpha=-1/(2M)$, and if the fusion angle is $\rho$, then the angle parameter
$u$ of the produced particle is $u=\pi/2-\rho/2$. The fusion of a kink of mass
$M$ and a particle with $\alpha=-1/(2M)$ produces a kink of mass $M$. In the
diagrammatic representation there are essentially two types of vertices: the
kink-kink-particle and the three-particle vertices.

To a supersymmetric boson-fermion  multiplet
$W_P(\Theta)$  with  definite rapidity  and mass $m=2M \cos (\rho/2)$ we
assign the following (not ordered) set of rapidities:
$$L[W_P(\Theta)]=\{ \Theta-i\rho/2, \Theta +i\rho/2\}\ , $$  
where it is not required that $\Theta \pm i\rho /2$ be in the physical
strip. The elements of the set are the rapidities of those kink multiplets
which fuse into the boson-fermion multiplet $W_P(\Theta)$.  $L[W_P(\Theta)]$
and $M$ determines $W_P(\Theta)$ uniquely. The set $L[W_K(\Theta)]=\{ \Theta
\} $ is assigned to a kink multiplet $W_K(\Theta)$. In terms of these sets the
fusion rule of two boson-fermion multiplets takes the form  
\begin{equation}
\{\Theta_{1},\Theta_{2}\}+\{\Theta_{3},\Theta_{1}\pm
i\pi\}\rightarrow\{\Theta_{2},\Theta_{3}\}\ ,
\label{eq.fr1}
\end{equation}
 where $\Theta_{1},\Theta_{2},\Theta_{3}$
are appropriate complex rapidities. Similarly, the fusion of a kink
and a particle takes the form 
\begin{equation}
\{\Theta_{1}\}+\{\Theta_{2},\Theta_{1}\pm
i\pi\}\rightarrow\{\Theta_{2}\}\ .
\label{eq.fr2}
\end{equation}
A kink-kink fusion  takes
the form 
\begin{equation}
\{\Theta_{1}\}+\{\Theta_{2}\}\rightarrow\{\Theta_{1},\Theta_{2}\}\ .
\label{eq.fr3}
\end{equation}
In these fusions the set of rapidities
corresponding to the final state is obtained in the following way:
the disjoint union of the two sets of rapidities corresponding to
the fusing particles/kinks is formed and the pair of rapidities differing
by $\pm i\pi$ is deleted (if there is any such pair). It is important that we allow here
and further on that a set contains certain elements several times,
i.e.\ the elements of the sets we consider have multiplicity. Such sets denoted
by $L[\dots]$ will be used in the boundary case as well and they will be
called labeling sets.

The rules (\ref{eq.fr1})-(\ref{eq.fr3}) follow from the fact that $S_{KK}$ is
bijective (of rank 6) in the physical strip but is of rank 3 at $\Theta=i\pi$.
The bijectivity of the fusion tensor $f_{KK}^{K^{(2)}}$ together with the bootstrap
equations and the symmetry properties of  $f_{KK}^{K^{(2)}}$ implies that a
multiplet of
two-kink-states $W_{K^{(2)}}(\Theta)$ has the same scattering and transformation
properties as the multiplet of two-kink states $(\hat{f}_{KK}^{K^{(2)}})^{-1}(W_{K^{(2)}}(\Theta))$. 
Similar statement applies if we use the boson-fermion states. 
The
kinks can be considered as elementary states, whereas boson-fermions or
two-kink-states as composite states (mentioned at the end of Section
\ref{sec.fst}) constituted by two kinks with fixed rapidity difference.

In summary, the supersymmetric factors are characterized by a single mass
parameter $M$ which is the common mass of the kinks, and each
particle multiplet has a mass $m\le 2M$ and a parameter $0<u\le \pi/2$ 
so that $m=2M\sin(u)$ (see (\ref{eq.u})).
The fusion rules satisfy the constraint 
$u_{ij}^{k}\in\{ u_{i}+u_{j},\pi-u_{i}+u_{j},u_{i}+\pi-u_{j}\}$
for the fusion angle of a $particle_{i}+particle_{j}\rightarrow particle_{k}$
fusion (this constraint is equivalent to (\ref{eq.fr1})), $u_{k}=\pi/2-u_{ij}^{k}/2$ for a $kink_{i}+kink_{j}\rightarrow particle_{k}$
fusion (which is equivalent to (\ref{eq.fr3})), and $u_{ij}^{k}=\pi/2+u_{i}$ for a $particle_{i}+kink_{j}\rightarrow kink_{k}$
fusion (this constraint is equivalent to (\ref{eq.fr2})).
These constraints are nontrivial, because  the fusion angle is not restricted
kinematically in general by the masses of the fusing particles.
The supersymmetric factors are
$S_{PP}(\Theta,u_i,u_j,M)$, $S_{KK}(\Theta)$, $S_{PK}(\Theta,M,u_i)$, $S_{KP}(\Theta,M,u_i)$ and
$f_{PP}^P(u_i,u_j,u_k,M)$, $d_P^{PP}(u_i,u_j,u_k,M)$, $f_{KK}^P(u_k)$, $d_{P}^{KK}(u_k)$, $f_{KP}^K(u_j)$, $d_{K}^{PK}(u_j)$,
$f_{PK}^K(u_i)$, $d_{K}^{KP}(u_i)$. (Assuming that the boson-fermion
representation is used, not the  two-kink-representation.)

\subsection{Supersymmetric reflection matrix factors}
\label{sec.ssrmf}
\markright{\thesubsection.\ \ SUPERSYMMETRIC REFLECTION MATRIX FACTORS}

We consider one-particle kink reflection matrix factors $R_{K}$  on ground state boundary
first. As the left and right RSOS labels should be conserved, $\{ R_{K}\}_{K_{1\hlf}}^{K_{0\hlf}}(\Theta)=\{ R_{K}\}_{K_{0\hlf}}^{K_{1\hlf}}(\Theta)=0$
must hold, i.e.\ $R_{K}$ is diagonal. The general solution of the boundary Yang-Baxter equation,
unitarity condition and crossing equation without imposing supersymmetry
is \cite{AK1,Chim} 
\begin{equation}
\notag
\{ R_{K}\}_{K_{0\hlf}}^{K_{0\hlf}}(\Theta)=(1+A\sinh(\Theta/2))M(\Theta)\ ,
\end{equation}
\begin{equation}
\notag
\{ R_{K}\}_{K_{1\hlf}}^{K_{1\hlf}}(\Theta)=(1-A\sinh(\Theta/2))M(\Theta)\ ,
\end{equation}
where $M(\Theta)$ is restricted by unitarity and crossing symmetry.
After imposing the boundary supersymmetry condition one finds that
in the $(+)$ case \cite{BPT,Nep2}
\begin{equation}
\notag
\{ R_{K}^{(+)}\}_{K_{0\hlf}}^{K_{0\hlf}}(\Theta)=\{
R_{K}^{(+)}\}_{K_{1\hlf}}^{K_{1\hlf}}(\Theta)=2^{-\Theta/(\pi
  i)}P(\Theta),\ \qquad A=0\ .
\end{equation}
(A formula for $P(\Theta)$ can be found in the Appendix.) In the $(-)$ case there
are two distinct solutions for a given $\gamma$ corresponding to the
two values of the sign $e$:
\begin{equation}
\{ R_{K,e}^{(-)}\}_{K_{0\hlf}}^{K_{0\hlf}}(\Theta)
=(\cos\frac{\xi}{2}+ei\sinh\frac{\Theta}{2})K(\Theta-i\xi)K(i\pi-\Theta-i\xi)2^{-\Theta/(\pi
  i)}P(\Theta)\ ,\label{refl1}
\end{equation}
\begin{equation}
\{
R_{K,e}^{(-)}\}_{K_{1\hlf}}^{K_{1\hlf}}(\Theta)
=(\cos\frac{\xi}{2}-ei\sinh\frac{\Theta}{2})K(\Theta-i\xi)K(i\pi-\Theta-i\xi)2^{-\Theta/(\pi
  i)}P(\Theta)\ ,\label{refl2}
\end{equation}
where $\gamma=-2\sqrt{M}\cos\frac{\xi}{2}$ and $0\leq\xi\leq\pi$,
$M$ is the kink mass. 
(\ref{refl1})
and (\ref{refl2}) are invariant under $\xi\leftrightarrow-\xi$.
These reflection amplitudes are minimal, they have no poles and zeroes
in the physical strip. The sign $e$ seems to have a correspondence here with the
$0$ and $1$ RSOS vacua. It should be noted that $R_{K}^{(+)}(\Theta)$
is independent of $\gamma$. Furthermore, as symmetry under
${\Gamma}_B$ requires $R_{K_{0\hlf}}^{K_{0\hlf}}(\Theta)=R_{K_{1\hlf}}^{K_{1\hlf}}(\Theta)$,
the $R_{K}^{(+)}(\Theta)$ is automatically ${\Gamma}_B$-symmetric,
although this is not required a priori. On the other hand, $R_{K,e}^{(-)}(\Theta)$
are not ${\Gamma}_B$-symmetric. However, $\{ R_{K,e}^{(-)}\}_{K_{0\hlf}}^{K_{0\hlf}}(\Theta)=-\{ R_{K,e}^{(-)}\}_{K_{1\hlf}}^{K_{1\hlf}}(\Theta)$
if $\gamma=0$ ($A\rightarrow\infty$). We also remark that $\{ R_{K}\}_{K_{1\hlf}}^{K_{1\hlf}}(\Theta)/\{ R_{K}\}_{K_{0\hlf}}^{K_{0\hlf}}(\Theta)$
is determined by the supersymmetry condition, i.e.\ if we impose the
condition of invariance under supersymmetry, then we do not need to
solve the Yang-Baxter equation.

We determined the general solution of the Yang-Baxter equation for
the boundary supersymmetric particle reflection matrix factor (on the ground state
boundary). We imposed the supersymmetry condition first. 
The
resulting forms of the reflection amplitude in the $(+)$ and $(-)$
cases are  
\begin{multline}
R_{P}^{(+)}(\Theta)=
Z^{(+)}(\Theta)\frac{1}{\sqrt{m}}  \\
\times  \left(\begin{array}{cc}
(X^{(+)}(\Theta)+e\gamma Y^{(+)}(\Theta))c(\frac{\Theta}{2}-\frac{i\pi}{4}) & \sqrt{m}Y^{(+)}(\Theta)c(\Theta)\\
\sqrt{m}Y^{(+)}(\Theta)c(\Theta) & (X^{(+)}(\Theta)-e\gamma
Y^{(+)}(\Theta))c(\frac{\Theta}{2}+\frac{i\pi}{4})\end{array}\right),
\end{multline}
\begin{multline}
R_{P}^{(-)}(\Theta)=
Z^{(-)}(\Theta)\frac{1}{\sqrt{m}}  \\
\times   \left(\begin{array}{cc}
(X^{(-)}(\Theta)+e\gamma Y^{(-)}(\Theta))c(\frac{\Theta}{2}+\frac{i\pi}{4}) & i\sqrt{m}Y^{(-)}(\Theta)c(\Theta)\\
-i\sqrt{m}Y^{(-)}(\Theta)c(\Theta) & (X^{(-)}(\Theta)-e\gamma
Y^{(-)}(\Theta))c(\frac{\Theta}{2}-\frac{i\pi}{4})\end{array}\right),
\end{multline}
where $c$ stands for $\cosh$ and $X$, $Y$ and $Z$ are functions
not determined by supersymmetry. Now two cases can be distinguished
depending on whether ${\Gamma}_B$ is a symmetry or not: in the
first case, which is the ${\Gamma}_B$-symmetric case, $Y(\Theta)\equiv0$,
$X(\Theta)$ can be absorbed into the prefactor, and the structure
of the reflection amplitude is completely determined and does not
contain free parameters:\[
R_{P1}^{(\pm)}(\Theta)=\frac{1}{\sqrt{m}}ZX^{(\pm)}(\Theta)\left(\begin{array}{cc}
\cosh(\frac{\Theta}{2}\mp\frac{i\pi}{4}) & 0\\
0 & \cosh(\frac{\Theta}{2}\pm\frac{i\pi}{4})\end{array}\right).\]
 This case is discussed in \cite{SM}, the explicit form of $ZX^{(\pm)}$
can be found in the Appendix, see also \cite{BPT,SM,AK2}.
We checked that the boundary Yang-Baxter equation for incoming
particles of arbitrary masses is satisfied by this reflection amplitude.
$R_{P1}^{(\pm)}(\Theta)$ can also be obtained from $R_{K}^{(+)}(\Theta)$
and $R_{K}^{(-)}(\Theta)$ at $\gamma=0$ by bootstrap \cite{BPT,AK2}.

In the second case, when ${\Gamma}_B$ is not conserved, $Y(\Theta)$
is not identically zero, and it can be absorbed into the prefactor,
so one free function $y^{(\pm)}(\Theta)=X^{(\pm)}(\Theta)/Y^{(\pm)}(\Theta)$
remains in the reflection amplitude, which is to be determined by
the boundary Yang-Baxter equation. To obtain $y^{(\pm)}(\Theta)$
we solved the Yang-Baxter equation first in the case when the conserved
quantum numbers introduced in (\ref{eq.fackb})  have the same values for the two incoming particles.
Although the boundary Yang-Baxter equation is quadratic in general,
in this case it is inhomogeneous linear in the variables $y^{(\pm)}(\Theta_{1})$
and $y^{(\pm)}(\Theta_{2}).$ The coefficient of the quadratic term
$y^{(\pm)}(\Theta_{1})y^{(\pm)}(\Theta_{2})$ vanishes precisely because
$R_{P1}^{(\pm)}(\Theta)$ satisfies the Yang-Baxter equation.
The Yang Baxter equation consists of 16 scalar equations in our case.
Some of them are trivial (0=0), and the remaining $n$ equations are
of the form\[
a_{q}^{(\pm)}(\Theta_{1},\Theta_{2})y^{(\pm)}(\Theta_{1})+b_{q}^{(\pm)}(\Theta_{1},\Theta_{2})y^{(\pm)}(\Theta_{2})+c_{q}^{(\pm)}(\Theta_{1},\Theta_{2})=0,\quad q=1..n.\]
 It is possible to choose two inequivalent equations from this set.
Two such equations can be solved for the numbers $y^{(\pm)}(\Theta_{1})$
and $y^{(\pm)}(\Theta_{2})$. The solution turns out to be of the
form $y^{(\pm)}(\Theta_{1})=g^{(\pm)}(\Theta_{1}),$ $y^{(\pm)}(\Theta_{2})=g^{(\pm)}(\Theta_{2})$
(for general coefficients
$d_{q_{1}},e_{q_{1}},f_{q_{1}};d_{q_{2}},e_{q_{2}},f_{q_{2}}$ instead of 
$a_{q_{1}},b_{q_{1}},c_{q_{1}};a_{q_{2}},b_{q_{2}},c_{q_{2}}$
it would be of the form $y(\Theta_{1})=g_{1}(\Theta_{1},\Theta_{2}),$
$y(\Theta_{2})=g_{2}(\Theta_{1},\Theta_{2})$, which does not define
a function $y(\Theta)$), where $g^{(\pm)}$ is a function that depends
also on $m$, $\alpha$, $\gamma$, but has no other parameters. Consequently,
the reflection amplitude depends on the conserved quantum numbers introduced
in (\ref{eq.fackb})
through these parameters only. We checked that the solution obtained
in this way satisfies the other $n-2$ equations as well. In the next
step we checked that the solutions $R_{P2,e}^{(\pm)}(\Theta)$ satisfy
the Yang-Baxter equation for incoming particles of different masses
as well. The two functions $y^{(+)}(\Theta)$ and $y^{(-)}(\Theta)$
have very similar form.

The solutions that we obtained can be brought to the following form:
\begin{align}
\notag
&\{ R_{P2,e}^{(\pm)}\}_{b}^{b}(\Theta)=A_{+}^{(\pm)}(\Theta)&& \{ R_{P2,e}^{(\pm)}\}_{f}^{f}(\Theta)=A_{-}^{(\pm)}(\Theta)\\
&
\{ R_{P2,e}^{(\pm)}\}_{b}^{f}(\Theta)=\pm B^{(\pm)}(\Theta)&& \{
R_{P2,e}^{(\pm)}\}_{f}^{b}(\Theta)=B^{(\pm)}(\Theta)\notag
\end{align}
\begin{multline}
\notag
A_{\pm}^{(-)}(\Theta)=\tilde{Z}^{(-)}(\Theta)\left\{ \cosh\left(\frac{\Theta}{2}\right)\left(\frac{\gamma^{2}}{4M}-\left[\sin^{2}\left(\frac{\rho}{4}\right)+\sinh^{2}\left(\frac{\Theta}{2}\right)\right]\right)\right.\\
\left.\mp
  i\sinh\left(\frac{\Theta}{2}\right)\left(\frac{\gamma^{2}}{4M}+\left[\sin^{2}\left(\frac{\rho}{4}\right)+\sinh^{2}\left(\frac{\Theta}{2}\right)\right]\right)\right\} 
\end{multline}
\begin{multline}
\notag
A_{\pm}^{(+)}(\Theta)=\tilde{Z}^{(+)}(\Theta)\left\{ -i\cosh\left(\frac{\Theta}{2}\right)\left(\frac{\gamma^{2}}{4M}-\left[\sin^{2}\left(\frac{\rho}{4}\right)-\cosh^{2}\left(\frac{\Theta}{2}\right)\right]\right)\right.\\
\left.\pm\sinh\left(\frac{\Theta}{2}\right)\left(\frac{\gamma^{2}}{4M}+\left[\sin^{2}\left(\frac{\rho}{4}\right)-\cosh^{2}\left(\frac{\Theta}{2}\right)\right]\right)\right\} 
\end{multline}
\begin{equation}
\notag
B^{(\pm)}(\Theta)=\tilde{Z}^{(\pm)}(\Theta)\frac{e\gamma}{2\sqrt{M}}\sqrt{\cos(\rho/2)}\sinh(\Theta),
\end{equation}
where
\begin{equation}
m=2M\cos(\frac{\rho}{2}),\quad\frac{\rho}{2}=\frac{\pi}{2}-u,\quad\alpha=-\frac{1}{2M},\label{eq.defrho}
\end{equation}
$0\leq\rho<\pi$, $e=\pm1$ in the $(-)$ case and $e=1$ in the
$(+)$ case. Note that $R_{P2,e}^{(\pm)}$ depends on two parameters:
$\gamma^{2}/M$ and $\rho$ only. $R_{P2,e}^{(-)}$ has the same structure
as the particle reflection amplitude obtained in \cite{BPT} for
the case of the boundary supersymmetric sine-Gordon model from the
kink reflection amplitude by bootstrap. Consequently, there is no
need now to solve the crossing and unitarity equations for $\tilde{Z}^{(-)}(\Theta)$,
we take it from \cite{BPT}. We determined $\tilde{Z}^{(+)}(\Theta)$
using the unitarity and crossing equations and exploiting the fact
that these equations take a similar form for $\tilde{Z}^{(-)}(\Theta)$.
Explicit formulae for these prefactors can be found in the Appendix.

In the $(+)$ case we introduce the parameter $\xi$ in the following
way: 
$\gamma=-2\sqrt{M}i\sin(\xi/2)$, 
$\xi \in [-\pi,\pi]$.
It should be noted, however, that if $\xi \ne 0$, then the condition
$Q_B^\dagger=Q_B$ is violated. If we choose $\xi$ so that $\gamma \in \RR$,
then, as we can see from  
the formula for $\tilde{Z}^{(+)}(\Theta)$,  the condition
5) in Section \ref{sec.fstb} requiring that the poles should be in $i\RR$  is
not satisfied. We also note that if $\xi=0$, then
$R_{P2,e}^{(+)}(\Theta)=R_{P1}^{(+)}(\Theta)$.

To summarize, we have the supersymmetric reflection matrix factors
\begin{equation}
\begin{array}{ll}
R_K^{(+)}(\Theta)\qquad &  \\ 
R_{K,e}^{(-)}(\Theta,\xi)\qquad &  \xi
\in [0,\pi],\ \ e=\pm 1\\  
R_{P1}^{(\pm)}(\Theta)\qquad &   \\
R_{P2,e}^{(-)}(\Theta, M, \rho, \xi)\qquad & M>0,\ \ 0\le \rho< \pi,\ \  \xi
\in [0,\pi],\ \  e=\pm 1\\
R_{P2,e}^{(+)}(\Theta, M, \rho, \xi)\qquad & M>0,\ \ 0\le \rho< \pi,\ \ \xi
\in [-\pi,\pi],\ \   e=1\ . 
\end{array}
\end{equation}

The same set of kink and particle reflection matrix factors can be obtained
by solving the Yang-Baxter equations without imposing the boundary
supersymmetry condition \cite{AK1,AK2,Chim}. The supersymmetry
condition relates the parameters of the reflection matrix factors obtained
in this way to the parameters of the representations of the supersymmetry
algebra. The results described above show that the task of solving
the Yang-Baxter equations is greatly simplified if one imposes the
supersymmetry condition first.

\subsection{Properties of the ground state reflection matrix factors}
\label{sec.property1}
\markright{\thesubsection.\ \ PROPERTIES OF THE GROUND STATE REFLECTION FACTORS}

In this section various important properties of the ground state reflection
matrix factors are collected. 

$R_{P2,e}^{(\pm)}$ are not symmetric with respect to ${\Gamma}_B$,
$R_{P1}^{(\pm)}$ are symmetric with respect to ${\Gamma}_B$. 
$R_{P1}^{(\pm)}$
and $R_{P2,e}^{(\pm)}$ do not satisfy the Yang-Baxter equation together.
It is also important to note that $\lim_{\gamma\to
  0}R_{P2,e}^{(\pm)}=R_{P1}^{(\pm)}$. We shall assume that $\gamma \ne 0$ when
we mention $R_{P2,e}^{(\pm)}$.

The second boundary bootstrap equation 
applied to the $kink+kink\rightarrow particle$ bulk fusion determines
reflection matrix factors for boson-fermions on ground state boundary. They turn
out \cite{BPT,AK2} to be the same as those obtainable by solving
the boundary Yang-Baxter, crossing and unitarity equations. In terms
of the reflection matrix factors 
\begin{equation}
R_{K}^{(+)}+R_{K}^{(+)}\rightarrow R_{P1}^{(+)}
\end{equation}
and 
\begin{equation}
R_{K,e}^{(-)}(\xi) +R_{K,e}^{(-)}(\xi) \rightarrow R_{P2,e}^{(-)}(\xi)
\end{equation}
with
appropriate values of the parameters. Similarly, it can be checked
that the second boundary bootstrap equation is also satisfied for the
$particle+particle\rightarrow particle$, $kink+particle\rightarrow kink$
fusions with the reflection matrix factors
\begin{align}
R_{P1}^{(\pm)}+R_{P1}^{(\pm)} &\rightarrow R_{P1}^{(\pm)}\\
R_{P2,e}^{(\pm)}(\xi)+R_{P2,e}^{(\pm)}(\xi) &\rightarrow
R_{P2,e}^{(\pm)}(\xi)\\
R_{K}^{(+)}+R_{P1}^{(+)} &\rightarrow R_{K}^{(+)}\\
R_{K,e}^{(-)}(\xi)+R_{P2,e}^{(-)}(\xi) &\rightarrow R_{K,e}^{(-)}(\xi)
\end{align}
respectively. These relations are nontrivial, although it is clear
that they are satisfied up to CDD factors. It is remarkable that 
$R_{P2,e}^{(+)}$
cannot be obtained by bootstrap from kink reflection matrix factors, while
the other particle reflection matrix factors $R_{P1}^{(\pm)}$ and $R_{P2,e}^{(-)}$
can be obtained in this way.

$R_{K}^{(+)}(\Theta)$ is bijective (of rank two) in the physical strip. $\{ R_{K,+1}^{(-)}\}_{K_{0\hlf}}^{K_{0\hlf}}(\Theta)$
and\\
$\{ R_{K,-1}^{(-)}\}_{K_{1\hlf}}^{K_{1\hlf}}(\Theta)$ have a
zero at $\Theta=i(\pi-\xi)$, so $R_{K,e}^{(-)}$ is of rank one at
this angle. This zero is in the physical strip if $\pi>\xi>\pi/2$,
any other zeroes of the kink amplitudes are outside the physical strip.

Consequently, the relations $R_{K}^{(+)}+R_{K}^{(+)}\rightarrow R_{P1}^{(+)}$,
and $R_{K,e}^{(-)}+R_{K,e}^{(-)}\rightarrow R_{P2,e}^{(-)}$ together
with the bijectivity of the kink-kink fusion tensor and $S_{KK}$ imply
that $R_{P1}^{(+)}$ is of rank two (bijective) and has no poles in
the physical strip, and $R_{P2,e}^{(-)}(\Theta)$ is also bijective
for generic values of $\Theta,$ but it is of rank one if $\Theta=i(\pi-\xi\pm\rho/2)$.
It is possible for these angles to be in the physical strip and on
the imaginary axis.  $(\pi-\xi-\rho/2)>-\pi/2$ holds,
so if $(\pi-\xi-\rho/2)$ is negative, then there is a pole in the
physical strip at $i(\rho/2+\xi-\pi)$ because of unitarity. If $(\pi-\xi-\rho/2)>0$,
then $R_{P2,e}^{(-)}(\Theta)$ has no poles and zeroes in the physical
strip, and within the physical strip it is of rank 1 if and only if
$\Theta=i(\pi-\xi\pm\rho/2)$. We therefore impose the following condition
on $\xi$:
\begin{equation}
\pi-\xi\geq\rho/2\ .\label{eq.feltetel0}
\end{equation}
 $R_{P1}^{(-)}$ is also bijective and has no poles in the physical
strip. It can be verified by direct calculation that $R_{P2,e}^{(+)}(\Theta)$
is degenerate (of rank two) at $\Theta=i(\pi-\xi\pm\rho/2)$ and
$\Theta=i(\pi+\xi\pm\rho/2)$, therefore
  the condition (\ref{eq.feltetel0}) reads in this case as
\begin{equation}
 \pi-|\xi|\geq\rho/2\ .\label{eq.feltetel0 vesszo}
 \end{equation}
(Note that the relation between $\xi$ and $\gamma$ is different
in the $(+)$ and $(-)$ cases). 

\begin{table}
\caption{Degeneracy properties of reflection factors \hspace{7cm}}
\vspace{0.2cm}
\begin{tabular*}{16cm}{@{\extracolsep\fill}lp{13.5cm}}
\hline
$R_K^{(+)}(\Theta)$: & bijective (rank two)\\
$R_{K,e}^{(-)}(\Theta)$: & degenerate (rank one) at $\Theta=i(\pi-\xi)$, which is in the
physical strip if $\pi>\xi>\pi/2$.\\
$R_{P1}^{(\pm)}(\Theta)$: &  bijective (rank two)\\
$R_{P2,e}^{(-)}(\Theta)$: & degenerate (rank one) at $\Theta=i(\pi-\xi\pm\rho/2)$, one
or both of these angles can be in the physical strip. $\pi-\xi\ge \rho/2$ is
necessary and sufficient for   $R_{P2,e}^{(-)}$ not to have any poles in the
physical strip.\\
$R_{P2,e}^{(+)}(\Theta)$: & degenerate (rank one) at $\Theta=i(\pi-\xi\pm\rho/2)$, 
$\Theta=i(\pi+\xi\pm\rho/2)$. 
 Some
 of these angles can be in the physical strip. $\pi-|\xi|\ge \rho/2$ is
 necessary and sufficient for   $R_{P2,e}^{(+)}$ not to have any poles in the
 physical strip.
\\
\hline
\end{tabular*}
\end{table}

A particular boundary scattering theory is characterized by the $M$ parameter 
 of the underlying bulk theory and by the sign $(+)$ or $(-)$, and
also by the parameters $\xi$ and $e$. Considering the properties of the
 supersymmetric factors the following three cases can be distinguished:

\begin{enumerate}
\item  The
   boundary co-multiplication is $\Delta_B^{+}$, the theory may contain kinks
   as well as particles, the supersymmetric reflection matrix
   factors on ground state boundary are $R_K^{(+)}$ and  $R_{P1}^{(+)}$. 
 \item  The 
    boundary co-multiplication is $\Delta_B^{+}$, the theory may contain
    particles only, the supersymmetric reflection matrix
    factor on ground state boundary is $R_{P2,e}^{(+)}$. $\gamma \ne 0$ is assumed.
\item   The 
   boundary co-multiplication is $\Delta_B^{-}$, the theory may contain kinks
   as well as particles, the supersymmetric reflection matrix
   factors on ground state boundary are $R_{K,e}^{(-)}$ and  $R_{P2,e}^{(-)}$.
   In this case we allow $\gamma =0$.
\item   The
   boundary co-multiplication is $\Delta_B^{-}$, the theory may contain particles
   only, the supersymmetric reflection matrix
   factor on ground state boundary is  $R_{P1}^{(-)}$. In this case it is
   assumed that $\gamma \ne 0$. 
\end{enumerate}

(\ref{eq.feltetel0})  
imply that if $R_{P2,e}^{(-)}$
describes the ground state reflections of the particles in a theory (and
$\gamma \ne 0$),
then 
\begin{equation}
\pi-|\xi|\geq\rho_{max}/2\ ,\qquad\rho_{max}=\max_{i}(\rho_{i})\ ,\label{eq.feltetel
  00}
\end{equation}
where $i$ runs over all particles in a particular theory, is necessary and sufficient for
all $R_{P2,e}^{(-)}$ and  $R_{P2,e}^{(+)}$ factors (in a particular theory) not to have  poles in the physical strip.

\subsection{Higher level supersymmetric boundary states}
\label{sec.hlssbs}
\markright{\thesubsection.\ \ HIGHER LEVEL SUPERSYMMETRIC BOUNDARY STATES}

\subsubsection{Cases 1 and 3}
\label{sec.rules13} 

In case 1.\ and 3.\ of Section \ref{sec.property1} 
the supersymmetric part of any boundary multiplet can be labeled as
\begin{equation}
\label{eq.lab13}
W(\nu_1,\nu_2,\dots,\nu_n,B_{\hlf} )
\end{equation}
where
\begin{equation}
\label{eq.00}
\pi>\nu_1>\nu_2>\dots>\nu_n>0
\end{equation}
\begin{equation}
\nu_i+\nu_j\ne \pi\qquad \forall i,j=1\dots n,\ i\ne j\ . 
\end{equation}
We assume here and below that $\nu_i\ne \pi-\xi\ \forall i$. The special situation
when $\pi-\xi$ is also allowed and the situation when equalities are also
allowed in (\ref{eq.00}) will be discussed after the description of the
general case.

$W(\nu_1,\nu_2,\dots,\nu_n,B_{\hlf} )$ stands for a linear space which is
 spanned by the states belonging to the (supersymmetric part of the) multiplet,
\begin{equation}
v=W(\nu_1,\nu_2,\dots,\nu_n,B_{\hlf} )=W_{K}\otimes W_{K} \otimes \dots \otimes W_{K}
\otimes W(B_{\hlf})\ , 
\end{equation}
where $W_K$ stands for the internal space of the kink representation of mass $M$. 
The second equality is, strictly speaking, an equality up to an isomorphism.
The space $v$, taking into consideration the kink adjacency conditions, has dimension 
\begin{equation}
\dim v= 2^{\left\lceil n/2\right\rceil }\ .
\end{equation}
The representation of $\mathcal{A}_B$ on $v$ is
\begin{equation}
K(\nu_1,\nu_2,\dots,\nu_n,B_{\hlf})=K_M(i\nu_1)\times K_M(i\nu_2) \times \dots \times K_M(i\nu_n) \times B_{\hlf}\ .
\end{equation}
Strictly speaking, this equality is an equivalence of representations, the
intertwining map being the isomorphism mentioned above.
We assign a labeling set to a boundary multiplet in the following way:
\begin{equation}
L[ W(\nu_1,\nu_2,\dots,\nu_n,B_{\hlf} ) ]= \{i\nu_1, i\nu_2, \dots , i\nu_n\}\ . 
\end{equation}

The reflection factors on $v$ have no poles and zeroes on the imaginary
axis in the physical strip if and only if 
\begin{equation}
\nu_{i}<\pi-\rho_{max}/2\qquad\forall i=1\dots n\ .\label{eq.feltetel 4}
\end{equation}
(Note that $\rho$ was defined in (\ref{eq.defrho}).)
Let $p+v\to y$ be a boundary fusion where $p$ is either a kink or a particle
 multiplet with appropriate rapidity and $v$ and $y$ are boundary multiplets,
$L[v]=\{i\nu_1,\dots,i\nu_n \}$.
$L[y]$ can be obtained form $L[p]$ and $L[v]$:\\
If $p$ is a kink multiplet, $L[p]=\{iw\}$,   and $w+\nu_i \ne \pi  ,\ i=1\dots n$,  then 
\begin{equation}
\{iw \} + \{ i\nu_1, i\nu_2, \dots , i\nu_n   \} \to 
\{ i\nu_1, \dots , i\nu_{k}, iw, i\nu_{k+1}, \dots,    i\nu_n   \}\ ,
\end{equation}
if $w+\nu_k=\pi$, then
\begin{equation}
\{iw \} + \{ i\nu_1, i\nu_2, \dots , i\nu_n   \} \to 
\{ i\nu_1, \dots , i\nu_{k-1},  i\nu_{k+1}, \dots,    i\nu_n   \}\ .
\end{equation}
If $p$ is a particle multiplet, $L[p]=\{ iw_1, iw_2 \}$,  and $|w_1|>|w_2|$ and $|w_1| +\nu_i \ne \pi ,\
 i=1\dots n$,  
$|w_2|+\nu_i \ne \pi ,\ i=1\dots n$,   then
\begin{equation}
\{iw_1, iw_2 \} + \{ i\nu_1, i\nu_2, \dots , i\nu_n   \} \to 
\{ i\nu_1, \dots , i\nu_{k}, i|w_1|, i\nu_{k+1}, \dots, i\nu_{l}, i|w_2|,
i\nu_{l+1}, \dots ,   i\nu_n   \},\ 
\end{equation}
if 
$|w_1|+\nu_k=\pi$, 
$|w_2|+\nu_i\ne \pi  ,\ i=1\dots n$, then
\begin{equation}
\{iw_1, iw_2 \} + \{ i\nu_1, i\nu_2, \dots , i\nu_n   \} \to 
\{ i\nu_1, \dots , i\nu_{k-1}, i\nu_{k+1}, \dots, i\nu_{l}, i|w_2|,
i\nu_{l+1}, \dots ,   i\nu_n   \},\ 
\end{equation}
if
$|w_1|+\nu_i \ne \pi ,\ i=1\dots n$, 
$|w_2|+ \nu_l=\pi $, then
\begin{equation}
\{iw_1, iw_2 \} + \{ i\nu_1, i\nu_2, \dots , i\nu_n   \} \to 
\{ i\nu_1, \dots , i\nu_{k}, i|w_1|, i\nu_{k+1}, \dots, i\nu_{l-1}, 
i\nu_{l+1}, \dots ,   i\nu_n   \},\ 
\end{equation}
if
$|w_1|+\nu_k=\pi$, 
$|w_2|+\nu_l=\pi$, then 
\begin{equation}
\{iw_1, iw_2 \} + \{ i\nu_1, i\nu_2, \dots , i\nu_n   \} \to 
\{ i\nu_1, \dots , i\nu_{k-1}, i\nu_{k+1}, \dots, i\nu_{l-1}, 
i\nu_{l+1}, \dots ,   i\nu_n   \}.\ 
\end{equation}

In other words,  $L[y]$ is obtained in the following way:
we form the union $b=L[p]\cup L[v]$ (the elements may have multiplicities),
replace $iw_1$, $iw_2$ by $i|w_1|$, $i|w_2|$, and remove all the pairs of
 elements $i\Theta_1,i\Theta_2$ satisfying $\Theta_1+\Theta_2=\pi$.  This rule is analogous to the bulk fusion rules, but the amplitudes
$i\Theta$ and $-i\Theta$ ($\Theta\in\RR$) are identified.

\begin{figure}
\begin{center}
\includegraphics[scale=0.7]{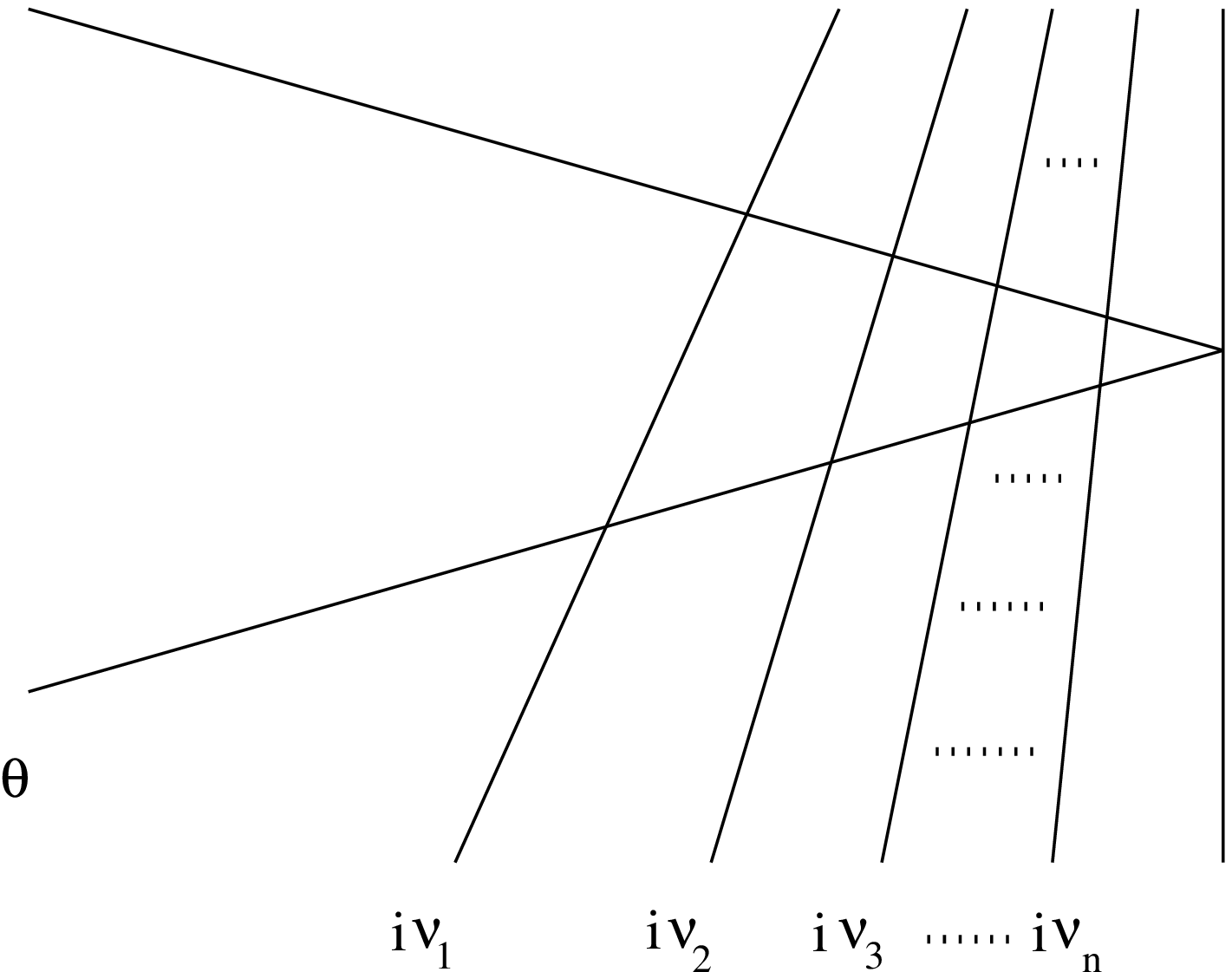}
\caption{\label{fig.multi}Graphical representation for (\ref{eq.y1})-(\ref{eq.y4})  }
\end{center}
\end{figure}

The reflection matrix factor of a particle or kink on the boundary multiplet\\
$W(\nu_1,\nu_2,\dots,\nu_n,B_{\hlf})$ is 
\begin{equation}
\label{eq.y1}
R_{XW(\nu_1,\nu_2,\dots,\nu_n,B_{\hlf})}(\Theta)=
U_1U_2\dots U_n R T_nT_{n-1} \dots T_1
\end{equation}
where
\begin{align}
&T_k = \underbrace{I\otimes \dots \otimes I}_{k-1}\otimes S_{XK}(\Theta-i\nu_k)
\otimes \underbrace{I \otimes \dots \otimes I}_{n-k}
\otimes I\\
&R =  \underbrace{I\otimes \dots \otimes I}_{n} \otimes R_X(\Theta)\\
\label{eq.y4}
&U_k =  \underbrace{I\otimes \dots \otimes I}_{k-1}\otimes S_{KX}(\Theta+i\nu_k)
\otimes \underbrace{I \otimes \dots \otimes I}_{n-k}
\otimes I 
\end{align}
and
$X$ stands either for $K$ (kink) or $P$ (particle). A graphical illustration
is given in Figure \ref{fig.multi}.

We consider now the special case when kink rapidities $i(\pi-\xi)$ are also
allowed and case 3.\ of Section  \ref{sec.property1} applies. The speciality of
$i(\pi-\xi)$ is that at this rapidity the ground state reflection factor
$R_{K,e}^{(-)}$ is degenerate. The rules above are modified in the following
way: there is no change in the rules for the labeling sets, but if the
labeling set contains $i(\pi-\xi)$, i.e.\ $L= \{i\nu_1, i\nu_2, \dots,
i(\pi-\xi), \dots , i\nu_n\}$, then the linear space of the multiplet is
denoted as 
$W(\nu_1,\nu_2,\dots,\nu_n,B_{0} )$ or $W(\nu_1,\nu_2,\dots,\nu_n,B_{1})$ and 
$v=W(\nu_1,\nu_2,\dots,\nu_n,B_{0} )=W_{K}\otimes W_{K} \otimes \dots \otimes W_{K}
\otimes W(B_{0})$, $v=W(\nu_1,\nu_2,\dots,\nu_n,B_{1} )=W_{K}\otimes W_{K} \otimes \dots \otimes W_{K}
\otimes W(B_{1})$,
where  $W(B_{0})$ or   $W(B_{1})$ is the one-dimensional
space spanned by 
$\ket{B_0}$
or  
$\ket{B_1}$, 
which are the boundary states created by the fusion of $K_{0\hlf}(i(\pi-\xi))$
and $\ket{B_{\hlf}}$ of $K_{1\hlf}(i(\pi-\xi))$ and $\ket{B_{\hlf}}$, respectively.
 `$0$' should be taken in these
and the following
formulae if $e=-1$, `$1$' if $e=+1$.   The dimension of $v$ is $2^{\left\lceil n/2\right\rceil
-1}$. 
$W(B_0)$ or $W(B_1)$ is invariant with respect to the action of $\mA_B$, the
representation on $W(B_0)$ or $W(B_1)$ is denoted by
$B_0$ or $B_1$. The representation of  $\mA_B$ on $v$ is $K(i\nu_1)\times K(i\nu_2) \times \dots \times
K(i\nu_n) \times B_{0}$
or
$K(i\nu_1)\times K(i\nu_2) \times \dots \times K(i\nu_n) \times B_{1}$. 
The formulae (\ref{eq.y1})-(\ref{eq.y4}) apply with the modification that
$R_X(\Theta)$ should be replaced by
the reflection matrix factor on $W(B_0)$ or $W(B_1)$, which can be calculated
by the bootstrap equation.

The only modification
if equalities between the $\nu$'s (i.e.\ in (\ref{eq.00})) are also allowed is
that the labeling sets can contain certain elements with multiplicities
greater than 1. If $i(\pi-\xi)$ has multiplicity greater than 1 in the
labeling set, then only one copy of it is absorbed to form the label $B_{0}$ or $B_{1}$.

It can be verified that the labeling (\ref{eq.lab13}) of the boundary multiplets
is unambiguous, i.e.\ if two multiplets have different labels, then they have
different transformation and scattering properties. This can be seen from the
dimensions of the multiplets and from the singularity properties of the
reflection factors on them.

It is also possible that (\ref{eq.feltetel 4}) is not satisfied for
certain bound states, but the poles of the supersymmetric reflection
factors on them are canceled by zeroes of the corresponding non-supersymmetric
reflection factor, so the poles of the supersymmetric reflection factors
do not introduce new singularity. The fusion rules above apply in
this case as well.

We remark that the above characterization of the supersymmetric factors for
higher level boundary bound states can be interpreted as a characterization in
terms of `boundary particle groups'. In this interpretation we can say, in
particular, that the supersymmetric parts of
the boundary bound states can be described as supersymmetric boundary particle
groups in which the boundary is in the ground state.    

We refer the reader to \cite{sajat1} for a detailed justification of the statements in
this section. Here we remark only that the degeneracy properties of the
two-particle S-matrix factors, the bulk supersymmetric fusion rules  and the
degeneracy properties of the one-particle ground state reflection
matrix factors  play very important role.

It should be noted that we have not assumed particular statistical
properties of the particles.

As we mentioned in Section \ref{sec.sssf}, the boson-fermion representation can be
replaced  by the two-kink-representation without difficulty.

\subsubsection{Cases 2 and 4}
\label{sec.rules24}

In cases 2.\ and 4.\ of Section \ref{sec.property1} 
the supersymmetric part of any boundary multiplet can be labeled as
\begin{equation}
\label{eq.form1}
W(\nu_1,\nu_2,\dots,\nu_n,B_{\hlf} )
\end{equation}
where $n$ is even,
\begin{equation}
\label{eq.form2}
\pi>\nu_1>\nu_2>\dots>\nu_{n-1}>0,\quad \nu_{n}>-\pi
\end{equation}
\begin{equation}
\label{eq.form3}
\nu_{n-1}>|\nu_{n}|
\end{equation}
\begin{equation}
\label{eq.form4}
\nu_i+\nu_j\ne \pi\qquad \forall i,j=1\dots n,\ i\ne j
\end{equation}
\begin{equation}
\label{eq.form5}
\nu_i\ne -\nu_n\qquad \forall i=1\dots n-1\ .
\end{equation}
We assume here and below that $\nu_i\ne \pi-|\xi|\ \forall i$. 
 The special situation
 when $\pi-|\xi|$ is also allowed and the situation when equalities are also
 allowed in (\ref{eq.form2}), (\ref{eq.form3}) will be discussed after the description of the
 general case.

\begin{equation}
v=W(\nu_1,\nu_2,\dots,\nu_n,B_{\hlf} )=W_{K}\otimes W_{K} \otimes \dots \otimes W_{K}
\otimes W(B_{\hlf})\ .
\end{equation}
The dimension of $v$ is
\begin{equation}
\dim v= 2^{\left\lceil n/2\right\rceil }.\ 
\end{equation}
The representation on $v$ is
\begin{equation}
K(\nu_1,\nu_2,\dots,\nu_n,B_{\hlf})=K(i\nu_1)\times K(i\nu_2) \times \dots \times K(i\nu_n) \times B_{\hlf}\ .
\end{equation}
We assign a labeling  set to a boundary multiplet as in the previous section:
\begin{equation}
L[ W(\nu_1,\nu_2,\dots,\nu_n,B_{\hlf} ) ]= \{i\nu_1, i\nu_2, \dots , i\nu_n\}\ . 
\end{equation}

The reflection factors on $v$ have no poles and zeroes on the imaginary
axis in the physical strip if and only if (\ref{eq.feltetel 4}) is satisfied.

Let $p+v\to y$ be a boundary fusion where $p$ is a particle
 multiplet with appropriate rapidity and $v$ and $y$ are boundary multiplets,
$L[p]=\{iw_1, iw_2 \}$,  $L[v]= \{
i\nu_1, i\nu_2, \dots , i\nu_n   \}$. $L[y]$ is obtained in the following way:
we form the union $b=L[p]\cup L[v]$ (the elements may have multiplicities),
then delete (zero, two or four) elements in a few steps applying  the
 following algorithm: 
1.\ if $\Theta-\Theta_2=i\pi$ for two elements, then they are removed, 2.\ the
 sign of any even number of
 elements of $b$ can be changed freely, 3.\ sign changes are done until no more
 deletions can be done. Finally if $b$ is not yet of the form (\ref{eq.form1})-(\ref{eq.form5}), then we bring it to this
form by changing the sign of an even number (usually two) of  appropriate  elements of $b$. The $b$ obtained in this way
equals to $L[y]$.

The formulae (\ref{eq.y1}) - (\ref{eq.y4}) apply in this case as well, $X$
should be replaced by $P$.

 We consider now the special case when kink rapidities $i(\pi-|\xi|)$ are also
 allowed and case 2.\ of Section  \ref{sec.property1} applies. The speciality of
  $i(\pi-|\xi|)$ is related to the degeneracy properties of the reflection factor
 $R_{P2}^{(+)}$. The rules above are modified in the following
 way:  if the
 labeling set contains $i(\pi-|\xi|)$, i.e.\ $L= \{i\nu_1, i\nu_2, \dots,
 i(\pi-|\xi|), \dots , i\nu_n\}$, then the linear space of the multiplet is
 denoted as 
 $W(\nu_1,\nu_2,\dots,\nu_{n-1},B_{\hlf}' )$  and 
 $v=W(\nu_1,\nu_2,\dots,\nu_{n-1},B_{\hlf}' )=W_{K}\otimes W_{K} \otimes \dots \otimes W_{K}
 \otimes W(B_{\hlf}')$, where  $W(B_{\hlf}')$ is the one-dimensional
 space of the
 boundary state arising in the fusion of a particle composed of the kink
 multiplets $K(i(\pi-|\xi|)$ and $K(i\nu_n)$   and
 $\ket{B_{\hlf}}$. The dimension of $v$ is $2^{\left\lceil n/2\right\rceil
 -1}$. 
 The representation on $v$ is $K(i\nu_1)\times K(i\nu_2) \times \dots \times
 K(i\nu_{n-1}) \times B_{\hlf}'$.
 The formulae (\ref{eq.y1})-(\ref{eq.y4}) apply with the modification that
 $R_P(\Theta)$ should be replaced by the (particle) reflection factor on
 $B_{\hlf}'$ which can be computed using the bootstrap equation.

The modification in the above rules
if equalities between the $\nu$'s (in (\ref{eq.form2}), (\ref{eq.form3})) are also allowed is
that the labeling sets can contain certain elements with multiplicities
greater than 1. 
 If $i(\pi-|\xi|)$ has multiplicity greater than 1 in the
 labeling set, then only one copy of it is absorbed to form the label $B_{\hlf}'$.
If the last element $\nu_n$  is negative, it can have
multiplicity 1 only.  

It can be verified that the labeling (\ref{eq.form1}) of the boundary multiplets
is unambiguous, i.e.\ if two multiplets have different labels, then they have
different transformation and scattering properties.

The remarks made at the end of the previous section apply to the present
section as well. 
\vspace{4mm}
\begin{center}
-----------------------
\end{center}
\vspace{4mm}

We consider now the situation when a boundary state (or multiplet) in the non-supersymmetric
theory arises as a bound state in more than one way. In this case the question
that arises is whether the supersymmetric parts obtained by the rules of
this section 
give the same result in each
way of creating the boundary bound state.
%  Affirmative answer is necessary for
% the applicability of the ansatz described in Sections \ref{sec.ansatz} and \ref{sec.ansatzb}.  
In the non-supersymmetric theory the diagrams
corresponding to the different ways of creation can usually be transformed into each other
by shifting lines using the bootstrap and
Yang-Baxter equations and by splitting or merging  vertices. It is natural
to expect that these transformations are usually applicable to the
supersymmetric parts as well, and so the answer to the question above is
positive. This should nevertheless be checked in specific models. 
If the answer is negative, then one has to take direct sums of the
multiplets (\ref{eq.lab13}) or (\ref{eq.form1}) as supersymmetric parts of the
boundary multiplets.

% ************************************* scat14.tex vege *****************************************

\section{Examples}
\label{sec.ex..}
\markright{\thesection.\ \ EXAMPLES}

In this section we present examples for the application of the fusion rules
written down in Section \ref{sec.hlssbs}. The sinh-Gordon model and the free particle
will be discussed only briefly, restricting mainly to the dimension of
the boundary multiplets. The reason for this is that in these cases not all
the assumptions that we have made are satisfied. The sine-Gordon model and the
$a_2^{(1)}$ and $a_4^{(1)}$ affine Toda models will be 
discussed in more detail.

\subsection{Boundary sine-Gordon model}
\markright{\thesubsection.\ \ BOUNDARY SINE-GORDON MODEL}

It was argued in \cite{Skly,GZ} that the boundary version of sine-Gordon model
which has the action
\begin{equation}
S=\int_{-\infty }^{\infty }dt\int_{-\infty }^{0}dx\ {\cal {L}}_{SG}-\int_{-\infty}^{\infty}dt\ V_{B}(\Phi_{B})\ ,
\end{equation}
where
\begin{equation} 
{\cal {L}}_{SG}=\frac{1}{2}(\partial _{\mu }\Phi )^{2}-\frac{m^{2}}{\beta
  ^{2}}(1-\cos (\beta \Phi ))
\end{equation}
is the Lagrangian density of the sine-Gordon model,
 $ \Phi (x,t) $ is a real scalar field, $ \beta $ is a real
dimensionless coupling constant and $ \Phi _{B}(t)=\Phi (x,t)|_{x=0}$,
preserves the integrability of the bulk theory if the boundary potential
is chosen as 
\begin{equation}
V_{B}(\Phi_{B})=M_{0}\left( 1-\cos \left( \frac{\beta }{2}(\Phi_{B}-\phi_{0})\right) \right)\ ,
\end{equation}
 where $ M_{0} $ and $ \phi_{0} $ are free parameters. As a
result of the boundary potential the scalar field satisfies the boundary
condition
\begin{equation}
\label{hatfel}
\partial_{x}\Phi |_{x=0}=-M_{0}\frac{\beta}{2}\sin \left( \frac{\beta}{2}(\Phi_{B}-\phi_{0})\right)\ .
\end{equation}

The particle spectrum of the bulk sine-Gordon theory (SG) contains
a soliton $s$ and an anti-soliton $\bar{s}$ of mass $M$ and
the breathers $B_{n}$ of mass $m_{n}=2M\sin(u_{n})$, where $u_{n}=\pi n/(2\lambda)$,
$n=1,\dots,[\lambda]$,  $\lambda=\frac{8\pi}{\beta^2}-1$. $M$ can be expressed
in terms of $\beta, m$. 
The breathers are self-conjugate, and the conjugate of $s$ is $\bar{s}$.
The fusions (given in the form {\it process, (fusion angle)}) are the
following: $s+\bar{s}\rightarrow B_{n}$, $(\pi-2u_{n})$; $B_{n}+B_{m}\rightarrow B_{n+m}$,
$(u_{n}+u_{m})$ provided $n+m\leq[\lambda]$; and the crossed versions
of these rules. 
These rules are consistent with the following decomposition (see (\ref{eq.fack})):
$\langle s,\bar{s} \rangle \oplus \langle b_1 \rangle \oplus \dots \oplus 
\langle b_{[\lambda]} \rangle$, where $\langle v, \dots \rangle$ denotes the
linear space spanned by the vectors $v, \dots$.   
One can associate kink representations
to $\langle s,\bar{s} \rangle$ and particle or `two-kink-representations'
with $\alpha=-1/(2M)$ to the breathers (see also (\ref{eq.svvvv})). The labeling sets corresponding to
the supersymmetric parts $W_s$, $W_n$ of $\langle s,\bar{s} \rangle$, 
$\langle B_{n} \rangle$ are $L[W_s(\Theta)]=\{\Theta\}$,
$L[W_n(\Theta)]=\{\Theta+i(\pi/2-u_{n}),\Theta-i(\pi/2-u_{n})\}$.

The boundary sine-Gordon model (BSG)  
has the boundary spectrum containing the states $\ket{n_{1},n_{2},\dots,n_{k}}$,
where $n_{1},n_{2},\dots,n_{k}$ are nonnegative integers satisfying
the condition $\pi/2\geq\nu_{n_{1}}>w_{n_{2}}>\nu_{n_{3}}>\dots\geq 0$,
where $\nu_{n}=\eta/\lambda-u_{2n+1}$, $w_{n}=\pi-\eta/\lambda-u_{2n-1}$,
and $0<\eta\leq\frac{\pi}{2}(\lambda+1)$ is a boundary parameter which can be
expressed in terms of $\phi_0$, $M_0$, $\beta$, $m$ (see \cite{sajat1}).
The fusion rules \cite{sajat1,Mattsson} are listed in Table \ref{tab.bsg}.

\begin{table}
\caption{\label{tab.bsg}Fusion rules of the boundary sine-Gordon model\hspace{6cm}}
\vspace{2mm}
\begin{tabular*}{16cm}{@{\extracolsep\fill}lccl}
\hline 
Initial state&
 Particle&
 Rapidity&
 Final state\tabularnewline
\hline
$|n_{1},...,n_{2k}\rangle$&
 $s,\bar{s}$&
 $i\nu_{n}$&
 $|n_{1},...,n_{2k},n\rangle$\tabularnewline
$|n_{1},...,n_{2k-1}\rangle$&
 $s,\bar{s}$&
 $iw_{n}$&
 $|n_{1},...,n_{2k-1},n\rangle$\tabularnewline
$|n_{1},...,n_{2k},n_{2k+1},...\rangle$&
 $B_{n}$&
 $i\frac{1}{2}(\nu_{l}-w_{n-l})$&
 $|n_{1},...,n_{2k},l,n-l,n_{2k+1},...\rangle$\tabularnewline
$|n_{1},...,n_{2k-1},n_{2k},...\rangle$&
 $B_{n}$&
 $i\frac{1}{2}(w_{l}-\nu_{n-l})$&
 $|n_{1},...,n_{2k-1},l,n-l,n_{2k},...\rangle$\tabularnewline
$|n_{1},...,n_{2k},...\rangle$&
 $B_{n}$&
 $i\frac{1}{2}(\nu_{-n_{2k}}-w_{n+n_{2k}})$&
 $|n_{1},...,n_{2k}+n,...\rangle$\tabularnewline
$|n_{1},...,n_{2k-1},...\rangle$&
 $B_{n}$&
 $i\frac{1}{2}(w_{-n_{2k-1}}-\nu_{n+n_{2k-1}})$&
 $|n_{1},...,n_{2k-1}+n,...\rangle$ \tabularnewline
\hline
\end{tabular*}
\end{table}

The first and second columns contain the initial boundary state and the
incoming bulk particle, the fusion angle times $i$ is shown in the third column and the final
state is shown in the fourth column.

Because of the presence of the kinks, only the cases 1 and 3 of Section \ref{sec.property1}
can apply.

In the decomposition (\ref{eq.fackb}) every subspace is one-dimensional and is
spanned by the boundary bound states introduced above.  

The first two lines show that the whole boundary spectrum can be generated
by kinks. Correspondingly, we associate to the BSG state $\ket{n_{1},n_{2},\dots,n_{k}}$
the supersymmetric part $W({\nu_{n_{1}},w_{n_{2}},\nu_{n_{3}},\dots,B_{\hlf}})$
(using the notation introduced earlier). Now we have to verify whether the fusion
rules given in the 3-6th lines are also valid for these supersymmetric
parts. This is easily done  using the rules given in Section \ref{sec.rules13}.
Let us consider the 3rd line first. Let $v=W({\nu_{n_{1}},\dots,w_{n_{2k}},\nu_{n_{2k+1}},\dots,B_{\hlf}})$,
$p=W_P(i\hlf(\nu_{l}-w_{n-l}))$ with mass parameter $u_{n}$, and $w_{2k}>\nu_{l}>w_{n-l}>\nu_{2k+1}$
and $p+v\rightarrow y$. In this case $L[p]=\{ i\nu_{l},-iw_{n-l}\}$,
so $L[y]=\{ i\nu_{n_{1}},\dots,iw_{n_{2k}},i\nu_{n_{2k+1}},\dots,i\nu_{l},iw_{n-l}\}$.
The 4th line is similar. Turning to the 5th line, let $v=W({\nu_{n_{1}},\dots,w_{n_{2k}},\dots,B_{\hlf}})$,
$p=W_P(i\hlf(\nu_{-n_{2k}}-w_{n+n_{2k}}))$ with mass parameter $u_{n}$, and $p+v\rightarrow y$.
Now $L[p]=\{ i\nu_{-n_{2k}},-iw_{n+n_{2k}}\}$, and because of $\nu_{-n_{2k}}+w_{n_{2k}}=\pi$
we have $L[y]=\{ i\nu_{n_{1}},\dots,iw_{n+n_{2k}},\dots\}$. The 6th
line is similar to the 5th line.

Condition (\ref{eq.feltetel 4}) is clearly satisfied for all boundary bound
states, so the supersymmetric factors of the reflection matrix blocks
do not have poles on the imaginary axis in the physical strip.

The Lagrangian function  and also the one-particle reflection matrix blocks and two-particle
S-matrix blocks of the BSG model contain two bulk and two boundary parameters.
A Lagrangian function for the boundary supersymmetric 
sine-Gordon model was written down in \cite{Nep1} (it can also be found in
\cite{BPT}), this Lagrangian function also contains four parameters only. The supersymmetric
reflection matrix 
constructed above (and in   \cite{BPT}), however, contains the
parameter $\gamma$ (see (\ref{gr st rep})) in addition, therefore one can expect that
$\gamma$ can be related to the other four parameters. This relation  is not yet
known. In particular,  it
is not known whether any of the angles $\nu_{n}$ and $w_{n}$
coincides with $\pi-\xi$ or not. The supersymmetric  transformation and scattering
properties and in particular the dimensions of the boundary bound state multiplets are modified by such a coincidence.

We remark that this section confirms  the results of \cite{BPT}.

\subsection{Boundary sinh-Gordon model\label{sec BShG}}
\markright{\thesubsection.\ \ BOUNDARY SINH-GORDON MODEL}

 The classical equation of
 motion for the sinh-Gordon  field $\Phi$, which is a real field,
 is
 \begin{equation}
 \label{eq.shg}
 \partial_t^2\Phi-\partial_x^2\Phi+
 \frac{\sqrt{8}m^2}{\beta}\sinh(\sqrt{2}\beta\Phi)=0\ ,
 \end{equation}
 where $m$ and $\beta$ are parameters.

 The sinh-Gordon model (ShG) can be restricted to the left half-line
 $-\infty\leq x\leq 0$
 without losing integrability by imposing the
  boundary condition
 \begin{equation}
  \left. \partial_x\Phi\right|_0=\frac{\sqrt{2}m}{\beta}
  \left(\epsilon_0e^{-\frac{\beta}{\sqrt{2}}\Phi(0,t)}-
 \epsilon_1e^{\frac{\beta}{\sqrt{2}}\Phi(0,t)}\right)\ ,
 \end{equation}
 where $\epsilon_0$ and $\epsilon_1$ are two additional parameters \cite{GZ,
 MacI}.

Some quantities of the sinh-Gordon and the  boundary sinh-Gordon (BShG) model
can be obtained from those of the sine-Gordon and boundary sine-Gordon model
by analytic continuation of their coupling constant $\beta$  to the imaginary axis.

The particle spectrum of the ShG model contains only one self-conjugate
particle $P$ with a two-particle S-matrix and reflection
matrix that can  be obtained from the corresponding matrices
of the first SG breather ($B_{1}$) by the analytic continuation mentioned
above. 
The BShG model also contains a series of finitely many boundary bound states $b_{n}$ found in
\cite{CorDel} (see also \cite{CorTaor}).
These states correspond to the BSG states $\ket{\nu_{0},w_{n}}$, $n=1\dots $
which are precisely those states that can be generated using $B_{1}$
only. $b_{0}$ is the ground state. The boundary  fusion rules are listed in
Table \ref{tab.bshg}. The parameter $\eta$ is determined by
$\epsilon_0$, $\epsilon_1$, $\beta$. For a more detailed
description we refer the reader to \cite{CorDel,CorTaor}.

\begin{table}
\caption{\label{tab.bshg}Fusion rules of the boundary sinh-Gordon model\hspace{6cm}}
\vspace{2mm}
\begin{tabular*}{16cm}{@{\extracolsep\fill}cccc}
\hline 
Initial state&
 Particle&
 Rapidity&
 Final state\\
\hline
$b_{n}$&
 $P$&
 $i(\frac{\eta}{\lambda}-\frac{\pi}{2}+\frac{\pi}{\lambda}n)$&
 $b_{n+1}$ \\
\hline
\end{tabular*}
\end{table}

Some quantities (the Lagrangian densities and the scattering and reflection
matrices, for example) of the supersymmetric BShG model are also obtained by the analytic continuation
above  \cite{AN}. This, however, implies $\lambda<0$ and  $M<0$, $u<0$ in the formula $m=2M\sin(u)$ for the particle
mass, i.e.\ the values of these parameters are not
in the range that we have considered. One consequence of this,
for example, is that the supersymmetric factor of the two-particle
S-matrix has a pole in the physical strip (canceled by
a zero of the non-supersymmetric factor) \cite{AN}.  Applying the rules of
Section \ref{sec.rules13} (cases 1 and 3) formally  we can nevertheless obtain 
 supersymmetric parts  $W_{n}$ for the states $b_{n}$, and  corresponding
reflection matrix factors. In this way the multiplets $W_{n}$ turn out to be two-dimensional in the supersymmetric BShG model if the value of $\xi$ is generic.

\subsection{Free particle on the half-line\label{sec.free part}}
\markright{\thesubsection.\ \ FREE PARTICLE ON THE HALF-LINE}

The supersymmetric factors of the scattering and ground state reflection
matrices of a massive real free boson can be obtained by taking the limit $\alpha\rightarrow0$,
which implies $M\rightarrow\infty$, $u\rightarrow0$, $\rho\rightarrow\pi$;
in the $(-)$ case $\xi\rightarrow\pi$, $R_{P2,e}^{(-)}\rightarrow R_{P1}^{(-)}$,
and in the $(+)$ case $\xi\rightarrow0$, $R_{P2,e}^{(+)}\rightarrow R_{P1}^{(+)}$.
The same result can be obtained by solving the Yang-Baxter equations.
$S_{PP}$ is also bijective in the physical strip. 

The dimension
of a boundary bound state multiplet created from the ground state in $n$ steps will be $2^{n}$ irrespectively of
the particular boundary fusion rules. The reflection matrix factors on
the boundary bound states are also free from physical strip poles
for any values of the fusion angles. However, if some value of the
boundary fusion angles occurs several times and we take into consideration the statistical
properties of the particles, then the dimensions of the boundary bound state multiplets
are modified.
  In
particular, if all the boundary fusion angles are the same (see \cite{CorTaor}),
then the all excited boundary bound state multiplets are 2-dimensional. 
This is consistent with the multiplicities
obtained by taking the zero bulk coupling limit of the boundary sinh-Gordon
model.

The boundary condition for the  real scalar field $\Phi$ of mass $m$ considered in
\cite{CorTaor} is
\begin{equation}
\notag
\left.\partial_x\Phi\right|_0=-m\lambda \Phi\ ,
\end{equation} 
where $\lambda$ is a coupling constant. The spectrum of boundary  states
consists of an 
infinite tower of states $b_n$, $n=1,2,\dots$, the boundary fusion rules are
$P+b_n \to b_{n+1}$, and the fusion angle is independent of $n$ ($P$ denotes
the bulk boson). In this case the dimension of the supersymmetric parts of
the $b_n$-s is 2.

\subsection{Boundary affine Toda field theories}
\markright{\thesubsection.\ \ BOUNDARY AFFINE TODA FIELD THEORIES}

As in \cite{DeliusGand}, only $a_n^{(1)}$ Toda theory will be considered.
The classical equation of motion for the $n$-component bosonic
field $\phi$ of  $a_n^{(1)}$  Toda theory is
\begin{equation}
\partial_t^2{\phi}-\partial_x^2{\phi}+\frac{m^2}{\beta}\sum_{i= 0}^n
\alpha_ie^{\beta\alpha_i\cdot \phi}= 0\ .
\label{eq.phieom}
\end{equation}
The $\alpha_i$, $i=1,\dots,n$ are the simple roots of the Lie algebra
$a_n=\text{sl}_{n+1}$ and $\alpha_0$ is minus the highest root.
The coupling constant
$\beta$ could be removed from the
equations of motion by rescaling the field and
therefore the coupling
constant plays a role only in the quantum theory.
We shall use the notation
$$B=\frac{1}{2\pi} \frac{\beta^2}{1+\frac{\beta^2}{4\pi}}\ .$$

It was discovered  \cite{corri94,bowco95} that
this equation of motion
can be restricted to the left half-line $x<0$ without
losing integrability if one imposes a boundary condition at
$x= 0$ of the form
\begin{equation}
\label{eq.bcond}
\left.\beta\,\partial_x\phi+\sum_{i= 0}^n C_i\,\alpha_i
e^{\beta\alpha_i\cdot\phi/2}\right|_{x= 0}= 0\ ,
\end{equation}
where the boundary parameters $C_i$
satisfy either
\begin{equation}
C_i = 0\;, \quad (i=0,1,\dots,n)\ ,
\end{equation}
which is the Neumann boundary condition, or
\begin{equation}
C_i = \pm 1\;, \quad (i=0,1,\dots,n)\ , 
\label{eq.boundcond}
\end{equation}
which will be denoted as the (++...-- --...) boundary
conditions.

\subsubsection{Boundary $a_{2}^{(1)}$ affine Toda field theory}

The bulk spectrum of this model contains two particles 1 and 2 of
equal mass $m_{1}=m_{2}$. Their fusion rules are $1+1\rightarrow2$
$(\pi/3)$ and $2+2\rightarrow1$ $(\pi/3)$. The anti-particle 
of 1 is 2. 
These rules are easily seen to be consistent with
the decomposition (see (\ref{eq.fack})) consisting of a single term $\langle
1,2 \rangle$ and with
 associating
supersymmetric particle representation to  $\langle
1,2 \rangle$ with $\alpha=-1/(2M)$,
$m_{1}=m_{2}=2M\sin(\pi/3)$ \cite{HolMav}. The labeling set for the
supersymmetric part of 1 and 2 denoted by $W_{12}$ is $L[W_{12}(\Theta)]=\{ \Theta-\frac{i\pi}{6}, \Theta+\frac{i\pi}{6}  \}$.
$M$ is determined by $\beta$ and $m$.

The boundary version of this model with any of the solitonic boundary conditions
$(+--)$, $(-+-)$ or $(--+)$  contains
the boundary states $b_{n,m}$ for all $n,m\in\ZZ$, $n+m\geq0$,
$-\frac{1}{2B}-\hlf<n,m<\frac{1}{2B}+\hlf$, and the states $b_{-n,n}$
and $b_{n,-n}$ for all $n\in\ZZ$, $0\leq\frac{3}{2B}+\hlf$, as described in \cite{DeliusGand}. We consider only generic values
of $B$ and only the domain $0<B<1$, as in \cite{DeliusGand}. $b_{0,0}$
is the ground state. The fusion rules are shown in Table \ref{tab.a21}.

\begin{table}
\caption{\label{tab.a21}Fusion rules of the boundary $a_2^{(1)}$ affine Toda
  field theory\hspace{6cm}}
\vspace{2mm}
\begin{tabular*}{16cm}{@{\extracolsep\fill}lccl}
\hline 
Initial state &
 Particle&
 Rapidity&
 Final state\tabularnewline
\hline
$b_{n,m}$&
 1&
 $i\frac{\pi}{6}-i\frac{\pi}{6}B(2n+1)$&
 $b_{n+1,m}$\tabularnewline
$b_{n,m}$&
 2&
 $i\frac{\pi}{6}-i\frac{\pi}{6}B(2m+1)$&
 $b_{n,m+1}$\tabularnewline
$b_{-n,n}$, $0\leq n$&
 1&
 $i\frac{\pi}{2}-i\frac{\pi}{6}B(2n+1)$&
 $b_{-n-1,n+1}$\tabularnewline
$b_{n,-n}$, $0\leq n$&
 2&
 $i\frac{\pi}{2}-i\frac{\pi}{6}B(2n+1)$&
 $b_{n+1,-n-1}$ \tabularnewline
\hline
\end{tabular*}
\end{table}

The decomposition (\ref{eq.fackb}) is $\oplus_{n,m} \langle b_{n,m}, b_{m,n}
\rangle$, where $n \ge m$.
The supersymmetric part associated to $\langle b_{n,m}, b_{m,n}
\rangle$ is denoted by $W_{n,m}$.
We assume, for the sake of simplicity, that the cases 1 and 3 of Section \ref{sec.property1} apply.
It is straightforward to verify that the fusion rules above imply
that the  labeling sets $L[W_{n,m}]$
are the following:
\begin{multline}
L[W_{n,m}]=
Abs \{ i(\frac{\pi}{3}-\frac{\pi}{6}B(2j+1)),\ 
i(\frac{\pi}{6}B(2j+1)),\  i(\frac{\pi}{3}-\frac{\pi}{6}B(2l+1)),\\
i(\frac{\pi}{6}B(2l+1))\,|\, j=0\dots n-1,\ l=0\dots m-1\}
\quad \mbox{if}\quad  n,m\geq 0,
\end{multline}
\begin{multline}
L[W_{n,-n}]=Abs \{
i(\frac{2\pi}{3}-\frac{\pi}{6}B(2j+1)),\ i(\frac{\pi}{3}-\frac{\pi}{6}B(2j+1))\,|\,
j=0\dots n-1\}
\end{multline}
\begin{multline}
L[W_{n,m}]=Abs \{ i(\frac{2\pi}{3}-\frac{\pi}{6}B(2j+1)),\ 
i(\frac{\pi}{3}-\frac{\pi}{6}B(2l+1)),\  i(\frac{\pi}{6}B(2r+1))\,|\,
j=0\dots (-m-1),\\ l=0\dots n-1,\ r=(-m)\dots n-1\}\quad
\mbox{if}\quad m<0\ , 
\end{multline}
where $Abs(ix)=i|x|$ if $x\in \RR$, and the limit for the running variables given on the left-hand side
of the dots is always assumed to be lower than or equal to the limit given on the
right-hand side, if this condition is not satisfied, then the set of allowed
values is empty. These conventions apply below as well.
The dimension of $W_{n,m}$ is $2^{n+m}$ if $n,m\geq0$ and
$2^{n}$ if $m<0$, provided that there is no
coincidence between $\pi-\xi$ and the rapidities above. $\pi-\rho_{max}/2=5\pi/6$,
so condition (\ref{eq.feltetel 4}) is satisfied and the supersymmetric
reflection matrix factors do not have poles on the imaginary
axis in the physical strip.

\subsubsection{Boundary $a_{4}^{(1)}$ affine Toda field theory}

The bulk spectrum of this model contains four particles 1, 2, 3, 4
of mass $m_{1}=m_{4}=2M\sin(\pi/5)$, $m_{2}=m_{3}=2M\sin(2\pi/5)$. 
The fusion rules are $a+b\rightarrow c$,
where either $c=a+b$ or $c=a+b-5$. The corresponding fusion angles
are $\frac{\pi}{5}(a+b)$ if $c=a+b$ and $\frac{\pi}{5}(10-a-b)$
if $c=a+b-5$. The anti-particle of 1 and 2 is 4 and 3. These
rules are easily seen to be consistent with 
the decomposition (see (\ref{eq.fack})) $\langle 1,4 \rangle \oplus  
\langle 2,3 \rangle$ and with 
associating supersymmetric
particle representations to $\langle 1,4 \rangle$ and $\langle 2,3 \rangle$
 with $\alpha=-1/(2M)$ \cite{HolMav}.
The labeling sets corresponding to the supersymmetric parts of
 $\langle 1,4 \rangle$ and $\langle 2,3 \rangle$
denoted by $W_{14}$ and $W_{23}$ are
$L[W_{14}(\Theta)]=\{ \Theta-i\frac{3\pi}{10},
\Theta+i\frac{3\pi}{10} \} $,
$L[W_{23}(\Theta)]=\{ \Theta-i\frac{\pi}{10},
\Theta+i\frac{\pi}{10} \} $.

The boundary version of this model described in \cite{DeliusGand} has
two classes of inequivalent solitonic boundary conditions to which
different boundary spectra belong. The first class contains the boundary
conditions $(-+++-), (+++--), (++--+), (--+++)$ and $(+--++)$,
the second class contains the boundary conditions $(- + + - +)$,
$(- + - + +)$, $(+ + - + -)$, $(- + - - -)$, $(+ - + + -)$, $(+ -
+ - +)$, $(- - + - -)$, $(- - - + -)$, $(- - - - +)$
and $(+ - - - -)$.

If the boundary condition belongs to the first class, then there are
the boundary states $b_{n_{2},n_{3}}$, $n_{2},n_{3}\in\ZZ$, $n_{2}+n_{3}\geq0$,
$-\frac{1}{2B}-\hlf<n_{2},n_{3}<\frac{1}{2B}+\hlf$, and $b_{n,-n}$
and $b_{-n,n}$ for all $n\in\ZZ$, $0\leq n<\frac{5}{2B}+\hlf$,
where $B$ is a parameter of the bulk model. Generic values of $B$
are considered in the domain $0<B<1$. The fusion rules, which are
analogous to those of the $a_{2}^{(1)}$ model, are listed in Table \ref{tab.a41f}.

\begin{table}
\caption{\label{tab.a41f}Fusion rules of the $a_4^{(1)}$ affine Toda
  field theory with first class boundary conditions}
\vspace{2mm}
\begin{tabular*}{16cm}{@{\extracolsep\fill}lccl}
\hline 
Initial state&
 Particle&
 Rapidity&
 Final state\tabularnewline
\hline
$b_{n_{2},n_{3}}$&
 2&
 $i\frac{\pi}{10}-i\frac{\pi}{10}B(2n_{2}+1)$&
 $b_{n_{2}+1,n_{3}}$\tabularnewline
$b_{n_{2},n_{3}}$&
 3&
 $i\frac{\pi}{10}-i\frac{\pi}{10}B(2n_{3}+1)$&
 $b_{n_{2},n_{3}+1}$\tabularnewline
$b_{-n,n}$, $0\leq n$&
 1&
 $i\frac{\pi}{2}-i\frac{\pi}{10}B(2n+1)$&
 $b_{-n-1,n+1}$\tabularnewline
$b_{n,-n}$, $0\leq n$&
 4&
 $i\frac{\pi}{2}-i\frac{\pi}{10}B(2n+1)$&
 $b_{n+1,-n-1}$ \tabularnewline
\hline
\end{tabular*}
\end{table}

The decomposition (\ref{eq.fackb}) is $\oplus_{n_2,n_3} \langle
b_{n_2,n_3},b_{n_3.n_2}  \rangle $, where $n_2 \ge n_3$.
The supersymmetric part associated to $ \langle
b_{n_2,n_3},b_{n_3.n_2}  \rangle $ is denoted by
$W_{n_{2},n_{3}}$. 
Again, we assume for the sake of simplicity that the cases 1 and 3 of Section \ref{sec.property1} apply.
We
have
\begin{multline}
L[W_{n_{2},n_{3}}]=Abs \{ i(\frac{\pi}{5}-\frac{\pi}{10}B(2j+1)),\
i(\frac{\pi}{10}B(2j+1)),\ i(\frac{\pi}{5}-\frac{\pi}{10}B(2l+1)),\\ i(\frac{\pi}{10}B(2l+1))\,|\, j=0\dots n_{2}-1,l=0\dots n_{3}-1\}
\quad \mbox{if}\quad n_{2},n_{3}\geq 0,
\end{multline}
\begin{multline}
L[W_{n,-n}]=Abs \{ i(\frac{4\pi}{5}-\frac{\pi}{10}B(2j+1)),\ i(\frac{\pi}{5}-\frac{\pi}{10}B(2j+1))\,|\, j=0\dots n-1\},
\end{multline}
\begin{multline}
L[W_{n_{2},n_{3}}]=Abs \{ i(\frac{4\pi}{5}-\frac{\pi}{10}B(2j+1)),\
i(\frac{\pi}{5}-\frac{\pi}{10}B(2l+1)),\ i(\frac{\pi}{10}B(2r+1))\,|\, \\
j=0\dots (-n_{3}-1),\ l=0\dots n_{2}-1,\ r=-n_{3}\dots n_{2}-1\}
\quad
\mbox{if}\quad n_{3}<0.
\end{multline}

The dimension of $W_{n_{2},n_{3}}$ is $2^{n_{2}+n_{3}}$ if $n_{2},n_{3}\geq0$ and
 $2^{n_{2}}$ if $n_{3}<0$, provided that
there is no coincidence between $i(\pi-\xi)$ and the rapidities above.

If $B$ is sufficiently small, then $W_{1,-1},W_{2,-2},\dots $
violate condition (\ref{eq.feltetel 4}), so the supersymmetric factor
of the reflection matrix of particle 1 on these states have poles
on the imaginary axis in the physical strip.

If the boundary condition belongs to the second class, then there are
the boundary states $b_{n_{1},n_{2},n_{3},n_{4}}$, $n_{1},n_{2},n_{3},n_{4}\in\ZZ$,
$n_{1}+n_{2}\geq0$, $n_{2}+n_{3}\geq0$, $n_{3}+n_{4}\geq0$. The ground state
is $b_{0,0,0,0}$.
 The
fusion rules are listed in Table \ref{tab.a41s}.

\begin{table}
\caption{\label{tab.a41s}Fusion rules of the $a_4^{(1)}$ affine Toda
  field theory with second class boundary conditions}
\vspace{2mm}
\begin{tabular*}{16cm}{@{\extracolsep\fill}lccl}
\hline 
Initial state&
 Particle&
 Rapidity&
 Final state\tabularnewline
\hline
$b_{n_{1},n_{2},n_{3},n_{4}}$&
 1&
 $i\frac{\pi}{10}-i\frac{\pi}{10}B(2n_{1}+1)$&
 $b_{n_{1}+1,n_{2},n_{3},n_{4}}$\tabularnewline
$b_{n_{1},n_{2},n_{3},n_{4}}$&
 2&
 $i\frac{\pi}{10}-i\frac{\pi}{10}B(2n_{2}+1)$&
 $b_{n_{1},n_{2}+1,n_{3},n_{4}}$\tabularnewline
$b_{n_{1},n_{2},n_{3},n_{4}}$&
 3&
 $i\frac{\pi}{10}-i\frac{\pi}{10}B(2n_{3}+1)$&
 $b_{n_{1},n_{2},n_{3}+1,n_{4}}$\tabularnewline
$b_{n_{1},n_{2},n_{3},n_{4}}$&
 4&
 $i\frac{\pi}{10}-i\frac{\pi}{10}B(2n_{4}+1)$&
 $b_{n_{1},n_{2},n_{3},n_{4}+1}$\tabularnewline
$b_{-n_{2},n_{2},n_{3},n_{4}}$, $0\leq n_{2}$&
 1&
 $i\frac{3\pi}{10}-i\frac{\pi}{10}B(2n_{2}+1)$&
 $b_{-n_{2}-1,n_{2}+1,n_{3},n_{4}}$\tabularnewline
$b_{n_{1},n_{2},n_{3},-n_{3}}$, $0\leq n_{3}$&
 4&
 $i\frac{3\pi}{10}-i\frac{\pi}{10}B(2n_{3}+1)$&
 $b_{n_{1},n_{2},n_{3}+1,n_{3}-1}$ \tabularnewline
\hline
\end{tabular*}
\end{table}

The decomposition (\ref{eq.fackb}) is $\oplus_{n_1,n_2,n_3,n_4} \langle
   b_{n_1,n_2,n_3,n_4} , b_{n_4,n_2,n_3,n_1}, b_{n_1,n_3,n_2,n_4} , b_{n_4,n_3,n_2,n_1}  
   \rangle $, where $n_1 \le n_4$, $n_2 \ge n_3$, and $b_{k,l,m,n}=0$ if the
   condition $k+l \ge 0$ and $l+m \ge 0$ and $m+n \ge 0$ is not satisfied. 
The supersymmetric part associated to $b_{n_1,n_2,n_3,n_4}$ is denoted by $W_{n_1,n_2,n_3,n_4}$.
Assuming, for the sake of simplicity, that the cases 1 and 3 of Section \ref{sec.property1} apply,
using our rules  we get
\begin{multline}
L[W_{n_{1},n_{2},n_{3},n_{4}}]=Abs \{
i(\frac{2\pi}{5}-\frac{\pi}{10}B(2l_{1}+1)),\
i(\frac{\pi}{5}+\frac{\pi}{10}B(2l_{1}+1)),\
i(\frac{2\pi}{5}-\frac{\pi}{10}B(2l_{4}+1)),\\
i(\frac{\pi}{5}+\frac{\pi}{10}B(2l_{4}+1)),\
i(\frac{\pi}{5}-\frac{\pi}{10}B(2l_{2}+1)),\ i(\frac{\pi}{10}B(2l_{2}+1)),\
i(\frac{\pi}{5}-\frac{\pi}{10}B(2l_{3}+1)),\\ i(\frac{\pi}{10}B(2l_{3}+1))\,|\,
l_{i}=0\dots n_{i}-1\}\quad 
\mbox{if}\quad n_{1},n_{2},n_{3},n_{4}\geq 0\ ,
\end{multline}
\begin{multline}
L[W_{n_{1},n_{2},n_{3},n_{4}}]=Abs \{
i(\frac{\pi}{5}-\frac{\pi}{10}B(2l_{3}+1)),\ i(\frac{\pi}{10}B(2l_{3}+1)),\
i(\frac{2\pi}{5}-\frac{\pi}{10}B(2l_{4}+1)),\\
i(\frac{\pi}{5}+\frac{\pi}{10}B(2l_{4}+1)),\ 
i(\frac{6\pi}{10}-\frac{\pi}{10}B(2j_{1}+1)),\ i(\frac{\pi}{10}B(2j_{2}+1)),\\
i(\frac{\pi}{5}-\frac{\pi}{10}B(2j_{3}+1))\,|\, l_{3}=0\dots n_{3}-1,\
l_{4}=0\dots n_{4}-1,\ j_{2}=0\dots n_{2}-1,\\ j_{1}=0\dots (-n_{1}-1),\ j_{3}=-n_{1}\dots n_{2}-1\}\quad
\mbox{if}\quad n_{3},n_{4},n_{2}\geq 0, n_{1}<0\ ,
\end{multline}
\begin{multline}
L[W_{n_1,n_2,n_3,n_4}]=Abs \{ 
i\frac{\pi}{10}B(2l_1+1),\ 
i(\frac{6\pi}{10}-\frac{\pi}{10}B(2l_1+1)),\
i\frac{\pi}{10}B(2l_4+1),\\ 
i(\frac{6\pi}{10}-\frac{\pi}{10}B(2l_4+1)),\
i(\frac{\pi}{5}-\frac{\pi}{10}B(2l_2+1)),\ 
i\frac{\pi}{10}B(2l_2+1),\\
i(\frac{\pi}{5}-\frac{\pi}{10}B(2l_2+1)),\ 
i\frac{\pi}{10}B(2l_2+1)\,|\, l_1=0\dots (-n_1-1),\ l_4=0\dots (-n_4-1),\\ 
l_2=-n_1\dots n_2-1,\ 
l_3=-n_4\dots n_3-1\}
\quad 
\mbox{if}\quad n_1,n_4<0\ .
\end{multline}

The dimension  of $W_{n_{1},n_{2},n_{3},n_{4}}$ is $2^{n_{1}+n_{2}+n_{3}+n_{4}}$
if $n_{1},n_{2},n_{3},n_{4}\geq0$,
$2^{n_2+n_3+n_4}$ if $n_1<0$, $n_2,n_3,n_4 \ge 0$
and $2^{n_{2}+n_{3}}$ if $n_{1},n_{4}<0$ provided that there is
no coincidence between $\pi-\xi$ and the rapidities above. Condition
(\ref{eq.feltetel 4}) is satisfied in this case, so the supersymmetric
factors of the reflection matrix blocks do not have poles on the imaginary
axis in the physical strip.

% ********************************* example10.tex vege *****************************

\section{Discussion}
\markright{\thesection.\ \ DISCUSSION}

We  studied the boundary supersymmetric bootstrap programme in a special
framework in which the blocks of the full two-particle  S-matrix and the
blocks of the full one-particle reflection matrix take a factorized
form. We assumed that the ground state is a singlet with RSOS label $\hlf$ and
the bulk particles transform in the kink and boson-fermion representations.

We introduced the boundary supersymmetry algebra  in
 the framework proposed by \cite{DeliusGeorge,DeliusMacKay}, which requires that
 the boundary supersymmetry algebra  be a co-ideal of the bulk
 supersymmetry algebra. 
It is a remarkable feature of the boundary supersymmetry algebra  that it admits
 essentially two possible boundary co-multiplications --- the corresponding two
 cases are denoted by $(+)$ and
 $(-)$. The two co-multiplications lead to different supersymmetric
ground state reflection matrix factors. We found that these factors are essentially
the same as those given in \cite{AK1,AK2,Chim}. Although the
two co-multiplications  appear to play symmetric role algebraically,
the corresponding kink reflection matrix factors turn out to be significantly
different. A further important difference between the two cases is
that in the $(+)$ case the boson-fermion reflection matrix factor can be
obtained by bootstrap from the kink reflection matrix factor only at special
values of its parameters \cite{AK2}. We also found that the kink
and boson-fermion reflection factors can be degenerate at particular
rapidities depending on a parameter $\gamma$ of the ground state
representation.

We presented supersymmetric boundary fusion rules by which the representations
and reflection matrix factors for excited boundary bound states can be easily
determined in specific models. The main difficulty of the problem of finding
such rules 
is to handle the degeneracies of the boundary fusion tensors that
occur at particular rapidities (resulting from the degeneracies of
the one-particle reflection matrix factors). These degeneracies are closely
related to the degeneracies of the bulk two-particle S-matrix factors
and of the ground state one-particle reflection matrix factors. 
We found
that the boundary fusion rules are analogous to the bulk rules of \cite{HolMav,Sch},
and that it is useful to characterize the boson-fermion multiplets
by their constituent kinks. The kink representation appears to be an
elementary object. 

For the sake of simplicity we assumed that the two-particle S-matrix
factors and ground state reflection matrix factors  have no
poles and overall zeroes on the imaginary axis in the physical strip and there is no interplay between the poles and zeroes of the supersymmetric
and non-supersymmetric factors of the S-matrix and reflection matrix.
We found, regarding the boundary part, that the main restriction on the applicability of the described
framework follows from this condition.

We applied our rules to the  sine-Gordon model, 
to the $a_{2}^{(1)}$ and $a_{4}^{(1)}$ affine Toda field theories,
to the free particle and to the sinh-Gordon model. We found that 
in the case of the $a_{4}^{(1)}$ model
with first class boundary condition the supersymmetric
reflection matrix factors on some excited boundary states have poles in the
physical strip.

It is a further problem to consider the axioms in 10) of Section
\ref{sec.fstb}. Initial steps in this direction 
were made in \cite{SM}.
The possible Coleman-Thun diagrams could be considered as well,
although we think that it is less likely that they yield further
constraints. 
Writing down formulae for the fusion and decay tensors is also a
task for the future.  

At the present stage it is an open problem to tell how large class of non-supersymmetric 
theories
can  be supersymmetrized under the assumption that the bulk particles transform
in the kink and boson-fermion representation and the ground state is a singlet
with label $1/2$, and, generally, how uniquely  supersymmetric theories  are characterized by the one-particle representations and the
ground state representation occurring in them.

In general,  representations beyond the kink and boson-fermion
for the bulk particles and other (possibly non-singlet) ground state
representations are also relevant to some models. 
We think that  at least two  lessons can be learned from our investigation
which are relevant for cases with other representations:
the first one is that the 
degeneracy properties of the supersymmetric ground state reflection matrix factor and the  supersymmetric two-particle
S-matrix factor have very important role in the fusion rules, the second one
is that one
should find  `elementary' representations for which the
mentioned degeneracies show a simple pattern and from which one can build the
other representations of interest by multiplication.  

% ******************************* disc4.tex vege ****************************

\section{Appendix}
\label{sec.app..}
\markright{\thesection.\ \ APPENDIX}

$$
G^{[i,j]}(\Theta)=R^{[i,j]}(\Theta)R^{[i,j]}(\pi i-\Theta)
$$

\begin{multline}
\notag
R^{[i,j]}(\Theta)=\frac{1}{\Gamma(\frac{\Theta}{2\pi
    i})\Gamma(\frac{\Theta}{2\pi
    i}+\hlf)}\prod_{k=1}^{\infty}\frac{\Gamma(\frac{\Theta}{2\pi
    i}+\Delta_{1}+k+1)\Gamma(\frac{\Theta}{2\pi
    i}-\Delta_{1}+k)}{\Gamma(\frac{\Theta}{2\pi
    i}+\Delta_{1}+k-\hlf)\Gamma(\frac{\Theta}{2\pi
    i}-\Delta_{1}+k+\hlf)}\times\\
\times\frac{\Gamma(\frac{\Theta}{2\pi
    i}+\Delta_{2}+k-\hlf)\Gamma(\frac{\Theta}{2\pi
    i}-\Delta_{2}+k-\hlf)}{\Gamma(\frac{\Theta}{2\pi
    i}+\Delta_{2}+k)\Gamma(\frac{\Theta}{2\pi i}-\Delta_{2}+k)}\ ,
\end{multline}
where 
$\Delta_{1}=(u_{i}+u_{j})/(2\pi)$, $\Delta_{2}=(u_{i}-u_{j})/(2\pi)$.

$$
K(\Theta)=\frac{1}{\sqrt{\pi}}\prod_{k=1}^{\infty}\frac{\Gamma(k-\frac{1}{2}+\frac{\Theta}{2\pi i})\Gamma(k-\frac{\Theta}{2\pi i})}{\Gamma(k+\frac{1}{2}-\frac{\Theta}{2\pi i})\Gamma(k+\frac{\Theta}{2\pi i})}$$

$$
P(\Theta)=\prod_{k=1}^{\infty}\left.\left[\frac{\Gamma(k-\frac{\Theta}{2\pi
      i})^{2}}{\Gamma(k-\frac{1}{4}-\frac{\Theta}{2\pi
      i})\Gamma(k+\frac{1}{4}-\frac{\Theta}{2\pi
      i})}\right/ \{\Theta\leftrightarrow-\Theta\}\right]
$$

$$
ZX^{(+)}(\Theta)=\sqrt{m}P(\Theta+i\rho/2)P(\Theta-i\rho/2)\sqrt{2}K(2\Theta)2^{-\Theta/(i\pi)}\
,
$$
 where $m=2M\cos(\frac{\rho}{2})$, $0\leq\rho<\pi$, $M=-1/\alpha$.

\begin{align}
\notag
&\tilde{Z}^{(-)}(\Theta)=K(2\Theta)2^{-\Theta/(i\pi)}F(\Theta-i\rho/2)F(\Theta+i\rho/2)\\
\notag
&F(\Theta)=P(\Theta)K(\Theta+i\xi)K(\Theta-i\xi)\ ,
\end{align}
where $\gamma=-2\sqrt{M}\cos\frac{\xi}{2}$.

$$
ZX^{(-)}(\Theta)=\sqrt{m}\tilde{Z}^{(-)}(\Theta)\frac{1}{\sqrt{2}}(\cos(\rho/2)-\cosh(\Theta))\
,
$$
where $\xi=\pi$. 

\begin{align}
\notag
&\tilde{Z}^{(+)}(\Theta)=iK(2\Theta)2^{-\Theta/(i\pi)}F(\Theta-i\rho/2)F(\Theta+i\rho/2)U(\Theta)\\
\notag
&F(\Theta)=P(\Theta)K(\Theta+i\xi)K(\Theta-i\xi)\ ,
\end{align}
where $\gamma=-2\sqrt{M}i\sin(\xi/2)$,
$$
U(\Theta)=f(\Theta)/f(-\Theta)
$$
$$
\frac{U(i\Theta)}{U(i\Theta-i\pi)}=-\frac{\cos(\Theta)-\cos(\rho/2)}{\cos(\Theta)+\cos(\rho/2)}
$$
$$
f(\Theta)=\prod_{k=1}^{\infty}\frac{\Gamma(\frac{\rho}{4\pi}-\frac{\Theta}{2\pi
    i}-\frac{1}{2}+k)\Gamma(\frac{\rho}{4\pi}+\frac{\Theta}{2\pi
    i}+k)\Gamma(-\frac{\rho}{4\pi}-\frac{\Theta}{2\pi
    i}-\frac{1}{2}+k)\Gamma(-\frac{\rho}{4\pi}+\frac{\Theta}{2\pi
    i}+k)}{\Gamma(\frac{\rho}{4\pi}-\frac{\Theta}{2\pi
    i}+k)\Gamma(\frac{\rho}{4\pi}+\frac{\Theta}{2\pi
    i}+\frac{1}{2}+k)\Gamma(-\frac{\rho}{4\pi}-\frac{\Theta}{2\pi
    i}+k)\Gamma(-\frac{\rho}{4\pi}+\frac{\Theta}{2\pi i}+\frac{1}{2}+k)}\ .
$$

% ********************************** appendix2.tex vege *******************************

\chapter{Truncation effects in the boundary flows of the Ising model on a strip} 
\label{sec.chap3}
\markboth{CHAPTER \thechapter.\ \  TRUNCATION EFFECTS IN ISING BOUNDARY FLOWS}{}

\section{On the definition of field theories on the strip}
\label{sec.def}
\markright{\thesection.\ \ FIELD THEORY ON THE STRIP}

By quantum field theory on the strip  $[0,L]\times \RR$ we generally mean a
collection of the following elements: a Hilbert space $\mH_B$ of states, a
Hamiltonian operator $H_B$ acting on the Hilbert space, a set of fields
$\Phi_B(x,t)$ which satisfy the time evolution equation 
\begin{equation}
\Phi_B(x,t)=\exp(iH_Bt)\Phi_B(x,0)\exp(-iH_Bt)\ ,
\end{equation}
a set of fields $\phi_B(t,0)$ defined on the left boundary $\{ 0 \}
\times \RR$ which satisfy the time evolution equation
\begin{equation}
\phi_B(0,t)=\exp(iH_Bt)\phi_B(0,0)\exp(-iH_Bt)
\end{equation}
and a
set of fields $\phi_B(t,L)$ defined on the right boundary $\{ L \}
\times \RR$ which satisfy the time evolution equation 
\begin{equation}
\phi_B(L,t)=\exp(iH_Bt)\phi_B(L,0)\exp(-iH_Bt)\ .
\end{equation} 
The factor
$[0,L]$ in  $[0,L]\times \RR$  represents the space, $\RR $ represents the time. 
Fields defined on one of the boundaries only are often called boundary fields.
We assume that the strip  $[0,L]\times \RR$ is equipped with Minkowski metric.

New boundary theories can be obtained from old ones by perturbing the
Hamiltonian operator:
\begin{equation}
\hat{H}_B=H_B+H_{I,B}\ ,
\end{equation}
where the perturbation $H_{I,B}$ is
often a sum of various kinds of terms.  A bulk
term in the perturbation takes the form 
\begin{equation}
H_{I,B}=g_1\int_{0}^L \ dx\  \Psi_B(x,0)\ ,
\end{equation}
a
boundary perturbation takes the form 
\begin{equation}
H_{I,B}=g_2\psi_B(L,0)\quad \mbox{or}\quad
H_{I,B}=g_3\psi_B(0,0)\ ,
\end{equation}
where $\Psi_B(x,t)$,  $\psi_B(L,t)$ and  $\psi_B(0,t)$ are
certain fields of the unperturbed theory and $g_1,g_2,g_3$ are coupling constants.

The fields of the perturbed theory, which are distinguished by a hat, are defined by the equations
\begin{align}
\label{eq.int1}
\hat{\Phi}_B(x,t) & =\exp(i\hat{H}_Bt)\Phi_B(x,0)\exp(-i\hat{H}_Bt)\\
\hat{\phi}_B(0,t) & =\exp(i\hat{H}_Bt)\phi_B(0,0)\exp(-i\hat{H}_Bt)\\
\label{eq.int3}
\hat{\phi}_B(L,t) & =\exp(i\hat{H}_Bt)\phi_B(L,0)\exp(-i\hat{H}_Bt)\ .
\end{align}
In particular, 
\begin{equation}
\label{eq.111}
\hat{\Phi}_B(x,0)=\Phi_B(x,0)\ ,\quad \hat{\phi}_B(0,0)=\phi_B(0,0)\ ,\quad
\hat{\phi}_B(L,0)=\phi_B(L,0)\ .
\end{equation}

The calculations of Section \ref{sec.eel} are done in this perturbed
Hamiltonian operator framework.

We remark that to define a quantum (and classical) field theory on a strip
it is usual in the literature to take a classical
Lagrangian function of the form
\begin{equation}
\label{eq.blagr}
\mathcal{L}=\int_0^L  \mathcal{L}_{bulk}(x)\ dx\
+\mathcal{L}_{b,1}(x=0)+\mathcal{L}_{b,2}(x=L)\ ,
\end{equation}
which contains a bulk term and boundary terms (the notation of the dependence
of $ \mathcal{L}_{bulk}$,
$\mathcal{L}_{b,1}$ and $\mathcal{L}_{b,2}$ on the fields is suppressed).  
The classical equations of motion are
derived from (\ref{eq.blagr}) via a variational
principle.
These equations of motion usually
consist of  bulk partial differential equations and certain boundary
conditions. Finally a quantization is carried out (see for example
\cite{GZ,C1,C2}).

It is also possible to remain in the framework of
the standard Lagrangian formalism without boundaries on the full Minkowski
space, in this case the presence of boundaries
is taken into consideration in the Lagrangian function by step functions and
Dirac-deltas (see \cite{Corrigan1}).   
One can also start with a classical Hamiltonian function (see e.g.\
\cite{G,BDG}) instead of a Lagrangian function.

Boundary conditions usually play an important role in the literature in the definition
of boundary field theories.

We remark finally that as far as we know, a complete, systematic and definitive discussion of the definition and
basic formalism of
 classical and quantum field theory with boundaries cannot be found in the
 literature.

\section{Unitary representations of the Virasoro algebra}
\label{sec.repvir}
\markright{\thesection.\ \ UNITARY REPRESENTATIONS OF THE VIRASORO ALGEBRA}

The Virasoro algebra with central charge $c$ is generated by the
elements $L_n$, $n\in \ZZ$ and the identity element $I$ and the relations
\begin{equation}
\label{Virasoro}
[L_n,L_m]=(n-m)L_{n+m}+\frac{c}{12}(n^3-n)\delta_{n+m,0}I\ .
\end{equation}
A unitary highest weight representation is characterized by the following
properties: there exists a unitary highest weight vector $\ket{h}$ which
satisfies the relations 
\begin{equation}
\label{eq.l1}
L_n\ket{h}=0\qquad \forall n>0
\end{equation}
\begin{equation}
\label{eq.l3}
L_0\ket{h}=h\ket{h}\ ,
\end{equation} 
and the scalar product is related to the generators by the following relation:
\begin{equation}
\label{eq.l2}
L_n^\dagger =L_{-n}\ .
\end{equation}

A highest weight unitary representation $M(c,h)$ is uniquely specified by the numbers
$c$ and $h$. $h$ is called the weight of the representation. The positive definiteness of the scalar product and the above requirements do not
permit any values for $(c,h)$. For $0\le c<1$ only the following  discrete set
is allowed:
\begin{equation}
c(m)=1-\frac{6}{m(m+1)}
\end{equation}
\begin{equation}
h_{p,q}(m)=\frac{[(m+1)p-mq]^2-1}{4m(m+1)}
\end{equation}
where $m,p,q$ are integers, $m=2,3,4,5,6,\dots $, 
$1\le p<m$, $1\le q<m+1$. The pair of numbers $(p,q)$ is called Kac-label.  Each
weight appears twice since $h(p,q)=h(m-p,m+1-q)$.

We restrict to conformal field theories the  Hilbert space of which is a
 finite sum
 $\mH=\oplus_i M(c,h_i)$, where $(c,h_i)$ belongs to the above discrete series.

A construction for these representations is the following:
we take a vector $\ket{h}$ with the properties $\braket{h}{h}=1$, (\ref{eq.l1}), (\ref{eq.l3}),
and we take the vectors   
\begin{equation}
\label{eq.vbas}
L_{-n_1}L_{-n_2}\dots L_{-n_k}\ket{h}\qquad n_1\ge n_2\ge \dots  \ge n_k>0 
\end{equation}
which are assumed to be linearly independent. 
These vectors (including $\ket{h}$) form a basis of a representation $V(c,h)$ of the Virasoro
algebra. They are also eigenvectors of $L_0$ with eigenvalue
\begin{equation}
h+n_1+n_2+\dots +n_k\ .
\end{equation}
We say that the level of the vector $L_{-n_1}L_{-n_2}\dots L_{-n_k}\ket{h}$ is
 $n_1+n_2+\dots +n_k$. The space spanned by the eigenvectors of level $i$ is
 denoted by $V_i(c,h)$.  
 $V(c,h)$ can be written as
\begin{equation}
V(c,h)=\oplus_{i=0}^{\infty} V_i(c,h)\ .
\end{equation}
A
unique scalar product is determined on the representation $V(c,h)$ by  (\ref{eq.l1}),
(\ref{eq.l3}), (\ref{eq.l2}) and (\ref{Virasoro}), in particular the scalar
product of any two basis vectors can be calculated recursively using these
formulae. The strategy for this calculation is to use the commutation
relations  (\ref{Virasoro}) to move $L$-s with positive index to the right and
$L$-s with negative index to the left. This scalar product is not guaranteed
to be positive definite for general values of $(c,h)$. For the above discrete values of $(c,h)$ it is
positive semidefinite, there is a unique maximal subspace $V^{null}(c,h)$
which is orthogonal to the whole $V(c,h)$.  $V^{null}(c,h)$ decomposes as
\begin{equation}
 V^{null}(c,h)=\oplus_{i=0}^{\infty} V^{null}_i(c,h)
\end{equation}
where  $V^{null}_i(c,h)=0$ may hold for some values of $i$.
The unitary irreducible representation $M(c,h)$  is obtained by
quotienting  out  $V^{null}(c,h)$,
\begin{equation}
M(c,h)=V(c,h) /  V^{null}(c,h) = \oplus_{i=0}^{\infty}\ ( V_i(c,h) /
V_i^{null}(c,h))\ .
\end{equation}
The corresponding projection is denoted by $q: V(c,h)\to M(c,h)$.
The scalar product on $M(c,h)$ is defined in the standard way, i.e.\ by the
requirement that $\braket{v}{w}=\braket{q(v)}{q(w)}$. 

\subsection{Primary fields} 
\label{sec.repvir1}

A (chiral) Virasoro primary field of weight $l$ is a field $\Phi(z)$, $z\in\CC$ that
satisfies the property
\begin{equation}
\label{primary}
[L_m,\Phi(z)]=l(m+1)z^m\Phi(z)+z^{m+1}\frac{\partial\Phi}{\partial z}(z)\ .
\end{equation}
The matrix elements of a primary field satisfy the relation 
\begin{equation}
\label{primary2}
\brakettt{A}{\Phi(z)}{B}=\frac{\brakettt{A}{\Phi(1)}{B}}{z^{l+h_B-h_A}}\ ,
\end{equation}
where $l$ is the weight of $\Phi(z)$, $\ket{A}$ and $\ket{B}$ are eigenvectors
of $L_0$ with eigenvalues $h_A$, $h_B$. 

Equations   (\ref{eq.l1}),
(\ref{eq.l3}), (\ref{eq.l2}), (\ref{Virasoro}),  (\ref{primary}),
(\ref{primary2}) determine $\frac{\brakettt{A}{\Phi(1)}{B}}{\brakettt{h_1}{\Phi(1)}{h_2}}$ uniquely, where
$\ket{A}$, $\ket{B}$ are vectors of the form (\ref{eq.vbas}) with $h_1$ and
$h_2$, respectively. It is assumed here that $\brakettt{h_1}{\Phi(1)}{h_2} \ne 0$, if
$\brakettt{h_1}{\Phi(1)}{h_2} = 0$, then $\brakettt{A}{\Phi(1)}{B}=0$ $\forall\  \ket{A}\in V(c,h_1)$,
$\ket{B}\in V(c,h_2)$.   
 $\frac{\brakettt{A}{\Phi(1)}{B}}{\brakettt{h_1}{\Phi(1)}{h_2}}$ 
can be calculated recursively
using the equations   (\ref{eq.l1}),
(\ref{eq.l3}), (\ref{eq.l2}), (\ref{Virasoro}),  (\ref{primary}),
(\ref{primary2}), the strategy of the calculation being again to use
(\ref{Virasoro}) and   (\ref{primary})  to move $L$-s with positive index to the right and
$L$-s with negative index to the left.

Although the matrix elements  $\frac{\brakettt{A}{\Phi(1)}{B}}{\brakettt{h_1}{\Phi(1)}{h_2}}$,  where $\ket{A}\in V(c,h_1)$,
$\ket{B}\in V(c,h_2)$, are determined uniquely, they determine the matrix
elements 
   $\frac{\brakettt{q(A)}{\Phi(1)}{q(B)}}{\brakettt{h_1}{\Phi(1)}{h_2}}$  
unambiguously only if   $\brakettt{A}{\Phi(1)}{B}=0$ whenever  
 $\ket{A}\in V^{null}(c,h_1)$ or 
$\ket{B}\in V^{null}(c,h_2)$. This condition restricts the value of $l$ if
$c,h_1,h_2$ are fixed. If $l$ has other value, then  
 $\brakettt{A}{\Phi(1)}{B}=0$   $\forall\ \ket{A}\in M(c,h_1)$ and
$\ket{B}\in M(c,h_2)$. 

The permitted values of $l$ are
given by the Verlinde fusion rule.
If $h_1=h_{r,s}(m)$ and $h_2=h_{p,n}(m)$, then the permitted values  are
$h_{k,j}(m)$, where the numbers $k$ and $j$ take the values given by the following equations:
\begin{align}
\label{eq.frules1}
& k  = 1+|r-p|\ \dots\ k_{max},\quad  k+r+p=1\ mod\ 2   \\
& j  = 1+|s-n|\ \dots\ j_{max},\quad j+s+n=1\ mod\ 2   \\
& k_{max}=min(r+p-1,\ 2m-1-r-p)\\
\label{eq.frules2}
& j_{max}=min(s+n-1,\ 2m+1-s-n)\ .
\end{align}
It is  usual to write the Verlinde fusion rules as
\begin{equation}
h_{a}(m) \times h_{b}(m) = \sum_c n_{ab}^c h_c(m),\ 
\end{equation}
where the Verlinde fusion numbers $n_{ab}^c$ are either 0 or 1;  $a,b,c$
stand for Kac labels, e.g.\ $a=(r,s)$, $b=(p,n)$, $c=(k,j)$. The summation is
done over the possible Kac labels modulo the relation $(p,q) \sim (m-p,m+1-q)$.
 The values of $c$ for which
$n_{ab}^c$ is 1 are determined by (\ref{eq.frules1})-(\ref{eq.frules2}).

We refer the reader to the literature on conformal field theory  (e.g.\ \cite{FMS}) for
the description of descendant fields.

\subsection{c=1/2}
\label{sec.repvir2}

The value of the central charge that is relevant for the model (\ref{eq.yyy}) is
$c=1/2$, this  corresponds to $m=3$. The values of the weights $h_{p,q}(3)$ are
\begin{align}
& h_{1,1}(3)=0 && h_{1,2}(3)=1/16 && h_{1,3}(3)=1/2\\
& h_{2,1}(3)=1/2 && h_{2,2}(3)=1/16 && h_{2,3}(3)=0\ , 
\end{align}
i.e.\ there are three different unitary highest weight representations of the
Virasoro algebra, the weights of these representations are $0,1/2,1/16$.
The dimensions of the subspaces of definite level of these  representations
are listed  in Table \ref{tab.lev} (taken from \cite{CH}) up to level 15.
\begin{table}
\caption{\label{tab.lev}Dimensions of levels of $c=1/2$ unitary representations of the
  Virasoro algebra}
%\begin{center}
\vspace{2mm}
\begin{tabular*}{16cm}{@{\extracolsep\fill}llcccccccccccccccc}
\hline
&\vline\ Level & 0 & 1 & 2 & 3 & 4 & 5 & 6 & 7 & 8 & 9 & 10 & 11 & 12 & 13 &
14 & 15 \\
\hline
$h=0$ & \vline\ dimension & 1 & 0 & 1 & 1 & 2 & 2 & 3 & 3 & 5 & 5 & 7 & 8 & 11
& 12 & 16 & 18 \\
$h=1/2$  & \vline\ dimension & 1 & 1 & 1 & 1 & 2 & 2 & 3 & 4 & 5 & 6 & 8 & 9
& 12 & 14 & 17 & 20 \\
$h=1/16$  & \vline\ dimension & 1 & 1 & 1 & 2 & 2 & 3 & 4 & 5 & 6 & 8 & 10 &
12 & 15 & 18 & 22 & 27\\
\hline 
\end{tabular*}
%\end{center}
\end{table}

The fusion rule (\ref{eq.frules1})-(\ref{eq.frules2})  written in the usual form is
\begin{alignat}{6}
& 0 \times 0  && = 0\qquad && 0 \times 1/2 &&= 1/2\qquad && 0\times 1/16 &&= 1/16\\
& 1/2 \times 1/2 && = 0\qquad && 1/2 \times 1/16 && = 1/16\qquad && 1/16 \times 1/16 &&= 0+1/2
\end{alignat}
in this case.

\section{Conformal field theory on the strip}
\label{sec.CFTstrip}
\markright{\thesection.\ \ CONFORMAL FIELD THEORY ON THE STRIP}

In this section we discuss some basic properties of boundary conformal field
theory on the strip $[0,L]\times \RR$. We restrict to those elements which are
necessary for our work.

The Hilbert space is a sum of irreducible
unitary highest weight representations of the Virasoro algebra
(\ref{Virasoro})  with  common
central charge $c$. 
We restrict to  the cases when the Hilbert space is a finite sum of the Virasoro algebra
representations belonging to the unitary minimal series described in Section (\ref{sec.repvir}). 

The Hamiltonian operator of the theory is $\frac{\pi}{L}L_0$, where $L_0$ is the $L_0$ element of the Virasoro
algebra. A term proportional to the identity operator may also be added.

One can have right and left moving primary fields $\Phi_R(x,t)$,
$\Phi_L(x,t)$  which depend only on
$t-x$ or $t+x$, respectively, and satisfy the equations
\begin{align}
\label{eq.pri1}
[L_n,\Phi_R(x,t)] & =  D_n(l)\Phi_R(x,t)\\
\label{eq.pri2}
[L_n,\Phi_L(x,t)] & =  \bar{D}_n(l)\Phi_L(x,t)\ .
\end{align}
$l$ is the weight of these fields, and  
$D_n(l)$ and $\bar{D}_n(l)$ are operators on functions on the strip:
\begin{align}
\label{eq.d1}
D_n(l) & =   \frac{-iL}{2\pi}e^{i\frac{n\pi}{L}(t-x)}\partial_-  
+lne^{i\frac{n\pi}{L}(t-x)}\\
\label{eq.d2}
\bar{D}_n(l) & =  \frac{-iL}{2\pi}e^{i\frac{n\pi}{L}(t+x)}\partial_+  
+lne^{i\frac{n\pi}{L}(t+x)}\ ,
\end{align}
where $\partial_-=\partial_t-\partial_x$,
$\partial_+=\partial_t+\partial_x$.

A boundary primary field of weight $l$ on the left boundary is a field 
$\Phi_B(0,t)$ that is
defined on the left boundary and satisfies
\begin{alignat}{2}
\label{eq.prib1}
[L_n,\Phi_B(0,t)] & = D_n^0(l)\Phi_B(0,t) && =
[ \frac{-iL}{\pi}e^{i\frac{n\pi}{L}t}\partial_t
+lne^{i\frac{n\pi}{L}t}]\Phi_B(0,t)\ .
\end{alignat}
Similarly, 
a boundary primary field of weight $l$ on the right boundary is a field 
$\Phi_B(L,t)$ that is
defined on the right boundary and satisfies
\begin{alignat}{2}
\label{eq.prib2}
[L_n,\Phi_B(L,t)] & =  D_n^L(l)\Phi_B(L,t) && =
(-1)^n[ \frac{-iL}{\pi}e^{i\frac{n\pi}{L}t}\partial_t
+lne^{i\frac{n\pi}{L}t}]\Phi_B(L,t)\ .
\end{alignat}

The restriction of left or right moving primary fields of weight $l$ to any one of the boundaries will be  boundary
fields of weight  $l$.

$D_n(l)$, $\bar{D}_n(l)$, $D^0_n(l)$ and $D^L_n(l)$ satisfy the identities
\begin{align}
[D_n(l), \bar{D}_m(l)] & =  0\\{}
[D_n(l),D_m(l)] & =  -(n-m)D_{n+m}(l)\\{}
[\bar{D}_n(l),\bar{D}_m(l)] & =  -(n-m)\bar{D}_{n+m}(l)\\{}
[D^0_n(l),D^0_m(l)] & =  -(n-m)D^0_{n+m}(l)\\{}
[D^L_n(l),D^L_m(l)] & =  -(n-m)D^L_{n+m}(l)
\end{align}
which are compatible with (\ref{Virasoro}) and 
(\ref{eq.pri1}), (\ref{eq.pri2}), 
(\ref{eq.prib1}), (\ref{eq.prib2}).

The matrix elements of the right and left moving primary fields and boundary
primary fields of weight $l$ satisfy the equations
\begin{align}
\label{eq.v1}
\brakettt{A}{\Phi_R(x,t)}{B} & =  \brakettt{A}{\Phi_R(0,0)}{B}\exp[(h_A-h_B)\frac{i\pi}{L}(t-x)]\\
\brakettt{A}{\Phi_L(x,t)}{B} & = 
\brakettt{A}{\Phi_L(0,0)}{B}\exp[(h_A-h_B)\frac{i\pi}{L}(t+x)]\\
\brakettt{A}{\Phi_B(0,t)}{B} & =  \brakettt{A}{\Phi_B(0,0)}{B}
\exp[(h_A-h_B)\frac{i\pi}{L}t]\\
\label{eq.v2}
\brakettt{A}{\Phi_B(L,t)}{B} & =  \brakettt{A}{\Phi_B(L,0)}{B}
\exp[(h_A-h_B)\frac{i\pi}{L}t]\ ,
\end{align}
where $\ket{A}$ and $\ket{B}$ are eigenstates of $L_0$ with eigenvalues $h_A$
and $h_B$.

Introducing the new variables 
$z=e^{i\frac{\pi}{L}(t-x)}$, $\bar{z}=e^{i\frac{\pi}{L}(t+x)}$,
$y=e^{i\frac{\pi}{L}t}$ and 
the operators 
$\hat{\Phi}_R(z)=z^{-l}\Phi_R(z)$, 
$\hat{\Phi}_L(\bar{z})=\bar{z}^{-l}\Phi_L(\bar{z})$,
$\hat{\Phi}_B(0,y)=y^{-l}\Phi_B(0,y)$, 
$\hat{\Phi}_B(L,y)=y^{-l}\Phi_B(L,y)$, 
and 
allowing $z$, $\bar{z}$
and $y$ to take any values from $\CC$, 
we have
\begin{align}
[L_n,\hat{\Phi}_R(z)] & =  ( z^{n+1}\partial_z  
+l(n+1)z^n) \hat{\Phi}_R(z)\\
[L_n,\hat{\Phi}_L(\bar{z})] & =  ( \bar{z}^{n+1}\partial_{\bar{z}} 
+l(n+1)\bar{z}^n) \hat{\Phi}_L(\bar{z}) \\
[L_n,\hat{\Phi}_B(0,y)] & =   (y^{n+1}\partial_y  
+l(n+1)y^n) \hat{\Phi}_B(0,y)\\
[L_n,\hat{\Phi}_B(L,y)]  & =  (-1)^n[ y^{n+1}\partial_{y} 
+l(n+1)y^n] \hat{\Phi}_B(L,y)
\end{align}
\begin{align}
\brakettt{A}{\hat{\Phi}_R(z)}{B} & =  \brakettt{A}{\hat{\Phi}_R(1)}{B}z^{h_A-h_B-l}\\
\brakettt{A}{\hat{\Phi}_L(\bar{z})}{B} & = 
\brakettt{A}{\hat{\Phi}_L(1)}{B}\bar{z}^{h_A-h_B-l}  \\
\brakettt{A}{\hat{\Phi}_B(0,y)}{B} & =  \brakettt{A}{\hat{\Phi}_B(0,1)}{B}
y^{h_A-h_B-l}\\
\brakettt{A}{\hat{\Phi}_B(L,y)}{B} & =  \brakettt{A}{\hat{\Phi}_B(L,1)}{B}
y^{h_A-h_B-l}\ ,
\end{align}
which are the standard formulae for chiral primary fields in conformal field
theory.

\subsection{Boundary conditions of the minimal models}

Certain conformal field theories on the strip can be obtained by imposing so called
conformal boundary conditions on conformal minimal models. J.\ Cardy investigated
this construction and introduced a certain type of elementary boundary conditions often
called 'Cardy boundary conditions'.
A complete classification of possible Cardy boundary conditions of minimal
models has been given in \cite{BPPZ1,BPPZ2,BPZ}; see also \cite{Watts2}. 
 For the A-type unitary minimal models  these boundary conditions are in one-one correspondence with the
irreducible representations of the Virasoro algebra belonging to the unitary
discrete series and so we can label both boundary
conditions and representations from the same set, e.g.\ the set of Kac
labels.

The Hilbert space of a model with boundary condition $\alpha$ on the left-hand
side and boundary condition $\beta$ on the right-hand side decomposes into the
following sum of Virasoro algebra representations:
\begin{equation}
\mH_{\alpha\beta}=\oplus_c\ n_{c\alpha}^\beta R_c
\end{equation}
where $n_{c\alpha}^\beta$ are Verlinde fusion numbers (see Section
\ref{sec.repvir1}). $R_c$ denotes the representation belonging to the label $c$.

In particular, if the boundary condition is $0$ on one side and $h$ on the
other side, then the Hilbert space will consist of the single representation
$0\times h=h$. The possible primary boundary fields are $h\times h$. In the
case that we will investigate
$c=1/2$, $h=1/16$, and we have $1/16 \times 1/16 = 0 + 1/2$, i.e.\ two nonzero boundary
primary fields exist (up to normalization), one has weight $0$ (this is the constant identity field), the other has weight $1/2$. 

The $1/16$ Cardy boundary condition is also known as the `free' boundary
condition, the $0$ and $1/2$ boundary conditions are known as `fixed up' and
`fixed down' boundary conditions \cite{Cardy1}. These boundary conditions
correspond to letting the boundary spins free or to fix them in the two
possible directions.

\subsection{Boundary flows}
\label{sec.bflows}

The  flows we shall consider are trajectories of the form  
\begin{equation}
\label{eq.flow}
H=g_0(t)H_0+g_1(t)H_1+g_2(t)H_2+\dots 
\end{equation}
in the space of possible Hamiltonian operators, where  the $g_i$
coupling constants are functions of a parameter $t$. Often $g_0(t)=1$, $t$
varies form $0$ to $\infty$, and in the simplest situations there is only one
perturbing term on the right-hand side of (\ref{eq.flow}) and $g_1(t)=t$.

In the context of perturbed conformal field theory on a strip a boundary 
renormalization group (RG)
flow is a trajectory of the form
\begin{equation}
\label{eq.rgflow1}
H=\frac{\pi}{L}L_0+g_1L^{-l_1}\phi_1(t=0)+g_2L^{-l_2}\phi_2(t=0)+\dots 
\end{equation} 
where the length of the line segment $L$ plays the role of the parameter of
the trajectory, $\phi_1,\phi_2,\dots $ are boundary perturbations with
definite weights $l_1,l_2,\dots $. 
Such flows are associated with changing the width of the strip.

A perturbation is called relevant if $l<1$, irrelevant if $l>1$, marginal if $l=1$.
Relevant operators generate RG flows away from  boundary conformal field
theory, irrelevant operators  generate RG
flows into  boundary conformal field
theory at small values of $L$.  In case of marginal  operators a
more detailed investigation is required to determine whether they generate
flows away from or into   boundary conformal field
theory or whether they generate a flow within the space of boundary conformal
field theories.

A massless RG flow is an RG flow in which nontrivial conformal symmetry is
restored in the endpoint (e.g.\ $L\to \infty$). Values of $L$ at which the
theory described by (\ref{eq.rgflow1})  
has conformal
symmetry are called fixed points.    

In the simplest case when there is only one perturbation, e.g.\ in the case of
(\ref{eq.yyy}), one can keep $L$
fixed and use the coupling constant instead as a parameter of the flow.

As we mentioned in the Introduction, a problem of interest, e.g.\ in string
theory, is to find the possible RG flows, i.e.\ to find the triplets 
consisting of a BCFT (boundary conformal field theory) which is the starting point of the flow, a boundary
perturbation which generates the flow and which is usually a relevant
boundary operator of the starting BCFT, and a boundary BCFT which is the
endpoint of the flow.

One can also include bulk perturbation terms into $H$, in this chapter, however,
we shall consider boundary flows only.
In Chapter \ref{sec.chap4} we shall see an example for renormalization group
flows of models defined on the cylinder, where the perturbations will be bulk perturbations.

\section{The Truncated Conformal Space Approach}
\label{sec.tcsades}
\markright{\thesection.\ \ THE TRUNCATED CONFORMAL SPACE APPROACH}

The method called Truncated Conformal Space Approach (TCSA) is a numerical
method for the 
computation of the eigenvalues and eigenvectors of  Hamiltonian operators of
perturbed conformal field theories.

It was introduced by Yurov and Al.\ Zamolodchikov in
1990 in \cite{Yurov-Zam} to study bulk perturbations of conformal field
theories, and it was
first applied to boundary perturbations by Dorey et al in \cite{Watts3}. A
modified version 
applied to a perturbed massive free field theory can be found in \cite{YZ2}.  
In this section we  describe the TCSA method in a general form that is
not tied to conformal field theory.

The idea is to compute the matrix of the Hamiltonian operator in a suitable
basis, to take a finite
corner of it and to calculate the eigenvectors and eigenvalues numerically.    

A detailed description of the method is the following:
let us assume that we want to obtain the eigenvalues and eigenvectors of a
Hamiltonian operator of the form $H_0+\lambda H_I$, where $\lambda$ is a
coupling constant,  $H_0$ is a Hamiltonian operator with known discrete
spectrum $E_0<E_1<E_2<\dots $ and the eigenspaces $\mH_i$ belonging to the eigenvalues
are finite dimensional. We also assume that a basis (which is not necessarily
orthonormal) of each eigenspace is
given, the elements of which are denoted by  $v_i^n$, $n=1\dots \dim \mH_i$ for
$\mH_i$. We assume that the matrix elements $h_{I,ij}^{nm}=\brakettt{v_i^n}{H_I}{v_j^m}$ of
$H_I$ and the scalar product matrix $g^{nm}_{ij}=\braket{v_i^n}{v_j^m}$ are
also known (in particular, $g$ is block-diagonal: $g^{nm}_{ij}=0$ if $i\ne j$). 

The entries of the matrix of $H_I$ with respect to the basis $\{ v_i^n \} $ are determined by
the above data in the following way: 
\begin{equation}
H_I\ket{v_j^m}=\sum_{i=0}^\infty \sum_{n=1}^{\dim \mH_i} \tilde{h}_{I,ij}^{nm}
\ket{v_i^n}\ ,
\end{equation} 
where $\tilde{h}_{I,ij}^{nm}$ denotes the entries of the matrix of $H_I$ we are looking
for,
\begin{equation}
h_{I,kj}^{lm}=
\brakettt{v_k^l}{H_I}{v_j^m}= \sum_{i=0}^\infty \sum_{n=1}^{\dim \mH_i}
\tilde{h}_{I,ij}^{nm}\braket{v_k^l}{v_i^n}=  \sum_{i=0}^\infty \sum_{n=1}^{\dim \mH_i} g_{ki}^{ln}
\tilde{h}_{I,ij}^{nm}\ ,
\end{equation}
so
\begin{equation}
\tilde{h}_{I,sj}^{rm}=  \sum_{k=0}^\infty \sum_{l=1}^{\dim \mH_k}
(g^{-1})_{sk}^{rl} h_{I,kj}^{lm}\ ,
\end{equation}
where $g^{-1}$ is the inverse matrix of $g$: 
\begin{equation}
 \sum_{k=0}^\infty \sum_{l=1}^{\dim \mH_k}
 (g^{-1})_{sk}^{rl}g_{ki}^{ln}=\delta_{si}\delta_{rn}\ .
\end{equation}
$g^{-1}$ is block-diagonal, the block $(g^{-1})_{ii}^{nm}$ (where $i$ is fixed)
is the inverse of
the block $g_{ii}^{nm}$. The blocks of $g^{-1}$ have finite sizes $\dim \mH_1,
\dim \mH_2,\dots $, so they can be calculated numerically by a computer.  

The entries of the matrix of the operator
\begin{equation}
H^{TCSA}(n_c)=P_{n_c}(H_0+\lambda H_I)P_{n_c}|_{\mH(n_c)}
\end{equation}
are
\begin{equation}
\tilde{h}_{0,ij}^{nm}+\lambda \tilde{h}_{I,ij}^{nm}=
E_i\delta_{ij}\delta_{nm}+\lambda \tilde{h}_{I,ij}^{nm}=
E_i\delta_{ij}\delta_{nm}+ \lambda \sum_{k=0}^\infty \sum_{l=1}^{\dim \mH_k}
(g^{-1})_{ik}^{nl} h_{I,kj}^{lm}\ ,
\end{equation}
where $i,j\le n_c$, $P_{n_c}$ is the orthogonal projector onto the finite
dimensional subspace $\mH(n_c)=\oplus_{i=0}^{n_c} \mH_i$ and
$\tilde{h}_{0,ij}^{nm}=E_i\delta_{ij}\delta_{nm}$ are the entries of the
matrix  of $H_0$
(as well as of $P_{n_c}H_0P_{n_c}|_{\mH(n_c)}$ for $i,j \le n_c$).  
$n_c$ is called truncation level.
Using the property that  $g^{nm}_{ij}=0$ if $i\ne j$ 
we can also write
\begin{equation}
\label{eq.jhjh}
\tilde{h}_{0,ij}^{nm}+\lambda \tilde{h}_{I,ij}^{nm}=
E_i\delta_{ij}\delta_{nm}+ \lambda \sum_{k=0}^{n_c} \sum_{l=1}^{\dim \mH_k}
(g^{-1})_{ik}^{nl} h_{I,kj}^{lm}\ .
\end{equation}
The second term on the right-hand side is just a product of two finite matrices which are
obtained by simply taking the $\dim \mH (n_c) \times \dim \mH (n_c) $ corner of the
infinite matrices $g^{-1}$ and $h_I$. Therefore the right-hand side of
(\ref{eq.jhjh}) can be calculated and diagonalized numerically  by a computer.    
In this way we obtain the eigenvalues of $P_{n_c}(H_0+\lambda H_I)P_{n_c}|_{\mH(n_c)}$ and the expansion coefficients
of its eigenvectors with respect to the basis $\{ v_i^n$, $i\le n_c \}$. These
eigenvalues and expansion coefficients approximate those of $H_0+\lambda H_I$.  The
approximation is generally better and applies to more eigenvectors if $n_c$
is larger. One also expects that the approximation gets worse as $\lambda$ is increased.

We remark that the above truncation method also serves as a regularization for
possible ultraviolet divergences. 

\subsection{TCSA for perturbed conformal field theory on the strip}

In this section we describe the application of the TCSA to a quantum field
theory on a strip with the Hamiltonian operator
\begin{equation}
H=\frac{\pi}{L}L_0+\lambda L^{-l}\Phi(t=0)\ ,
\end{equation}
where $\frac{\pi}{L}L_0$ is the Hamiltonian operator of a conformal field
theory on the strip with Hilbert space $\mH=\oplus_jM(c,h_j)$, 
$\Phi$ is a boundary primary field of weight $l$ of this conformal
field theory, $\lambda $ is a coupling constant and $L$ is the width of the strip.

The Hilbert space at truncation level $n_c$ will be 
$\mH(n_c)=\oplus_j( \oplus_{i=0}^{n_c} M_i(c,h_j) )$.
A basis for $M_i(c,h)$ is obtained in the following way:
we take the vectors
\begin{equation}
\label{eq.tcsab}
L_{-n_1}L_{-n_2}\dots L_{-n_k}\ket{h}\qquad n_1\ge n_2\ge \dots  \ge n_k>0\qquad
n_1+n_2+ \dots + n_k = i 
\end{equation}
which are linearly independent and form a basis of $V_i(c,h)$.
We compute the scalar product matrix as described in Section \ref{sec.repvir}, 
this calculation can be done by computer. Then we use a simple linear
algebraic algorithm (which is not specific to conformal field theory, and which we do not describe
here) to eliminate elements from the above basis so that the image of the
remaining vectors by the mapping $q$ (see Section \ref{sec.repvir}) will constitute a
basis for $M_i(c,h)$. The elimination is also done by computer.    

The matrix elements of a chiral primary field $\Psi(z)$ between the basis
vectors can also be calculated algorithmically by
computer
as described in Section \ref{sec.repvir1} if all the matrix elements
$\brakettt{h_1}{\Psi(1)}{h_2}$ between highest weight states are
known. Therefore the matrix elements of $\Phi(t=0)$ can also be calculated.

In summary, a TCSA calculation consists of the following steps:
\begin{enumerate}
\item
Fixing the data specifying the model under investigation (the representations
appearing in the decomposition of the
Hilbert space into irreducible representations of the Virasoro algebra,
weight of the perturbation, normalizations).  
\item
Fixing the truncation level $n_c$.
\item
Taking a basis  for $V_i(c,h_j)$ for all values of $i$ and $j$ and   
computing the inner product matrix for the generated basis.
\item
Generating a basis for $M_i(c,h_j)$  for all values of $i$ and $j$ as
described above, storing $\tilde{h}_0$ (which is a diagonal matrix with entry
$\frac{\pi}{L}(h_j+i)$ for a basis vector in $M_i(c,h_j)$), storing the inner product matrix $g$
for $\mH(n_c)$
 and computing its inverse. 
\item
Computing $h_I$, i.e.\ the matrix of the matrix elements of the perturbing
primary field $\Phi(t=0)$
between the basis vectors of $\mH(n_c)$.
\item 
Computing the product matrix $g^{-1} h_{I}$.
\item
Computing the eigenvalues and eigenvectors of the matrix  $\tilde{h}_0+\lambda
L^{-l}g^{-1}h_I$, which is the matrix of $H^{TCSA}(n_c)$, at the  values of
$\lambda$ which are of interest.
\end{enumerate}
These steps can obviously be extended to the case when the perturbation
consists of more than one
terms.   

An application of the TCSA is to locate  RG fixed points and identify the representation of the Virasoro
algebra at these points.  
When one deals with
spectra, 
the representations of the Virasoro algebra 
can be identified from the dimensions of the energy eigenspaces at various
levels (see Section \ref{sec.repvir2}).

\section{Exact spectrum}
\label{sec.eel}
\markright{\thesection.\ \ EXACT SPECTRUM}

In this section we present a quantum field theoretic description of the
model (\ref{eq.yyy}) introduced in Chapter \ref{sec.introduction}. 
Regarding that the TCSA consists in the (approximate) diagonalization of the
operator (\ref{eq.yyy}) given by its matrix elements,
the main goal is to realize this Hamiltonian operator explicitly in the field
theoretic framework and to calculate its
spectrum. This is what determined the choice of the approach that we take in this
section. In particular, the framework for our calculation will be the one
introduced in   Section \ref{sec.def} as the  perturbed Hamiltonian
operator framework, because  in this framework the perturbed  Hamiltonian
operator structure, on which the TCSA is based, is explicit. In a
calculation using the 
Bethe-Yang equations (which we also present, nevertheless) or in the approach of \cite{C1} where  boundary
conditions play central role  the link with the TCSA formulation would not be
entirely obvious. Furthermore, the formulation presented in this section is
also suitable for Rayleigh-Schr\"odinger perturbation theory and for the treatment of the mode
truncated version in the subsequent sections.  
We do not consider the classical level of the field theoretic model, mainly
because it is irrelevant to our problem. We remark that  the same model with massive
unperturbed part (see e.g.\ \cite{C2}) could be studied  along the same lines.

\subsection{Distributions on closed line segments}
\markright{\thesubsection.\ \ DISTRIBUTIONS ON CLOSED LINE SEGMENTS}

\noindent
We shall use distributions on the closed line
segment (interval) $[0,L]\subset \RR$. The necessary formulae for the
Dirac-delta $\delta(x)$ and step function $\Theta(x)$ distributions on this
interval are the following:

\begin{alignat}{2}
&
\int_0^L\delta(x-a)f(x)\ dx=f(a) &&\quad\quad \mbox{if}\quad a\in (0,L)\\
&
\int_0^L\delta(x-a)f(x)\ dx=\frac{1}{2}f(0) &&\quad\quad \mbox{if}\quad a=0\\
&
\int_0^L\delta(x-a)f(x)\ dx  =\frac{1}{2}f(L) &&\quad\quad \mbox{if}\quad a=L\\
&
\int_0^L\delta(x-a)f(x)\ dx  =0 &&\quad\quad \mbox{if}\quad a\not\in [0,L]\ ,
\end{alignat}
where $f$ is a function defined on $[0,L]$, and $x\in [0,L]$.
\begin{alignat}{2}
&
\Theta(x-a)=0 &&\quad\quad \mbox{if}\quad  x<a\\
&
\Theta(x-a)=1 &&\quad\quad \mbox{if}\quad x\ge a\ ,
\end{alignat}
where $a \in \RR$, 
\begin{alignat}{2}
&
\partial_x \Theta(x-L)=2\delta(x-L)&&\\
&
\partial_x \Theta(x)=0&&\\
&
\partial_x \Theta(x-a)=\delta(x-a) && \quad\quad \mbox{if}\quad a\in (0,L)\\
&
\partial_x \Theta(x-a)=0 && \quad\quad \mbox{if}\quad a \not\in [0,L]\ ,  
\end{alignat}
where $x \in [0,L]$. 

\begin{equation}
\label{eq.di1}
\sum_{k\in \frac{\pi\ZZ}{L}}\exp[ik(x-x')]=2L\delta(x-x')
\end{equation}
\begin{equation}
\label{eq.di2}
\sum_{k\in \frac{\pi\ZZ}{L}}\exp[ik(x+x')]=2L[\delta(x+x')+\delta(x+x'-2L)]\ ,
\end{equation}
where $x,x' \in [0,L]$.

\subsection{The free model}
\label{sec.freemod}
\markright{\thesubsection.\ \ EXACT SPECTRUM - THE FREE MODEL}

The defining constituents of the unperturbed model are the following:
two  fermion fields $\Psi_1(x,t)$ and $\Psi_2(x,t)$ and a fermionic  operator $A_2(t)$ with the anticommutators
\begin{align}
\label{ac1}
\{ \Psi_1(x,t),\Psi_1(y,t)\} & =   4L\delta(x-y)\\
\{ \Psi_2(x,t),\Psi_2(y,t)\} & =  4L\delta(x-y)\\
\{ \Psi_1(x,t),\Psi_2(y,t)\} & =  -4L[\delta(x+y)+\delta(x+y-2L)]\\
\{ A_2(t), \Psi_1(x,t) \} & =  0\\
\{ A_2(t), \Psi_2(x,t) \} & =  0\\
\{ A_2(t),A_2(t) \} & =  2\  \label{ac2}
\end{align}
and the relations
\begin{equation}
\Psi_1(x,t)^\dagger =\Psi_1(x,t)\qquad \Psi_2(x,t)^\dagger =\Psi_2(x,t)\qquad
A_2(t)^\dagger =A_2(t)\ ,
\end{equation}
the Hamiltonian operator
\begin{equation}
H_0=-\frac{i}{8L}\int_0^L\ dx\ \Psi_1(x,0)\partial_x \Psi_1(x,0)+\frac{i}{8L}\int_0^L\
dx\ \Psi_2(x,0)\partial_x \Psi_2(x,0)\ ,
\end{equation}
the equations of motion
\begin{equation}
\label{mozg1}
\frac{d}{dt}A_2(t)=[iH_0,A_2(t)]=0\ , 
\end{equation}
\begin{multline}
\label{mozg2}
\partial_t\Psi_1(x,t)=[iH_0,\Psi_1(x,t)]=\\ =-\partial_x
\Psi_1(x,t)+\frac{1}{2}[\Theta(-x)+\Theta(x-L)][-\partial_x\Psi_2(x,t)+\partial_x\Psi_1(x,t)]+\\
+\frac{1}{2}[-\Psi_1(0,t)-\Psi_2(0,t)]\delta(x)+\frac{1}{2}[\Psi_1(L,t)+\Psi_2(L,t)]\delta(x-L)\ ,
\end{multline}
\begin{multline}
\label{mozg3}
\partial_t\Psi_2(x,t)=[iH_0,\Psi_2(x,t)]=\\=\partial_x
\Psi_2(x,t)+\frac{1}{2}[\Theta(-x)+\Theta(x-L)][-\partial_x\Psi_2(x,t)+\partial_x\Psi_1(x,t)]+\\
+\frac{1}{2}[\Psi_1(0,t)+\Psi_2(0,t)]\delta(x)+\frac{1}{2}[-\Psi_1(L,t)-\Psi_2(L,t)]\delta(x-L)\ .
\end{multline}
The fermion fields, which can be regarded as one-component real fermion fields with
zero mass,  have the following expansion:
\begin{align}
\Psi_1 (x,t) & =  \sum_{k\in\frac{\pi}{L}\ZZ,\ \
  k>0}\sqrt{2}[a(k)e^{ik(t-x)}+a^\dagger (k)e^{-ik(t-x)}]+A_1\\
\Psi_2 (x,t) & =  \sum_{k\in\frac{\pi}{L}\ZZ,\ \
  k>0}\sqrt{2}[-a(k)e^{ik(t+x)}-a^\dagger (k)e^{-ik(t+x)}]-A_1
\end{align}
\begin{align}
\label{a1}
a(k)^\dagger &=a(-k) &  A_1^\dagger &=A_1\\ 
\{ a(k_1),a(k_2)\} &= \delta_{k_1,-k_2} &
\{ a(k),A_1 \} &= 0\\
\{A_2,A_1 \} &=  0 &
\{ A_1,A_1 \} &=  2\\
\{ A_2,a(k)\} &=0 &
\label{a2}
\end{align}
The fermion fields and $A_2$ are dimensionless.

A unitary representation for the above operator algebra is defined by the
following formulae:
an orthonormal basis of the Hilbert space is
\begin{equation}
\label{eq.freebasis}
\ket{a(k_1)a(k_2)a(k_3)\dots a(k_n) u}\qquad 
\ket{a(k_1)a(k_2)a(k_3)\dots a(k_n) v}\ ,
\end{equation}
where $k_1>k_2>k_3>\dots >k_n>0$, $k_i \in \frac{\pi \ZZ}{L}$ and $n\ge 0$,
\begin{align}
A_1\ket{a(k_1)a(k_2)a(k_3)\dots a(k_n) u} & =  (-1)^n \ket{a(k_1)a(k_2)a(k_3)\dots a(k_n) v}\\
A_1\ket{a(k_1)a(k_2)a(k_3)\dots a(k_n) v} &=  (-1)^n \ket{a(k_1)a(k_2)a(k_3)\dots a(k_n)
  u}\\
A_2\ket{a(k_1)a(k_2)a(k_3)\dots a(k_n) u} & =  (-1)^n (-i) \ket{a(k_1)a(k_2)a(k_3)\dots a(k_n) v}\\
A_2\ket{a(k_1)a(k_2)a(k_3)\dots a(k_n) v} & =  (-1)^n (+i)
\ket{a(k_1)a(k_2)a(k_3)\dots a(k_n) u}\ .
\end{align}
The condition  $k_1>k_2>k_3>\dots >k_n>0$, $k_i \in \frac{\pi \ZZ}{L}$ and $n\ge
0$ also applies below  if not stated otherwise. $a(k)\ket{u}=0$,
$a(k)\ket{v}=0$ if $k<0$. 

The Hamiltonian operator can be written as 
\begin{equation}
H_0=\sum_{k\in\frac{\pi}{L}\ZZ,\ \
  k>0}\ k[a(k)a^\dagger (k)]\ ,
\end{equation}
where an infinite constant is subtracted, by normal ordering for example, which is
not denoted explicitly.
The energy eigenvalue of a basis vector $\ket{a(k_1)a(k_2)a(k_3)\dots a(k_n)
  u}$ is
\begin{equation}
E_{\{ k_1,k_2,\dots ,k_n,u\}  }=\sum_{i=1}^n k_i\ ,
\end{equation} 
and the same applies if $u$ is replaced by $v$.

The following boundary conditions are satisfied by the fermion fields:
\begin{equation}
\brakettt{E_1}{\Psi_1(0,t)+\Psi_2(0,t)}{E_2}=0\qquad
\brakettt{E_1}{\Psi_1(L,t)+\Psi_2(L,t)}{E_2}=0\ ,
\end{equation}
where $\ket{E_1}$ and $\ket{E_2}$ are energy eigenstates.

A representation of the Virasoro algebra can also be defined on the Hilbert
space in the following way (which is well known essentially in conformal field
theory; see \cite{FMS,Ginsparg}):

\begin{equation}
a(0)=\frac{1}{\sqrt{2}}A_1
\end{equation}
\begin{equation}
\label{eq.ln1}
L_{N}=\frac{L}{2\pi}\sum_{k\in\frac{\pi}{L}\ZZ}
-ka(-k)a(k-\frac{N\pi}{L})
\end{equation}
where $N=1,2,3,4,\dots $,
\begin{equation}
L_{-N}=(L_{N})^\dagger
\end{equation}
\begin{equation}
\label{eq.ln2}
L_0=\frac{L}{2\pi}\sum_{k\in\frac{\pi}{L}\ZZ} [-k:a(-k)a(k):]+\frac{1}{16}\ ,
\end{equation}
where $::$ denotes the normal ordering for fermionic creation and annihilation
operators, $:a(-k)a(k):=a(-k)a(k)$ if $k<0$,  $:a(-k)a(k):=-a(k)a(-k)$ if $k>0$.
$L_N$, $N\in \ZZ$ will be the generators of the Virasoro algebra. They satisfy the relations
\begin{equation}
[L_N,L_M]=(N-M)L_{N+M}\qquad \mbox{if}\quad N+M\ne 0
\end{equation}
and
\begin{multline}
\label{eq.LNLN}
[L_N,L_{-N}]=\\ 
=\frac{L^2}{4\pi^2}\sum_{k\in\frac{\pi}{L}\ZZ}[-k(-2k+\frac{N\pi}{L})a(-k)a(k)+k(-2k+\frac{N\pi}{L})a(-k+\frac{N\pi}{L})a(k-\frac{N\pi}{L})]
\end{multline}
if $N>0$.
It can be verified that in the above representation of the creation and
annihilation operators the right-hand side of the equation (\ref{eq.LNLN}) is
equal to
\begin{equation}
2NL_0+\frac{1}{24}N(N-1)(N+1)\ ,
\end{equation}
and so the operators $L_N$ satisfy the (usual) relations of the generators of
the Virasoro algebra with central
charge $c=1/2$.  $A_2$ commutes with the $L_N$-s. $\ket{v}$ and $\ket{u}$ are highest weight states with weight
$1/16$ and the Hilbert space decomposes into two copies of the $M(c=1/2, h=1/16)$
unitary highest weight representation of the Virasoro algebra. The 
invariant subspace belonging to $u$  is spanned by the vectors 
$\ket{a(k_1)a(k_2)\dots a(k_n) u}$ with $n$ even and   
$\ket{a(k_1)a(k_2)\dots a(k_n) v}$ with $n$ odd. The 
invariant subspace belonging to $v$  is spanned by the vectors\\ 
$\ket{a(k_1)a(k_2)\dots a(k_n) u}$ with $n$ odd and   
$\ket{a(k_1)a(k_2)\dots a(k_n) v}$ with $n$ even. These subspaces will be called
$u$ and $v$ sectors.

The relation between $H_0$ and $L_0$ is 
\begin{equation}
H_0=\frac{\pi}{L}L_0-\frac{1}{16}\frac{\pi}{L}\ .
\end{equation}

The fields $\Psi_1$ and $\Psi_2$ can be written as
\begin{align}
\Psi_1(x,t) & =  \sqrt{2} \sum_{k\in \frac{\pi}{L}\ZZ} a(k) e^{ik(t-x)}\\
\Psi_1(z) & =  \sqrt{2} \sum_{n \in \ZZ} \tilde{a} (n) z^n \\
\Psi_2(x,t) & =  \sqrt{2}\sum_{k\in \frac{\pi}{L}\ZZ} -a(k) e^{ik(t+x)}\\
\Psi_2(\bar{z}) & =  \sqrt{2}\sum_{n\in\ZZ} -\tilde{a}(n) \bar{z}^n\ ,
\end{align}
where $\tilde{a}(n)=a(k)$, $k=n\frac{\pi}{L}$, $z=e^{i\frac{\pi}{L}(t-x)}$, $\bar{z}=e^{i\frac{\pi}{L}(t+x)}$.

The following commutation relations are satisfied:
\begin{align}
[L_N,\Psi_1(x,t)] & =
\frac{-iL}{2\pi}e^{i\frac{N\pi}{L}(t-x)}[\partial_t-\partial_x]\Psi_1(x,t)+\frac{N}{2}e^{i\frac{N\pi}{L}(t-x)}\Psi_1(x,t)\\
{}
[L_N,\Psi_1(z)] & =  z^{N+1}\frac{d\Psi_1}{dz}(z)+\frac{1}{2}Nz^N\Psi_1(z) \label{z1}\\ {}
[L_N,\Psi_2(x,t)] & = 
\frac{-iL}{2\pi}e^{i\frac{N\pi}{L}(t+x)}[\partial_t+\partial_x]\Psi_2(x,t)+\frac{N}{2}e^{i\frac{N\pi}{L}(t+x)}\Psi_2(x,t)\\ {}
[L_N,\Psi_2(\bar{z})] & = 
\bar{z}^{N+1}\frac{d\Psi_2}{d\bar{z}}(\bar{z})+\frac{1}{2}N\bar{z}^N\Psi_2(\bar{z})\
. \label{z2}
\end{align}
In the  equations (\ref{z1}) and  (\ref{z2}) the domains of $z$ and $\bar{z}$ are extended to the
whole complex plane and the differentiation with respect to $z$ and $\bar{z}$ is
the usual complex differentiation.

For 
$\epsilon(z)=\frac{\Psi_1(z)}{\sqrt{z}}$ we have 
\begin{equation}
[L_N,\epsilon(z)]=z^{N+1}\frac{d\epsilon}{dz}(z)+\frac{1}{2}(N+1)z^N\epsilon(z)
\end{equation}
and for $\bar{\epsilon}(\bar{z})=\frac{\Psi_2(\bar{z})}{\sqrt{\bar{z}}}$
\begin{equation}
[L_N,\bar{\epsilon}(\bar{z})]=\bar{z}^{N+1}\frac{d\bar{\epsilon}}{d\bar{z}}(\bar{z})+\frac{1}{2}(N+1)\bar{z}^N\bar{\epsilon}(\bar{z})\ ,
\end{equation}
i.e.\ $\epsilon(z)$ and  $\bar{\epsilon}(\bar{z})$ are chiral Virasoro primary
fields of weight $1/2$. 
The same relations apply to the fields $A_2\Psi_1$, $A_2\Psi_2$,
$A_2\epsilon$ and $A_2\bar{\epsilon}$, in particular $A_2\epsilon$ and
$A_2\bar{\epsilon}$ are also chiral primary fields of weight $1/2$.
$A_2\epsilon$ and
$A_2\bar{\epsilon}$ have zero matrix elements between the $u$ and $v$ sector,
whereas  $\epsilon(z)$ and  $\bar{\epsilon}(\bar{z})$ have zero matrix
elements within the $u$ and $v$ sectors. 

The  operator $A_2$ is an auxiliary operator and if it is  omitted, then it is possible to represent
the fields so that the Hilbert space is a single Virasoro module $(1/2,1/16)$.
It is the next section where the presence of $A_2$ will be really useful.

We remark that if we demand the equations of motion 
$\partial_t\Psi_1(x,t)=-\partial_x\Psi_1(x,t)$,
$\partial_t\Psi_2(x,t)=\partial_x\Psi_2(x,t)$ for $x \in [0,L]$, the fermionic
nature of the mode creating and annihilating operators and the 
boundary conditions 
\begin{equation}
\brakettt{E_1}{\Psi_1(0,t)+\Psi_2(0,t)}{E_2}=0\qquad
\brakettt{E_1}{\Psi_1(L,t)+\Psi_2(L,t)}{E_2}=0\ 
\end{equation}
or
\begin{equation}
\brakettt{E_1}{\Psi_1(0,t)-\Psi_2(0,t)}{E_2}=0\qquad
\brakettt{E_1}{\Psi_1(L,t)-\Psi_2(L,t)}{E_2}=0\ ,
\end{equation}
where $\ket{E_1}$ and $\ket{E_2}$ are energy eigenstates, then the $(c=1/2,h=1/16)$ representation 
of the Virasoro algebra can be defined on the Hilbert space  (without
$A_2$ and considering the simplest possibility). 
If we demand the boundary conditions 
\begin{equation}
\label{eq.hinf}
\brakettt{E_1}{\Psi_1(0,t)+\Psi_2(0,t)}{E_2}=0\qquad
\brakettt{E_1}{\Psi_1(L,t)-\Psi_2(L,t)}{E_2}=0\ 
\end{equation}
or
\begin{equation}
\brakettt{E_1}{\Psi_1(0,t)-\Psi_2(0,t)}{E_2}=0\qquad
\brakettt{E_1}{\Psi_1(L,t)+\Psi_2(L,t)}{E_2}=0\ ,
\end{equation}
then the representation
of the Virasoro algebra on the Hilbert space will be $(c=1/2,h=1/2)\oplus (c=1/2,h=0)$.
The anticommutation relations of the fields are slightly different in these
four cases.

\subsection{The perturbed model}
\label{sec.pertmod}
\markright{\thesubsection.\ \ EXACT SPECTRUM - THE PERTURBED MODEL}

\noindent
The Hamiltonian operator is
\begin{multline}
\label{eq.pertham}
H=-\frac{i}{8L}\int_0^L\  \Psi_1(x,0)\partial_x \Psi_1(x,0)\ dx\
+\frac{i}{8L}\int_0^L\ \Psi_2(x,0)\partial_x \Psi_2(x,0)\ dx\ +\\
+hiA_2(0)[\Psi_2(L,0)-\Psi_1(L,0)]\ ,
\end{multline}
where $h$ is a coupling constant of dimension $mass$.
The perturbing term\\
$hH_I=hiA_2(0)[\Psi_2(L,0)-\Psi_1(L,0)]$ has zero matrix elements
between vectors belonging to different sectors,
which means that $H$ can be
restricted to the $u$ and $v$ sectors separately.
These restrictions are denoted by $H|_u$ and $H|_v$.
$H_I$ is also  a primary boundary
field of weight $1/2$ with respect to the Virasoro algebra defined in the
previous section  taken at $t=0$. The matrix elements of
$H_I$, i.e.\ of $H_I|_u$ and $H_I|_v$, are uniquely determined by this property and
by the values of the matrix elements  $\brakettt{u}{H_I}{u}$ and
$\brakettt{v}{H_I}{v}$. It is easy to verify that
$2=\brakettt{u}{H_I}{u}=-\brakettt{v}{H_I}{v}$, and there exists an
intertwiner $Y$
of the Virasoro algebra representations on the $u$ and $v$ sectors so that $Yu=v$, 
$YH_0|_uY^{-1}=H_0|_v$. This also implies that $YH_I|_u Y^{-1}=-H_I|_v$. This
means finally that we can restrict to $0\le h \le \infty$, and the $u$ sector and
$H|_u$ will
correspond to the $h\ge 0$ case of (\ref{eq.yyy}) and to (\ref{eq.flow1}), the
$v$ sector and $H|_v$ will
correspond to the $h\le 0$ case of (\ref{eq.yyy}) and to (\ref{eq.flow2}). For $0\le h \le
\infty$ the operator 
(\ref{eq.pertham}) describes in the two sectors the two flows mentioned in the
Introduction. The precise relation between the $h$ in (\ref{eq.yyy}) in the Introduction and the
$h$ in (\ref{eq.pertham}) is the following: $h_{Introd}=2L^{1/2}h$ in the $u$ sector,
$h_{Introd}=-2L^{1/2}h$ in the $v$ sector.
Further on we shall assume that  $0<h<\infty$.

The eigenvalues of the Hamiltonian operator (\ref{eq.pertham}) are ultraviolet
divergent in perturbation theory; the divergence can be removed by adding
a term $ch^2I$ with appropriate value of the logarithmically divergent (as a
function of the cutoff energy) coefficient $c$. This means that
the differences of the eigenvalues of $H$ are not ultraviolet divergent. We shall
assume that the ground state energy is set to zero by the renormalization, and
the $ch^2I$ term will not be written explicitly.

The equations of motion are obtained by adding the following terms to the
right-hand side of the unperturbed equations of motion (\ref{mozg1}), (\ref{mozg2}), (\ref{mozg3}):
\begin{align}
[ihH_I,\Psi_1(x,t)]& =  
8LhA_2(t)\delta(x-L)\\ {}
[ihH_I,\Psi_2(x,t)] & = 
-8LhA_2(t)\delta(x-L) \\ {}
[ihH_I,A_2(t)] & =  2h(\Psi_2(L,t)-\Psi_1(L,t))\ .
\end{align}
These terms are linear in the fermion fields and $A_2$, so the equations of
motion for the perturbed theory are linear and by sandwiching these equations
between energy eigenstates we get a system of three first-order differential
equations for the expectation values of the fields and $A_2$.  
These expectation values can be assumed to take the form
\begin{align}
\label{eq.ans1}
\brakettt{E_1}{\Psi_1(x,t)}{E_2} & =   -e^{ik(t-x)}-\Theta(x-L)C_1(k)e^{ikt}\\
\label{eq.ans2}
\brakettt{E_1}{\Psi_2(x,t)}{E_2} & =  e^{ik(t+x)}-\Theta(x-L)C_2(k)e^{ikt}\\
\label{eq.ans3}
\brakettt{E_1}{A_2(t)}{E_2} & =  C_3(k)e^{ikt}\ ,
\end{align}
where $C_1(k)$, $C_2(k)$, $C_3(k)$ are finite constants, $\ket{E_1}$ and
$\ket{E_2}$ are eigenstates of $H$ and $k=E_1-E_2$. $\ket{E_1}$ and
$\ket{E_2}$ are not necessarily normalized to 1 here.

Substituting (\ref{eq.ans1})-(\ref{eq.ans3}) into the equations of motion we get algebraic equations for
$C_1(k)$, $C_2(k)$, $C_3(k)$, which have the following solution:

\begin{equation}
\label{spektr}
kL\tan(kL)=16L^2h^2
\end{equation}

\begin{alignat}{2}
\psi_1(k)(x,t) &= \brakettt{E_1}{\Psi_1(x,t)}{E_2} &&= -e^{ik(t-x)}-\Theta(x-L)i\sin(kL)e^{ikt}\\
\psi_2(k)(x,t) &= \brakettt{E_1}{\Psi_2(x,t)}{E_2} &&= e^{ik(t+x)}-\Theta(x-L)i\sin(kL)e^{ikt}\\
a_2(k)(t) &= \brakettt{E_1}{A_2(t)}{E_2} &&= -\frac{i\sin(kL)}{4Lh}e^{ikt}\ .
\end{alignat}
(\ref{spektr}) is the formula that determines the possible values of $k$ at a
given value of $h$ and $L$ and so the spectrum of $H$ up to an undetermined
additive overall constant. 

The assumption (\ref{eq.ans1}), (\ref{eq.ans2}) is motivated by the fact that
the equations of motion for $x \in (0,L)$ are $(\partial_t+\partial_x)\Psi_1(x,t)=0$,
$(\partial_t-\partial_x)\Psi_2(x,t)=0$. We remark that the equations of motion could also be
solved directly without making any assumptions on the form of the expectation values.

\begin{figure}[h]
\begin{tabular}{lr}
\includegraphics[clip=true,height=12cm]{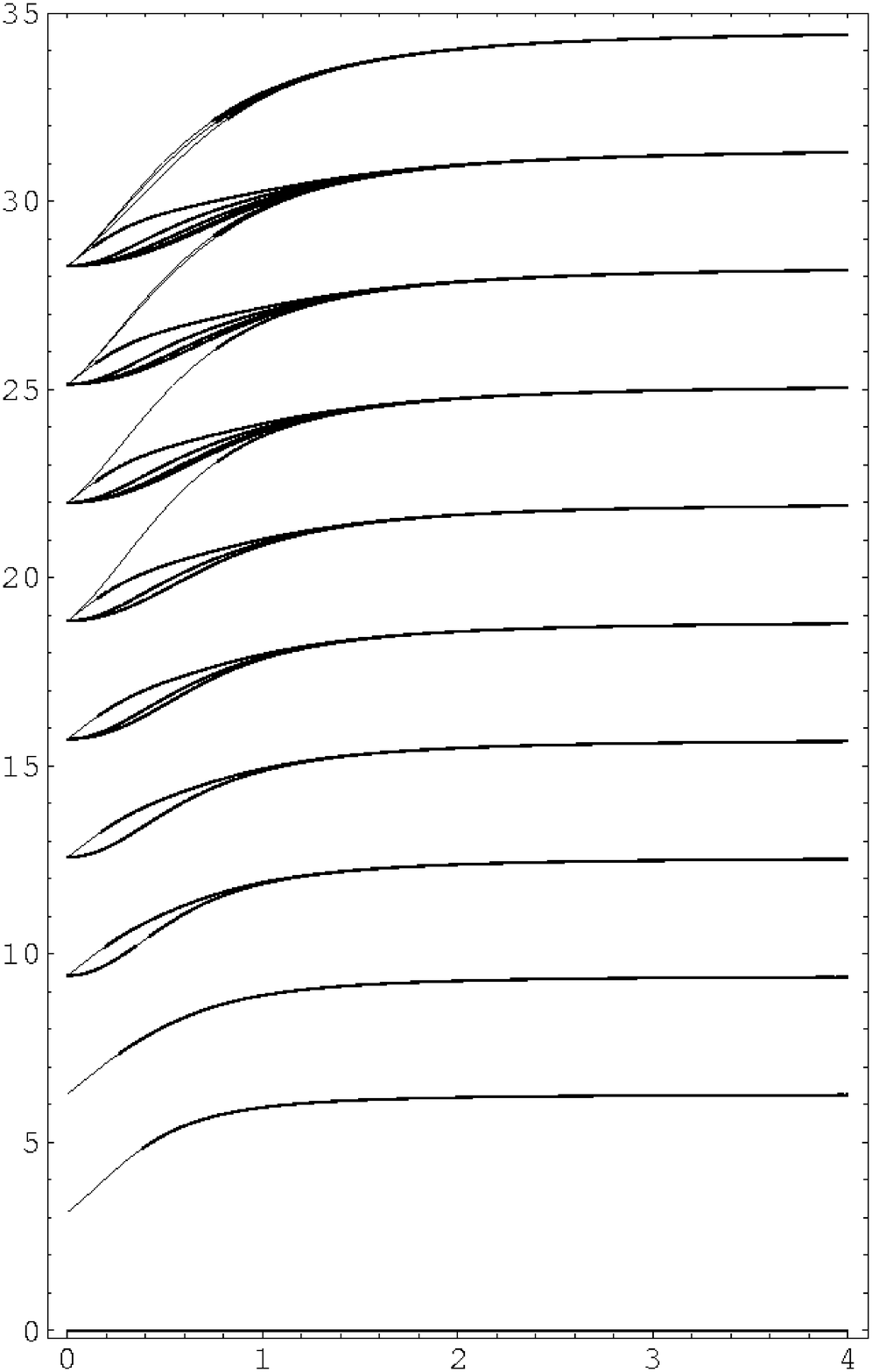}
&
\includegraphics[clip=true,height=12cm]{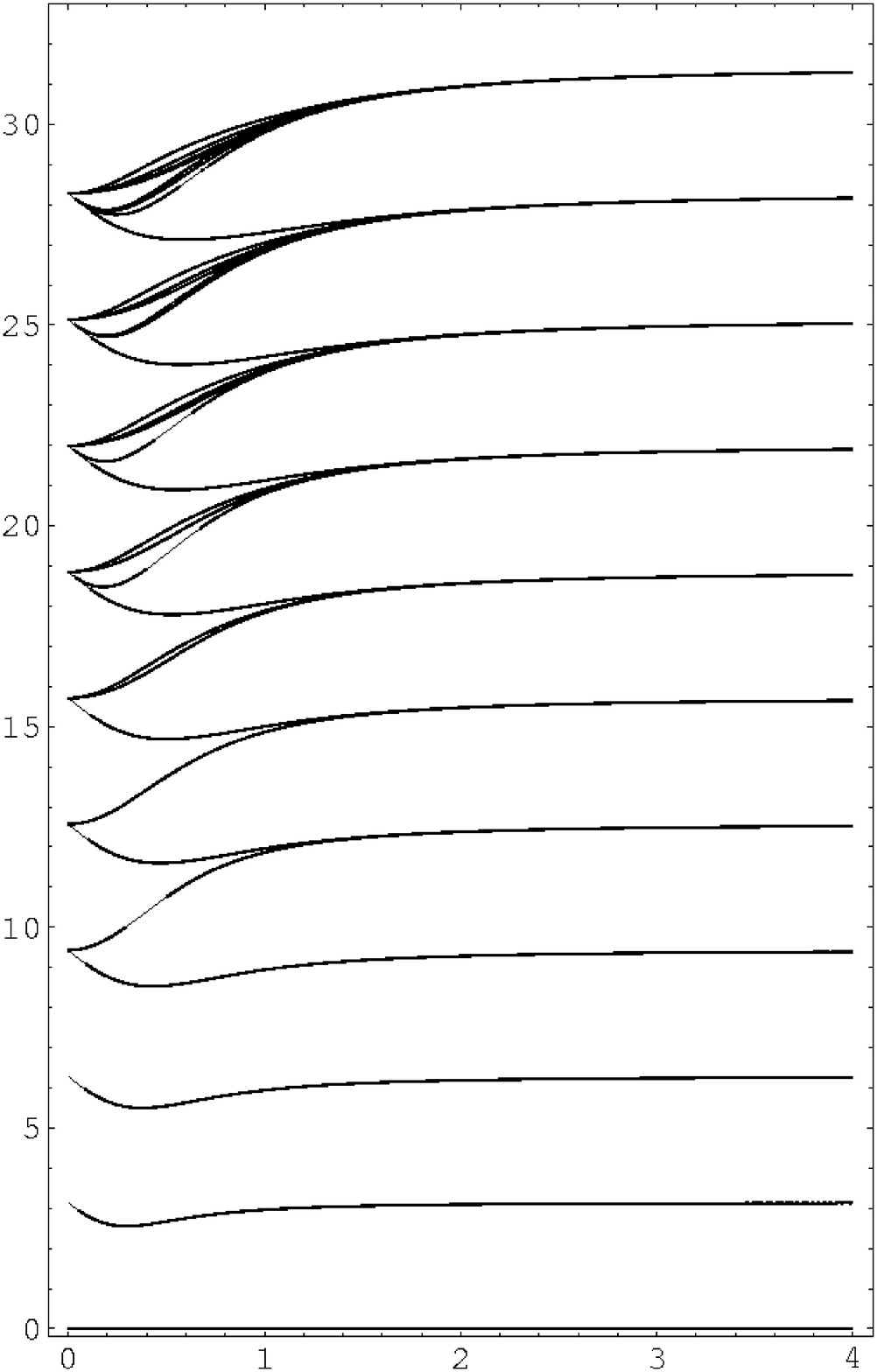}
\end{tabular}
\caption{\label{fig.1}Exact energy gaps in the $v$ and $u$ sectors as a function of $h$}
\end{figure}

Introducing the notation
\begin{equation}
n(k)=(\psi_1(k),\psi_2(k),a_2(k))
\end{equation}
the mode expansion of $(\Psi_1,\Psi_2,A_2)$ is
\begin{equation}
\label{eq.mode}
(\Psi_1,\Psi_2,A_2)=\sum_{k\in S} b(k)n(k)\ ,
\end{equation}
where the $b(k)$ are creation/annihilation operators and the summation is done
over the set $S$ of all real solutions of (\ref{spektr}). 

The anticommutation relation of the $b(k)$-s can be obtained in the following way:
we define a scalar product on the classical complex valued solutions of the
equations of motion:
\begin{equation}
\label{scal}
\braket{\psi_1,\psi_2,a_2}{\phi_1,\phi_2,b_2}=\int_0^L
[\psi^*_1\phi_1+\psi_2^*\phi_2] dx + 2La_2^*b_2\ .
\end{equation}
This product should be calculated at a fixed time. Using the equations of
motion it can be shown that the product is independent of this time. 

The essential properties of this scalar product are that it is defined by a
local expression and that the $n(k)$-s are orthogonal with respect to it:
\begin{equation}
\braket{n(k_1)}{n(k_2)}=\delta_{k_1-k_2,0}(2L+\frac{\sin(2k_1L)}{k_1})\ .
\end{equation}
The creation/annihilation operators can be expressed in the following way:
\begin{equation}
\label{eq.86}
\braket{n(k)}{(\Psi_1,\Psi_2,A_2)}=b(k)\braket{n(k)}{n(k)}
\end{equation}
Using the formula (\ref{scal}) and the anticommutation relations (\ref{ac1})-(\ref{ac2}) we get
\begin{equation}
\{ b(k_1),b(k_2)\} = \delta_{k_1+k_2,0}\frac{4Lk_1}{2Lk_1+\sin(2Lk_1)}\ .
\end{equation}
The equation 
\begin{equation}
b(k)^\dagger=b(-k)
\end{equation}
is also satisfied.

The expansion (\ref{eq.mode})  and (\ref{ac1})-(\ref{ac2}) imply that the following nontrivial
formulae hold:
\begin{align}
\label{eq.ortog}
\sum_{k\in S}f(k)\psi_1(k)(x,0)\psi_1(-k)(y,0)  & =   4L\delta(x-y)\\
\sum_{k\in S}f(k)\psi_2(k)(x,0)\psi_2(-k)(y,0)  & =  4L\delta(x-y)\\
\sum_{k\in S}f(k)\psi_1(k)(x,0)\psi_2(-k)(y,0)  & =  -4L[\delta(x+y)+\delta(x+y-2L)]\\
\sum_{k\in S}f(k)a_2(k)(0)\psi_1(-k)(x,0)   & =  0\\
\sum_{k\in S}f(k)a_2(k)(0)\psi_2(-k)(x,0)   & =  0\\
\sum_{k\in S}f(k)a_2(k)(0)a_2(-k)(0)   & =  2\ ,  \label{eq.ortog2}
\end{align}
where
\begin{equation}
f(k)=\frac{4Lk}{2Lk+\sin(2Lk)}\ .
\end{equation}
These formulae are generalizations of (\ref{eq.di1}), (\ref{eq.di2}).

Using the formulae
\begin{align}
a(k) & =  \frac{1}{2\sqrt{2}L}[ \int_0^L e^{ikx} \Psi_1(x,0)\ dx - \int_0^L
e^{-ikx} \Psi_2(x,0)\ dx ] \\
A_1 & =  \frac{1}{2L}[ \int_0^L  \Psi_1(x,0)\ dx - \int_0^L
 \Psi_2(x,0)\ dx ]
\end{align}
and (\ref{eq.mode}) and (\ref{eq.111}) we obtain the following relations:
\begin{align}
\label{eq.bg1}
a(k) & =  \frac{-1}{\sqrt{2}L}\sum_{k' \in S} b(k') \frac{\sin[(k-k')L]}{k-k'}\\
A_1 & =  \frac{-1}{L}\sum_{k' \in S} b(k') \frac{\sin[k'L]}{k'}\\
\label{eq.bg2}
A_2(0) & =  \sum_{k\in S} b(k) \frac{-i\sin(kL)}{4Lh}=-i\sum_{k\in S} b(k)
\frac{\sin(kL)}{\sqrt{kL\tan(kL)}}\ .
\end{align}
Using (\ref{eq.86}) we get the relation
\begin{equation}
\label{eq.bg3}
b(k)\left( 2L+\frac{\sin(2kL)}{k} \right) = \sqrt{2} \sum_{k'\in \frac{\pi}{L}\ZZ} a(k')
\frac{-2\sin[(k'-k)L]}{k'-k} + \frac{i\sin(kL)}{2h}A_2(0)\ .
\end{equation}
(\ref{eq.bg1})-(\ref{eq.bg3}) can be regarded as Bogoliubov transformation
formulae for this model.

We remark that 
the above scalar
product technique is also suitable for free fields on the half-line or in the
usual full Minkowski space without boundaries in arbitrary
spacetime dimensions. 

We also remark that we have not given a mathematically completely rigorous
proof  that (\ref{eq.mode}) satisfies (\ref{ac1})-(\ref{ac2}), but we think that this would be
possible.

The Hilbert space is
spanned by the orthogonal eigenstates
$\ket{b(k_1)b(k_2)b(k_3)\dots b(k_n) 0_h}$ where $k_1>k_2>\dots >k_n>0$, $\ket{0_h}$
is the ground state, which is unique, and  
$b(k)\ket{0_h}=0$ if $k<0$.
The eigenvalues of these states are 
\begin{equation}
E_{\{ k_1,k_2,\dots ,k_n\} }=\sum_{i=1}^n k_i\ .
\end{equation}
The eigenvector $\ket{b(k_1)b(k_2)\dots b(k_n) 0_h}$ belongs to the $v$ sector if
$n$ is even and to the $u$ sector if $n$ is odd. 
The first few energy gaps (i.e.\ energies relative to the lowest energy) within the two
sectors are shown in Figure \ref{fig.1} as functions of $h$ with $L=1$.

In the  $h\to 0$ limit 
\begin{eqnarray}
b(k(n,h)) & \to & -\sqrt{2}a(k(n,0))\qquad n\ge 1\\
b(k(0,h))+b(k(0,h))^\dagger & \to & -A_1\\
i(b(k(0,h))^\dagger -b(k(0,h))) & \to & A_2(0)\\
\ket{0_h} & \to & \ket{v}\ ,
\end{eqnarray}
where $k(n,h)$ is the $n$-th nonnegative root of (\ref{spektr}) as a function of $h$,
$k(n,0)=n\pi/L$, $n=0,1,2,\dots $.

The nonzero matrix elements of the fields are 
\begin{align}
\frac{\brakettt{P}{(\Psi_1(x,t),\Psi_2(x,t),A_2(t))}{Q}}{\sqrt{|\braket{P}{P}\braket{Q}{Q}|}}
& =  n(k)
(-1)^m \sqrt{f(k)}\\
\frac{\brakettt{Q}{(\Psi_1(x,t),\Psi_2(x,t),A_2(t))}{P}}{\sqrt{|\braket{P}{P}\braket{Q}{Q}|}}
& =  n(-k)
(-1)^m \sqrt{f(k)}\ ,
\end{align}
where $Q=\ket{b(k_1)b(k_2)\dots b(k_n) 0_h}$,
$P=\ket{b(k_1),\dots ,b(k_m),b(k),b(k_{m+1}),\dots ,b(k_n) 0_h}$,
\begin{equation}
\braket{Q}{Q}=\prod_{i=1}^n f(k_i)\ .
\end{equation}

The Hamiltonian operator can be written as 
\begin{equation}
H=\sum_{k\in S,\ k>0} \frac{k}{f(k)}b(k)b(-k)\ .
\end{equation}

The following formula can be written for $\ket{0_h}$:
\begin{equation}
N\ket{0_h}=\lim_{\alpha
  \to \infty} e^{-\alpha H}\ket{v}=\prod_{k\in S,\ k>0}(1-\frac{1}{f(k)}b(k)b(-k))\ket{v}
\end{equation}
where $N$ is a normalization factor. 
The second equation on the right-hand
side can be verified directly using the following formulae:
\begin{equation}
[b(k_1)b(-k_1),b(k_2)b(-k_2)]=0
\end{equation}
if $k_1 \ne k_2$ (and $k_1,k_2 >0$) and
\begin{equation}
(b(k)b(-k))^n=f(k)^{n-1}b(k)b(-k)\ .
\end{equation}

The following boundary conditions are satisfied:
\begin{align}
\label{eq.bc0}
\brakettt{E_1}{\Psi_1(0,t)+\Psi_2(0,t)}{E_2}& =  0\\
\label{eq.bc00}
\brakettt{E_1}{\Psi_1(L,t)+\Psi_2(L,t)}{E_2} & =  0
\end{align}
and
\begin{align}
\label{eq.bc1}
\lim_{x\to 0}\brakettt{E_1}{\Psi_1(x,t)+\Psi_2(x,t)}{E_2}& =  0\\
\label{eq.bc2}
\lim_{x\to
  L}\brakettt{E_1}{\partial_x\Psi_1(x,t)-\partial_x\Psi_2(x,t)}{E_2} & =  16Lh^2\brakettt{E_1}{\Psi_2(L,t)-\Psi_1(L,t)}{E_2} \\
\label{eq.kkkk1}
\lim_{h\to\infty}\lim_{x\to L}\brakettt{E_1}{\Psi_1(x,t)-\Psi_2(x,t)}{E_2} & =
 0\\
\label{eq.kkkk2}
\lim_{h\to\infty}\lim_{x\to L}\brakettt{E_1}{\Psi_1(x,t)+\Psi_2(x,t)}{E_2} &
\ne   0\ ,
\end{align}
where $\ket{E_1}$ and $\ket{E_2}$ are eigenstates of $H$. 
The boundary conditions (\ref{eq.bc0}) and (\ref{eq.bc00}) are the same as
those satisfied by the free fields and they are also in agreement with the
the definition (\ref{eq.int1})-(\ref{eq.int3}). On the other hand, (\ref{eq.bc1}) and
(\ref{eq.bc2}) are similar to the boundary conditions written down in \cite{GZ,CZ,C1}.
The equations (\ref{eq.kkkk1}), (\ref{eq.kkkk2}) show that in the 
$h: 0 \to \infty$ limit the boundary condition (\ref{eq.hinf}) is realized. From the
point of view of the boundary conditions one can say that the perturbation
$H_I$ induces a flow from the boundary condition
$\lim_{x \to L}\brakettt{E_1}{\Psi_1(x,t)+\Psi_2(x,t)}{E_2} =  0$ to the boundary
condition $\lim_{x\to L}\brakettt{E_1}{\Psi_1(x,t)-\Psi_2(x,t)}{E_2}  =
 0$ and the boundary condition on the left-hand side remains constant, which
 is in accordance with the literature (see e.g.\ \cite{GZ}).
The boundary condition $\brakettt{E_1}{\Psi_1(L,t)+\Psi_2(L,t)}{E_2} =  0$ is
called free spin boundary condition in the literature (e.g.\ \cite{GZ}) and 
$\brakettt{E_1}{\Psi_1(L,t)-\Psi_2(L,t)}{E_2}=0$ is called fixed spin boundary
condition. One also has the physical picture of these boundary conditions that at zero magnetic field
(i.e.\ $h=0$) the direction of the spin at the boundary is free, whereas at
infinite magnetic field it is completely fixed to one of the two directions
(depending on the sign of the boundary magnetic field).

We remark that from
(\ref{eq.bc1}) and (\ref{eq.bc2}) and the bulk equations of motion
$(\partial_t+\partial_x)\Psi_1=0$, $(\partial_t-\partial_x)\Psi_2=0$ the equation (\ref{spektr}) can be recovered.

In the 
$h: 0 \to \infty$ limit
\begin{equation}
k(n,h)\to
k(n,0)+\frac{1}{2}\frac{\pi}{L}=\frac{n\pi}{L}+\frac{1}{2}\frac{\pi}{L}
\end{equation}
and
\begin{equation}
\{ b(k_i),b(k_j) \} \to 2 \delta_{k_i+k_j,0}\ .
\end{equation}

It can be verified that in the $h\to \infty$ limit the $(c=1/2,h=0)$
representation of the Virasoro algebra can be introduced in the $v$ sector and
the $(c=1/2,h=1/2)$ representation can be introduced in the $u$ sector. One
can write   expressions (which are well known essentially, see \cite{FMS,Ginsparg}) for the
generators in terms of the $b(k)$-s similar to
(\ref{eq.ln1}), (\ref{eq.ln2}). Therefore $h:0\to \infty$ corresponds to the $(1/2,1/16)\to (1/2,0)$
flow in the $v$ sector and to the $(1/2,1/16)\to (1/2,1/2)$ flow in the $u$ sector. 
We remark that the easiest way to determine which representations are realized
in the $h\to \infty$ limit is to count the degeneracies of the first few
energy levels (separately in the two sectors) and compare the result with
Table \ref{tab.lev}.

We define the fields
\begin{align}
\Phi_1(x,t) & =  \sum_{k\in S} -b(k)e^{ik(t-x)} \\
\Phi_2(x,t) & =  \sum_{k\in S} b(k)e^{ik(t+x)}\ .
\end{align}

They satisfy the following equations: 
\begin{align}
(\Psi_1-\Psi_2)(L,t) & =  \frac{1}{16Lh^2}\partial_x (\Phi_2-\Phi_1)(L,t)\\
A_2(t) & =   -\frac{1}{8Lh}(\Phi_1+\Phi_2)(L,t)
\end{align}
\begin{equation}
\label{eq.h000}
H_0=-\frac{i}{8L}\int_0^L\ dx\ \Phi_1(x,0)\partial_x \Phi_1(x,0)+\frac{i}{8L}\int_0^L\
dx\ \Phi_2(x,0)\partial_x \Phi_2(x,0)\ -\frac{1}{2}hH_I\ .
\end{equation}
Note that the energies of the modes are not in $\frac{\pi}{L}\ZZ$, so one
cannot conclude that the sum of the first two terms in (\ref{eq.h000}) equals to $H$.

The above equations suggest how to describe the model discussed in this section
as a perturbation of the $h \to \infty$ limiting model. This description
will be given in the next section.

\subsection{Reverse description}
\label{sec.reverse}
\markright{\thesubsection.\ \ EXACT SPECTRUM - REVERSE DESCRIPTION}

In this section we propose the description of the model (\ref{eq.pertham}) as a perturbation of its $h \to
\infty$ limit. It should be noted that the meaning of the notation $H_0$ or $H_I$ etc.\ 
differs from that in the previous section. The precise correspondence between
the quantities in this section and in the previous section will be given
explicitly for  the coupling constant and for the spectrum. 

\subsubsection{The free model}

The fundamental objects of the  model at $h=\infty$  are two one-component real
fermion fields
$\Phi_1(x,t)$, $\Phi_2(x,t)$ with the anticommutation relations
\begin{align}
\{ \Phi_1(x,t),\Phi_1(y,t)\} & =   4L\delta(x-y)\\
\{ \Phi_2(x,t),\Phi_2(y,t)\} & =   4L\delta(x-y)\\
\{ \Phi_1(x,t),\Phi_2(y,t)\} & =   -4L[\delta(x+y)-\delta(x+y-2L)]
\end{align}
and reality property
\begin{equation}
\Phi_1(x,t)^\dagger =\Phi_1(x,t)\qquad \Phi_2(x,t)^\dagger =\Phi_2(x,t)\ .
\end{equation}
$\Phi_1$ and $\Phi_2$ are dimensionless.\\
The Hamiltonian operator is 
\begin{equation}
H_0=-\frac{i}{8L}\int_0^L\ dx\ \Phi_1(x,0)\partial_x \Phi_1(x,0)+\frac{i}{8L}\int_0^L\
dx\ \Phi_2(x,0)\partial_x \Phi_2(x,0)\ .
\end{equation}
The equations of motion are
\begin{multline}
\label{eq.rmo1}
\partial_t\Phi_1(x,t)=[iH_0,\Phi_1(x,t)]= -\partial_x\Phi_1(x,t) +\\
+\frac{1}{2}\delta(x-L)
[\Phi_1(L,t)-\Phi_2(L,t)]+\frac{1}{2}\delta(x) [-\Phi_1(0,t)-\Phi_2(0,t)]+\\
+\frac{1}{2} \Theta(-x) [\partial_x \Phi_1(x,t)-\partial_x \Phi_2(x,t)]
+\frac{1}{2} \Theta(x-L) [\partial_x \Phi_1(x,t)+\partial_x \Phi_2(x,t)]
\end{multline}
\begin{multline}
\label{eq.rmo2}
\partial_t\Phi_2(x,t)=[iH_0,\Phi_2(x,t)]= \partial_x\Phi_2(x,t) +\\
+\frac{1}{2}\delta(x-L)
[\Phi_1(L,t)-\Phi_2(L,t)]+\frac{1}{2}\delta(x) [\Phi_1(0,t)+\Phi_2(0,t)]+\\
+\frac{1}{2} \Theta(-x) [\partial_x \Phi_1(x,t)-\partial_x \Phi_2(x,t)]
+\frac{1}{2} \Theta(x-L) [-\partial_x \Phi_1(x,t)-\partial_x \Phi_2(x,t)]\ .
\end{multline}
The fermion fields have the following mode expansion:
\begin{align}
\Phi_1(x,t) & =  \sum_{k \in \frac{\pi}{L} \ZZ+\frac{\pi}{2L}} \sqrt{2} a(k)
  e^{ik(t-x)}\\
\Phi_2(x,t) & =  \sum_{k \in \frac{\pi}{L} \ZZ+\frac{\pi}{2L}} -\sqrt{2} a(k)
  e^{ik(t+x)}
\end{align}
\begin{equation}
\{ a(k_1),a^\dagger (k_2)\} =\delta_{k_1,k_2}\qquad a(k)^\dagger =a(-k)\ .
\end{equation}
An orthonormal basis for the Hilbert space is formed by the vectors
\begin{equation}
\ket{a(k_1)a(k_2)\dots a(k_n)0}\ ,
\end{equation}
where $k_i>0$, $k_i\in  \frac{\pi}{L} \ZZ+\frac{\pi}{2L}$, $n\ge 0$.
$a(k)\ket{0}=0$ if $k<0$.
The Hamiltonian operator can be written as  
\begin{equation}
H_0=\sum_{k>0,\ k \in \frac{\pi}{L} \ZZ+\frac{\pi}{2L}}\ k[a(k)a^\dagger (k)]
\end{equation}
after subtracting an infinite constant, by normal ordering for example, which is not
denoted explicitly.
The eigenvalue of the eigenvector $\ket{a(k_1)a(k_2)\dots a(k_n)0}$ is
\begin{equation}
E_{\{k_1,k_2,\dots ,k_n,0 \} }=\sum_{i=1}^n k_i\ . 
\end{equation}
The fermion fields satisfy the following boundary conditions:
\begin{equation}
\brakettt{E_1}{\Phi_1(0,t)+\Phi_2(0,t)}{E_2}=0\qquad
\brakettt{E_1}{\Phi_1(L,t)-\Phi_2(L,t)}{E_2}=0\ ,
\end{equation}
where $\ket{E_1}$, $\ket{E_2}$ are eigenstates of $H_0$. 

One can define a representation of the Virasoro algebra on the Hilbert space
in the same way as in Section \ref{sec.freemod} (see also \cite{FMS,Ginsparg}). This representation
is $(1/2,0)\oplus (1/2,1/2)$. The fields $\Phi_1(x,t)$ and $\Phi_2(x,t)$ can
be converted to right and left moving 
weight $1/2$ primary fields by multiplying them by a suitable simple exponential factor.

\subsubsection{The perturbed model}

The Hamilton operator is
\begin{multline}
H=-\frac{i}{8L}\int_0^L\ dx\ \Phi_1(x,0)\partial_x \Phi_1(x,0)+\frac{i}{8L}\int_0^L\
dx\ \Phi_2(x,0)\partial_x \Phi_2(x,0)\ +\\
+g[i(\Phi_1+\Phi_2)(L,0) \lim_{x\to L} \partial_x
(\Phi_2-\Phi_1)(x,0) ]\ ,  
\end{multline}
where $g$ is a dimensionless coupling constant. In the same way as in Section \ref{sec.pertmod}, the
perturbing term has zero matrix elements between vectors belonging to  different irreducible
representations. 

The limit prescription in the perturbing term is important, and the limit
$\lim_{x\to L}$ should be taken at the end of any calculation. 
It should  be assumed that $\lim_{x \to L} \delta (x-L)=0$, for example,
and similarly for the derivatives of $\delta (x-L)$.

It should be noted that the only
primary field in the $0$ or in the $1/2$ representation is the identity
operator, all other fields are descendant and non-relevant fields.

The equations of motion are obtained by adding the following terms to the
right-hand side of the unperturbed equations of motion (\ref{eq.rmo1}), (\ref{eq.rmo2}): 
\begin{align}
\label{eq.rint1}
[igH_I,\Phi_1(x,t)] & =  g8L\delta(x-L)\lim_{x\to L}
\partial_x(\Phi_2-\Phi_1)(x,t)\\ {}
\label{eq.rint2}
[igH_I,\Phi_2(x,t)] & =  g8L\delta(x-L)\lim_{x\to L}
\partial_x(\Phi_2-\Phi_1)(x,t)
\end{align}
where
\begin{equation}
H_I= [i(\Phi_1+\Phi_2)(L,0) \lim_{x\to L} \partial_x
(\Phi_2-\Phi_1)(x,0) ]\ .
\end{equation}
We remark that we used the following formulae in the computation of
(\ref{eq.rint1}), (\ref{eq.rint2}):
\begin{align}
\{ \Phi_1(x,t), \lim_{y\to L} \partial_y(\Phi_2-\Phi_1)(y,t) \} & =  0\\
\{ \Phi_2(x,t), \lim_{y\to L} \partial_y(\Phi_2-\Phi_1)(y,t) \} & =  0\ .
\end{align}
Similar steps to those in Section \ref{sec.pertmod} can now be taken to obtain
$\brakettt{E_1}{\Phi_1(x,t)}{E_2}$ and\\ 
$\brakettt{E_1}{\Phi_2(x,t)}{E_2}$. 
In the same way as in Section \ref{sec.pertmod}, the forms 
\begin{align}
\brakettt{E_1}{\Phi_1(x,t)}{E_2} & =  e^{ik(t-x)}+\Theta(x-L)D_1(k)e^{ikt}\\
\brakettt{E_1}{\Phi_2(x,t)}{E_2} & =  -e^{ik(t+x)}+\Theta(x-L)D_2(k)e^{ikt}
\end{align}
can be assumed, where 
$D_1(k)$ and $D_2(k)$ are finite constants, $\ket{E_1}$ and
$\ket{E_2}$ are eigenstates of $H$ and $k=E_1-E_2$. 

Solving the equations of motion for
$D_1(k)$, $D_2(k)$, we get
\begin{align}
D_2(k) & =  \cos(kL)\\
D_1(k) & =  -\cos(kL)
\end{align}
and
\begin{equation}
\label{spektrrev}
(kL)\tan(kL)=\frac{-1}{16g}\ .
\end{equation}
(\ref{spektrrev}) is the formula that determines the spectrum of $H$ up to an
overall additive constant. 
The eigenvalues of $H$ are 
\begin{equation}
k_1+k_2+\dots +k_n\ ,
\end{equation}
where $n \ge 0$, $k_i\ge 0$, $k_i\ne k_j$ if $i \ne j$, the $k_i$-s are
real roots of  (\ref{spektrrev}) and the lowest eigenvalue is assumed to be
set to zero by adding a constant which is not written explicitly.
The substitution
\begin{equation}
g=\frac{-1}{256 L^2h^2}
\end{equation}
converts (\ref{spektrrev}) into (\ref{spektr}). Thus  $g:0\to -\infty$ corresponds to the flow
$ (0 \to 1/16) \oplus (1/2 \to 1/16)$.

The following boundary conditions are satisfied:
\begin{align}
\label{eq.revbc0}
\brakettt{E_1}{\Phi_1(0,t)+\Phi_2(0,t)}{E_2}& =  0\\
\label{eq.revbc00}
\brakettt{E_1}{\Phi_1(L,t)-\Phi_2(L,t)}{E_2} & =  0
\end{align}
and
\begin{align}
\label{eq.revbc1}
\lim_{x\to 0}\brakettt{E_1}{\Phi_1(x,t)+\Phi_2(x,t)}{E_2}& =  0\\
\label{eq.revbc2}
\lim_{x\to
  L}\brakettt{E_1}{\partial_x\Phi_1(x,t)-\partial_x\Phi_2(x,t)}{E_2} & =  \frac{1}{16Lg}\brakettt{E_1}{\Phi_1(L,t)-\Phi_2(L,t)}{E_2} \\
\lim_{g\to\infty}\lim_{x\to L}\brakettt{E_1}{\Phi_1(x,t)+\Phi_2(x,t)}{E_2} & =
 0\\
\lim_{g\to\infty}\lim_{x\to L}\brakettt{E_1}{\Phi_1(x,t)-\Phi_2(x,t)}{E_2} &
\ne   0\ ,
\end{align}
where $\ket{E_1}$ and $\ket{E_2}$ are eigenstates of $H$.

In perturbation theory there are divergences if we take $x=L$ at the
beginning. However, if one allows $x$ to take general values, then in
Rayleigh-Schr\"odinger perturbation theory  for the differences of the energy
levels one can
expect to get sums
at any fixed order  which are possible to evaluate. We expect
that the evaluation yields, besides non-singular parts, $\Theta(x-L)$, $\delta (x-L)$ and its
derivatives, and taking  the $x \to L$
limit gives
finite result eventually.

\subsection{Bethe-Yang equations}
\markright{\thesubsection.\ \ EXACT SPECTRUM - BETHE-YANG EQUATION}

The Bethe-Yang equations can be used to give a description of the spectrum  of
models in finite volume which have factorized
scattering in their infinite volume limit. The Bethe-Yang equations for models
defined on a cylinder are exposed, for example, in \cite{ZamTBA,KM1,key-21}; 
for models defined  on a strip they
were written down in \cite{FeSa,BPTby}. It should be noted that the Bethe-Yang
equations usually
give approximate result only.  

In the case of the  model that we study the ingredients of the Bethe-Yang description are the
following: there is a single massless particle with fermionic statistics,  the
two-particle S-matrix
is a constant scalar $S(k)=-1$, where $k$ is the relative momentum. The
reflection matrix on the left-hand side can be read from (\ref{eq.bc1}), it is
$R_L(k)=-1$, the reflection matrix on the right-hand side can be read from
(\ref{eq.bc2}), it is 
\begin{equation}
R_R(k)=\frac{16Lh^2+ik}{16Lh^2-ik}\ .
\end{equation}

The transfer matrices for $N$-particle states are  scalars:
\begin{multline}
T_i(k_1,k_2,\dots,k_N)=\\
=R_L(k_i)R_R(k_i)\prod_{j,\ j\ne k}S(k_i+k_j)\prod_{j,\ j\ne
  k}S(k_i-k_j)=-R_R(k_i)\ , \qquad i=1\dots N\ ,
\end{multline}
where $k_1>k_2>\dots k_N \ne 0$. This very simple form is the consequence of
the simplicity of the S-matrix.
The   Bethe-Yang equations
for the momenta $k_1,k_2,\dots k_N$ of the $N$-particle   states take the form
\begin{equation}
\label{eq.BY2}
e^{2ik_iL}T_i(k_1,k_2,\dots,k_N)=e^{2ik_iL}\frac{ik_i+16Lh^2}{ik_i-16Lh^2}=1\
,\qquad i=1\dots N\ . 
\end{equation}    
The total energy of an $N$-particle state in the Bethe-Yang framework is 
\begin{equation}
E=\sum_{i=1}^N k_i\ .
\end{equation}
(\ref{eq.BY2}) can be rewritten as 
\begin{equation}
k_iL\tan(k_iL)=16L^2h^2\ , \qquad i=1\dots N\ ,
\end{equation}
which has the same form as (\ref{spektr}). This means that the Bethe-Yang
description reproduces the result of Section \ref{sec.pertmod} for the
spectrum exactly.

The `reverse' model is similar, one can read from (\ref{eq.revbc1}) and (\ref{eq.revbc2})
that 
\begin{equation}
R_L(k)=-1\ ,\qquad R_R(k)=\frac{1-ik16Lg}{1+ik16Lg}\ ,
\end{equation}
and the Bethe-Yang
equations for the momenta can be written as
\begin{equation}
k_iL\tan(k_iL)=\frac{-1}{16g}\ , \qquad i=1\dots N\ ,
\end{equation}
which has the same form as (\ref{spektrrev}), i.e.\  the result of Section
\ref{sec.reverse} for the spectrum is reproduced exactly.

\section{Exact spectrum in the Mode Truncated version}
\label{sec.modetr}
\markright{\thesection.\ \ EXACT MODE TRUNCATED SPECTRUM}

\subsection{The free model}
\label{sec.modetrfree}

Let $n_c$, called the truncation level, be a positive integer. The mode truncated version of the free model
described in Section \ref{sec.freemod} is the following:
\begin{align}
\{ \Psi_1(x,t),\Psi_1(y,t)\} & =   2[1+2\sum_{k\in\frac{\pi}{L} \{ 1\dots n_c \} } \cos(k(x-y)) ]\\
\{ \Psi_2(x,t),\Psi_2(y,t)\} & =   2[1+2\sum_{k\in\frac{\pi}{L} \{ 1\dots n_c \} } \cos(k(x-y)) ]\\
\{ \Psi_1(x,t),\Psi_2(y,t)\} & =  -2[1+2\sum_{k\in\frac{\pi}{L} \{ 1\dots n_c \} } \cos(k(x+y)) ]\\
\{ A_2(t), \Psi_1(x,t) \} & =  0\\
\{ A_2(t), \Psi_2(x,t) \} & =  0\\
\{ A_2(t),A_2(t) \} & =  2\ , 
\end{align}
\begin{equation}
\Psi_1(x,t)^\dagger =\Psi_1(x,t)\qquad \Psi_2(x,t)^\dagger =\Psi_2(x,t)\qquad
A_2(t)^\dagger =A_2(t)\ ,
\end{equation}

\begin{align}
\Psi_1(x,t) & =  \sum_{k\in\frac{\pi}{L} \{ 1\dots n_c \} }\sqrt{2}[a(k)e^{ik(t-x)}+a^+(k)e^{-ik(t-x)}]+A_1\\
\Psi_2(x,t) & =  \sum_{k\in\frac{\pi}{L}\{ 1\dots n_c   \}
}\sqrt{2}[-a(k)e^{ik(t+x)}-a^+(k)e^{-ik(t+x)}]-A_1\ .
\end{align}

The equations (\ref{a1})-(\ref{a2}) apply unchanged. 

The Hamiltonian operator is 
\begin{equation}
H_0=\sum_{k\in\frac{\pi}{L} \{ 1\dots n_c \} }\ k [a(k)a^+(k)]\ .
\end{equation}
The equations of motion are
\begin{alignat}{2}
\partial_t \Psi_1(x,t) &= [iH_0,\Psi_1(x,t)]    &&= -\partial_x \Psi_1(x,t)\\
\partial_t \Psi_2(x,t) &= [iH_0,\Psi_2(x,t)]   &&= \partial_x \Psi_2(x,t)\\
\partial_t A_2 (t)     &= [iH_0, A_2(t)]       &&= 0\ .
\end{alignat}
The Hilbert space and the energy eigenstates are similar to those in Section
\ref{sec.freemod}, but 
\begin{equation}
k\in \frac{\pi}{L} \{ -n_c,-n_c+1,\dots ,n_c-1,n_c\}
\end{equation}
applies instead of $k \in \pi\ZZ /L$. The Hilbert space is $2\times 2^{n_c}$ dimensional.

\subsection{The perturbed model}

The perturbed Hamiltonian operator is $H_0+hH_I$, where
\begin{equation}
H_I=iA_2(0)[\Psi_2(L,0)-\Psi_1(L,0)]\ .
\end{equation}

The equations of motion are
\begin{alignat}{2}
\partial_t \Psi_1(x,t) &=[i(H_0+hH_I),\Psi_1(x,t)] && =-\partial_x \Psi_1(x,t)-hA_2(t)C_1(x)\\
\partial_t \Psi_2(x,t) &=[i(H_0+hH_I),\Psi_2(x,t)] && =\partial_x \Psi_2(x,t)-hA_2(t)C_2(x)\\
\frac{d}{dt}A_2(t) &=[i(H_0+hH_I),A_2(t)] && =2h(\Psi_2(L,t)-\Psi_1(L,t))\ ,
\end{alignat}
where
\begin{alignat}{2}
C_1(x) & = \{ \Psi_1(x,t),\Psi_2(L,t)-\Psi_1(L,t) \} && =-4[1+2 \sum_{k\in\frac{\pi}{L} \{ 1\dots n_c \} } \cos (k(x+L)) ]\\
C_2(x) & = \{ \Psi_2(x,t),\Psi_2(L,t)-\Psi_1(L,t) \} && =-C_1(x)\ .
\end{alignat}
These equations are linear as in the non-truncated case, so sandwiching them
between energy eigenstates gives a system of 3 first-order linear partial
differential equations for the expectation values. The analogue of
(\ref{spektr}) can be obtained from these equations in the following way: 
we can eliminate $A_2$:
\begin{align}
\label{wr1}
\partial_t^2 \Psi_1(x,t) & =  -\partial_{xt}\Psi_1(x,t)-2h^2(\Psi_2(L,t)-\Psi_1(L,t))C_1(x)\\
\partial_t^2 \Psi_2(x,t) & =  \partial_{xt}\Psi_1(x,t)-2h^2(\Psi_2(L,t)-\Psi_1(L,t))C_2(x)
\label{wr2}
\end{align}
and introduce the functions $f_1(x)$, $f_2(x)$:
\begin{align}
\brakettt{E_1}{\Psi_1(x,t)}{E_2} & =  f_1(x)e^{ikt}\\
\brakettt{E_1}{\Psi_2(x,t)}{E_2} & =  f_2(x)e^{ikt}\ ,
\end{align}
where $k=E_1-E_2$ and the dependence of $f_1$ and $f_2$ on $k$ is not denoted
explicitly. Equations (\ref{wr1}) and (\ref{wr2}) give 
\begin{align}
-k^2f_1(x) & =  -ikf_1'(x)-2h^2(f_2(L)-f_1(L))C_1(x)\\
-k^2f_2(x) & =  ikf_2'(x)-2h^2(f_2(L)-f_1(L))C_2(x)\ .
\end{align}
We also have the boundary conditions
\begin{align}
f_1(0) & =  -f_2(0)\\
f_1(L) & =  -f_2(L)
\end{align}
so
\begin{align}
-k^2f_1(x) & =  -ikf_1'(x)+4h^2C_1(x)f_1(L)\\
-k^2f_2(x) & =  ikf_2'(x)-4h^2C_2(x)f_2(L)\ .
\end{align}
These are almost first-order inhomogeneous linear ordinary differential
equations, the difference is that the inhomogeneity depends on the unknown
functions. The method of undetermined coefficients can nevertheless be applied:
\begin{gather}
f_1(x)=C(x)e^{-ikx}\\
C'(x)=\frac{4h^2}{ik}e^{ikx}C_1(x)f_1(L)\\
C(x)=C(0)+\int_0^x\ dx'\ \frac{4h^2}{ik}e^{ikx'}C_1(x')f_1(L)=C(0)+I_1(x)\\
f_1(L)=C(L)e^{-ikL}= [C(0)+ \int_0^L\ dx'\
\frac{4h^2}{ik}e^{ikx'}C_1(x')f_1(L)  ]e^{-ikL}\\
C(0)=f_1(0)\ ,
\end{gather}

\begin{gather}
f_2(x)=D(x)e^{ikx}\\
D'(x)=\frac{4h^2}{ik}e^{-ikx}C_2(x)f_2(L)\\
D(x)=D(0)+\int_0^x\ dx'\ \frac{4h^2}{ik}e^{-ikx'}C_2(x')f_2(L)=D(0)+I_2(x)\\
f_2(L)=D(L)e^{ikL}=[D(0)+ \int_0^L\ dx'\
\frac{4h^2}{ik}e^{-ikx'}C_2(x')f_2(L)  ]e^{ikL}\\
D(0)=f_2(0)\ .
\end{gather}
In particular
\begin{align}
f_1(L)e^{ikL} & =  C(0)+I_1(L)\\
f_2(L)e^{-ikL} & =  D(0)+I_2(L)\ .
\end{align}
Using the boundary condition $C(0)+D(0)=0$:
\begin{equation}
I_1(L)+I_2(L)=f_1(L)e^{ikL}+f_2(L)e^{-ikL}\ ,
\end{equation}
and then the boundary condition $f_1(L)+f_2(L)=0$:
\begin{equation}
\int_0^L \ dx'
\frac{4h^2}{ik}[e^{ikx'}C_1(x')-e^{-ikx'}C_2(x')]=e^{ikL}-e^{-ikL}\ .
\end{equation}
Simplifying this equation we finally obtain 
\begin{equation}
\label{spektrmt}
16h^2[\frac{1}{k^2}+\sum_{k_0\in\frac{\pi}{L} \{ 1\dots n_c \} }
  \frac{2}{k^2-k_0^2}   ]=1\ ,
\end{equation}
which is the analogue of (\ref{spektr}) and determines the energy of the modes
as functions of $h$.

(\ref{spektrmt}) as an algebraic equation for $k$  has finitely many real
roots, and if $k$ is a root, then $-k$ is a root as well. All real roots converge
to finite values as $h \to \infty$ except for the pair with the largest
absolute value. This pair of roots diverges linearly as $h \to \infty$. This
pair has the largest absolute value already at $h=0$.  A 
 consequence of this behaviour is that the lower half of the spectrum, namely
 those states which do not contain the mode with the highest energy, remains
 finite as $h\to\infty$, whereas the higher half of the spectrum, i.e.\ the
 states which contain the mode with the highest energy, diverges linearly as $h
 \to \infty$ with a common slope. Here it is assumed that the ground state
 energy is set to zero. In the subsequent sections we shall consider the lower half of the spectrum. The
 look of this half as a function of $h$ is very similar to that shown in Figure \ref{fig.1}. 
We remark that it is not hard to see that the two halves of the
 spectrum are mirror symmetric with respect to a horizontal line, if the
 ground state energy is set appropriately.

Applying the formula 
\begin{equation}
1+\sum_{n=1}^\infty \frac{2k^2}{k^2-4n^2\pi^2}=\frac{k/2}{\tan (k/2)}
\end{equation}
we can easily verify that the limit of (\ref{spektrmt}) as $n_c\to \infty$ is (\ref{spektr}).

\section{Power series expansion of the energy levels}
\label{sec.RS}
\markright{\thesection.\ \ POWER SERIES EXPANSION}

The eigenvectors of $H_0$ suitable for Rayleigh-Schr\"odinger perturbation
theory are those introduced in (\ref{eq.freebasis}). Degenerate perturbation
theory has to be used.

The nonzero matrix elements of $H_I$ in Section \ref{sec.pertmod} are the following:
\begin{align}
\brakettt{Qu}{H_I}{Qu} & =  2\\
\brakettt{Qv}{H_I}{Qv} & =  -2\\
\brakettt{Qu}{H_I}{Pv} = \brakettt{Pv}{H_I}{Qu} & =  2\sqrt{2}(-1)^{n+m}(-1)^{kL/\pi}\\
\brakettt{Qv}{H_I}{Pu} =\brakettt{Pu}{H_I}{Qv} & =  -2\sqrt{2}(-1)^{n+m}(-1)^{kL/\pi}\ ,
\end{align}
where 
\begin{align}
P & =  a(k_1)a(k_2)\dots a(k_m)a(k)a(k_{m+1})\dots a(k_n) \\
Q & =  a(k_1)a(k_2)\dots a(k_n)\ .
\end{align}

We remark that certain perturbative calculations were also done in \cite{CSS}.

\paragraph{Non-truncated case:}
The eigenvalue of the state  starting from\\
$\ket{ a(\frac{N_1\pi}{L})
  a(\frac{N_2\pi}{L})\dots  a(\frac{N_r\pi}{L}) v}$ at $h=0$ is
\begin{multline}
E_{ \{ N_1,N_2,\dots ,N_r,v\} }(h) =
\frac{(N_1+N_2+\dots +N_r)\pi}{L}- 2h +\\
+( \frac{16L}{\pi}
(\frac{1}{N_1}+\frac{1}{N_2}+\dots +\frac{1}{N_r})  
-\sum_{n=1}^\infty \frac{8L}{n\pi} )h^2
+ ( \sum_{n=1}^\infty \frac{32L^2}{n^2\pi^2})h^3+ \dots 
\end{multline}
and the eigenvalue  of the state  starting form $\ket{ a(\frac{N_1\pi}{L})
  a(\frac{N_2\pi}{L})\dots  a(\frac{N_r\pi}{L}) u}$ at $h=0$ is
\begin{multline}
E_{ \{ N_1,N_2,\dots ,N_r,u\} }(h) =
\frac{(N_1+N_2+\dots +N_r)\pi}{L}+ 2h +\\
+( \frac{16L}{\pi}
(\frac{1}{N_1}+\frac{1}{N_2}+\dots +\frac{1}{N_r})  
-\sum_{n=1}^\infty \frac{8L}{n\pi} )h^2
+ (- \sum_{n=1}^\infty \frac{32L^2}{n^2\pi^2})h^3+ \dots\ . 
\end{multline}
In particular
\begin{align}
E_{\{ v\} }(h) & =  0-2h-\sum_{n=1}^\infty \frac{8L}{n\pi}h^2+ h^3
\sum_{n=1}^\infty \frac{32L^2}{n^2\pi^2}+\dots \\
E_{\{ u \} }(h) & =  0+2h-\sum_{n=1}^\infty \frac{8L}{n\pi}h^2 -h^3
\sum_{n=1}^\infty \frac{32L^2}{n^2\pi^2}+\dots\ . 
\end{align}
We note that the  coefficient of $h^2$ is ultraviolet divergent and should be
regularized. The TCSA and the mode truncation both provide a regularization.

The energy differences corresponding to the creation operators
$b(k(n,h))$ are the following: for $n=0$
\begin{equation}
\Delta E_0(h)=E_{\{ u \} }(h)-E_{\{ v \} }(h)=0+4h+0+h^3(- \sum_{n=1}^{\infty}
\frac{64L^2}{n^2\pi^2})+\dots 
\end{equation}
and for $n=N>0$
\begin{equation}
\Delta E_N=E_{\{N,u\} }(h)-E_{\{v\} }(h)=\frac{N\pi}{L}+0+ h^2
\frac{16L}{N\pi}+0+\dots\ .   
\end{equation}

Generally
\begin{multline}
E_{ \{ N_1,N_2,\dots ,N_r,v\} }(h)- E_{ \{ v\} }(h) =
\frac{(N_1+N_2+\dots +N_r)\pi}{L}+ 0\cdot h +\\
+ \frac{16L}{\pi}
(\frac{1}{N_1}+\frac{1}{N_2}+\dots +\frac{1}{N_r})h^2
+  0\cdot h^3+ \dots 
\end{multline}
\begin{multline}
E_{ \{ N_1,N_2,\dots ,N_r,v\} }(h)- E_{ \{ u\} }(h) =
\frac{(N_1+N_2+\dots +N_r)\pi}{L} -4h +\\
+ \frac{16L}{\pi}
(\frac{1}{N_1}+\frac{1}{N_2}+\dots +\frac{1}{N_r})  
 h^2
+ (\sum_{n=1}^{\infty} \frac{64L^2}{n^2\pi^2})  h^3+ \dots 
\end{multline}
\begin{multline}
E_{ \{ N_1,N_2,\dots ,N_r,u\} }(h)- E_{ \{ u\} }(h) =
\frac{(N_1+N_2+\dots +N_r)\pi}{L}+ 0\cdot h +\\
+ \frac{16L}{\pi}
(\frac{1}{N_1}+\frac{1}{N_2}+\dots +\frac{1}{N_r})  
h^2
+  0\cdot h^3+ \dots 
\end{multline}
\begin{multline}
E_{ \{ N_1,N_2,\dots ,N_r,u\} }(h)- E_{ \{ v\} }(h) =
\frac{(N_1+N_2+\dots +N_r)\pi}{L} +4h +\\
+ \frac{16L}{\pi}
(\frac{1}{N_1}+\frac{1}{N_2}+\dots +\frac{1}{N_r})h^2
-   (\sum_{n=1}^{\infty} \frac{64L^2}{n^2\pi^2})  h^3+ \dots\ .
\end{multline}

\paragraph{MT scheme:}
\begin{multline}
E_{ \{ N_1,N_2,\dots ,N_r,v\} }(h) =
\frac{(N_1+N_2+\dots +N_r)\pi}{L}- 2h +\\
+( \frac{16L}{\pi}
(\frac{1}{N_1}+\frac{1}{N_2}+\dots +\frac{1}{N_r})  
-\sum_{n=1}^{n_c} \frac{8L}{n\pi} )h^2
+ ( \sum_{n=1}^{n_c} \frac{32L^2}{n^2\pi^2})h^3+ \dots 
\end{multline}
\begin{multline}
E_{ \{ N_1,N_2,\dots ,N_r,u\} }(h) =
\frac{(N_1+N_2+\dots +N_r)\pi}{L}+ 2h +\\
+( \frac{16L}{\pi}
(\frac{1}{N_1}+\frac{1}{N_2}+\dots +\frac{1}{N_r})  
-\sum_{n=1}^{n_c} \frac{8L}{n\pi} )h^2
+ (- \sum_{n=1}^{n_c} \frac{32L^2}{n^2\pi^2})h^3+ \dots 
\end{multline}
\begin{align}
E_{\{ v\} }(h) & =  0-2h-\sum_{n=1}^{n_c} \frac{8L}{n\pi}h^2+ h^3
\sum_{n=1}^{n_c} \frac{32L^2}{n^2\pi^2}+\dots \\
E_{\{ u \} }(h) & =  0+2h-\sum_{n=1}^{n_c} \frac{8L}{n\pi}h^2 -h^3
\sum_{n=1}^{n_c} \frac{32L^2}{n^2\pi^2}+\dots 
\end{align}
\begin{equation}
\Delta E_0(h)=E_{\{ u \} }(h)-E_{\{ v \} }(h)=0+4h+0+h^3(- \sum_{n=1}^{n_c}
\frac{64L^2}{n^2\pi^2})+\dots 
\end{equation}
\begin{equation}
\Delta E_N=E_{\{N,u\} }(h)-E_{\{v\} }(h)=\frac{N\pi}{L}+0+ h^2 \frac{16L}{N\pi}+0+\dots   
\end{equation}
In these formulae $n_c$ is the truncation level introduced in Section \ref{sec.modetrfree}. 

Generally
\begin{multline}
\label{eq.pertmt1}
E_{ \{ N_1,N_2,\dots ,N_r,v\} }(h)- E_{ \{ v\} }(h) =
\frac{(N_1+N_2+\dots +N_r)\pi}{L}+ 0\cdot h +\\
+ \frac{16L}{\pi}
(\frac{1}{N_1}+\frac{1}{N_2}+\dots +\frac{1}{N_r})h^2
+  0\cdot h^3+ \dots 
\end{multline}
\begin{multline}
E_{ \{ N_1,N_2,\dots ,N_r,v\} }(h)- E_{ \{ u\} }(h) =
\frac{(N_1+N_2+\dots +N_r)\pi}{L} -4h +\\
+ \frac{16L}{\pi}
(\frac{1}{N_1}+\frac{1}{N_2}+\dots +\frac{1}{N_r})  
 h^2
+ (\sum_{n=1}^{n_c} \frac{64L^2}{n^2\pi^2})  h^3+ \dots 
\end{multline}
\begin{multline}
E_{ \{ N_1,N_2,\dots ,N_r,u\} }(h)- E_{ \{ u\} }(h) =
\frac{(N_1+N_2+\dots +N_r)\pi}{L}+ 0\cdot h +\\
+ \frac{16L}{\pi}
(\frac{1}{N_1}+\frac{1}{N_2}+\dots +\frac{1}{N_r})  
h^2
+  0\cdot h^3+ \dots 
\end{multline}
\begin{multline}
E_{ \{ N_1,N_2,\dots ,N_r,u\} }(h)- E_{ \{ v\} }(h) =
\frac{(N_1+N_2+\dots +N_r)\pi}{L} +4h +\\
\label{eq.pertmt2}
+ \frac{16L}{\pi}
(\frac{1}{N_1}+\frac{1}{N_2}+\dots +\frac{1}{N_r})h^2
-   (\sum_{n=1}^{n_c} \frac{64L^2}{n^2\pi^2})  h^3+ \dots\ . 
\end{multline}

\paragraph{TCS scheme:}
\begin{multline}
E_{ \{ N_1,N_2,\dots ,N_r,v\} }(h) =
\frac{(N_1+N_2+\dots +N_r)\pi}{L} -2h +\\
+( \frac{16L}{\pi}
(\frac{1}{N_1}+\frac{1}{N_2}+\dots +\frac{1}{N_r})  
-\sum_{n=1}^{n_m} \frac{8L}{n\pi} )h^2
+ ( \sum_{n=1}^{n_m} \frac{32L^2}{n^2\pi^2})h^3+ \dots 
\end{multline}
\begin{multline}
E_{ \{ N_1,N_2,\dots ,N_r,u\} }(h) =
\frac{(N_1+N_2+\dots +N_r)\pi}{L}+ 2h +\\
+( \frac{16L}{\pi}
(\frac{1}{N_1}+\frac{1}{N_2}+\dots +\frac{1}{N_r})  
-\sum_{n=1}^{n_m} \frac{8L}{n\pi} )h^2
+ (- \sum_{n=1}^{n_m} \frac{32L^2}{n^2\pi^2})h^3+ \dots 
\end{multline}
where $n_m=n_c-(N_1+N_2+\dots +N_r)$, $n_c$ is the conformal truncation level.
\begin{align}
E_{\{ v\} }(h) & =  0-2h-\sum_{n=1}^{n_c} \frac{8L}{n\pi}h^2+ h^3
\sum_{n=1}^{n_c} \frac{32L^2}{n^2\pi^2}+\dots \\
E_{\{ u \} }(h) & =  0+2h-\sum_{n=1}^{n_c} \frac{8L}{n\pi}h^2 -h^3
\sum_{n=1}^{n_c} \frac{32L^2}{n^2\pi^2}+\dots 
\end{align}
\begin{equation}
\Delta E_0(h)=E_{\{ u \} }(h)-E_{\{ v \} }(h)=0+4h+0+h^3(- \sum_{n=1}^{n_c}
\frac{64L^2}{n^2\pi^2})+\dots 
\end{equation}
\begin{equation}
\Delta E_N=E_{\{N,u\} }(h)-E_{\{v\} }(h)=\frac{N\pi}{L}+0+ h^2
(\frac{16L}{N\pi} +\sum_{n=n_m+1}^{n_c} \frac{8L}{n\pi}) +h^3  \sum_{n=n_m+1}^{n_c} \frac{32L^2}{n^2\pi^2} +\dots   
\end{equation}
Generally
\begin{multline}
\label{eq.perttcsa1}
E_{ \{ N_1,N_2,\dots ,N_r,v\} }(h)- E_{ \{ v\} }(h) =
\frac{(N_1+N_2+\dots +N_r)\pi}{L}+ 0 +\\
+( \frac{16L}{\pi}
(\frac{1}{N_1}+\frac{1}{N_2}+\dots +\frac{1}{N_r})  
+\sum_{n=n_m+1}^{n_c} \frac{8L}{n\pi} )h^2
-  \sum_{n=n_m+1}^{n_c} \frac{32L^2}{n^2\pi^2}h^3+ \dots 
\end{multline}
\begin{multline}
E_{ \{ N_1,N_2,\dots ,N_r,v\} }(h)- E_{ \{ u\} }(h) =
\frac{(N_1+N_2+\dots +N_r)\pi}{L} -4h +\\
+( \frac{16L}{\pi}
(\frac{1}{N_1}+\frac{1}{N_2}+\dots +\frac{1}{N_r})  
+\sum_{n=n_m+1}^{n_c} \frac{8L}{n\pi} )h^2
+  (\sum_{n=n_1}^{n_m} \frac{32L^2}{n^2\pi^2} + \sum_{n=1}^{n_c} \frac{32L^2}{n^2\pi^2})  h^3+ \dots 
\end{multline}
\begin{multline}
E_{ \{ N_1,N_2,\dots ,N_r,u\} }(h)- E_{ \{ u\} }(h) =
\frac{(N_1+N_2+\dots +N_r)\pi}{L}+ 0 +\\
+( \frac{16L}{\pi}
(\frac{1}{N_1}+\frac{1}{N_2}+\dots +\frac{1}{N_r})  
+\sum_{n=n_m+1}^{n_c} \frac{8L}{n\pi} )h^2
+  \sum_{n=n_m+1}^{n_c} \frac{32L^2}{n^2\pi^2}h^3+ \dots 
\end{multline}
\begin{multline}
E_{ \{ N_1,N_2,\dots ,N_r,u\} }(h)- E_{ \{ v\} }(h) =
\frac{(N_1+N_2+\dots +N_r)\pi}{L} +4h +\\
\label{eq.perttcsa2}
+( \frac{16L}{\pi}
(\frac{1}{N_1}+\frac{1}{N_2}+\dots +\frac{1}{N_r})  
+\sum_{n=n_m+1}^{n_c} \frac{8L}{n\pi} )h^2
-  (\sum_{n=n_1}^{n_m} \frac{32L^2}{n^2\pi^2} + \sum_{n=1}^{n_c}
\frac{32L^2}{n^2\pi^2})  h^3+ \dots\ . 
\end{multline}

We remark that the above formulae show that the truncated energy gaps
converge to the non-truncated energy gaps as $1/n_c$.

\section{Perturbative results}
\label{sec.pertres}
\markright{\thesection.\ \ PERTURBATIVE RESULTS}

The renormalized Hamiltonian operator is 
\begin{equation}
H^{r}=s_0(h,n_c)H_0+s_1(h,n_c)H_I\ ,
\end{equation}
where the functions $s_0(h,n_c)$ and $s_1(h,n_c)$ are determined by the renormalization
condition, $n_c$ is the truncation level. The renormalization condition in the present case is the following: the differences  of those  eigenvalues of
$H^{r}$ that are low compared to the truncation
level  should be equal to those of the truncated Hamiltonian operator
$H^t(n_c)$.  $H^t(n_c)$ is, in particular, the TCSA Hamiltonian operator
$H^{TCSA}(n_c)$, or the Hamiltonian operator $H^{MT}(n_c)$ of the mode
truncated model. This condition applies separately and independently within the $u$ and $v$
sectors, and we have in fact a   pair $s_0^u$, $s_1^u$ for the $u$ sector and another
pair 
 $s_0^v$, $s_1^v$ for the $v$ sector.

The renormalization condition is a very strong condition on $s_0$ and $s_1$
and generally we cannot expect that it can be satisfied. It is possible,
however, that it can be  satisfied in certain approximations.

\subsection{Mode Truncation Scheme}

Using the equations (\ref{eq.pertmt1})-(\ref{eq.pertmt2}) in Section \ref{sec.RS} we can obtain the following
results:

The renormalization conditions have a solution if the eigenfunctions are
expanded into a power series in $h$ and terms that are of order higher than 3 are omitted: 
\begin{align}
\label{eq.crv1}
s_0(h,n_c) & =  1+x_1h^2+O(h^4)\\
\label{eq.crv2}
s_1(h,n_c) & =  h+x_2h^3+O(h^4)
\end{align}
\begin{equation}
x_1  =  0 \qquad
x_2  =  \frac{1}{2}(S-S(n_c))L^2
\end{equation}
where
\begin{equation}
S=\sum_{n=1}^\infty \frac{32}{n^2\pi^2} \qquad
S(n_c)=\sum_{n=1}^{n_c} \frac{32}{n^2\pi^2}\ .
\end{equation}
This solution applies to both the $u$ and $v$ sectors and it is exact in $n_c$.
We remark that 
\begin{equation}
S-S(n_c) = \frac{32}{\pi^2 n_c}+O(1/n_c^2) \ .
\end{equation}

In the MT scheme we can obtain another result that is not perturbative in $h$:
doing power series expansion we obtain the formula 
\begin{equation}
\frac{1}{k^2}+\sum_{n=1}^{n_c} \frac{2}{k^2-\frac{\pi^2}{L^2} n^2} =
\frac{L}{k\tan(kL)}+\frac{2L^2}{\pi^2}+\frac{1}{n_c} +O(1/n_c^2)\ .
\end{equation}
Omitting the terms which are of second or higher order in $1/n_c$, equation (\ref{spektrmt})  takes the form
\begin{equation}
1=16h^2\left( \frac{L}{k\tan(kL)}+\frac{2L^2}{\pi^2}+\frac{1}{n_c} \right)
\end{equation}
or
\begin{equation}
kL\tan(kL)=16L^2h^2 \frac{1}{1-\frac{32L^2h^2}{\pi^2 n_c}}
\end{equation}
which takes the form of (\ref{spektr}) if 
\begin{equation}
h_{\mathrm{eff}}=\frac{h}{\sqrt{1-\frac{32L^2h^2}{\pi^2
      n_c}}}=h+16h^3\frac{L^2}{\pi^2}\frac{1}{n_c} +O(1/n_c^2)
\end{equation}
is introduced:
\begin{equation}
kL\tan(kL)=16L^2h_{\mathrm{eff}}^2\ .
\end{equation}
This means that rescaling by 
\begin{align}
&s_0(h,n_c)=1+O(1/n_c^2)\\
&s_1(h,n_c)=h+16h^3\frac{L^2}{\pi^2}\frac{1}{n_c}+O(1/n_c^2)
\end{align}
improves the convergence of the mode truncated energy gaps to the exact
energy gaps from order $1/n_c$
to order $1/n_c^2$, i.e.\ the difference between the energy gaps of
$s_0(h,n_c)H_0+s_1(h,n_c)H_I$ and the energy gaps of $(H_0+hH_I)^{MT}$ tends to zero
as $1/n_c^2$ for any fixed finite value of $h$, whereas the difference between the
energy gaps of   $H_0+hH_I$ and  $(H_0+hH_I)^{MT}$ tends to zero as $1/n_c$.

\subsection{TCS scheme}

Using the equations (\ref{eq.perttcsa1})-(\ref{eq.perttcsa2}) in Section \ref{sec.RS} we  obtain the following
results:

The renormalization conditions have a solution if the eigenfunctions are
expanded into a power series in $h$ and in $1/n_c$ and terms that are of order
higher than 3 in $h$ and 1 in $1/n_c$  are omitted: 
\begin{align}
\label{eq.crv3}
s_0(h,n_c) & =  1+x_1h^2+y_1h^3 + O(h^4)\\
\label{eq.crv4}
s_1(h,n_c) & =  h+y_2h^2+x_2h^3+O(h^4)
\end{align}
\begin{align}
x_1 & =  \frac{8L^2}{\pi^2 n_{c}}+O(1/n_{c}^2) &
y_1 & =  0 + O(1/n_c^2) \\
x_2 & =  \frac{1}{2}(S-S(n_c))L^2+O(1/n_{c}^2) &
y_2 & =  0 + O(1/n_c^2)\ . 
\end{align}
This solution applies to both sectors. 
If terms of order $1/n_c^2$ are also taken into consideration, then the
renormalization conditions do not have a solution.

Calculating $s_0(h,n_c)$ and $s_1(h,n_c)$ from the 
 three lowest energy levels gives the result
\begin{align}
s_0^u(h,n_c) & =  1+h^2\frac{8L^2}{(n_c-1)\pi^2}-h^3\frac{32L^3}{(n_c-1)^2n_c\pi^3} 
+O(h^4)\\
s_1^u(h,n_c) & =  h+\frac{2L}{(n_c^2-n_c)\pi}h^2+
[\frac{16L^2(1-5n_c+3n_c^2)}{2(n_c-1)^2n_c^2
  \pi^2 } +\frac{1}{2}(S-S(n_c))L^2] h^3 +O(h^4) 
\end{align}
in the $u$ sector and 
\begin{align}
s_0^v(h,n_c) & =  1+h^2\frac{8L^2}{(n_c-1)\pi^2}+h^3\frac{32L^3}{(n_c-1)^2n_c\pi^3} 
+O(h^4)\\
s_1^v(h,n_c) & =  
h-\frac{2L}{(n_c^2-n_c)\pi}h^2+
[\frac{16L^2(1-5n_c+3n_c^2)}{2(n_c-1)^2n_c^2
  \pi^2 } +\frac{1}{2}(S-S(n_c))L^2] h^3 +O(h^4)
\end{align}
in the $v$ sector. These formulae are exact in $n_c$. In this case $s_0(h,n_c)$ and $s_1(h,n_c)$
are given by the following formulae:
\begin{align}
\label{eq.s1}
&\frac{s_1}{s_0}(h)=\left( \frac{E_2-E_0}{E_1-E_0} \right)^{-1}\left(  \left(
  \frac{E_2-E_0}{E_1-E_0}\right)^{TCSA}(h) \right)  \\
\label{eq.s2}
&s_0(h)=\frac{(E_1-E_0)^{TCSA}(h)}{(E_1-E_0)(\frac{s_1}{s_0}(h))}\ ,
\end{align}
where the superscriptless quantities are the non-truncated ones. $E_0,E_1,E_2$
denote the three lowest energy eigenvalues.

The coefficient $y_1$ of $h^3$ in $s_0(h)$ and the coefficient $y_2$ of $h^2$ in $s_1(h)$
depend on which three energy levels one uses to calculate them, and the same
applies to that part of $x_1$ and $x_2$  which is beyond first order in $1/n_c$.

\section{Numerical results}
\label{sec.numres}
\markright{\thesection.\ \ NUMERICAL RESULTS}

In the numerical calculations described in this section the value of $L$ was
set equal to 1. This does not affect the generality of the results. 
In the calculations we used the same normalizations as in Section \ref{sec.eel}. 
 
\subsection{Mode Truncation Scheme}
\markright{\thesubsection.\ \ NUMERICAL RESULTS - MODE TRUNCATION SCHEME}

\begin{figure}[h]
\begin{tabular}{lr}
\includegraphics[clip=true,height=12cm]{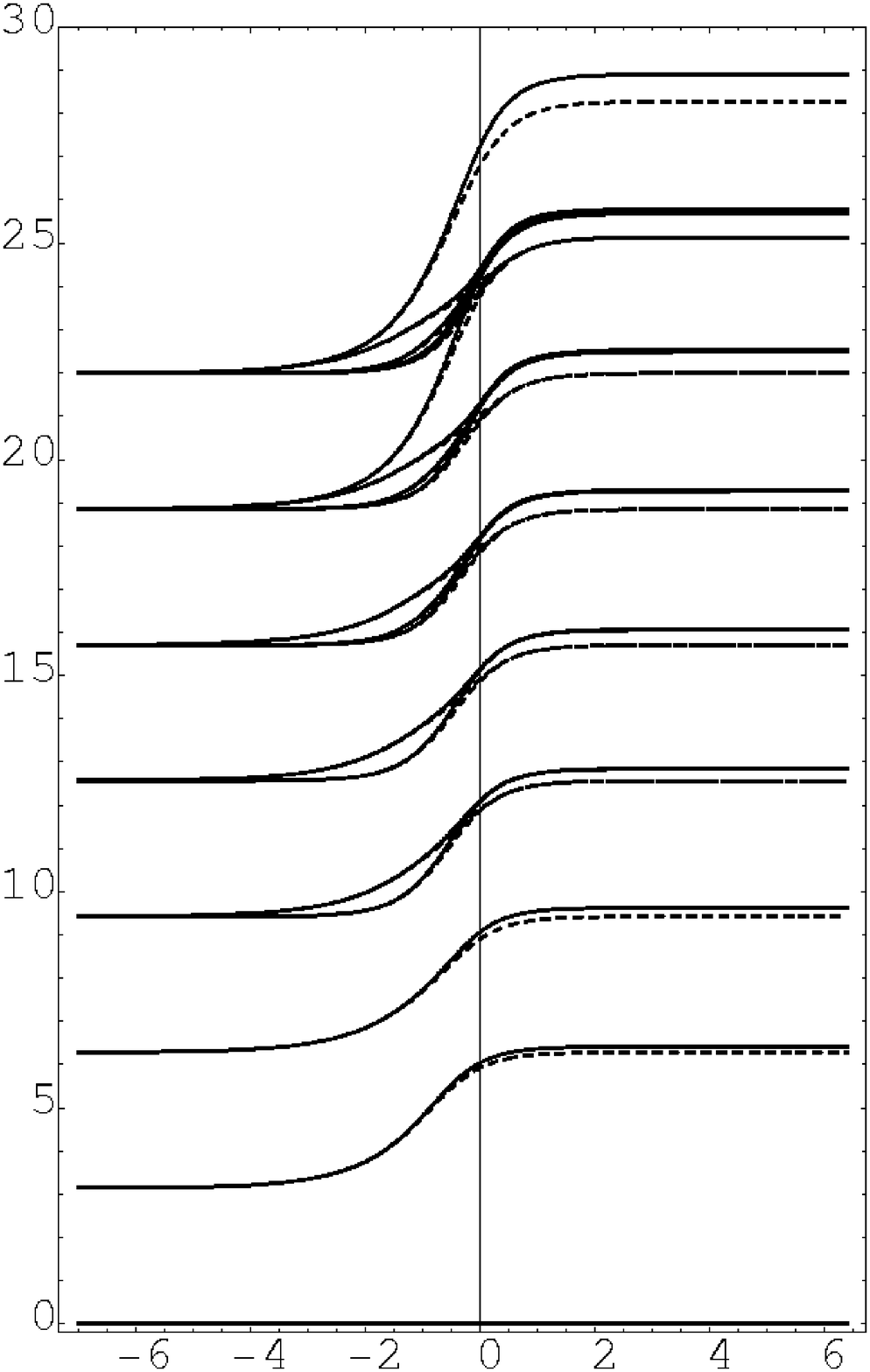}
&
\includegraphics[clip=true,height=12cm]{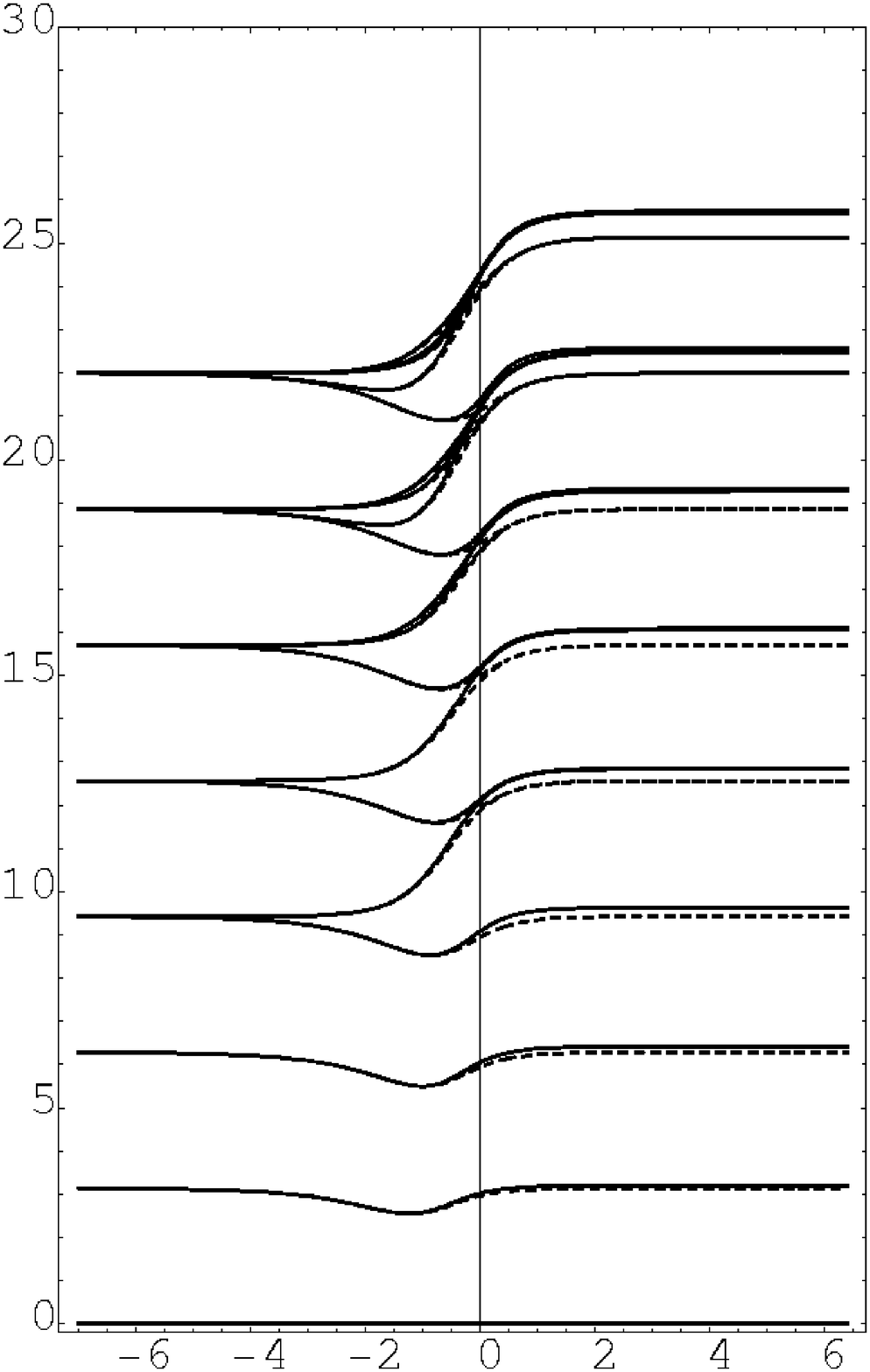}
\end{tabular}
\caption{\label{fig.mt1}Exact (dashed lines) and mode truncated (solid lines) energy
  gaps ($E_i-E_0$)
  in the $v$ and $u$ sectors respectively as a function of $\ln(h)$ at truncation level $n_c=9$}
\end{figure}

Figure \ref{fig.mt1} shows the exact and mode truncated spectra as a function of the
logarithm of the coupling constant. The truncation level is $n_c=9$, and the
dimension of the Hilbert space is 512 in each sector. It is remarkable that
there is a good qualitative agreement between the mode truncated and exact
spectra for all values of $h$.  Numerical values are listed in Table
\ref{tab.mt2} for the fifth energy gap $k(3,h)+k(0,h)$ of the $v$ sector and
in Table \ref{tab.mt4} for the fifth energy gap $k(4,h)-k(0,h)$ of the $u$
sector. The number of digits presented do not exceed the numerical precision.

\begin{figure}[!p]
\begin{tabular}{lr}
\includegraphics[clip=true,height=12cm]{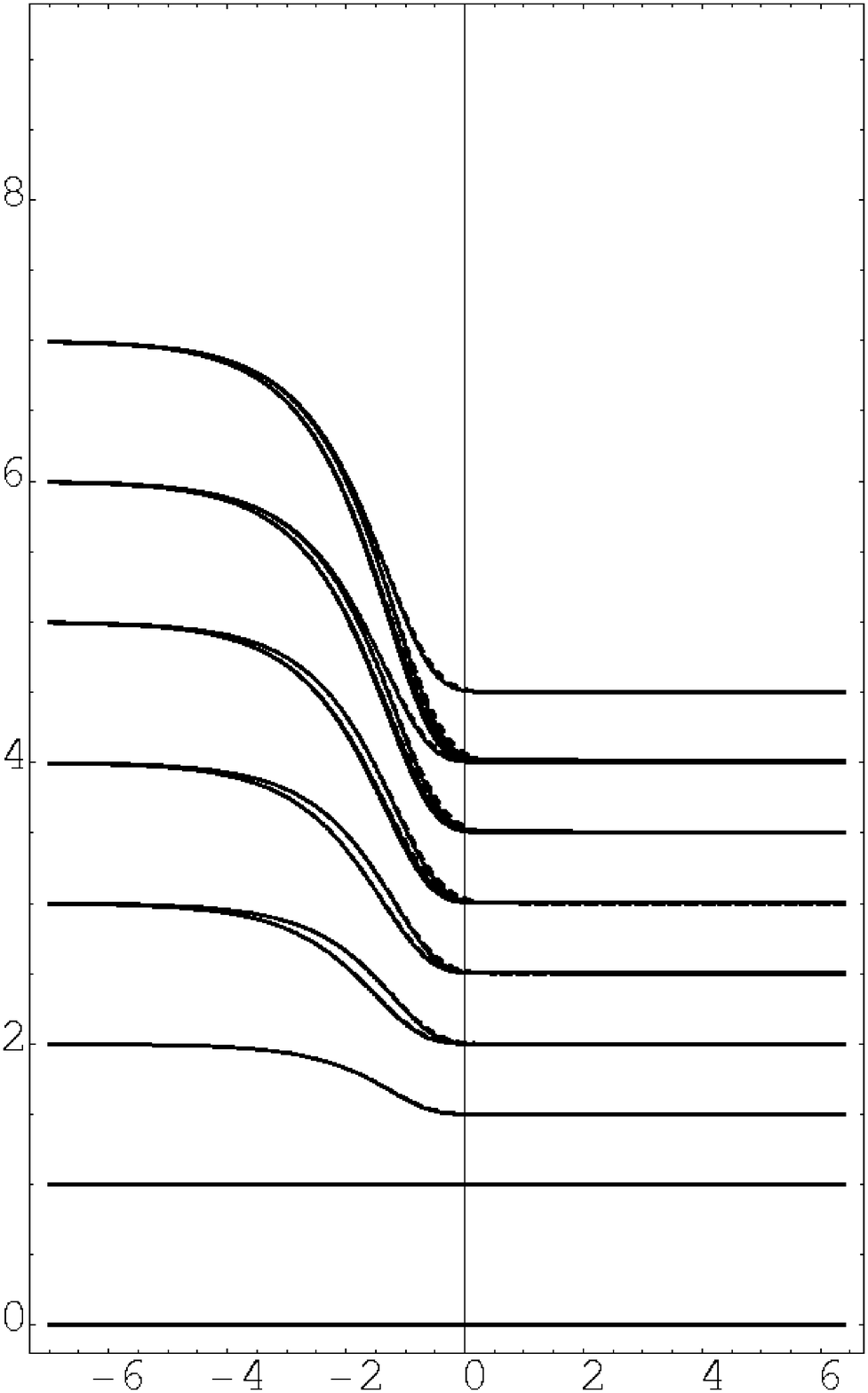}
&
\includegraphics[clip=true,height=12cm]{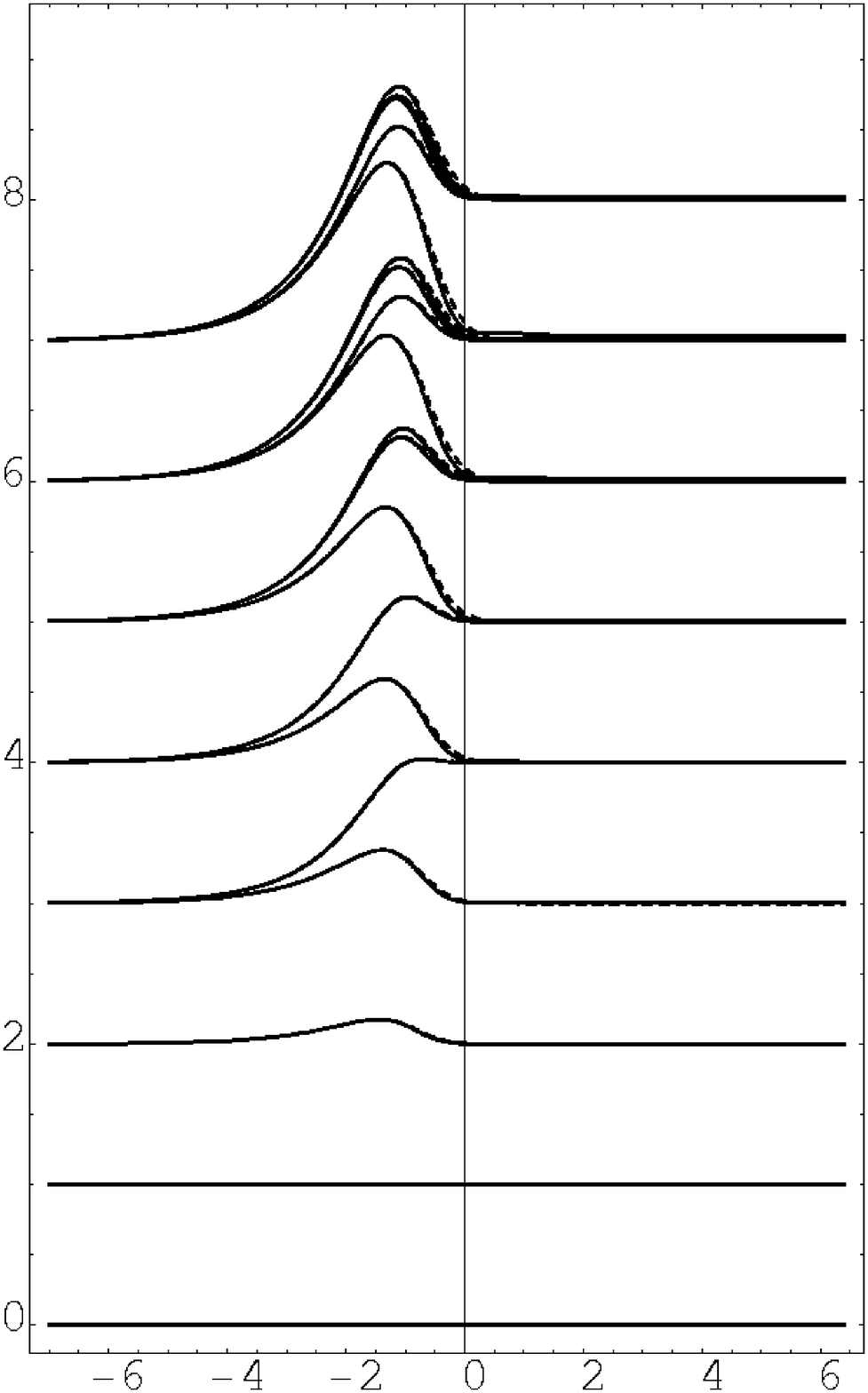}
\end{tabular}
\caption{\label{fig.mt2}Exact (dashed lines) and mode truncated (solid lines) normalized
  spectra
  in the $v$ and $u$ sectors respectively as a function of $\ln(h)$ at truncation level $n_c=9$}
\end{figure}

Figure \ref{fig.mt2}  shows the  same spectra,  but the  lowest gap is normalized to
$1$, i.e.\ the functions $\frac{E_i(h)-E_0(h)}{E_1(h)-E_0(h)}$ are shown. It is
remarkable that the agreement between exact and mode truncated spectra looks
considerably better than in the case of not normalized spectra.

\begin{figure}[t]
\begin{tabular}{lll}
\includegraphics[clip=true,height=4.5cm]{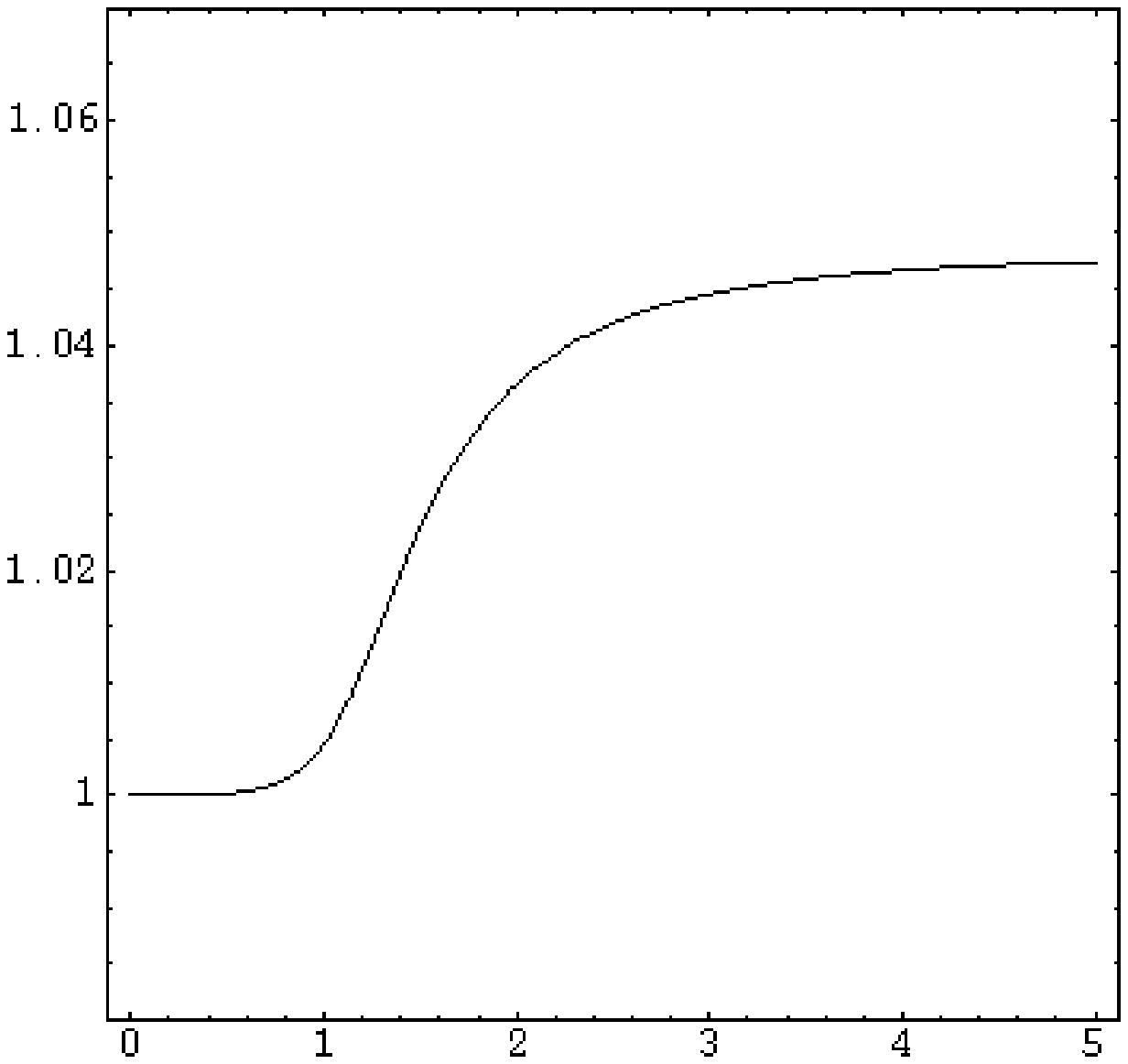}
&
\includegraphics[clip=true,height=4.5cm]{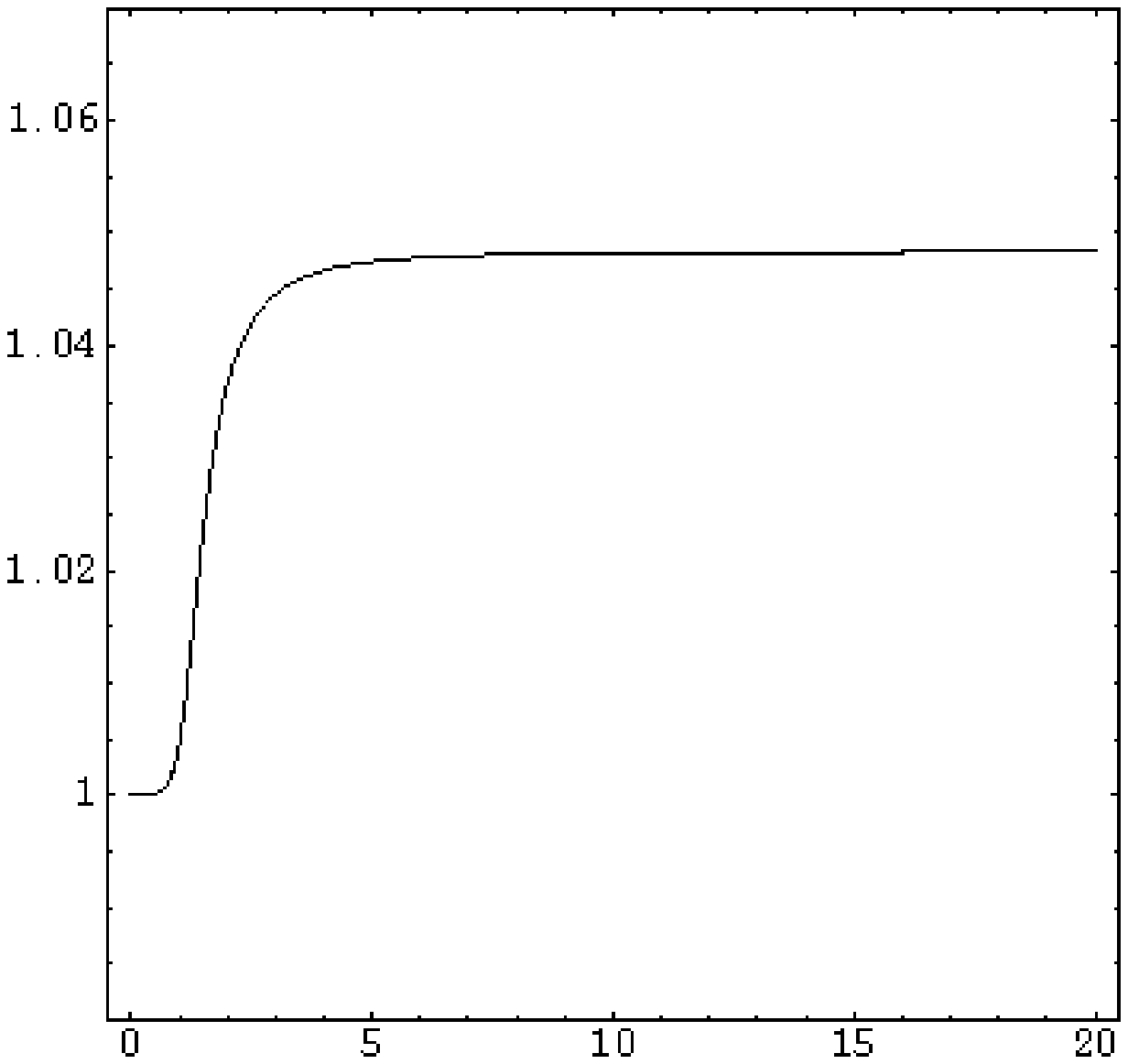}
&
\includegraphics[clip=true,height=4.5cm]{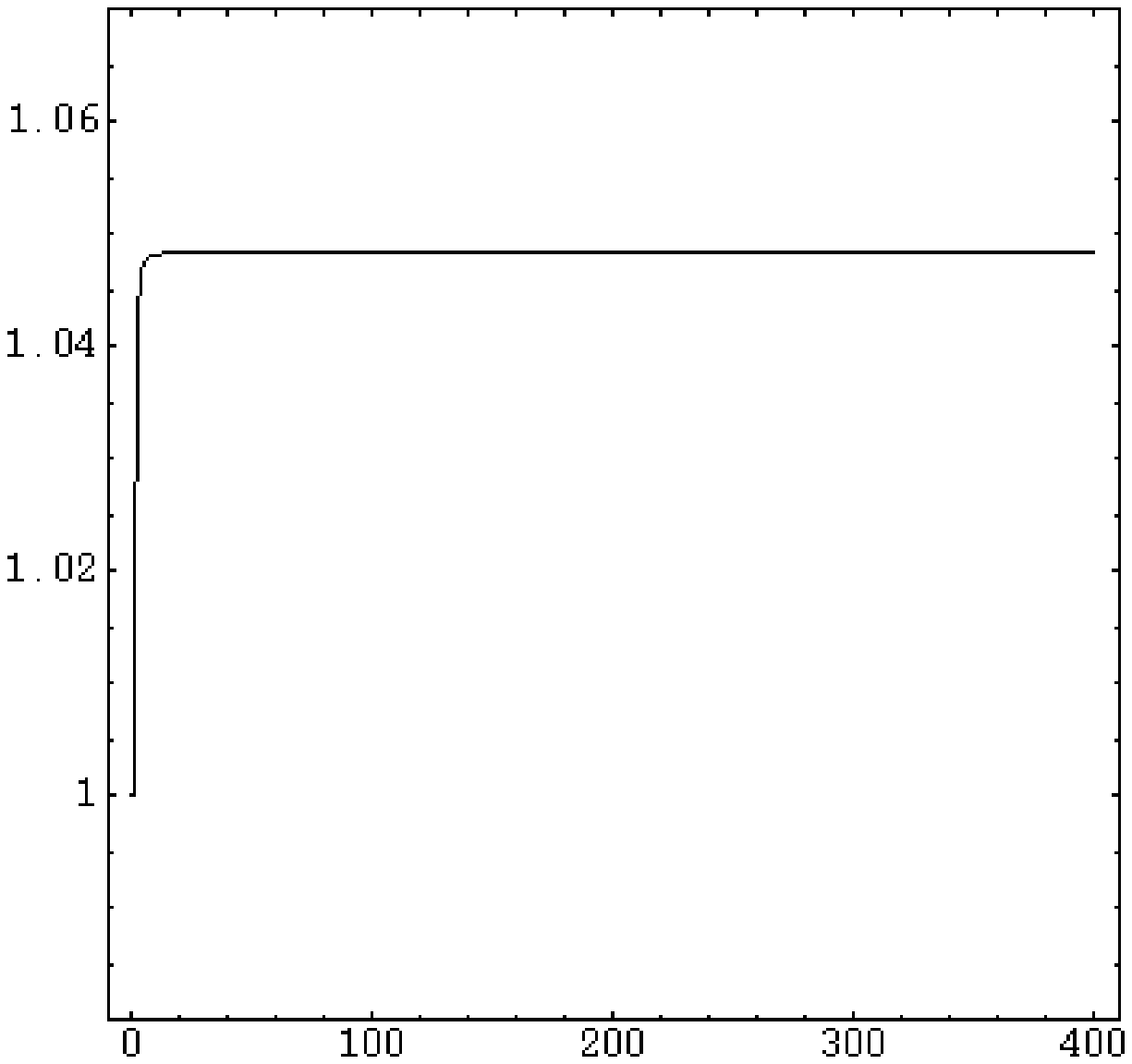}
\end{tabular}
\caption{\label{fig.mt3}The function $s_0(h)$ for the $v$ sector in the ranges $h\in [0,3]$, $h\in [0,20]$, $h\in [0,400]$ and
  $s_0\in [0.95,1.05]$ at truncation level $n_c=9$}
\end{figure}

\begin{figure}[h]
\begin{tabular}{lll}
\includegraphics[clip=true,height=4.5cm]{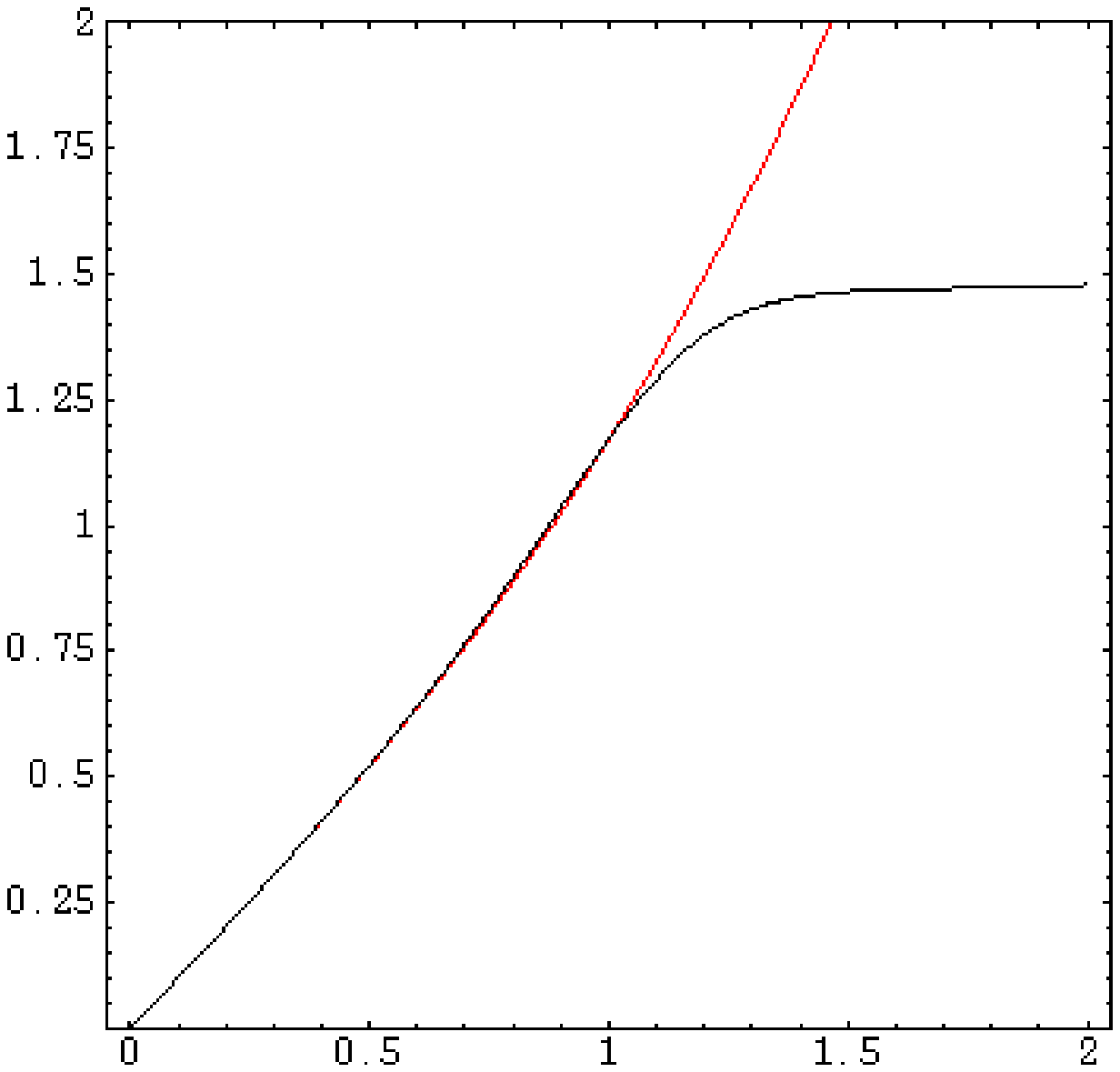}
&
\includegraphics[clip=true,height=4.5cm]{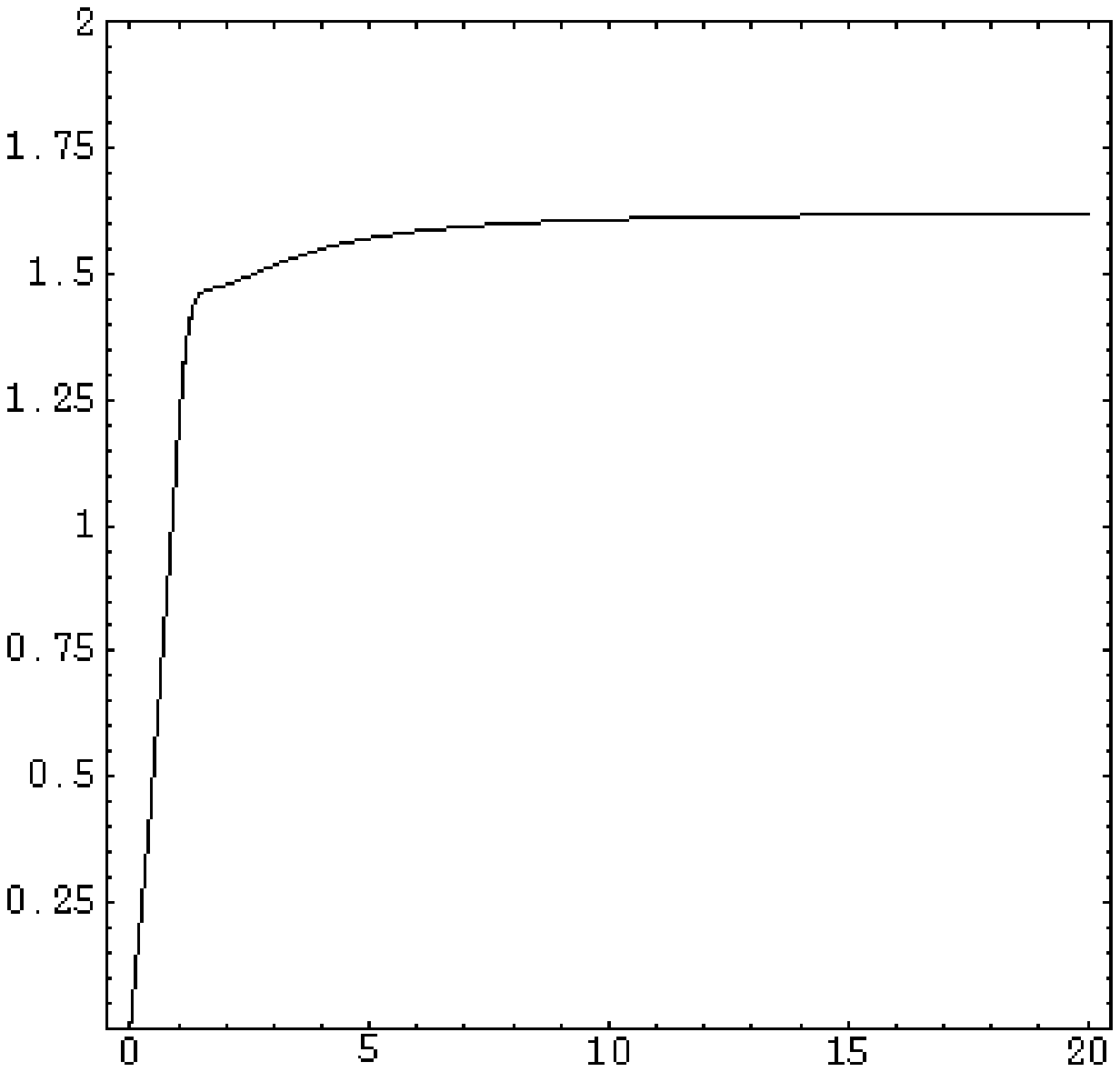}
&
\includegraphics[clip=true,height=4.5cm]{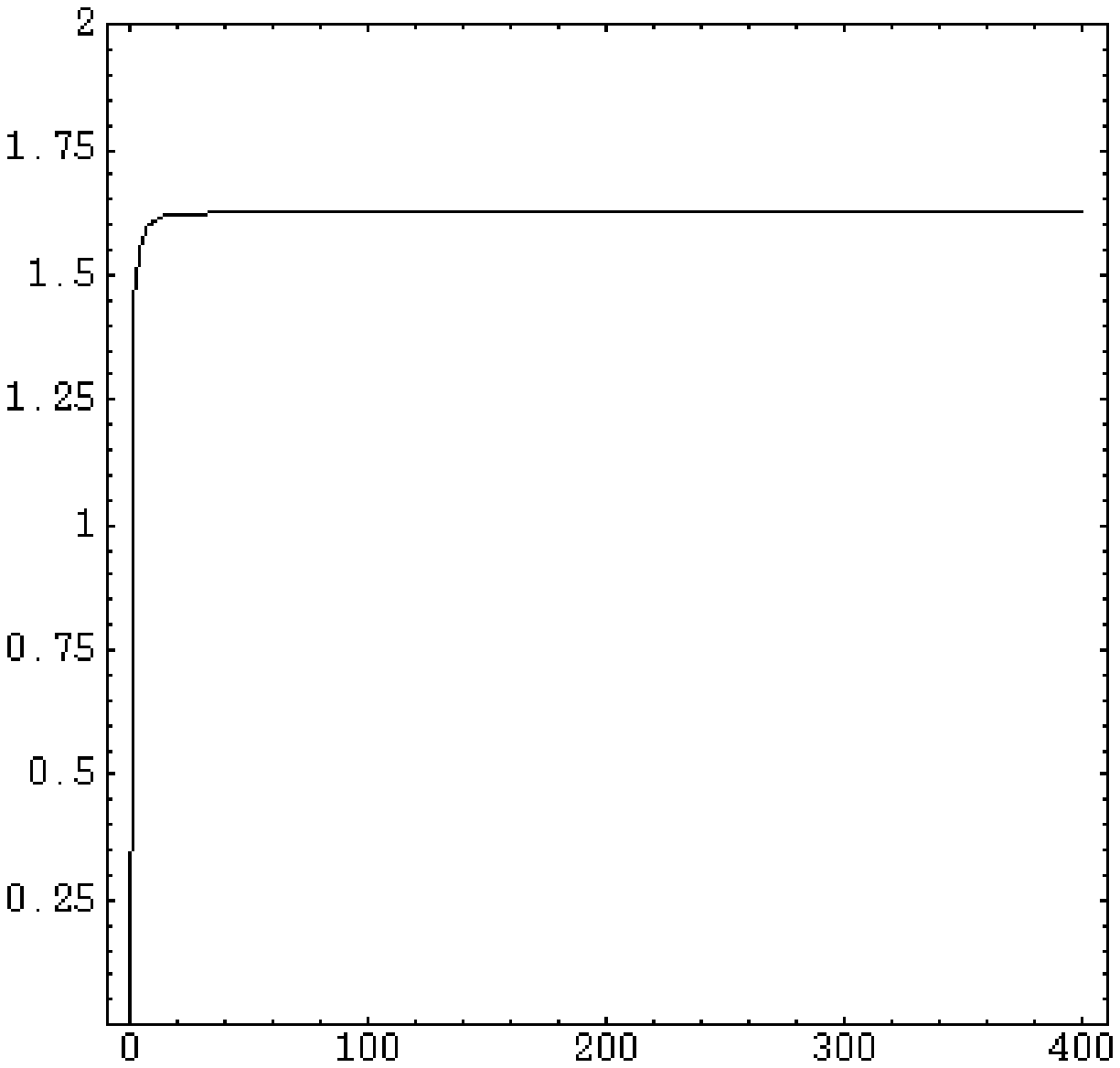}
\end{tabular}
\caption{\label{fig.mt4}The function $s_1(h)$ for the $v$ sector in the ranges $h\in [0,2]$, $h\in [0,20]$, $h\in [0,400]$ and
  $s_1\in [0,2]$ at truncation level $n_c=9$}
\end{figure}

\begin{figure}[!h]
\begin{tabular}{lll}
\includegraphics[clip=true,height=4.5cm]{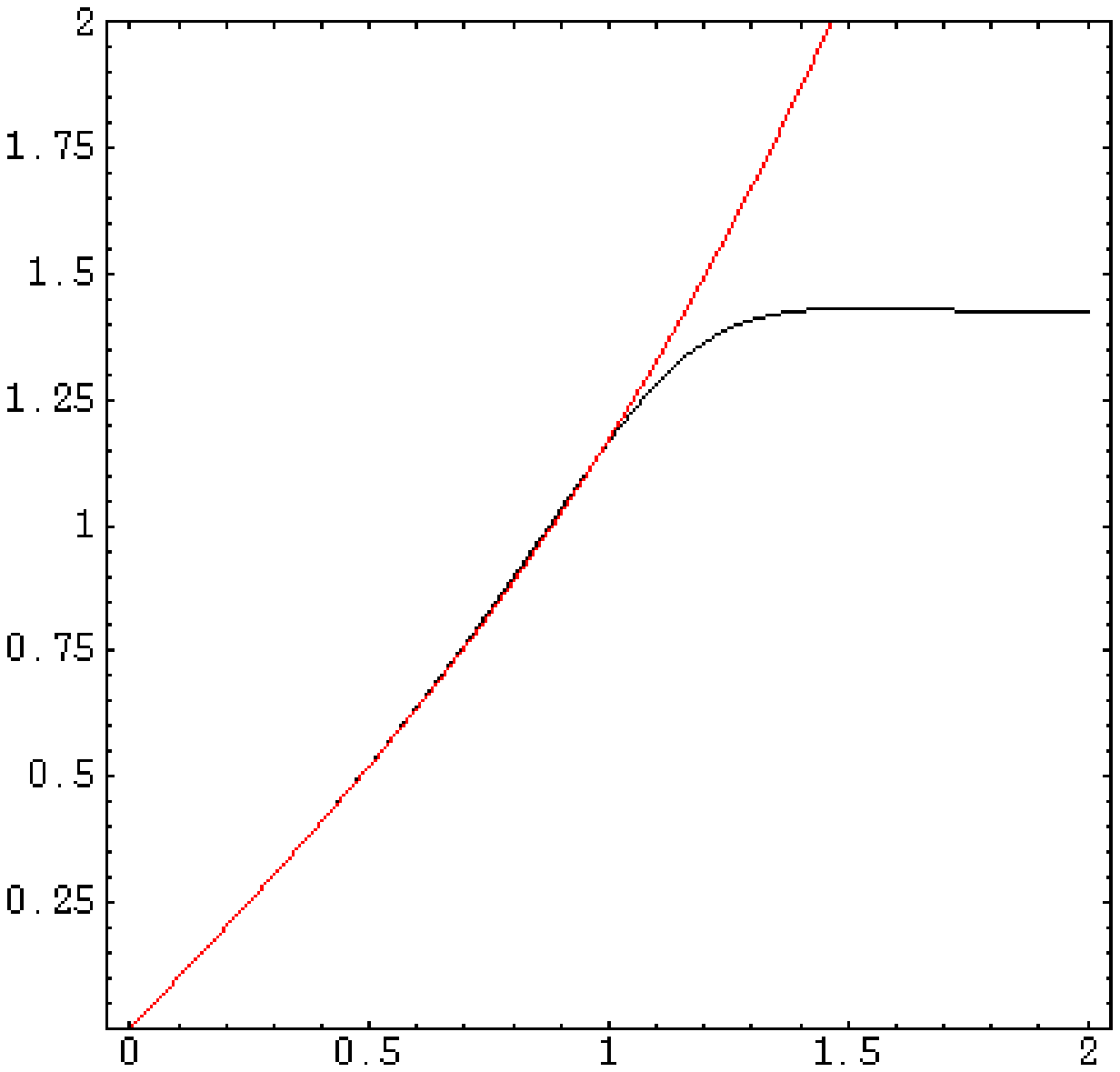}
&
\includegraphics[clip=true,height=4.5cm]{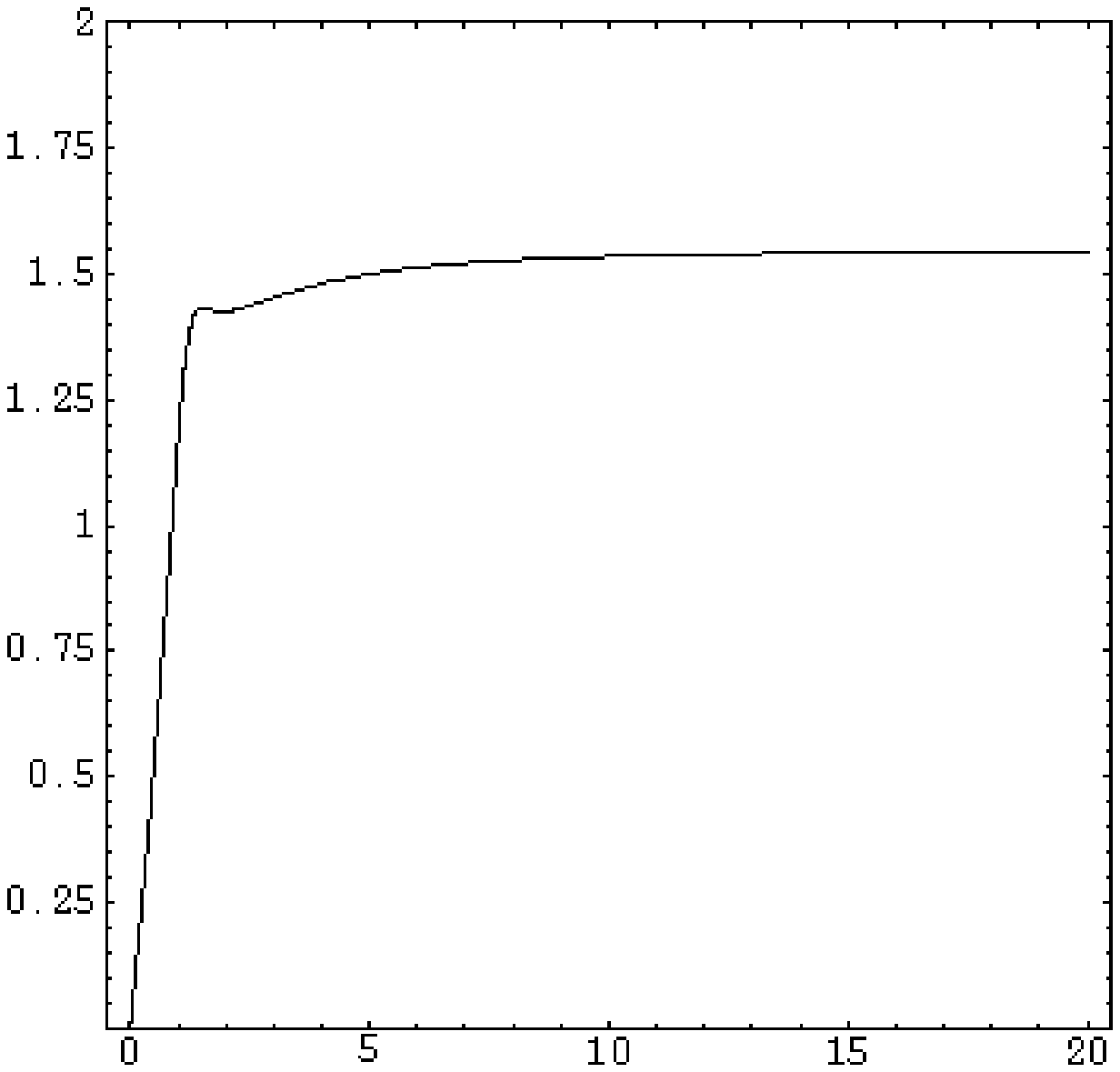}
&
\includegraphics[clip=true,height=4.5cm]{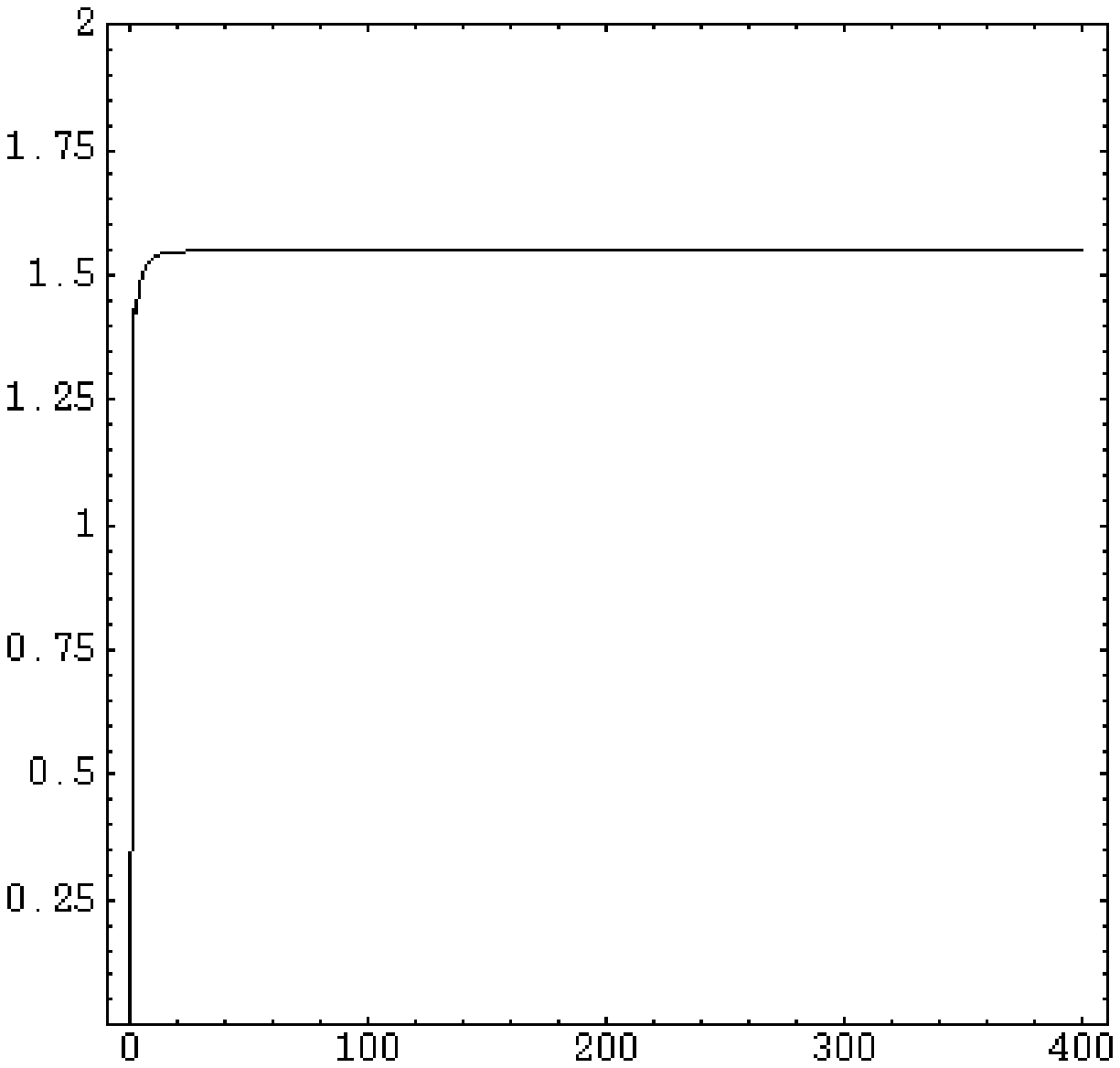}
\end{tabular}
\caption{\label{fig.mt5}The function $s_1(h)/s_0(h)$ for the $v$ sector in the ranges $h\in [0,2]$, $h\in [0,20]$, $h\in [0,400]$ and
  $s_1/s_0\in [0,2]$ at truncation level $n_c=9$}
\vspace{1cm}
\end{figure}

Figures \ref{fig.mt3}-\ref{fig.mt5}   show the functions $s_0(h)$, $s_1(h)$,
$s_1(h)/s_0(h)$ determined by the lowest three energy levels in the $v$ sector
via the formulae (\ref{eq.s1}), (\ref{eq.s2}) in
various ranges.
Figure \ref{fig.mt4} and \ref{fig.mt5} also show the curves given by 
(\ref{eq.crv1}) and (\ref{eq.crv2}) on the left-hand side  (red/grey line).
 $s_0(h)$ remains close to 1 and for large values of $h$ it
tends to a constant which can be expected to converge to 1 as $n_c \to \infty$. 
$s_1(h)$ also tends to a constant for large values of $h$ which can be
expected  to  increase to infinity as $n_c \to \infty$. 
The behaviour of $s_1(h)/s_0(h)$ and $s_1(h)$ are similar.

\begin{figure}[t]
\begin{tabular}{cc}
\includegraphics[clip=true,height=12cm]{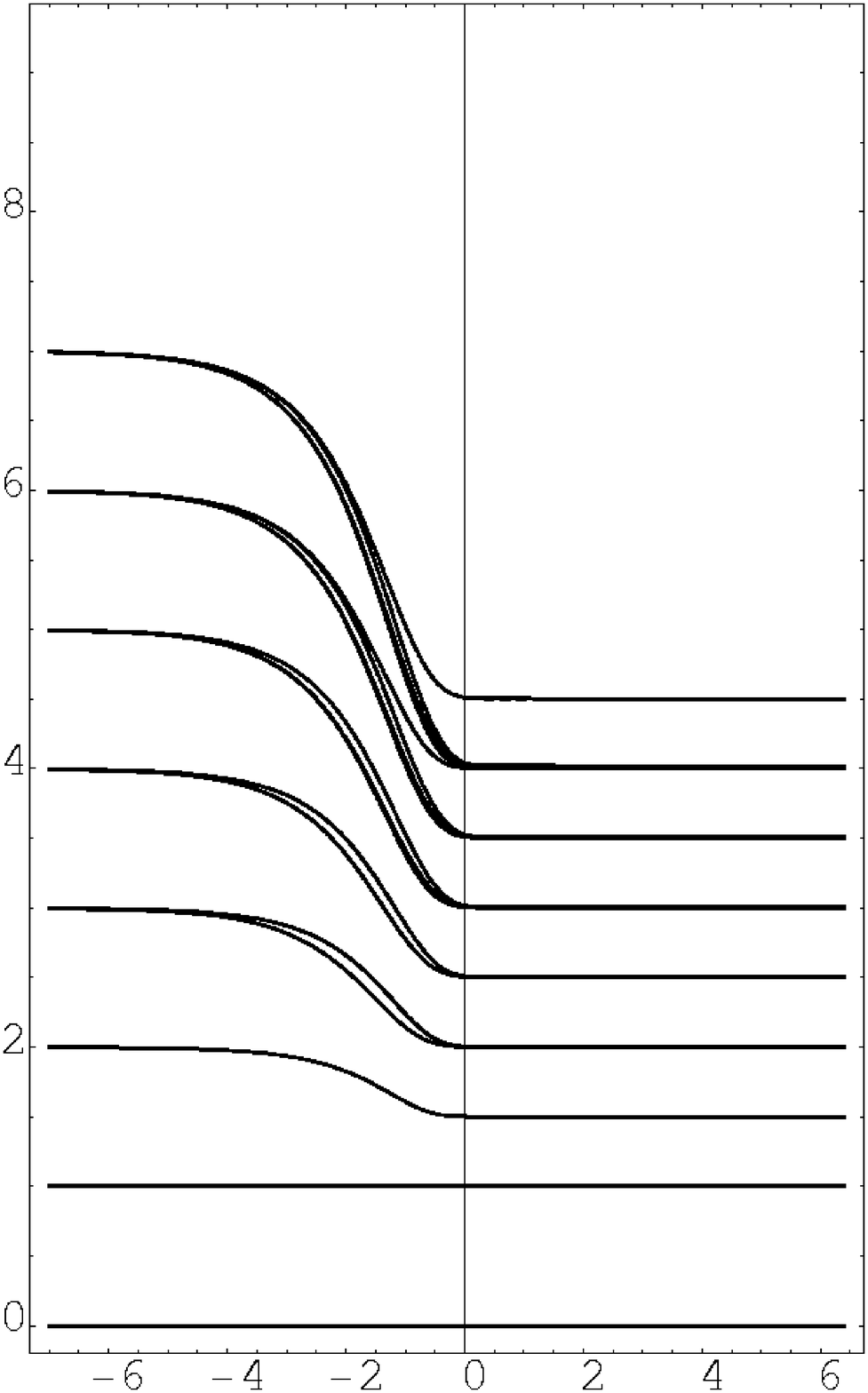}
&
\includegraphics[clip=true,height=12cm]{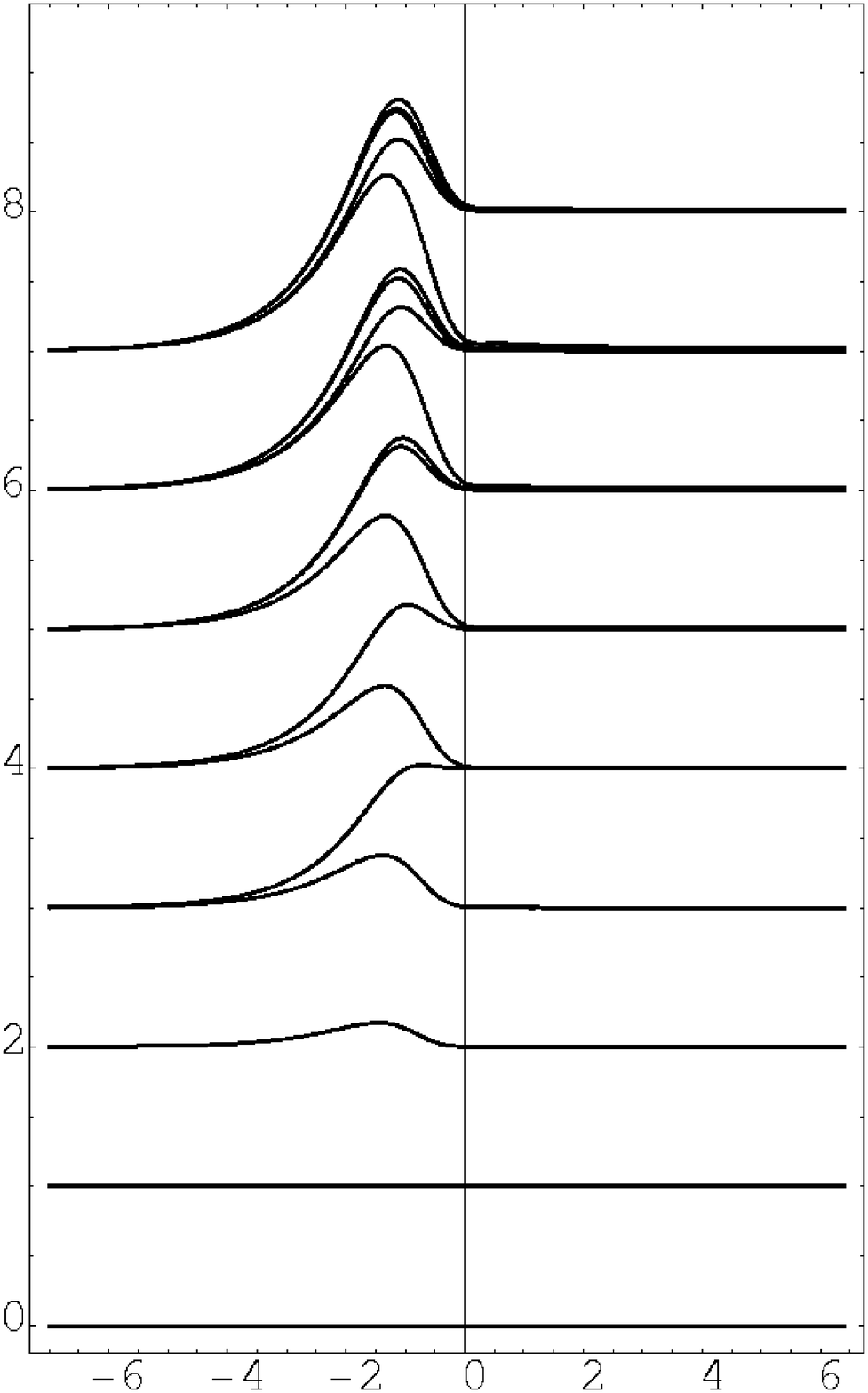}\\
a&b
\end{tabular}
\caption{\label{fig.mt6}The mode truncated (solid lines) and rescaled exact (dashed
  lines) normalized spectra in the $v$ and $u$ sectors respectively as a function of $\ln (h)$ at truncation level $n_c=9$}
\end{figure}

Figure  \ref{fig.mt6}.a  shows the normalized mode truncated spectrum and the normalized exact spectrum
rescaled by $s_0(h)$ and $s_1(h)$ in the $v$ sector. No
difference between the two is visible. In Table \ref{tab.mt1} values of the
fifth normalized energy gap $\frac{k(3,h)+k(0,h)}{k(1,h)+k(0,h)}$ of the $v$ sector are listed:
the values in the non-truncated case are listed in the first column, the
values in the mode truncated case are listed in the second column and the
rescaled non-truncated values are listed in the third column (which would
 be the same as the values in the second column if the renormalization could
 be satisfied exactly).

Figure  \ref{fig.mt6}.b  shows the normalized mode truncated spectrum and the
normalized exact spectrum
rescaled by $s_0(h)$ and $s_1(h)$ in the $u$ sector. The $s_0(h)$, $s_1(h)$
obtained in the $v$ sector were used for the rescaling, which corresponds to the assumption that
$s_0(h)$  and $s_1(h)$ are the same for both sectors. The difference  between the
mode truncated and rescaled exact spectra is not visible in the figure.

We also see from  Table \ref{tab.mt3} in which values of the
fourth normalized energy gap $\frac{k(3,h)-k(0,h)}{k(1,h)-k(0,h)}$ of the $u$
sector are listed  that the  rescaling together with the above
assumption works well.

We have not tried to calculate $s_0$ and $s_1$ for the $u$ sector because
$\frac{E_2-E_0}{E_1-E_0}(h)$ is not invertible in this case. One way to
circumvent this difficulty would be to use other energy levels $E_i,E_j,E_k$
for which $\frac{E_i-E_k}{E_j-E_k}(h)$ is invertible.

\clearpage

\clearpage

\begin{table}[p]
\caption{\label{tab.mt1}The normalized energy gap
  $\frac{k(3,h)+k(0,h)}{k(1,h)+k(0,h)}$  in the $v$ sector:  exact, MT $(n_c=9)$ 
   and rescaled  exact values}\vspace{3mm}
\begin{tabular}{@{}clll}
\hline
$\log(h)$ & Exact & MT & Rescaled exact \\
\hline
-7 & 2.997677 & 2.997677 & 2.997677  \\
-6 & 2.993681  & 2.993681 & 2.993681 \\
-5 & 2.982797  & 2.982797 & 2.982797\\
-4 & 2.953071  & 2.953068  & 2.953068 \\
-3 & 2.8719584  & 2.8719043  & 2.8719043\\
-2 & 2.66042177 & 2.6594328 & 2.6594328 \\
-1 & 2.25064420  & 2.24043886  & 2.24043637 \\
 0 & 2.00954942  & 2.00440950  & 2.00435155 \\
 1 & 2.00003474  & 2.00145821  & 2.00136637 \\
 2 & 2.00000008  & 2.00106850  & 2.00100783  \\
 3 & 2.0000000  & 2.0009872  & 2.0009201 \\
 4 & 2.0000000  & 2.0009756  & 2.0009202  \\
 5 & 2.0000000  & 2.0009740  & 2.0009187  \\
 6 & 2.0000000  & 2.0009738  & 2.0009185  \\
6.4 & 2.0000000  & 2.0009738  & 2.0009185 \\
\hline
\end{tabular}
\caption{\label{tab.mt2}The  energy gap
  $k(3,h)+k(0,h)$  in the $v$ sector:  exact and MT  $(n_c=9)$ 
   values}\vspace{3mm}
\begin{tabular}{@{}cll}
\hline
$\log(h)$ & Exact  & MT  \\
\hline
-7 & 9.428427 & 9.428427 \\
-6 & 9.434703 & 9.434703 \\
-5 & 9.451803 & 9.451804 \\
-4 & 9.498544 & 9.498549 \\
-3 & 9.626826 & 9.626912 \\
-2 & 9.972034 & 9.973705 \\
-1 & 10.746173 & 10.770531 \\
0 & 11.897034 & 12.101539 \\
1 & 12.461229 & 12.740942 \\
2 & 12.552003 & 12.833098 \\
3 & 12.564424 & 12.845456 \\
4 & 12.566107 & 12.847126 \\
5 & 12.566335 & 12.847352 \\
6 & 12.566366 & 12.847382 \\
6.4 & 12.566368 & 12.847385 \\
\hline
\end{tabular}
\end{table}

\begin{table}[p]
\caption{\label{tab.mt3}The normalized energy gap
  $\frac{k(3,h)-k(0,h)}{k(1,h)-k(0,h)}$  in the $u$ sector:  exact, MT  $(n_c=9)$ 
   and rescaled  exact values}\vspace{3mm}
\begin{tabular}{@{}clll}
\hline
$\log(h)$ & Exact & MT & Rescaled exact \\
\hline
-7 & 3.002321 & 3.002321 & 3.002321\\
-6 & 3.006305 & 3.006305 & 3.006305\\
-5 & 3.017105 & 3.017105 & 3.017105\\
-4 & 3.046201 & 3.046203 & 3.046203\\
-3 & 3.1225067 & 3.1225559 & 3.1225559\\
-2 & 3.29172833 & 3.29237422 & 3.29237405\\
-1 & 3.32029283 & 3.31272102 & 3.31271418\\
0  & 3.01716419 & 3.00801374 & 3.00788791\\
1  & 3.00006366 & 3.00269022 & 3.00249222\\
2  & 3.0000000 & 3.0019707 & 3.0018400\\
3  & 3.0000000 & 3.0018211 & 3.0017004\\
4  & 3.0000000 & 3.0017997 & 3.0016804\\
5  & 3.0000000 & 3.0017968 & 3.0016776\\
6  & 3.0000000 & 3.0017964 & 3.0016773\\
\hline
\end{tabular}
\caption{\label{tab.mt4}The  energy gap
  $k(3,h)-k(0,h)$  in the $u$ sector:  exact and MT  $(n_c=9)$ 
   values}\vspace{3mm}
\begin{tabular}{@{}cll}
\hline
$\log(h)$ & Exact & MT  \\
\hline
-7 & 9.421132 & 9.421132\\
-6 & 9.414873 & 9.414873\\
-5 & 9.397906 & 9.397906\\
-4 & 9.352150 & 9.352146\\
-3 & 9.231143 & 9.231064\\
-2 & 8.939485 & 8.938216\\
-1 & 8.545045 & 8.541635\\
0  & 8.939748 & 9.084020\\
1  & 9.345985 & 9.558387\\
2  & 9.414002 & 9.626803\\
3  & 9.423318 & 9.635922\\
4  & 9.424580 & 9.637153\\
5  & 9.424751 & 9.637320\\
6  & 9.424774 & 9.637342\\
\hline
\end{tabular}
\end{table}

\clearpage

\subsection{TCS scheme}
\markright{\thesubsection.\ \ NUMERICAL RESULTS - TCS SCHEME}

\begin{figure}
\begin{tabular}{lr}
\includegraphics[clip=true,height=12cm]{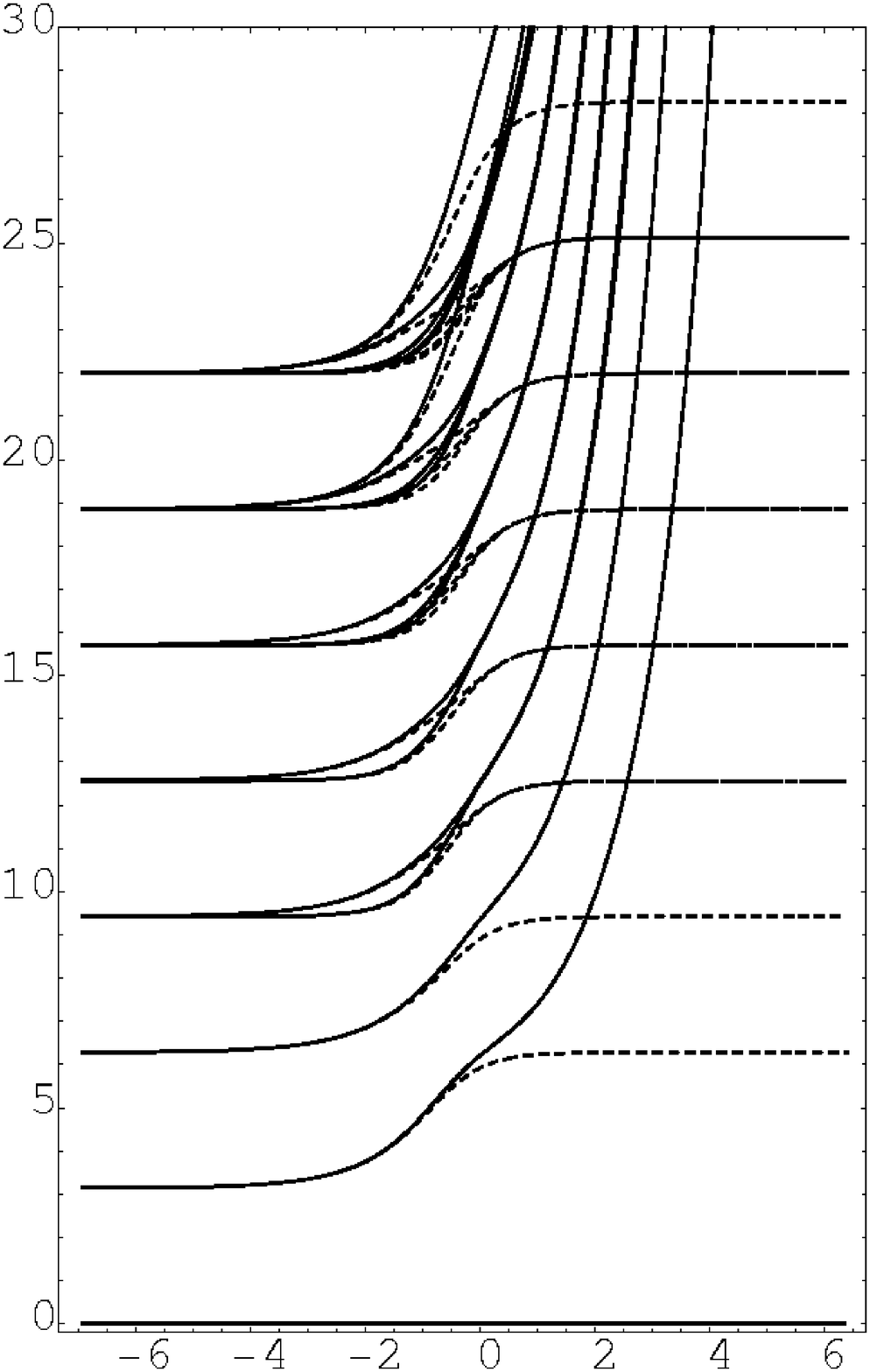}
&
\includegraphics[clip=true,height=12cm]{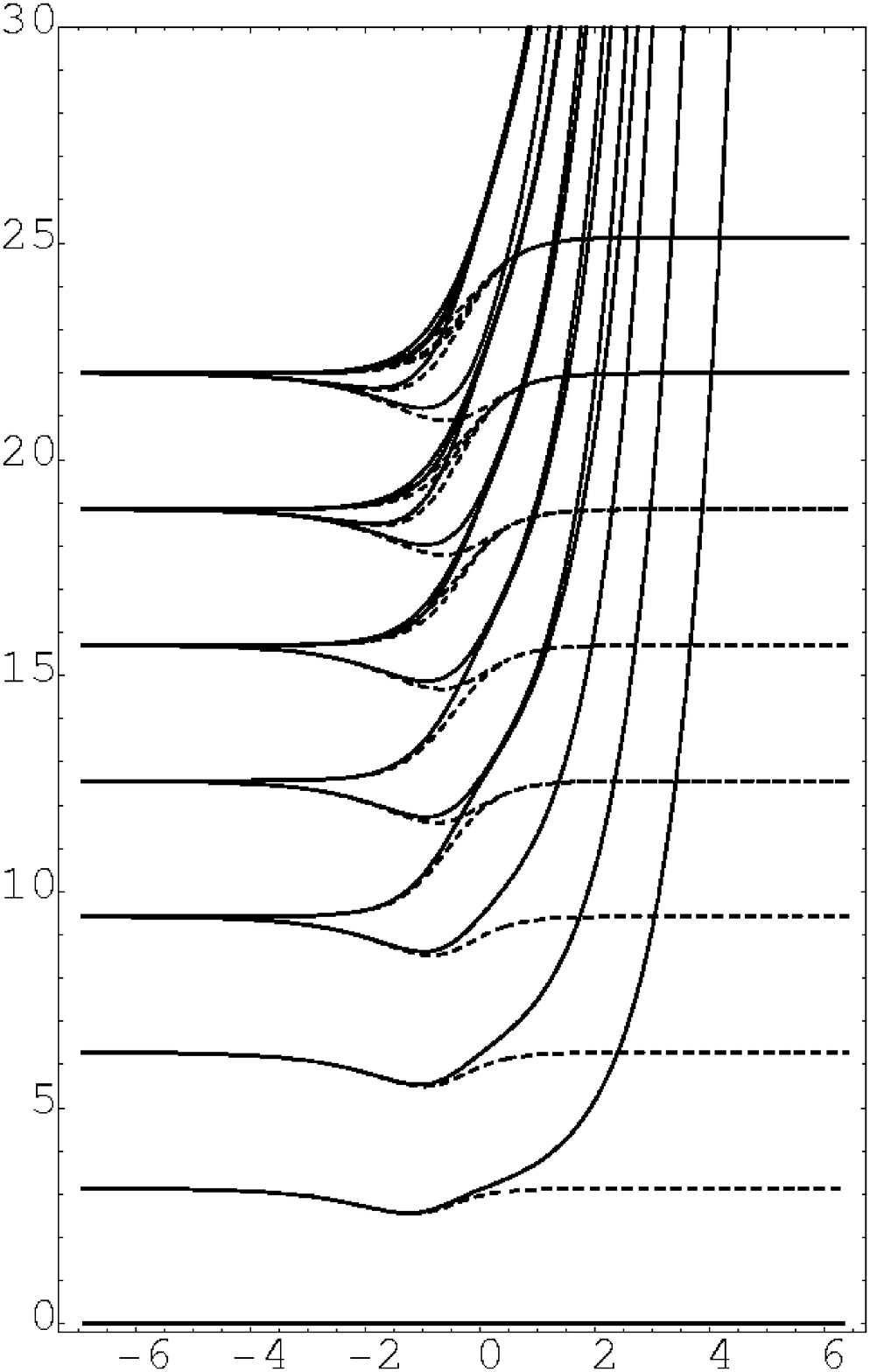}
\end{tabular}
\caption{\label{fig.tcsa1}Exact (dashed lines) and TCSA (solid lines) energy gaps ($E_i-E_0$)
  in the $v$ and $u$ sectors respectively as a function of $\ln(h)$ at
  truncation level $n_c=14$}
\end{figure}

Figure  \ref{fig.tcsa1} shows the exact and TCSA spectra as a function of the
logarithm of the coupling constant. The truncation level is $n_c=14$, and the
dimension of the Hilbert space is 110 in each sector. It is remarkable that
there is strong deviation between the TCSA and exact
spectra for large values of $h$. The behaviour of the TCSA energy gaps is
$E_i(h)-E_0(h) \propto h$ for large values of $h$.  Numerical values are listed in Table
\ref{tab.tcsa2} for the fifth energy gap $k(3,h)+k(0,h)$ of the $v$ sector and
in Table \ref{tab.tcsa4} for the fifth energy gap $k(4,h)-k(0,h)$ of the $u$
sector. The number of digits presented do not exceed the numerical precision.

\begin{figure}
\begin{tabular}{lr}
\includegraphics[clip=true,height=12cm]{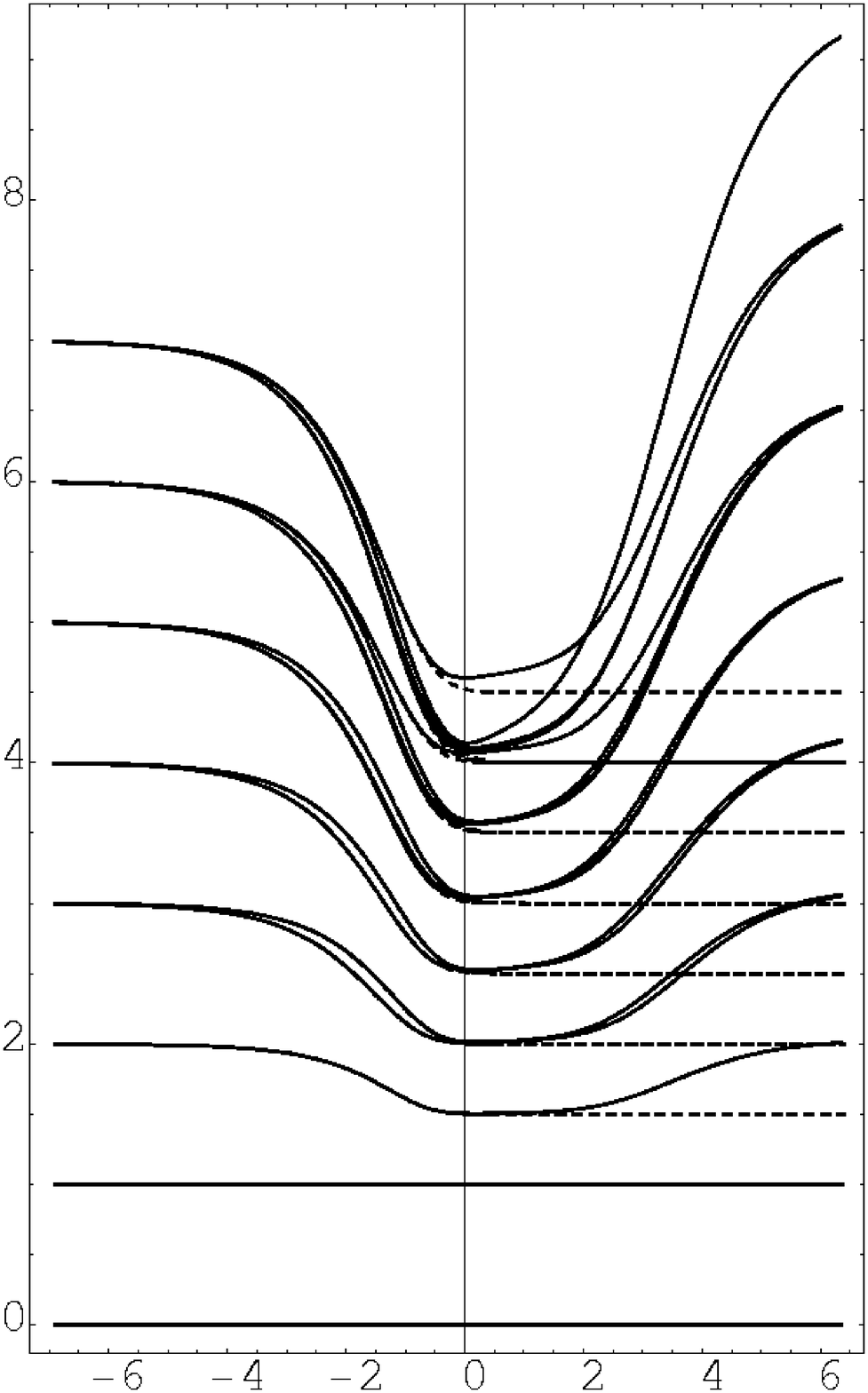}
&
\includegraphics[clip=true,height=12cm]{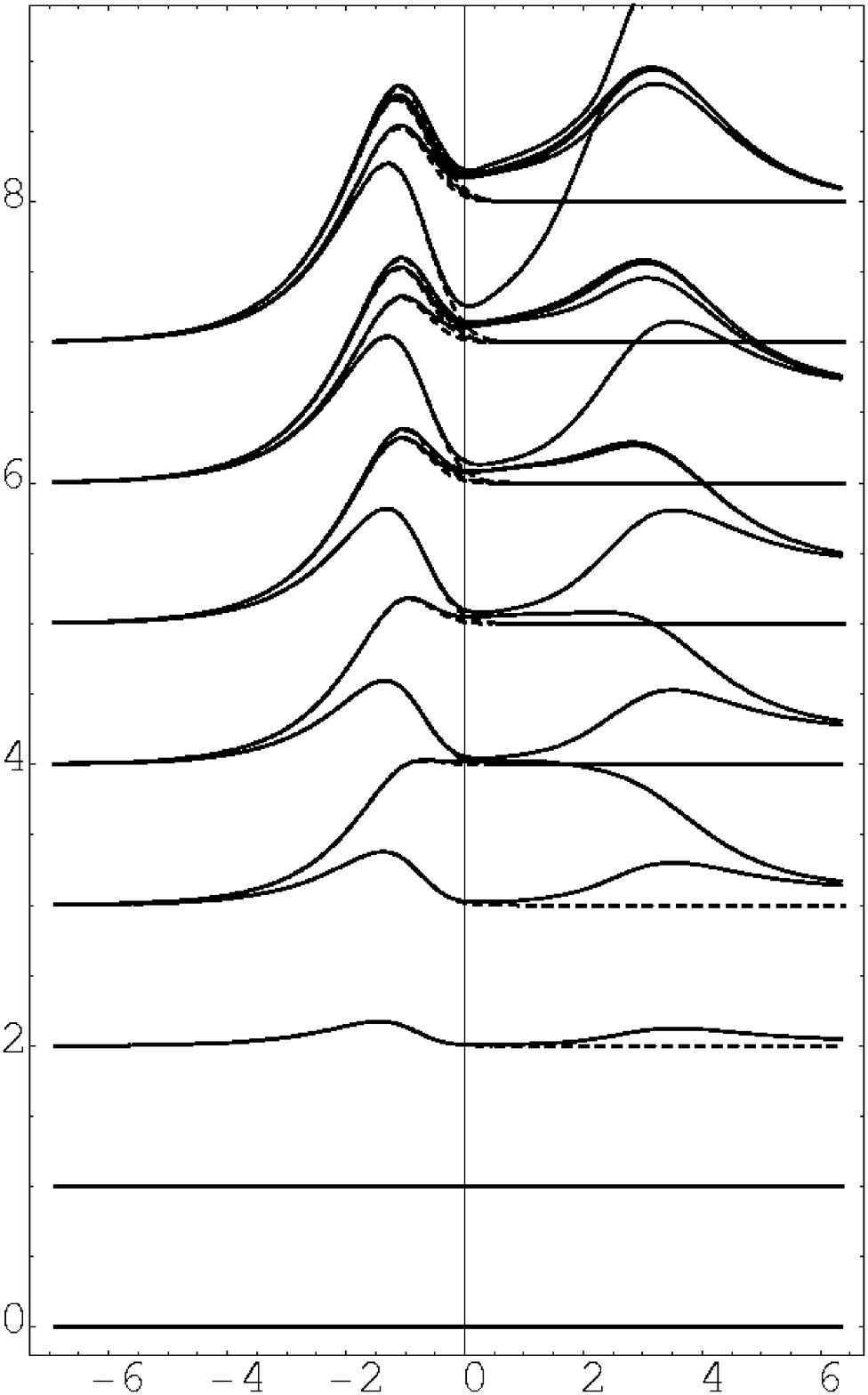}
\end{tabular}
\caption{\label{fig.tcsa2}Exact (dashed lines) and TCSA (solid lines) normalized spectra
  in the $v$ and $u$ sectors respectively as a function of $\ln(h)$ at
  truncation level $n_c=14$}
\end{figure}

\begin{figure}
\begin{tabular}{lr}
\includegraphics[clip=true,height=12cm]{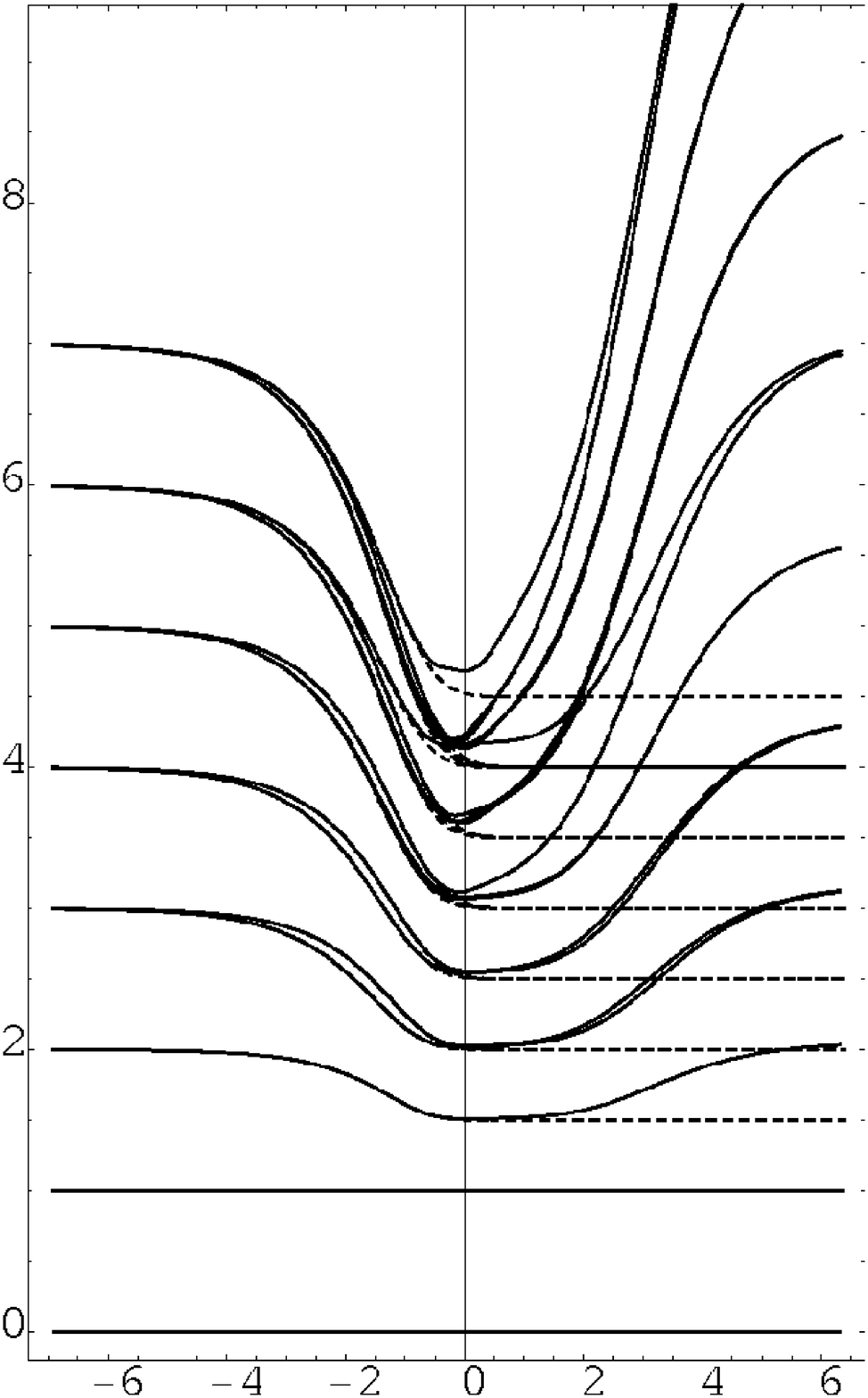}
&
\includegraphics[clip=true,height=12cm]{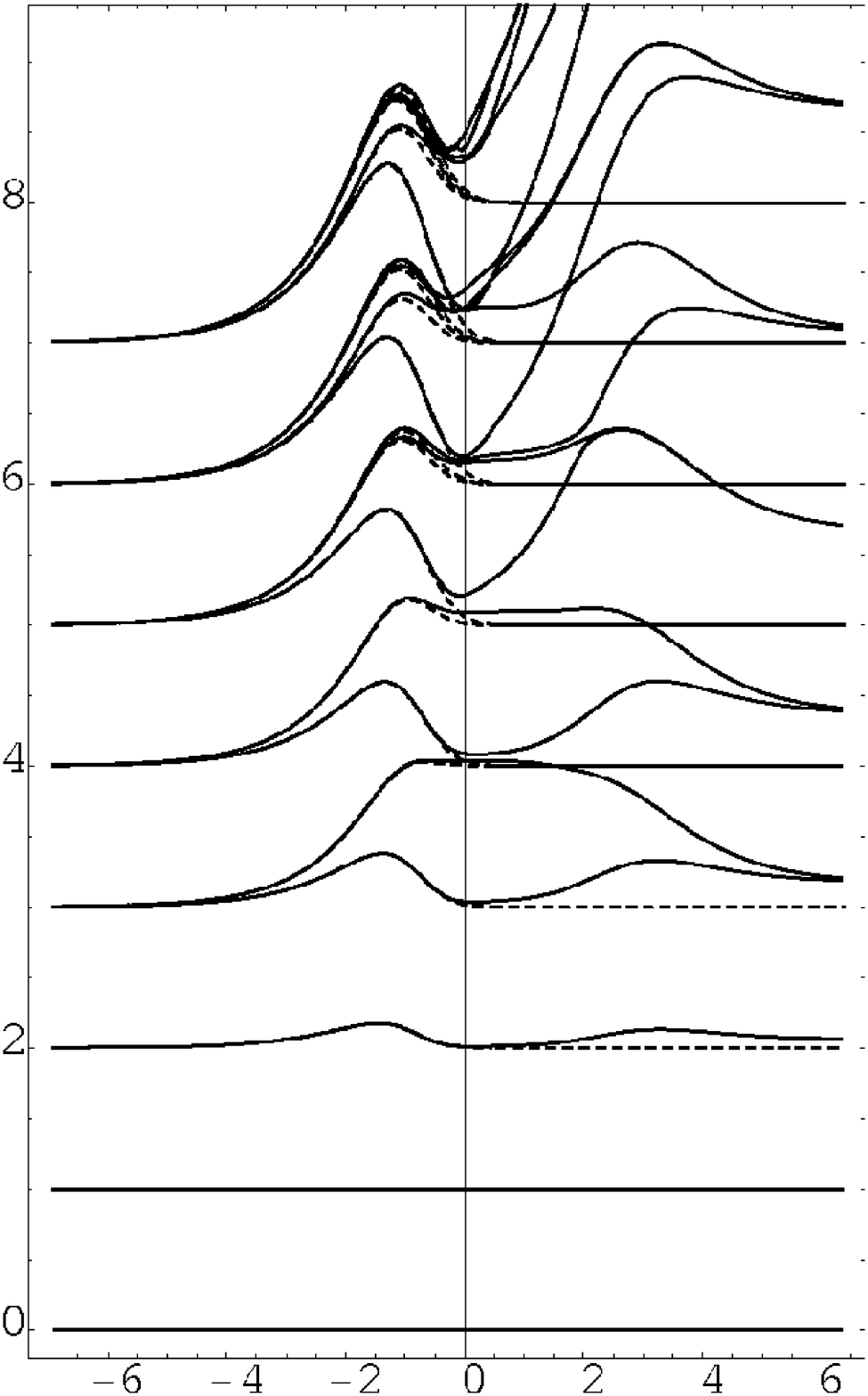}
\end{tabular}
\caption{\label{fig.tcsa2_10}Exact (dashed lines) and TCSA (solid lines) normalized spectra
  in the $v$ and $u$ sectors respectively as a function of $\ln(h)$ at
  truncation level $n_c=10$}
\end{figure}

Figure   \ref{fig.tcsa2} shows  the same spectra,  but the  lowest gap is normalized to
$1$, i.e.\ the functions $\frac{E_i(h)-E_0(h)}{E_1(h)-E_0(h)}$ are shown. It is
remarkable that the agreement between exact and TCSA spectra looks better 
than in the case of not normalized spectra.  The functions
$\frac{E_i(h)-E_0(h)}{E_1(h)-E_0(h)}$ have finite limit as  $h \to \infty$ and the
degeneracy pattern in this limit appears to  correspond to the $c=1/2,h=1/16$ representation of the
Virasoro algebra. This correspondence improves as $n_c$ is increased. As an
illustration of this improvement we show the spectra at $n_c=10$ in Figure  \ref{fig.tcsa2_10}.  
At any fixed finite value of $h$,
however, the TCSA data are expected to converge to the exact values as   $n_c \to
\infty$.  

\begin{figure}[!t]
\begin{tabular}{lll}
\includegraphics[clip=true,height=4.5cm]{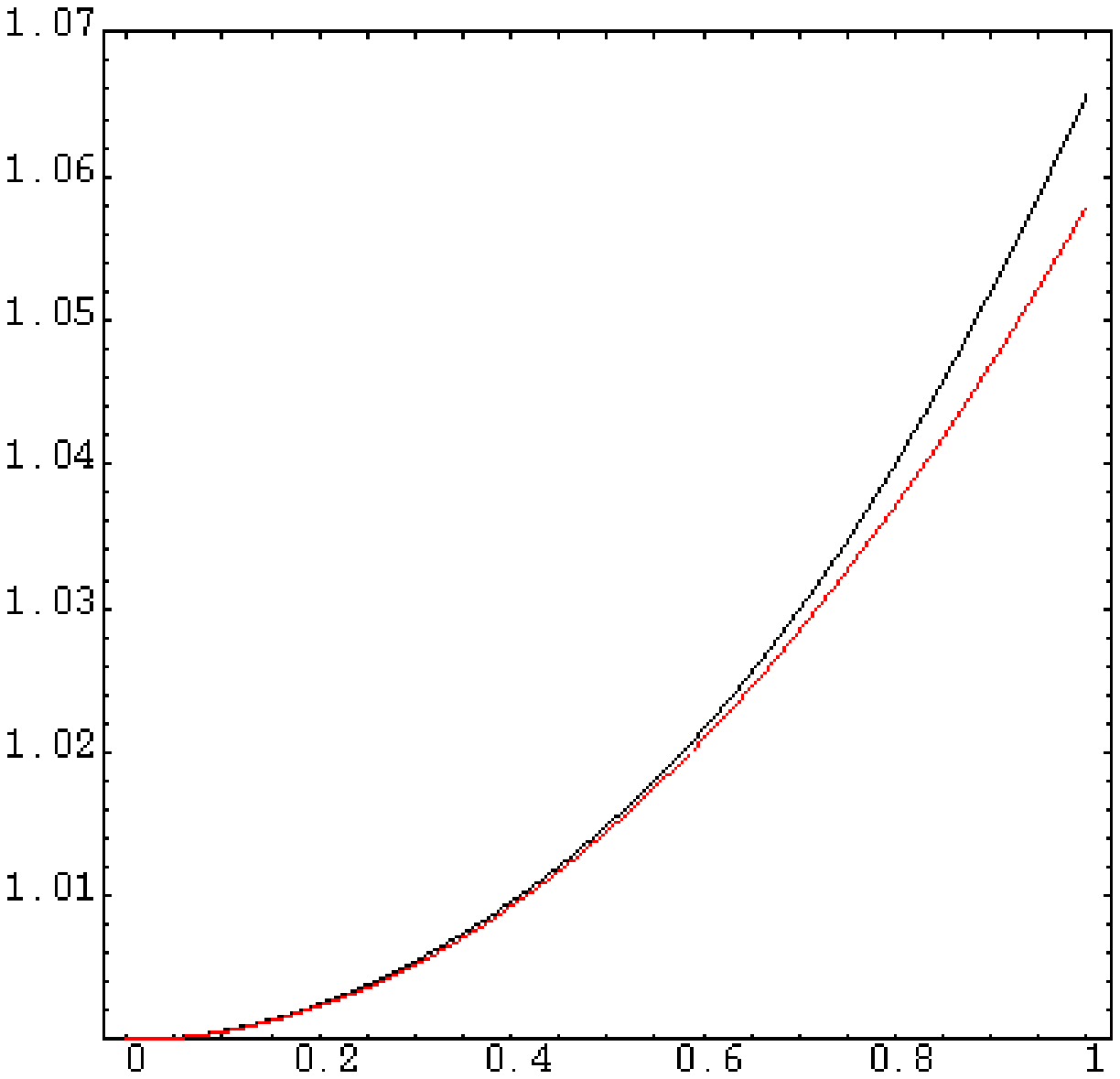}
&
\includegraphics[clip=true,height=4.5cm]{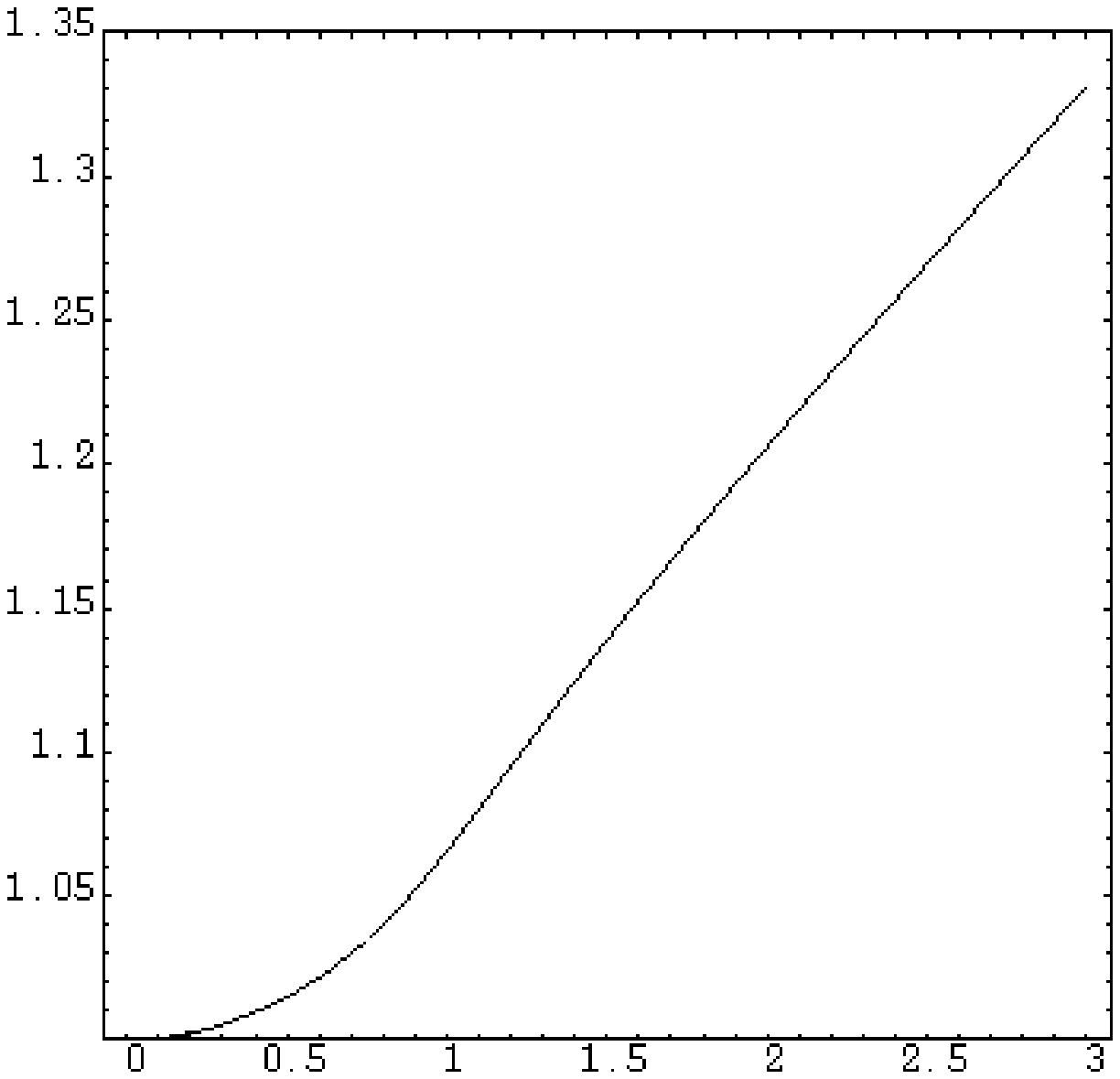}
&
\includegraphics[clip=true,height=4.5cm]{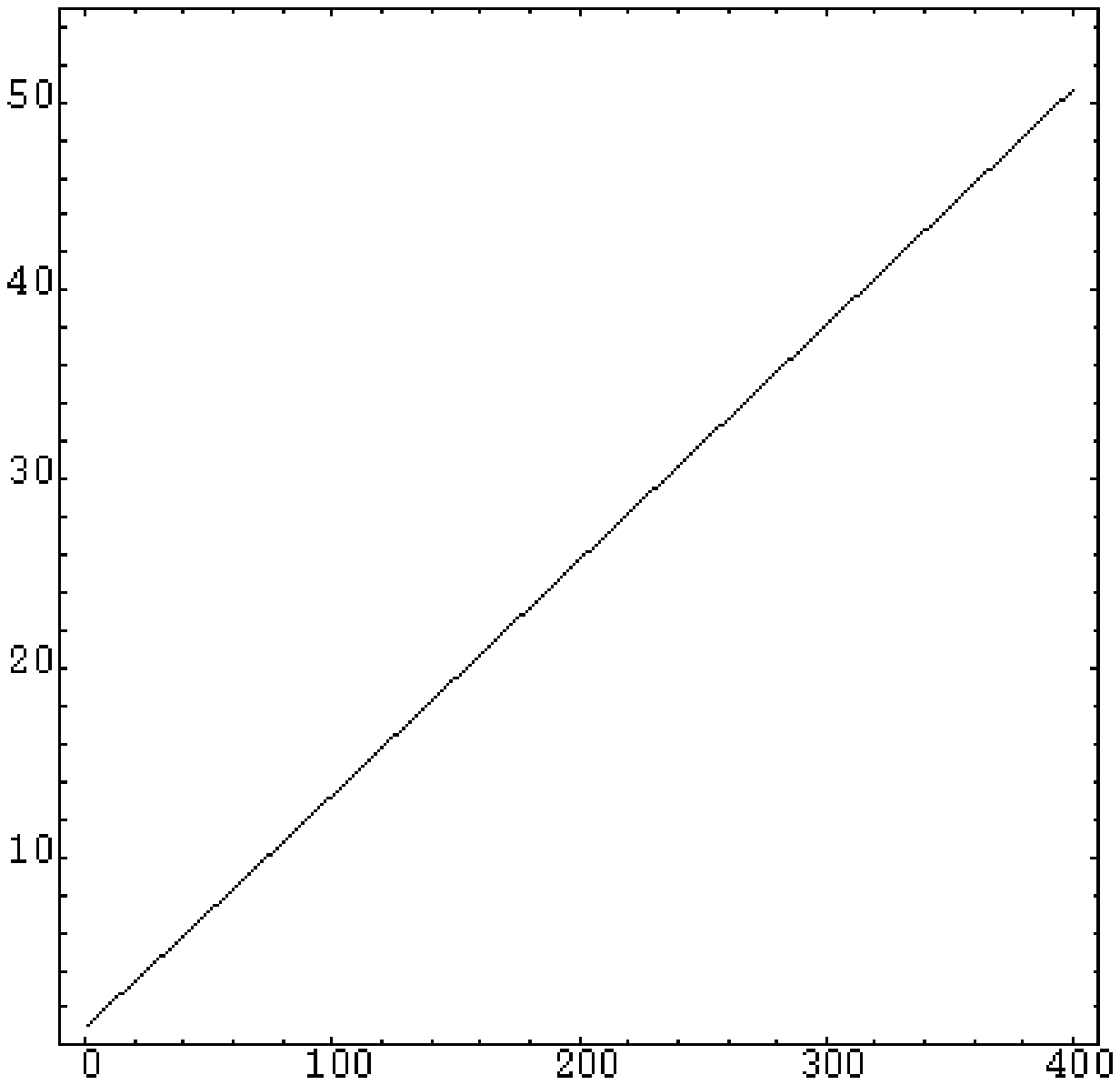}
\end{tabular}
\caption{\label{fig.tcsa3}The function $s_0(h)$ for the $v$ sector  in the ranges $h\in [0,1]$, $s_0\in [1,1.07]$; $h\in [0,3]$,
  $s_0\in [1,1.35]$; $h\in [0,400]$, $s_0\in [1,60]$ at truncation level $n_c=14$ }
\end{figure}

\begin{figure}[!h]
\begin{tabular}{lll}
\includegraphics[clip=true,height=4.5cm]{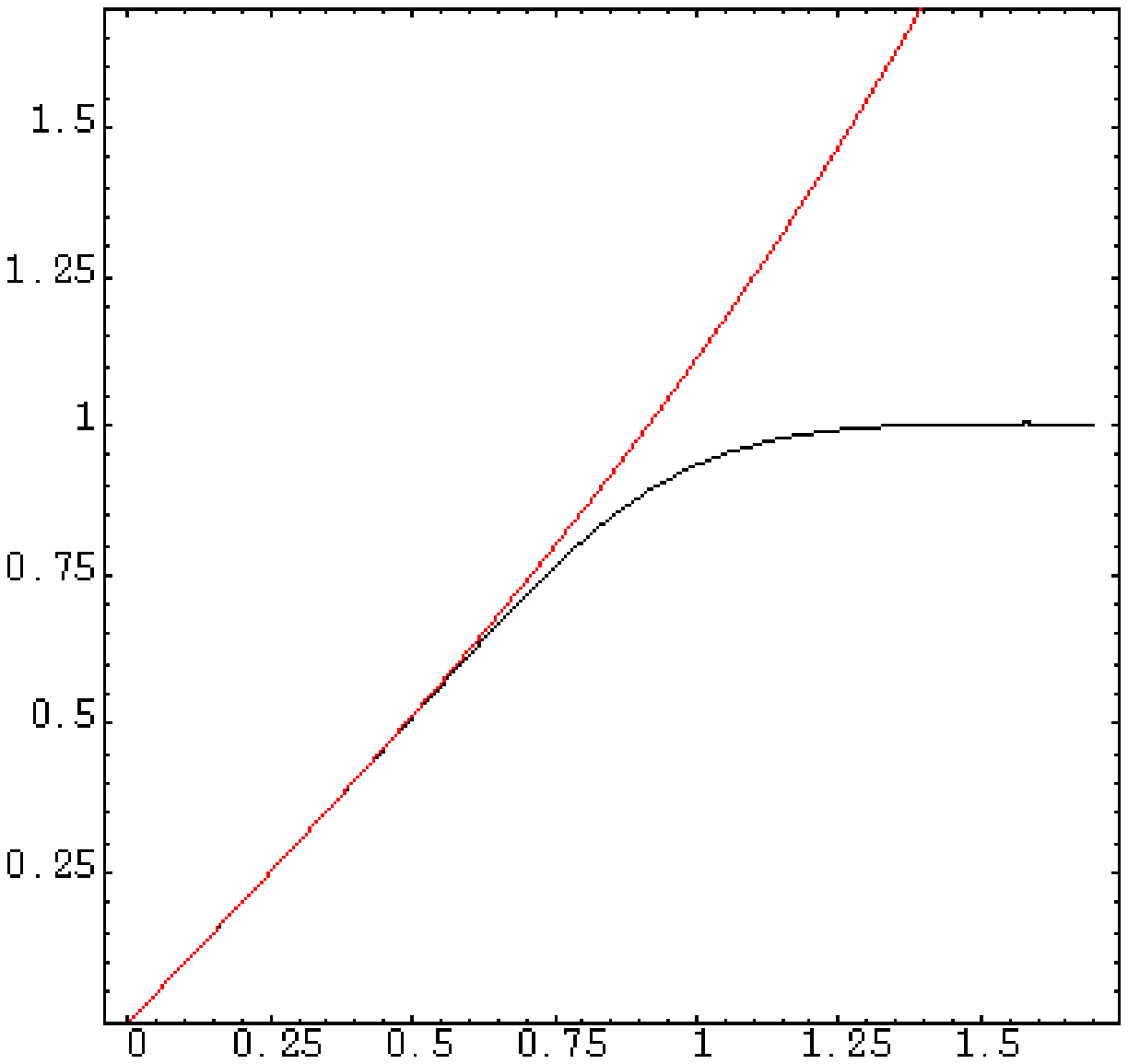}
&
\includegraphics[clip=true,height=4.5cm]{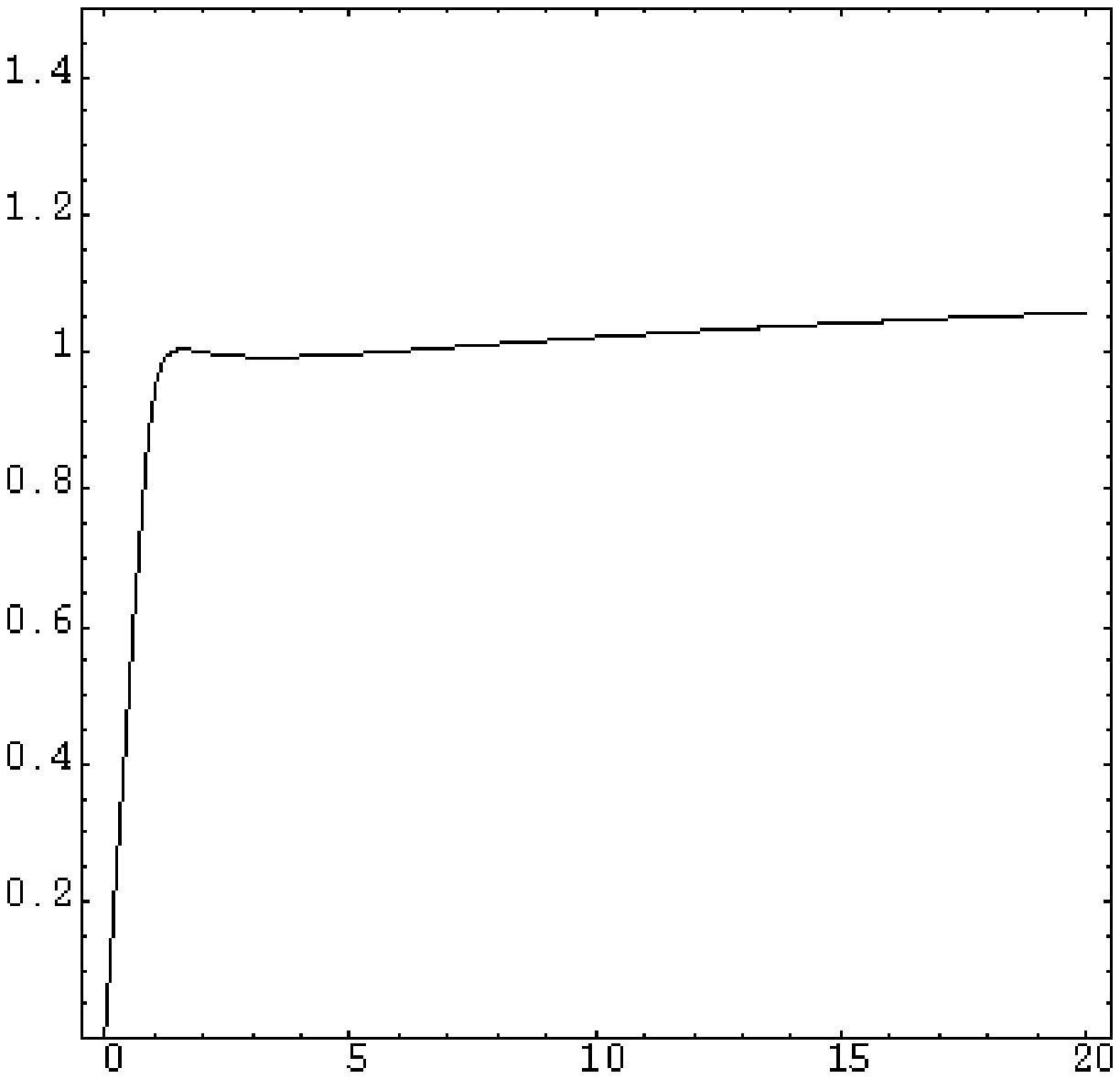}
&
\includegraphics[clip=true,height=4.5cm]{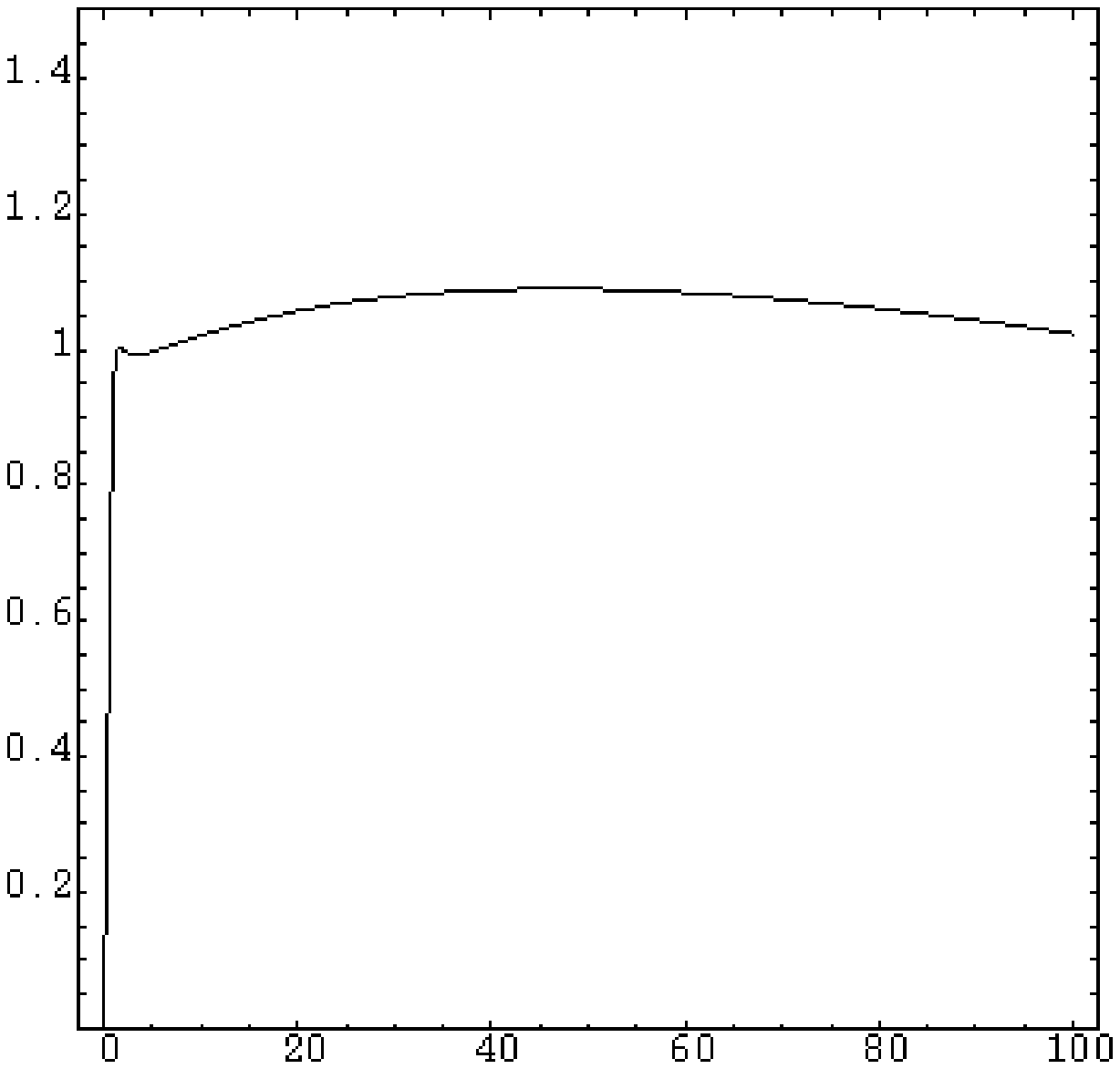}\\
\includegraphics[clip=true,height=4.57cm]{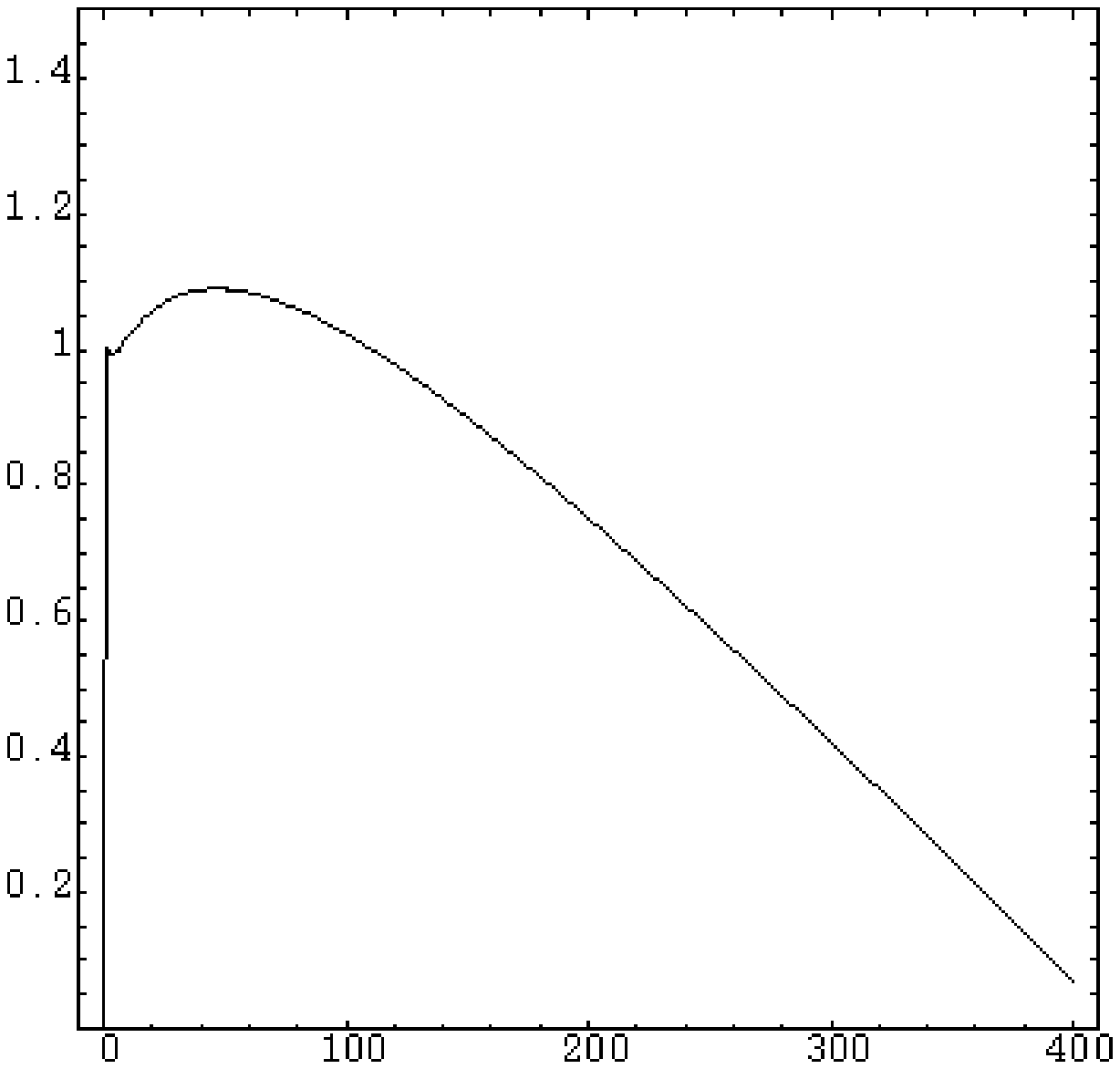} & 
\end{tabular}
\caption{\label{fig.tcsa4}The function $s_1(h)$ for the $v$ sector in the ranges $h\in [0,1.75]$, $s_1\in [0,1.7]$; $h\in [0,20]$,
  $s_1\in [0,1.5]$; $h\in [0,100]$, $s_1\in [0,1.5]$; $h\in [0,400]$, $s_1\in [0,1.5]$ at truncation level $n_c=14$  }
\end{figure}

\begin{figure}[!h]
\begin{tabular}{lll}
\includegraphics[clip=true,height=4.5cm]{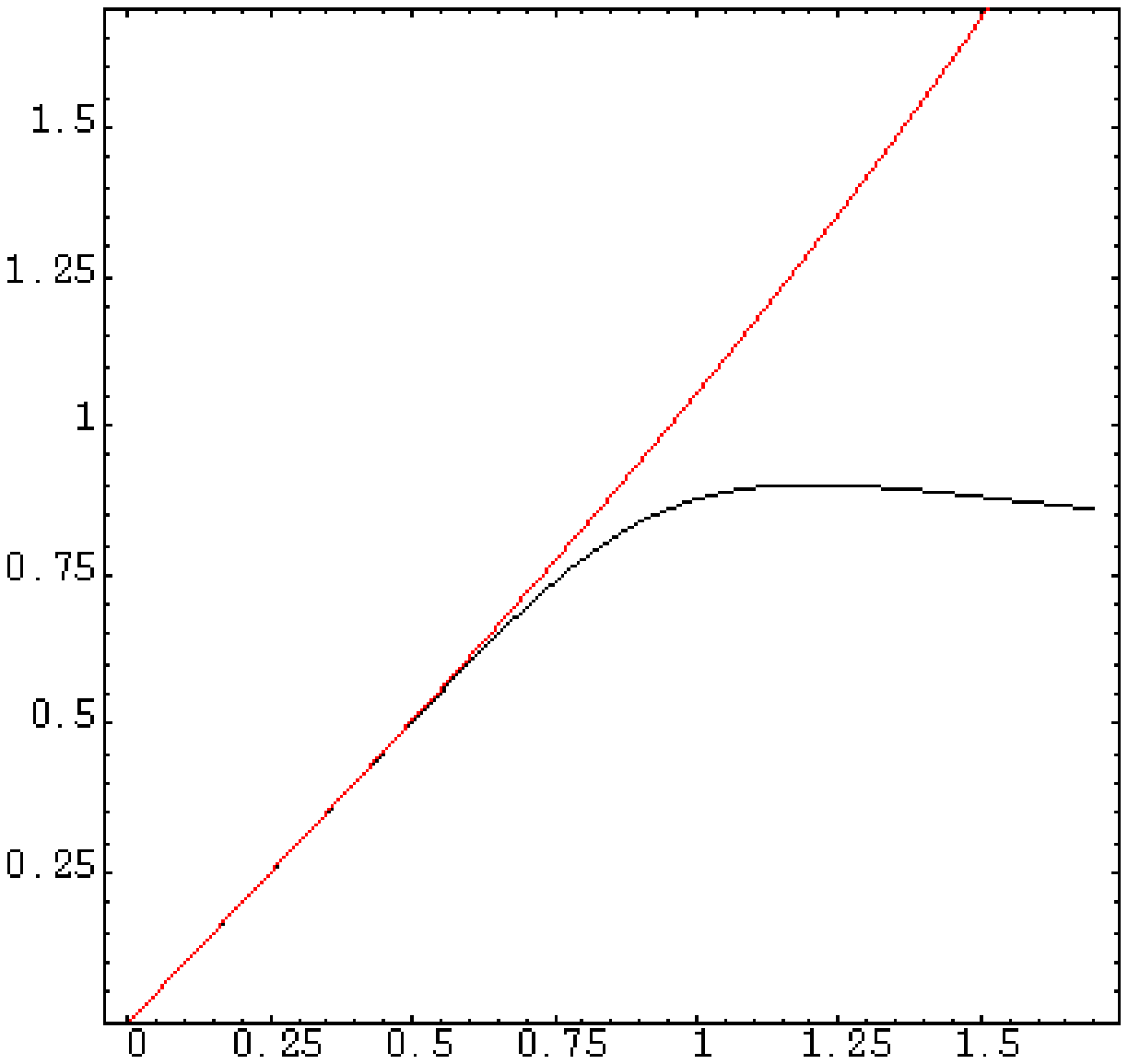}
&
\includegraphics[clip=true,height=4.5cm]{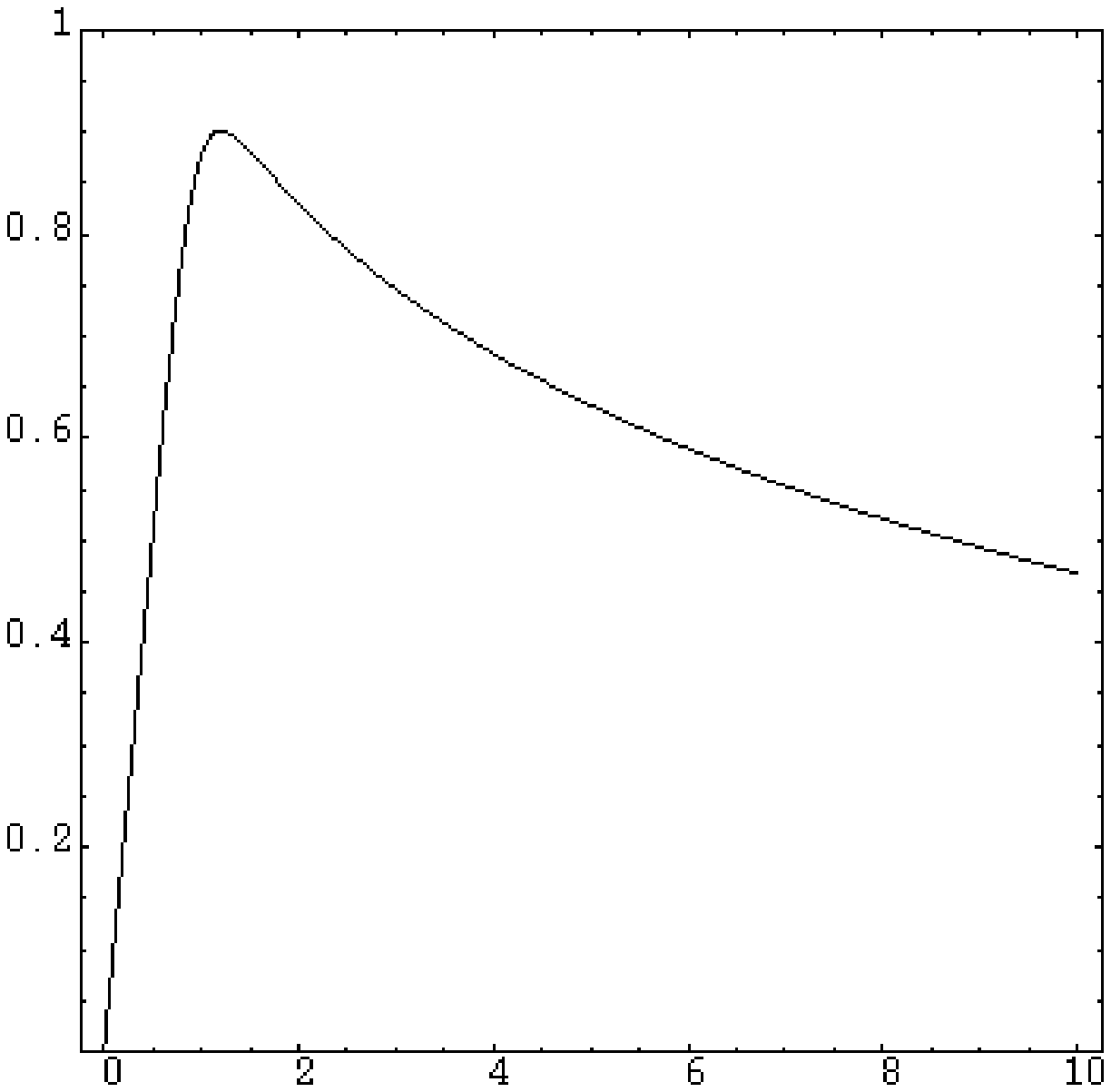}
&
\includegraphics[clip=true,height=4.5cm]{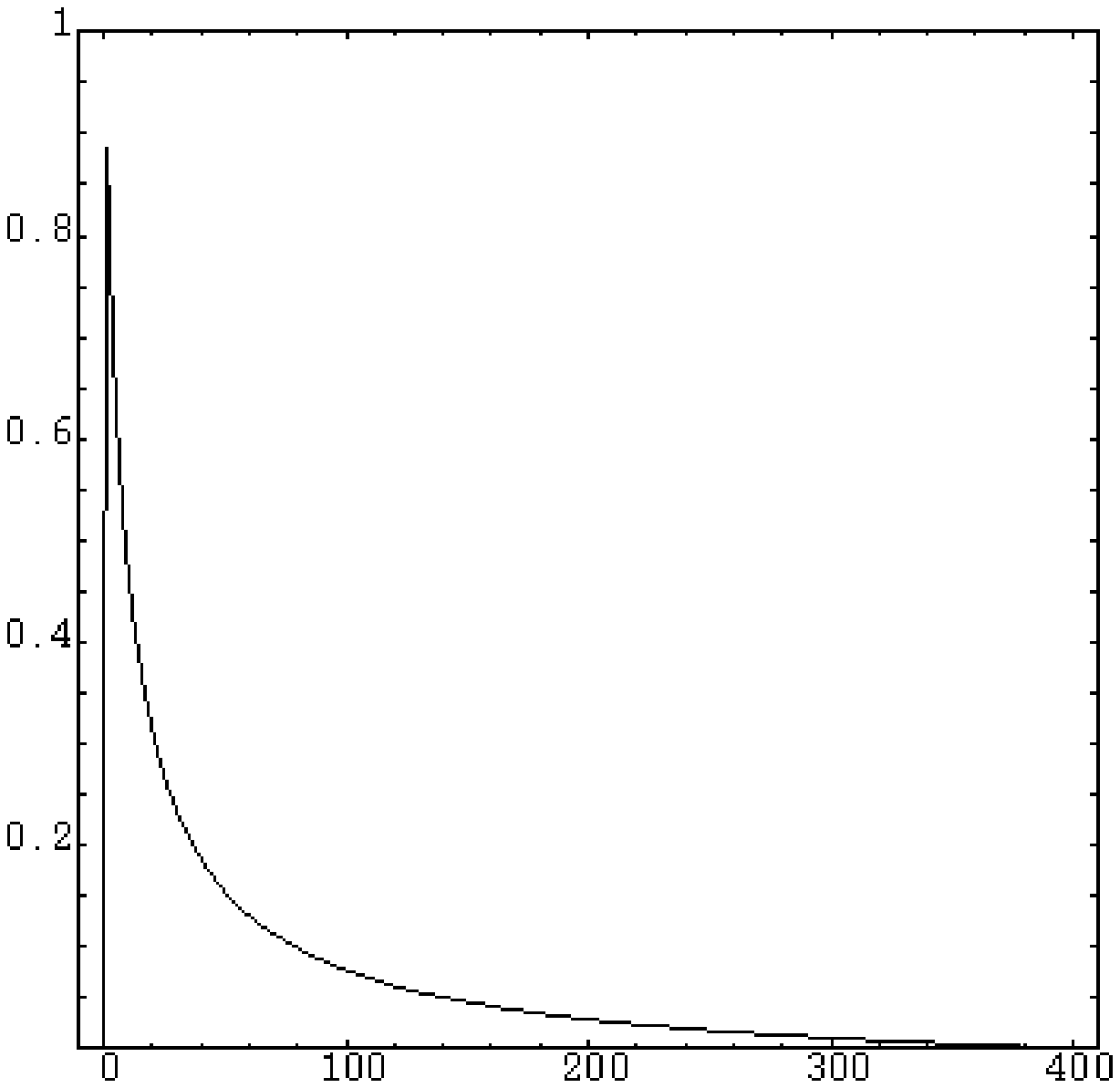}
\end{tabular}
\caption{\label{fig.tcsa5}The function $s_1(h)/s_0(h)$ for the $v$ sector  in the ranges $h\in [0,1.75]$, $s_1/s_0\in [0,1.75]$; $h\in [0,10]$,
  $s_1/s_0\in [0,1]$; $h\in [0,400]$, $s_1/s_0\in [0,1]$ at truncation level $n_c=14$ }
\end{figure}

Figures   \ref{fig.tcsa3}-\ref{fig.tcsa5} show the functions $s_0(h)$, $s_1(h)$, $s_1(h)/s_0(h)$ in
various ranges calculated in the same way as in the mode truncated case. 
The figures also show the curves given by 
(\ref{eq.crv3}) and (\ref{eq.crv4}) on the left-hand side  (red/grey line).
It is
remarkable that  $s_0(h)\propto h$  for large values of $h$.
Calculations at other values of $n_c$ show that the slope of $s_0(h)$
decreases as $n_c$ is increased and it can be expected to  converge  to 0 as $n_c \to \infty$. 
$s_1(h)$ appears to tend to a constant for moderately large values of $h$.
Calculations at other values of $n_c$ show that this constant increases as
$n_c$ is increased and it can be expected to 
converge to infinity as $n_c \to \infty$. 
For large values of $h$, $s_1(h)$
decreases.  
$s_1(h)/s_0(h)$ reaches a maximum at $h\approx 1.6$ and then decreases to
 zero. Calculations at other values of $n_c$ show that the maximum value and the value of $h$ where it is reached  increase as
 $n_c$ is increased and it can be expected that both values converge to
 infinity as $n_c \to \infty$. 

\begin{figure}[p]
\begin{tabular}{cc}
\includegraphics[clip=true,height=12cm]{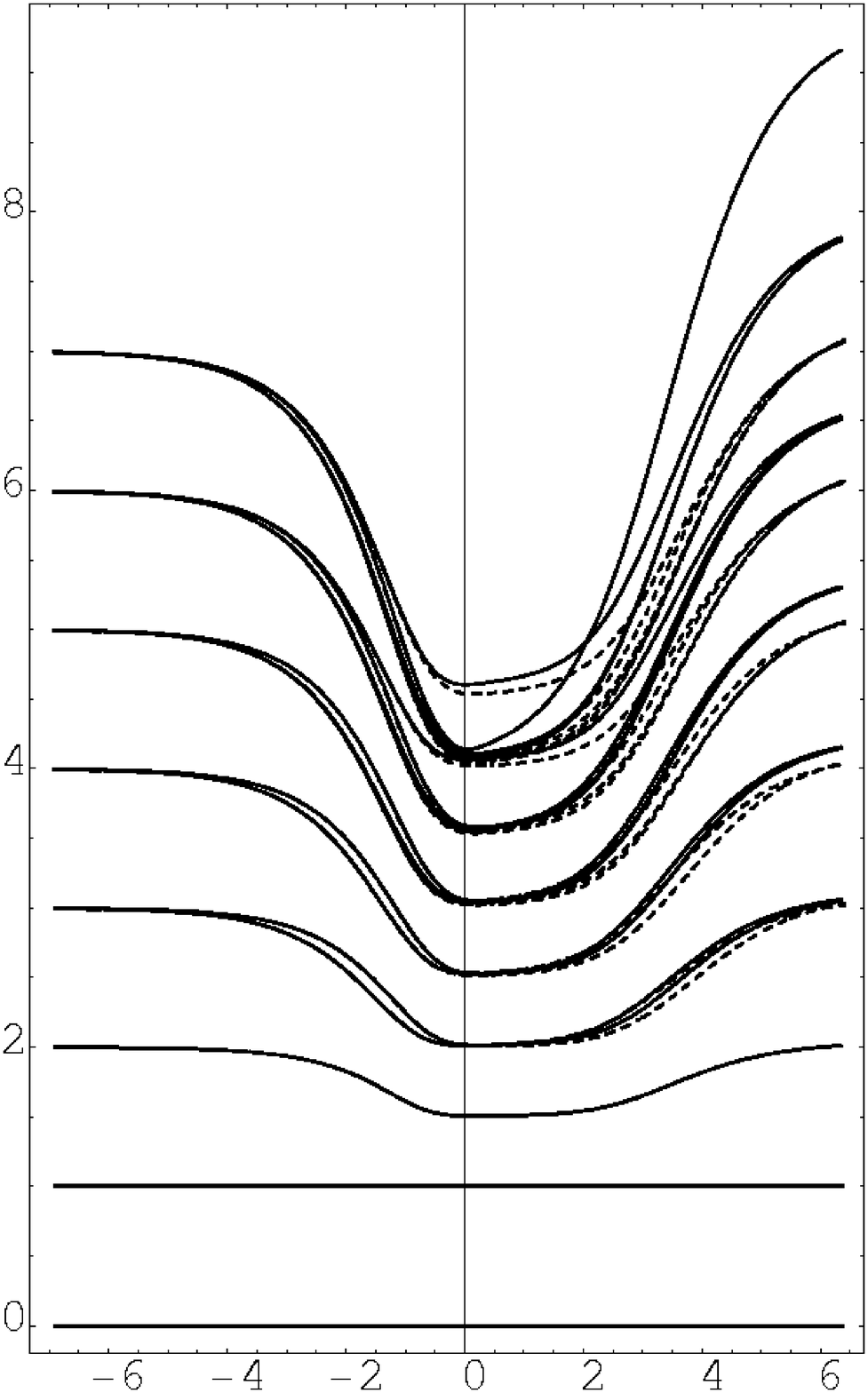}
&
\includegraphics[clip=true,height=12cm]{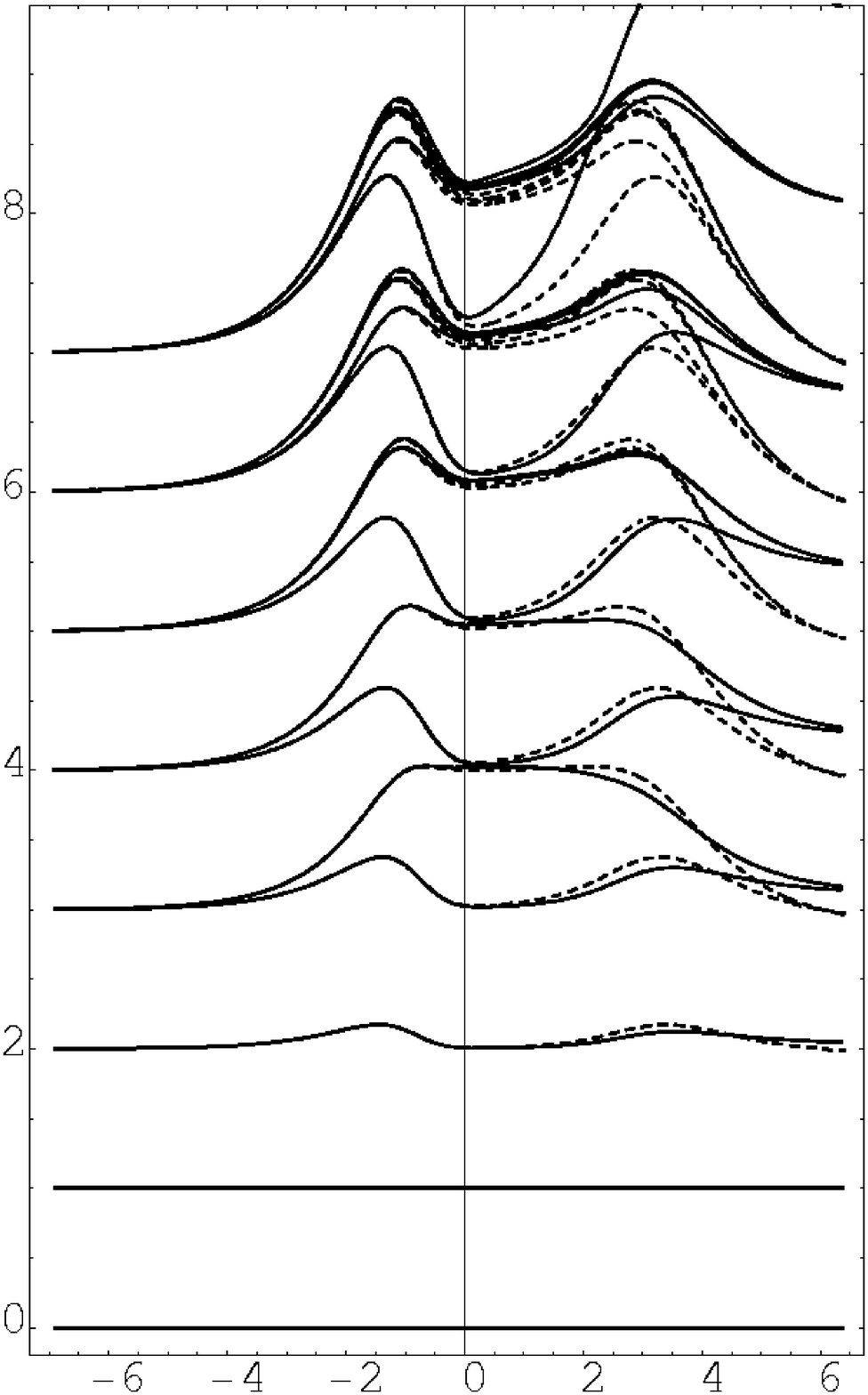}\\
a & b
\end{tabular}
\caption{\label{fig.tcsa6}The TCSA (solid lines) and rescaled exact (dashed
  lines) normalized spectra in the $v$ and $u$  sectors respectively as a
  function of $\ln (h)$ at truncation level $n_c=14$}
\end{figure}

Figure   \ref{fig.tcsa6}.a shows the normalized TCSA spectrum and the
normalized exact spectrum
rescaled by $s_0(h)$ and $s_1(h)$. They show  good
qualitative agreement. 
Values of the
fifth normalized energy gap $\frac{k(3,h)+k(0,h)}{k(1,h)+k(0,h)}$ of the $v$
sector are listed in Table \ref{tab.tcsa1} as in the mode truncated case. These data show that the
rescaling results significant improvement (which is especially noticeable if
$\ln(h)>-2$).

Figure  \ref{fig.tcsa6}.b  shows the normalized TCSA spectrum and the
normalized exact spectrum
rescaled by $s_0(h)$ and $s_1(h)$ in the $u$ sector. The $s_0(h)$, $s_1(h)$ functions
obtained in the $v$ sector were used for the rescaling as in the mode
truncated case. 
 Table \ref{tab.tcsa3} shows  values of the
fourth normalized energy gap $\frac{k(3,h)-k(0,h)}{k(1,h)-k(0,h)}$ of the $u$
sector. 
We see  from the table that the assumption that $s_0(h)$  and $s_1(h)$
are the same in both sectors  does not give very good result in this case, although the situation
might become better at higher truncation levels.

\
\newpage

\begin{table}[!h]
\caption{\label{tab.tcsa1}The normalized energy gap
  $\frac{k(3,h)+k(0,h)}{k(1,h)+k(0,h)}$  in the $v$ sector:  exact, TCSA  $(n_c=14)$
   and rescaled  exact values}\vspace{3mm}
\begin{tabular}{@{}clll}
\hline
$\log(h)$ & Exact  & TCSA   & Rescaled  exact \\
\hline
-6 & 2.993681 & 2.993681 & 2.993681    \\
-5 & 2.982797 & 2.982797 & 2.982797   \\
-4 & 2.953071 & 2.953074 & 2.953073   \\
-3 & 2.8719584 & 2.871974 & 2.871961  \\
-2 & 2.66042177 & 2.660322 & 2.660236    \\
-1 & 2.25064420 & 2.2488 & 2.24837    \\
0 & 2.00954942 & 2.01699 & 2.0177    \\
1 & 2.00003474 & 2.028321 & 2.031631   \\
2 & 2.00000008 & 2.101337 & 2.105516    \\
3 & 2.0000000 & 2.33433 & 2.32573  \\
4 & 2.0000000 & 2.670454 & 2.643480    \\
5 & 2.0000000 & 2.916318 & 2.879339   \\
6 & 2.0000000 & 3.037542 & 2.997132   \\
\hline   
\end{tabular} 

\caption{\label{tab.tcsa2}The  energy gap
  $k(3,h)+k(0,h)$  in the $v$ sector:  exact and TCSA  $(n_c=14)$
   values}\vspace{3mm}
\begin{tabular}{@{}cll}
\hline
$\log(h)$ & Exact  & TCSA  \\
\hline
-6 & 9.434703 & 9.434706 \\
-5 & 9.451803 & 9.451830 \\
-4 & 9.498544 & 9.498745 \\
-3 & 9.626826 & 9.628349 \\
-2 & 9.972034 & 9.983987 \\
-1 & 10.746173 & 10.840718  \\
0 & 11.897034 & 12.51855  \\
1 & 12.461229 & 15.0054  \\
2 & 12.552003 & 20.7975 \\
3 & 12.564424 & 36.1193  \\
4 & 12.566107 & 77.0368 \\
5 & 12.566335 & 187.694 \\
6 & 12.566366 & 488.292 \\
\hline
\end{tabular}
\end{table}

\
\newpage

\begin{table}[!h]
\caption{\label{tab.tcsa3}The normalized energy gap
 $\frac{k(3,h)-k(0,h)}{k(1,h)-k(0,h)}$  in the $u$ sector:  exact, TCSA  $(n_c=14)$ 
   and rescaled  exact values}\vspace{3mm}
\begin{tabular}{@{}clll}
\hline
$\log(h)$ & Exact  & TCSA  & Rescaled  exact \\
\hline
-7 & 3.002321 & 3.002321 & 3.002321\\
-6 & 3.006305 & 3.006305 & 3.006305\\
-5 & 3.017105 & 3.017105 & 3.017105\\
-4 & 3.046201 & 3.046206 & 3.046198\\
-3 & 3.1225067 & 3.122558 & 3.122504\\
-2 & 3.29172833 & 3.292196 & 3.291849\\
-1 & 3.32029283 & 3.31947 & 3.31865\\
0 & 3.01716419 &  3.0268 & 3.0315\\
1 & 3.00006366 &  3.03610 & 3.05522\\
2 & 3.0000000 &  3.11033 & 3.16799\\
3 & 3.0000000&  3.27063 & 3.36158\\
4 & 3.0000000 &  3.28320 & 3.30254\\
5 & 3.0000000 &  3.203134 & 3.115763\\
6 & 3.0000000 &  3.152931 & 3.002865\\
\hline   
\end{tabular} 

\caption{\label{tab.tcsa4}The  energy gap
   $k(3,h)-k(0,h)$  in the $u$ sector:  exact
   and TCSA  $(n_c=14)$ 
   values}\vspace{3mm}
\begin{tabular}{@{}cll}
\hline
$\log(h)$ & Exact  & TCSA  \\
\hline
-7 & 9.421132 &  9.421132\\
-6 & 9.414873 &  9.414877\\
-5 & 9.397906 &  9.397933\\
-4 & 9.352150 &  9.352345\\
-3 & 9.231143 &  9.232544\\
-2 & 8.939485 &  8.949181\\
-1 & 8.545045 &  8.61250\\
0 & 8.939748  &  9.4096\\
1 & 9.345985  &  11.3108\\
2 & 9.414002  &  16.0395\\
3 & 9.423318  &  29.8152\\
4 & 9.424580  &  69.5595\\
5 & 9.424751  &  179.6918\\
6 & 9.424774  &  480.0968\\
\hline
\end{tabular}
\end{table}

\clearpage 

\section{Scaling properties of the $s_1/s_0$, $s_1$ and  $s_0$ functions}
\label{sec.scaling}

In this section we describe results obtained mainly numerically for the
$n_c$-dependence of the  $s_1(n_c,h)/s_0(n_c,h)$,  $s_1(n_c,h)$ and
$s_0(n_c,h)$ functions. The value of $L$ is fixed.  $s_0(n_c,h)$ and
$s_1(n_c,h)$ are calculated from the three lowest energy levels. In the TCSA
case we considered $n_c=11,12,13,14$, in the MT
case $n_c=8,9,10,11$.

In the MT scheme
we found that the functions
\begin{align}
& n_c^{-1/2}\frac{s_1}{s_0}(n_c,h n_c^{1/2})\\
& n_c^{-1/2}s_1(n_c,h n_c^{1/2})\\
& n_c (s_0(n_c,h n_c^{1/2})-1)
\end{align}
are approximately independent of $n_c$. This is consistent with the
perturbative formulae (\ref{eq.crv1}), (\ref{eq.crv2}). It should be noted
that $s_0$ is very close to 1, therefore the scaling properties of $s_0$, $s_1$,
$s_1/s_0$ written down above are not inconsistent. 

In the TCSA scheme
we found that the functions 
\begin{align}
& n_c^{-\alpha}\frac{s_1}{s_0}(n_c,h n_c^{\beta})\\
& n_c^{-\alpha_1}s_1(n_c,h n_c^{\beta_1})\\
& s_0(n_c,h n_c^{\beta_2})
\end{align}
are approximately independent of $n_c$ with a suitable choice of the numbers
$\alpha$, $\beta$, $\alpha_1$, $\beta_1$,  $\beta_2$. The precision of this
independence, however, is not very high, especially for $s_1/s_0$ and $s_1$. We obtained
$\beta_2\approx 1/2$, which is consistent with the perturbative formula
(\ref{eq.crv3}). We also obtained that   $\alpha\approx\beta$ and
$\alpha_1\approx\beta_1$, which is consistent with (\ref{eq.crv3}) and (\ref{eq.crv4}), however the value of $\alpha$ and $\alpha_1$ appears
to be around $1/3$, which deviates from the expectation based on 
(\ref{eq.crv3}) and (\ref{eq.crv4}). Moreover, the value of  $\alpha$ and
$\alpha_1$ appears to depend on $n_c$,  and  on which three energy levels
$s_0$ and $s_1$ are calculated from. One can also obtain different numbers if
one takes different domains of the values of $h$
into consideration. The values one can obtain for $\alpha$ and $\alpha_1$ are
between $0.3$ and $0.5$. 

Generally one can expect that the functions   $s_1(n_c,h)/s_0(n_c,h)$,  $s_1(n_c,h)$ and
$s_0(n_c,h)$ can be written in the form
\begin{equation}
s(n_c,h)=n_c^{\gamma_1}F_1(hn_c^{\delta_1})+n_c^{\gamma_2}F_2(hn_c^{\delta_2})+\dots
\end{equation}
where $s(n_c,h)$ stands for  $s_1(n_c,h)/s_0(n_c,h)$,  $s_1(n_c,h)$ or
$s_0(n_c,h)$, and $\gamma_1$, $\delta_1$,  $\gamma_2$, $\delta_2$, \dots\ are
constants. Some of the functions $F_1$, $F_2$, \dots\ may depend on the method
used to calculate   $s_1(n_c,h)/s_0(n_c,h)$,  $s_1(n_c,h)$ and
$s_0(n_c,h)$. (The functions  $F_1$, $F_2$, \dots\ 
and the exponents are generally different for   $s_1/s_0$,  $s_1$ and
$s_0$, of course.)
In the MT scheme one term appears to
dominate clearly. For $s_0$ this term is the constant function 1, and the next
term also appears to be much larger than the further terms. In the TCSA scheme
it appears that there is one dominant term for  $s_0$,
and a dominant term cannot be isolated for $s_1/s_0$ and  $s_1$ at the values
of $n_c$ that we considered.

\section{Discussion}
\label{sec.disc2}
\markright{\thesection.\ \ DISCUSSION}

We have investigated the validity of the approach (\ref{eq.i2}) for the description
of truncation effects in TCSA spectra. Comparison  with a solvable truncation
method called mode truncation shows that the remarkably regular behaviour of the TCSA spectrum for large $h$ in the case
of the model (\ref{eq.yyy})  is not universal
(i.e.\ not independent of the truncation scheme). The numerical calculations show
that (\ref{eq.i2}) provides a good approximation of the truncated spectra in both the
TCSA and the mode truncation scheme. This is confirmed by perturbative
analytic 
calculations as well. The main difference between the mode truncated 
and TCSA spectra at large $h$ seems to be explicable through the different  behaviour of the function
$s_0(n_c,h)$ in the two schemes. Difference between the $s_0(n_c,h)$
functions appears also in perturbation theory. We
have shown analytically  that in the mode truncation scheme the
convergence of the truncated spectra to the exact spectra can be improved by one order in $1/n_c$ by
the  rescaling (\ref{eq.i2}). This has also been shown in the
TCSA scheme for low orders of perturbation theory in $h$.

We have also given a quantum field theoretic
discussion of the model (\ref{eq.yyy}). In particular we have discussed the change of the
boundary condition satisfied by the fermion fields as the coupling constant
(or external boundary magnetic field) is
increased. Such a change, which is emphasized in the literature, seems impossible 
naively --- in our formulation at least. The paradox is resolved by the
 phenomenon that the  fermion fields (more precisely their matrix
elements between energy eigenstates) develop a discontinuity at the boundary if the coupling
constant is nonzero.

It is still an open problem  to present  an  explanation of the validity of the
approach (\ref{eq.i2}). Within the framework of  (\ref{eq.i2}) the behaviour
of the TCSA spectrum at large $h$, in particular the second flow mentioned in
the Introduction, is explained by the behaviour of $s_0$ and $s_1$ at large
$h$: $s_0 \propto h$, $s_1$ is bounded from above, therefore $s_1/s_0$ tends to
zero. It is a further problem to give an analytic derivation of this behaviour
of $s_0$ and $s_1$.  
In the future we intend to investigate the scaling properties of $s_1/s_0$,
$s_1$ and $s_0$ and the TCSA and MT spectra
more thoroughly by taking higher values of $n_c$ \cite{TW}.
It would also be interesting to extend our results to other
perturbed boundary conformal minimal models, which show similar behaviour
numerically to the model that we have studied.  Certain results concerning
other minimal models and scaling properties already exist
\cite{FGPW,talk}; see also \cite{CLM}.
It is a further problem to
classify the possible behaviours of truncated spectra at large $h$ for various
truncation schemes.  Finally, the quantum field
theoretic description of the model (\ref{eq.yyy}) could be developed further
and extended to the massive case.

\chapter{
A nonperturbative study of phase transitions in the multi-frequency
sine-Gordon model}
\label{sec.chap4}
\markboth{CHAPTER \thechapter.\ \ PHASE TRANSITIONS IN THE MSG MODEL}{}

\section{\label{sec: ketto}The multi-frequency sine-Gordon model}
\markright{\thesection.\ \ THE MULTI-FREQUENCY SINE-GORDON MODEL}

In this section the definition and basic properties of the multi-frequency
sine-Gordon model (MSG) are described.

The action of the model is

\[
\mathcal{A}_{MSG}=\int\mathrm{d}t\int\mathrm{d}x\,\left(\frac{1}{2}\partial_{\mu}\Phi\partial^{\mu}\Phi-V(\Phi)\right),\]
 where \[
V(\Phi)=\sum_{i}^{n}\mu_{i}\cos(\beta_{i}\Phi+\delta_{i})\]
 is the potential, which contains $n$ cosine terms. $\Phi$ is a
real scalar field defined on the two-dimensional Minkowski space $\mathbb{R}^{1+1}$,
$\beta_{i}\in\mathbb{R}$ are the frequencies, $\beta_{i}\ne\beta_{j}$
if $i\ne j$, $\mu_{i}$ are the coupling constants (of dimension
mass$^{2}$ at the classical level) and $\delta_{i}\in\mathbb{R}$
are the phases in the terms of the potential.

Two cases can be distinguished according to the periodicity properties
of the potential. The first one is the rational case, when $V(\Phi)$
is a trigonometric polynomial: the ratios of the frequencies $\beta_{i}$
are rational and the potential is periodic. Let the period of the
potential be $2\pi r$ in this case. The target space of the field
$\Phi$ can be compactified: \[
\Phi\equiv\Phi+2kr\pi,\]
 where $k\in\mathbb{N}$ can be chosen arbitrarily. The model obtained
in this way is called the $k$-folded MSG. The well-known classical
sine-Gordon model corresponds to $n=1$, $k=1$.

The other case is the irrational one, when the potential is not periodic
and no such folding can be made. The irrational case is much more
complicated than the rational one, so we restrict our attention to
the rational case. We remark here only that although
$V(\Phi)$ always has a finite infimum, it does not necessarily admit
an absolute minimum. The potential $V(\Phi)$ can always be written
uniquely as a sum $V(\Phi)=V_{1}(\Phi)+V_{2}(\Phi)+\dots +V_{k}(\Phi)$,
where the terms $V_{1},\dots ,V_{k}$ are periodic but any sum of any
of these terms is not periodic. $V(\Phi)$ has an absolute minimum
if and only if $V_{1}(\Phi),\dots ,V_{k}(\Phi)$ have a common absolute
minimum. This occurs for special choice of the $\delta_{i}$, if the
values of $\beta_{i}$ are given. In particular, if $\beta_{i}/\beta_{j}$
are irrational for all $i\ne j$ and $\mu_{i}<0$ for all $i$, then
$V(\Phi)$ has a absolute minimum if and only if $\frac{\delta_{i}}{\beta_{i}}-\frac{\delta_{j}}{\beta_{j}}=\frac{2\pi b_{i}}{\beta_{i}}-\frac{2\pi b_{j}}{\beta_{j}}$
is satisfied with some numbers $b_{i}\in\ZZ$, which is equivalent
to the case $\delta_{i}=0$ for all $i$. See and \cite{DM} and \cite{BPTW}
for further remarks on the irrational case.

At the quantum level the theory can be regarded as a perturbed conformal
field theory: \[
\mathcal{A}_{MSG}=\mathcal{A}_{CFT}+\mathcal{A}_{pert}\ ,\]
 where\[
\mathcal{A}_{CFT}=\int\mathrm{d}t\int\mathrm{d}x\,\frac{1}{2}\partial_{\mu}\Phi\partial^{\mu}\Phi\
,\]
which is the action of the free scalar particle of zero mass, and\[
\mathcal{A}_{pert}=\int\mathrm{d}t\int\mathrm{d}x\,(-V(\Phi))=-\frac{1}{2}\int\mathrm{d}t\int\mathrm{d}x\,\sum_{i=1}^{n}(\mu_{i}e^{i\delta_{i}}V_{\beta_{i}}+\mu_{i}e^{-i\delta_{i}}V_{-\beta_{i}})\
,\]
where $V_{\omega}$ denotes the vertex operator \[
V_{\omega}=:e^{i\omega\Phi}:\ ,\]
which is a primary field with conformal dimensions \[
\Delta_{\omega}^{\pm}=\Delta_{\omega}=\frac{\omega^{2}}{8\pi}\]
 in the unperturbed (conformal) field theory. The upper index $\pm$
corresponds to the left/right conformal algebra and : : denotes the
conformal normal ordering. The dimensions of the coupling constants at the
quantum level are \[
[\mu_{i}]=(mass)^{2-2\Delta_{i}}\ ,\qquad\Delta_{i}\equiv\Delta_{\beta_{i}}\ .\]
 The perturbing operators are relevant only if 
\begin{equation}
\beta_{i}^{2}<8\pi\ ,
\label{relevance}
\end{equation}
 we restrict ourselves to this case. We also assume that \[
\beta_{i}^{2}<4\pi\ ,\]
 which is a necessary and sufficient condition for the model to be
free from ultraviolet divergencies in the perturbed conformal field
theory framework \cite{Kl-M,DM,BPTW}.

The model has a massgap in general, and it is clear that phase transitions
occur in the classical version of the model as the coupling constants
are tuned (assuming that $n>1$). It is also expected that there are
topologically charged solutions/states in the model \cite{DM}. We shall investigate the sector with zero topological
charge, which is sufficient for our purposes. We also restrict ourselves
to 1-folded models ($k=1$), as it is natural to expect that in infinite
volume a folding number $k\ne1$ results simply in a $k$-fold multiplication
of the spectrum corresponding to $k=1$.

\section{\label{sec: harom}The Truncated Conformal Space Approach for the
  multi-frequency sine-Gordon model}
\markright{\thesection.\ \ TCSA}

The following fields are primary fields in the folded free boson as a
conformal field theory: 
$$
V_{p,\bar{p}}(z,\bar{z})=:\exp[ip\phi_{CFT}(z)+i\bar{p}\bar{\phi}_{CFT}(\bar{z})]:\ .
$$
$V_{p,\bar{p}}(z,\bar{z})$ has 
conformal dimensions $\Delta^{+}=\frac{p^{2}}{8\pi}$, $\Delta^{-}=\frac{\bar{p}^{2}}{8\pi}$,
where $p=\frac{n}{r}+2\pi rm$, $\bar{p}=\frac{n}{r}-2\pi rm$, $n,m\in\ZZ$,
and the free boson field $\Phi_{CFT}$ is
$$
\Phi_{CFT}(x,t)=\phi_{CFT}(x-t)+\bar{\phi}_{CFT}(x+t)\ .
$$
$\mathcal{H}_{CFT}$ is spanned by the states $\ket{p,\bar{p}}=\lim_{z,\bar{z}\rightarrow0}V_{p,\bar{p}}(z,\bar{z})\ket{0}$
($\ket{0,0}\equiv\ket{0}$) and\\
$a_{n_{1}}\dots \bar{a}_{m_{1}}\dots \ket{p,\bar{p}}$,
where $a_{n_{i}}$ and $\bar{a}_{m_{i}}$ are creating operators of
Fourier modes of $\Phi_{CFT}$. The mode creating operators increase the
conformal weight by 1. 
The conformal generators
$L_{0}$ and $\bar{L}_{0}$ are diagonal in this basis. 
We refer the reader to \cite{FMS} for more details on the quantization of the
folded free boson in finite volume.

The basis of 
the TCSA Hilbert space
is obtained by taking those elements $\ket{v}$ of the basis above
which satisfy the truncation condition
$$
\frac{\bra{v}\frac{L}{2\pi}H_{CFT}\ket{v}}{\braket{v}{v}}<e_{cut}\ .
$$
$e_{cut}$ is the dimensionless upper conformal energy cutoff and
$L$ is the volume of space. We restrict ourselves to the sector with
zero topological charge ($p=\bar{p}$), this being the sector containing
the ground state(s) and the relevant information for the problem treated
in this chapter. 
We can also restrict to zero momentum states i.e.\ to states satisfying the condition
$(L_{0}-\bar{L}_{0})\ket{v}=0$. (The operator $L_{0}-\bar{L}_{0}$
commutes with $H$ and $H_{CFT}$ as well). We remark that $e_{cut}$
also serves as an ultraviolet cutoff.

The matrix elements of $H$ between two elements $\ket{a}$ and $\ket{b}$
of the basis of $\mathcal{H}_{CFT}$ above are given by 
\begin{multline}
\left(\frac{H}{M}\right)_{ab}=\frac{2\pi}{l}\left(L_{0}+\bar{L}_{0}-\frac{c}{12}\right)_{ab}\label{TCSA
  Hamilton}\\
+\frac{2\pi}{l}\sum_{j=1}^{n}sgn(\mu_{j})\kappa_{j}\left(\frac{M_{j}}{M}\right)^{x_{j}}\frac{l^{x_{j}}}{2(2\pi)^{x_{j}-1}}e^{i\delta_{j}}(V_{\beta_{j},\beta_{j}}(1,1))_{ab}\delta_{\Delta_{a}-\bar{\Delta}_{a},\Delta_{b}-\bar{\Delta}_{b}}\\
+\frac{2\pi}{l}\sum_{j=1}^{n}sgn(\mu_{j})\kappa_{j}\left(\frac{M_{j}}{M}\right)^{x_{j}}\frac{l^{x_{j}}}{2(2\pi)^{x_{j}-1}}e^{-i\delta_{j}}(V_{-\beta_{j},-\beta_{j}}(1,1))_{ab}\delta_{\Delta_{a}-\bar{\Delta}_{a},\Delta_{b}-\bar{\Delta}_{b}}\
  ,\end{multline}
where $M$ is a mass scale of the theory given below, $l=LM$ is
the dimensionless volume, $x_{j}=2-2\Delta_{j}$, $\Delta_{j}\equiv\Delta_{\beta_{j}}$;
$\Delta_{a}$, $\bar{\Delta}_{a}$, $\Delta_{b}$, $\bar{\Delta}_{b}$
are the conformal weights of the states $\ket{a}$ and $\ket{b}$,
$c$ is the central charge of the conformal theory ($c=1$ in the
present case), and we have made a replacement corresponding to 
$$
|\mu_{j}|=\kappa_{j}M_{j}^{x_{j}}\ .
$$
The `interpolating' mass scale $M$ is 
$$
M=\sum_{j}\eta_{j}M_{j}\ ,
$$
where 
\begin{equation}
\eta_{j}=\frac{|\mu_{j}|^{1/x_{j}}}{\sum_{i}|\mu_{i}|^{1/x_{i}}}\label{dless
  coupling}
\end{equation}
are the dimensionless coupling constants (of which only $n-1$ are
independent). (\ref{dless coupling}) implies that $\eta_{j}\in[0,1]$,
$\sum_{j}\eta_{j}=1$. $M$ depends smoothly on the $\eta_{j}$-s.
The precise expression for $\kappa$ is not essential for our problem,
we need only that $\kappa$ depends on $\Delta$ only and that it
is dimensionless. Following \cite{BPTW} we used the formula of \cite{Zam}:\[
\kappa_{j}=\frac{2\Gamma(\Delta_{j})}{\pi\Gamma(1-\Delta_{j})}\left(\frac{\sqrt{\pi}\Gamma(\frac{1}{x_{j}})}{2\Gamma(\frac{\Delta_{j}}{x_{j}})}\right)^{x_{j}}.\]
In the classical limit $\Delta_j=0$ and $x_j=2$.

The formula (\ref{TCSA Hamilton}) is written in terms of dimensionless
quantities, and the volume $(l)$ dependence of $H/M$ is also explicit.
We refer the reader to \cite{Yurov-Zam} and \cite{FRT4,BPTW,key-21,Ravanini} for further explanation of  (\ref{TCSA Hamilton}).
It should also be noted that the above basis for the Hilbert space is not the
one generated by the standard $L_N$ elements of the Virasoro algebra, which is used in
general for TCSA, but the one generated by the mode creating operators. In
this (orthogonal) basis the matrix elements
$(V_{\beta_{j},\beta_{j}}(1,1))_{ab}$ and
$(V_{-\beta_{j},-\beta_{j}}(1,1))_{ab}$ are relatively easy
to calculate.

It is clear from (\ref{TCSA Hamilton}) that TCSA gives an exact result
if $l\rightarrow 0$ (assuming (\ref{relevance})) and this limit of
the theory is the conformal theory (the massless free boson), and
the accuracy of the TCSA spectrum decreases at fixed $e_{cut}$ as
$l\rightarrow\infty$. For very large values of $l$ the $l$-dependence
of the spectrum of the TCSA Hamiltonian operator is power-like and it is determined
by the $l^{x_{j}-1}$ coefficients in (\ref{TCSA Hamilton}).
The TCSA Hamiltonian operator cannot be considered as good approximation for
these values of $l$. 

We denote the (dimensionless) energy levels of $H/M$ in volume $l$
by $e_{i}(l)$, $i=0,1,2,\dots$, and $e_{0}\leq e_{1}\leq e_{2}\leq \dots$
is assumed if not stated otherwise. We shall draw conclusions about the spectrum
at $l=\infty$ from the behaviour of the functions $e_{i}(l)$ for
low values of $i$ and moderately large values of $l$.

\section{\label{sec: negy}Phase structure in the classical limit }
\markright{\thesection.\ \ PHASE STRUCTURE IN THE CLASSICAL LIMIT}

\subsection{Phase structure of the two-frequency model in the classical limit}

The Lagrangian density takes the following form in the two-frequency
case: \[
\mathcal{L}=\frac{1}{2}\partial_{\mu}\Phi\partial^{\mu}\Phi-\mu\cos(\beta\Phi)-\lambda\cos(\alpha\Phi+\delta)\
,\]
 where 
\[
\frac{\beta}{\alpha}=\frac{n}{m}\ne1\ ,
\]
 $n$ and $m$ are coprimes (and the folding number equals to one).

The following proposition about the properties of $V(\Phi)$ can be used to
determine the phase structure:
\vspace{0.5cm}

Assume that $\mu,\lambda\ne0$. Then the following
three cases can be distinguished:

a.) If the function $V(\Phi)$ is symmetric with respect to the reflection
$\Phi\mapsto2\Phi_{0}-\Phi$, where $\Phi_{0}$ is a suitable constant, and
$n,m>1$, then $V$ has two absolute minima, which are mapped into
each other by the reflection. We remark that it depends on the value
of $\delta$ whether $V$ has this symmetry or not, and it is easy
to give a criterion for the existence of this symmetry in terms of
$n,m$ and $\delta$.

b.) If $V$ is symmetric with respect to a reflection as in case a.,
but $n=1$ or $m=1$, then $V$ has one or two absolute minima depending
on the values of $\mu$ and $\lambda$. In this case, assuming that
$n=1$, $V$ can be brought to the form \[
V(\Phi)=-|\mu|\cos(\beta\Phi)+|\lambda|\cos(m\beta\Phi)\]
 by an appropriate shift of $\Phi$. $V$ has two absolute minima
if $|\lambda/\mu|>1/m^{2}$, and one absolute minimum if $|\lambda/\mu|\leq1/m^{2}$.
The two absolute minima are mapped into each other by the reflection.
The second derivative of $V$ is nonzero at the minima if $|\lambda/\mu|\ne1/m^{2}$,
but it is zero if $|\lambda/\mu|=1/m^{2}$. In the latter case, the
fourth derivative of $V$ at the minimum is nonzero. The two minima
of $V$ merge and the value of the second derivatives of $V$ at the
two minima tends to zero as $|\lambda/\mu|$ approaches $1/m^{2}$
from above.

c.) If $V$ does not satisfy the requirements of a) and b), then $V$
has a single absolute minimum.
\vspace{0.5cm}

We omit the proof of this proposition, which is elementary, although
long and not completely trivial because of the arbitrariness of $n$
and $m$.

If $\mu$ or $\lambda$ equals to zero, then $V$ is periodic and
has $m$ or $n$ absolute minima, respectively. See \cite{key-21}
for a detailed investigation of these (integrable) limiting cases.

The phases of the classical model are determined by the behaviour
of the absolute minima of $V(\Phi)$ as the value of the coupling constants
vary. In particular, the proposition above implies that the phase structure
of the two-frequency model is the following:

The model exhibits an Ising-type second order phase transition at
the critical value \[
\eta_{c}=\frac{m}{1+m}\]
 of the dimensionless coupling constant $\eta=\sqrt{|\mu|}/(\sqrt{|\mu|}+\sqrt{|\lambda|})$
if $n=1$ and $V$ has the $\ZZ_{2}$-symmetry introduced above. This
critical point separates two massive phases with unbroken and spontaneously
broken $\ZZ_{2}$-symmetry. Equivalent statement can be made if $m=1$.
If $V$ is not symmetric, then there is only one massive phase with
nondegenerate ground state. If $m,n\ne1$ and $V$ is symmetric, then
there is one massive phase with doubly degenerate ground state (i.e.\
the reflection symmetry is spontaneously broken). In the limiting
cases $\eta=0$ and $\eta=1$ the model is massive and has spontaneously
(and completely) broken $\ZZ_{n}$ or $\ZZ_{m}$ symmetry.

\subsection{\label{sec 4.2}Phase structure of the three-frequency model in the
classical limit}

A complete description of the behaviour of the absolute minima of
$V$ for all values of the parameters becomes excessively difficult
in the three- and higher-frequency cases, so we restrict our attention
to particular values. The potential in the three-frequency case is
\begin{equation}
V(\Phi)=\mu_{1}\cos(\beta_{1}\Phi)+\mu_{2}\cos(\beta_{2}\Phi+\delta_{2})+\mu_{3}\cos(\beta_{3}\Phi+\delta_{3})\
 .\label{eq: 3freq pot}\end{equation}
 We choose the frequency ratios $3:2:1$, i.e.\  
\[
\beta_{1}=\beta\ ,\qquad\beta_{2}=\frac{2}{3}\beta\ ,\qquad\beta_{3}=\frac{1}{3}\beta\
.
\]
 This three-frequency model has a tricritical point if and only if
$\delta_{2}=\delta_{3}=0$ (and also in a few equivalent cases), in
this case $V$ is symmetric with respect to the reflection $\Phi\mapsto-\Phi$.
In the tricritical point the absolute minimum of $V$ can be located
only at $0$ or $\pi$. The two cases are equivalent, we consider
the case when the location of the absolute minimum is $0$. The tricritical
point in this case is located at\[
\frac{\mu_{1}}{\mu_{2}}=-\frac{1}{6}\ ,\qquad\frac{\mu_{1}}{\mu_{3}}=\frac{1}{15}\
.\]
 In this point $V^{(6)}(0)\ne0$. (The upper index $^{(6)}$ denotes
the sixth derivative with respect to $\Phi$.) $V^{(6)}(0)>0$ requires
$\mu_{1},\mu_{3}<0$ and $\mu_{2}>0$. We restrict ourselves to this
domain and to the values $\delta_{2}=\delta_{3}=0$.

\begin{figure}
\begin{center}\includegraphics[%
  scale=0.7]{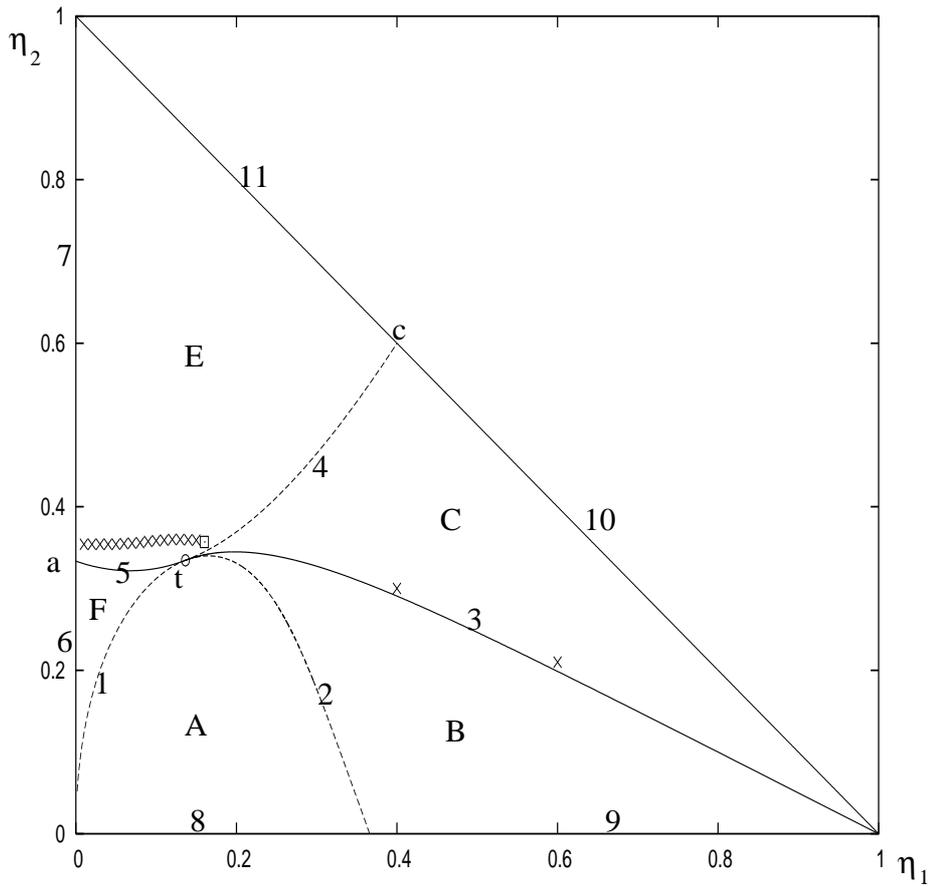}\end{center}

\caption{\label{threefreq phasediag}Phase diagram of the classical three-frequency
sine-Gordon model at $\beta_{1}/\beta_{2}/\beta_{3}=3/2/1$, $\delta_{1}=\delta_{2}=\delta_{3}=0$,
$\mu_{1},\mu_{3}<0$ and $\mu_{2}>0$. The crosses and the square correspond to certain
quantum theory values described in Section \ref{sec: het}.}
\end{figure}

\begin{figure}
\begin{center}\includegraphics[%
  height=0.17\textwidth]{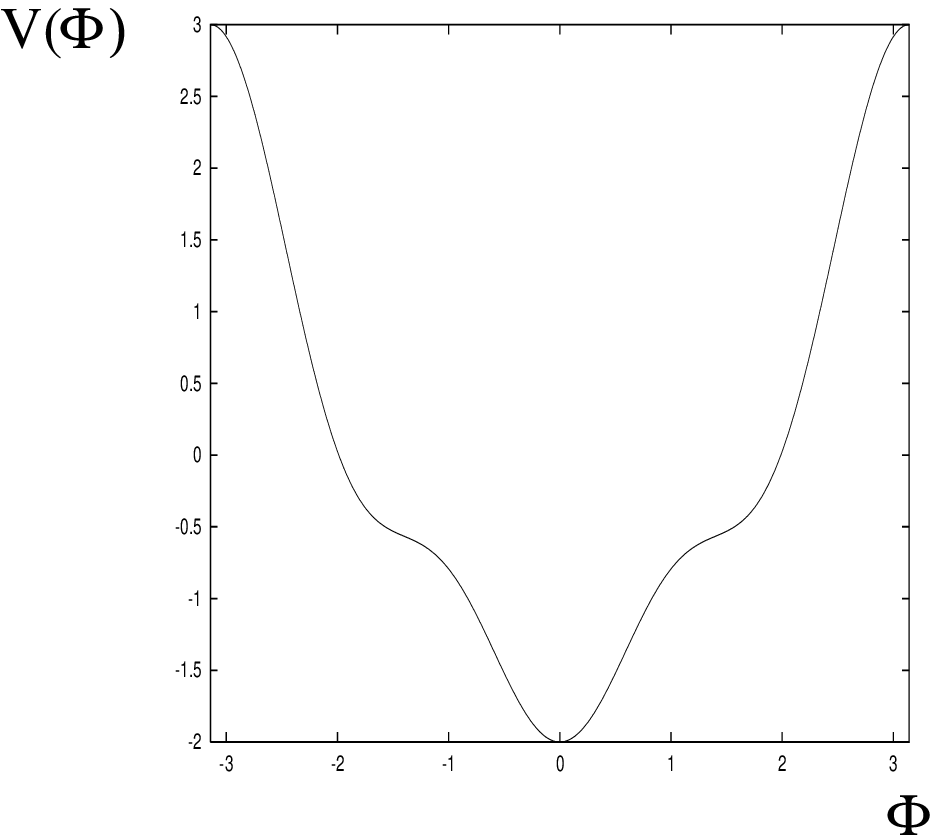}\includegraphics[%
  height=0.17\textwidth]{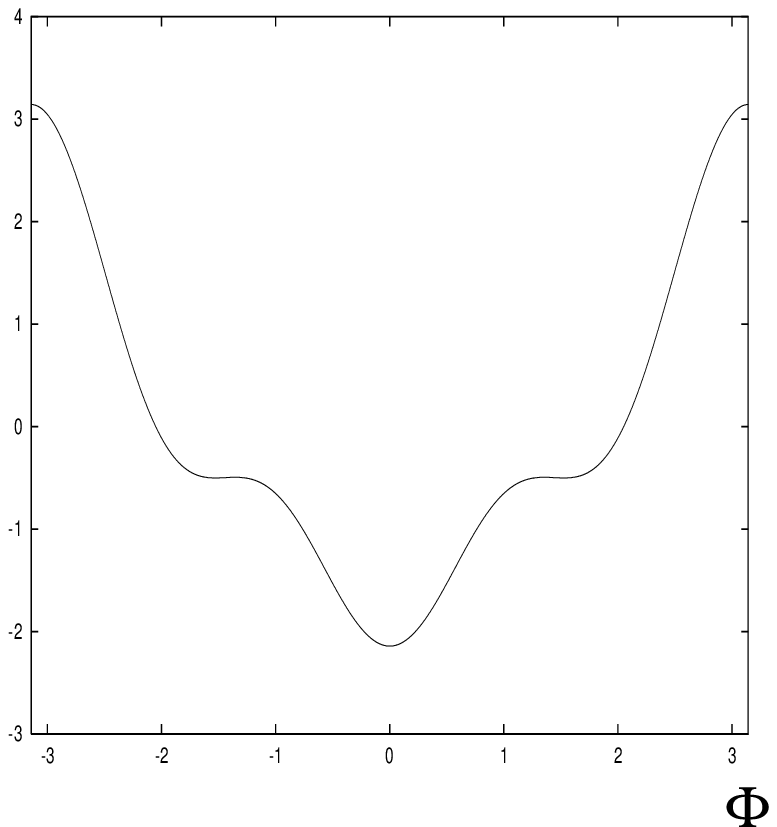}\includegraphics[%
  height=0.17\textwidth]{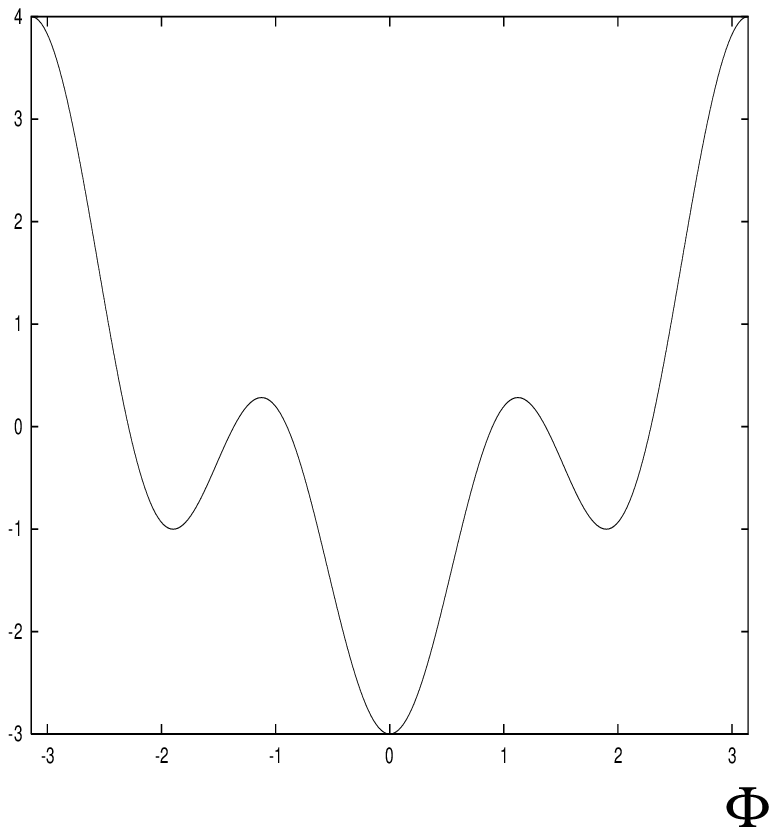}\includegraphics[%
  height=0.17\textwidth]{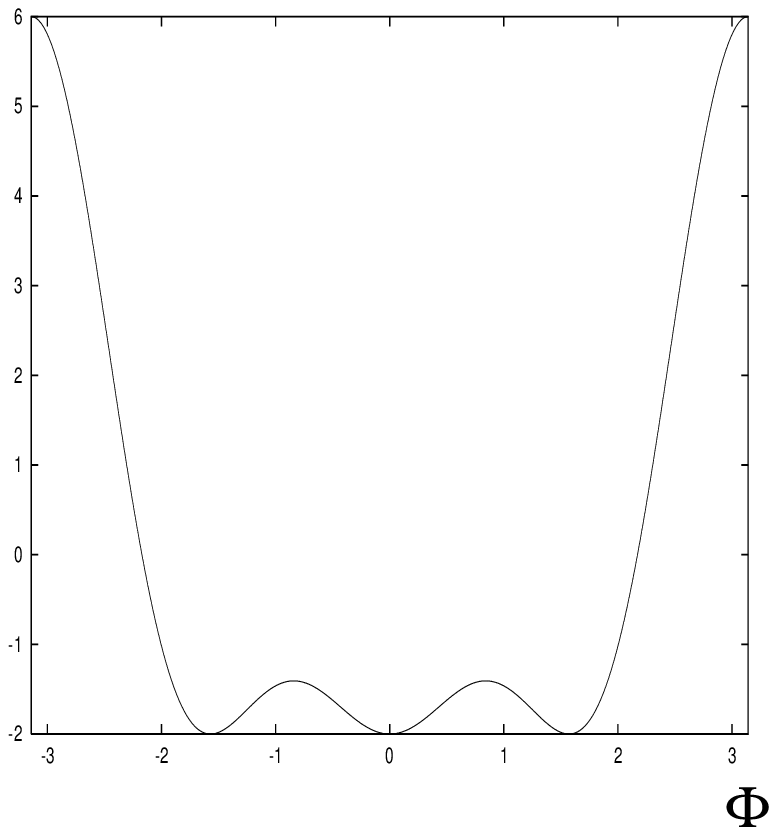}\end{center}

\vspace{-0.5cm} \hspace{3.7cm}1,6,8,A,F\hspace{1.5cm}2\hspace{2.2cm}B\hspace{2.3cm}3

\begin{center}\includegraphics[%
  height=0.17\textwidth]{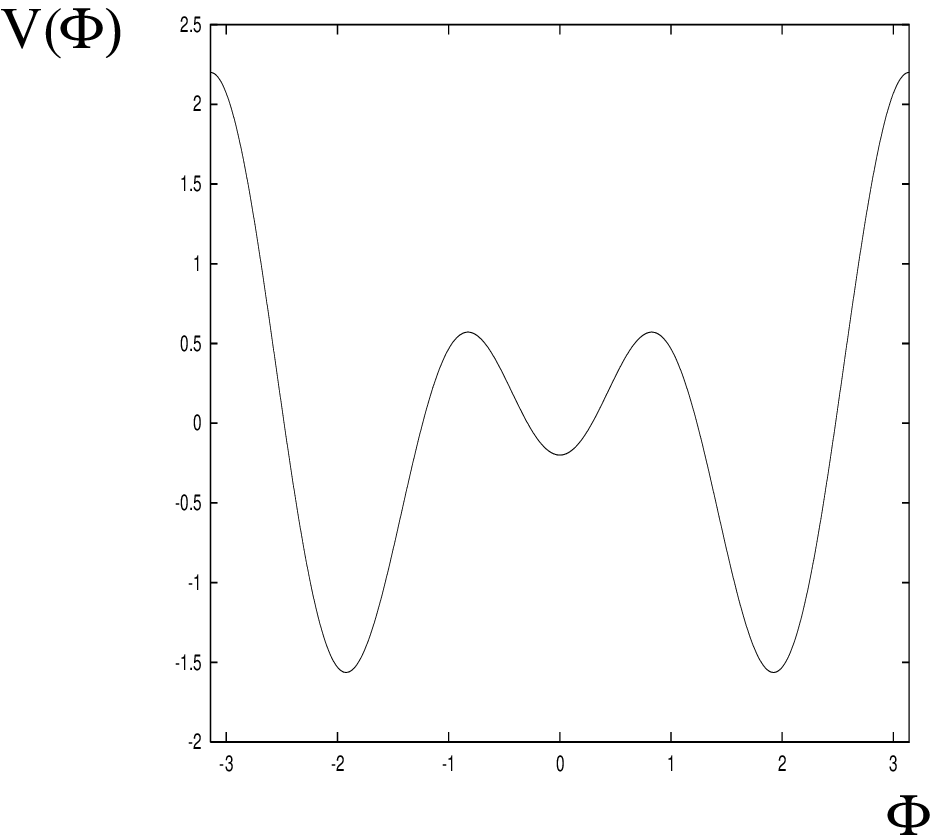}\includegraphics[%
  height=0.17\textwidth]{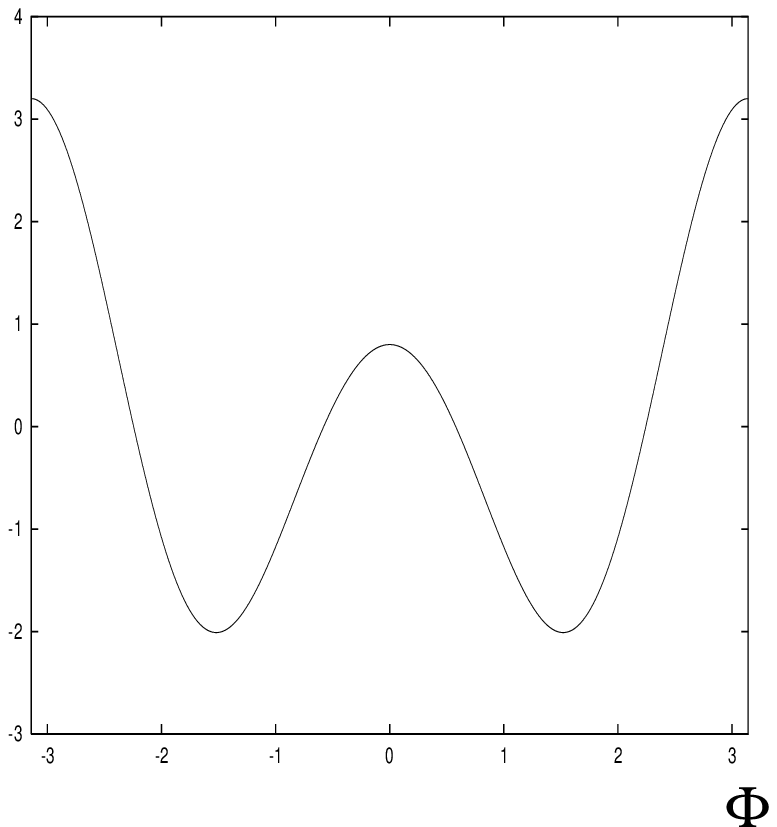}\includegraphics[%
  height=0.17\textwidth]{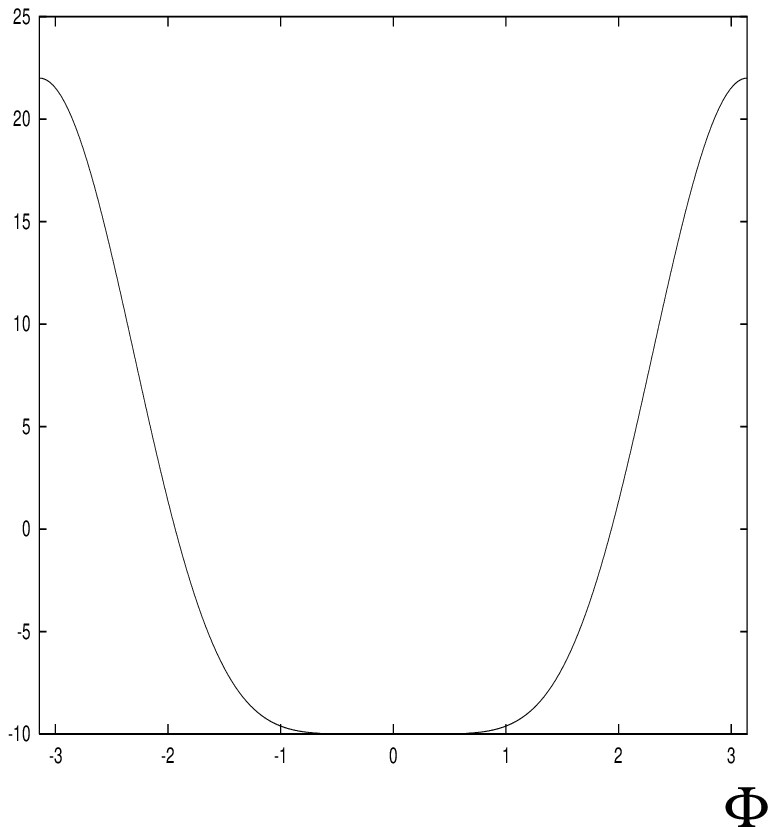}\end{center}

\vspace{-0.5cm} \hspace{5.7cm}C\hspace{1.8cm}4,E,7\hspace{1.6cm}a,5,t

\caption{\label{tpotfigs}Characteristic shapes of the potential }
\end{figure}

The phase diagram is shown in Figure \ref{threefreq phasediag}. The
points of the diagram correspond to the values of the pair $(\eta_{1},\eta_{2})$
of dimensionless parameters. The allowed values constitute the left
lower triangle, the straight line joining $(0,1)$ and $(1,0)$ corresponds
to $\eta_{3}=0$. The tricritical point is denoted by $t$, it is
located at \[
\left(\frac{1}{1+\sqrt{6}+\sqrt{15}}\
,\frac{\sqrt{6}}{1+\sqrt{6}+\sqrt{15}}\right)\approx(0.1365,0.3345)\ .\]
 At $t$ $V$ has one single and absolute minimum (at $\Phi=0$).
Phase transition occurs when the lines $5$ and $3$ shown in the
phase diagram are crossed. Second order Ising-type phase transition
occurs on $5$ and first order phase transition occurs on $3$. The
domain $A\cup B\cup F$ corresponds to a massive $\ZZ_{2}$-symmetric
phase (with unique ground state). The domain $E\cup C$ corresponds
to a massive phase with spontaneously broken $\ZZ_{2}$-symmetry.
Characteristic shapes of the potential in the various domains and
on the various lines of the phase diagram can be seen in Figure \ref{tpotfigs}.
Data applying to the quantum case are also shown in Figure \ref{threefreq phasediag},
they will be explained in subsequent sections.

\subsection{$n$-frequency model in the classical limit }

Let us take the $n$-frequency model with $\beta_{i}=i\beta$, $i=1\dots n$,
and $\delta_{i}=0$. In this case there exist unique values of $\mu_{i}/\mu_{1}$,
$i=2 \dots n$ so that $V(x)$ has a single global minimum at $x=0$ and
has no other local minima, and $V''(0)=0$, $V''''(0)=0$, ... ,$V^{(2n)}(0)=0$
also hold. The values of $\mu_{i}/\mu_{1}$ are determined by the
latter equations. The point corresponding to these values of $\mu_{i}/\mu_{1}$
is an $n$-fold multi-critical point in the phase space. The neighbourhood
of this multi-critical point contains $m$-fold multi-critical points
for any integer $0<m<n$.

These statements can be proved using well known properties of analytic
functions and the fact that $V$ is a trigonometric polynomial. We
omit the details of the proof.

\section{\label{sec: ot}Signatures of 1st and 2nd order phase transitions
in finite volume}
\markright{\thesection.\ \ SIGNATURES OF PHASE TRANSITIONS}

The considerations in this section apply to the quantum case.

The behaviour of the spectrum is governed by the $l\rightarrow0$
limiting conformal field theory for small values of $l$, so $e_{n}(l)-e_{0}(l)\sim1/l$.
Massive phases in infinite volume are characterized by the existence
of a massgap and the behaviour $\lim_{l\rightarrow\infty}(e_{n}(l)-e_{0}(l))=C_{n}$,
where $C_{n}\geq0$ are constants. $C_{n}=0$ if $0\leq n\leq d$
and $C_{n}>0$ if $n>d$, if the ground state has $d$-fold degeneracy
in infinite volume. In a phase with spontaneously broken symmetry
the spectrum is degenerate in the $l\rightarrow\infty$ limit, in
finite volume the degeneracy is lifted (at least partially) due to
tunneling effects. The resulting energy split between the degenerate
vacua vanishes exponentially as $l\rightarrow\infty$.

In the critical points (in infinite volume) the massgap vanishes and
the Hilbert space contains a sector that corresponds to the conformal
field theory specifying the universality class of the critical point.
We consider this sector in the following discussion. In finite but
large volume near the critical point this sector of the theory can
be regarded as the $l\rightarrow\infty$ limiting conformal theory
perturbed by some irrelevant and relevant operators. The corresponding
TCSA Hamiltonian operator takes the (generic) form 
\begin{equation}
H=\frac{2\pi}{L}\left((L_{0})_{IR}+(\bar{L}_{0})_{IR}-\frac{c_{IR}}{12}+\sum_{\psi}\frac{g_{\psi}L^{2-2\Delta_{\psi}}}{(2\pi)^{1-2\Delta_{\psi}}}\psi(1,1)\right)\
 ,\label{IR Hamilton}
\end{equation}
 where the $\psi$ are the perturbing fields of weight
 $(\Delta_\psi,\Delta_\psi)$, $g_\psi$ are constants.  
 This picture (proposed in \cite{BPTW})  gives the
following volume dependence of energy levels (in the first order of
conformal perturbation theory):
\begin{equation}
e_{\Psi}(l)-e_{0}(l)=\frac{2\pi}{l}(\Delta_{IR,\Psi}^{+}+\Delta_{IR,\Psi}^{-})+\sum_{\psi}A_{\Psi}^{\psi}l^{1-2\Delta_{\psi}}\
 ,\label{l fugges}
\end{equation}
 where $\Delta_{IR,\Psi}^{+}$ and $\Delta_{IR,\Psi}^{-}$ are the
conformal weights of the state $\Psi$ in the $l\rightarrow\infty$
limiting CFT, $A_{\Psi}^{\psi}$ are constants that also depend on
the particular energy eigenstate $\Psi$. The presence of irrelevant
perturbations ($1-2\Delta_{\psi}<-1$) is due to the finiteness of
the volume, whereas the presence of the relevant perturbations ($1-2\Delta_{\psi}>-1$)
is caused by the deviation of the parameters from the critical
value.
The effect of the truncation is not taken into consideration in  (\ref{IR Hamilton}).

The location of the critical points can be determined using the criterion
of vanishing massgap. A more precise method that also allows the determination
of the universality class of the critical point (i.e.\ the $l\rightarrow\infty$
limiting CFT) is the following: we make an assumption that the critical
point is in a certain universality class. This assumption predicts
the set of $\psi$-s, the values of $\Delta_{IR,\Psi}^{+}$ and $\Delta_{IR,\Psi}^{-}$,
and the values of the $\Delta_{\psi}$-s in (\ref{l fugges}). We
take leading terms of the series on the r.h.s.\ of (\ref{l fugges})
and determine the value of the $(\Delta_{IR,\Psi}^{+}+\Delta_{IR,\Psi}^{-})$-s
and of the $A_{\Psi}^{\psi}$-s by fitting to the TCSA energy data obtained at
several values of $l$.
The magnitude of the $A_{\Psi}^{\psi}$-s corresponding to the relevant
perturbations measures the deviation from the critical point, so if
the assumption on the universality class is right, then by tuning
the coupling constants one should be able to find a (critical) value
at which these $A_{\Psi}^{\psi}$-s are small,
the TCSA data are described well by (\ref{l fugges}) (terms from
higher orders of perturbation theory can also be included if necessary)
in a reasonably large interval of the values of $l$, and the values
of the $(\Delta_{IR,\Psi}^{+}+\Delta_{IR,\Psi}^{-})$-s obtained from
the TCSA data agree with the assumption with good precision. The interval
where (\ref{l fugges}) describes the TCSA data well is called the
scaling region. This region may be (and in fact is) different for
different energy levels. We remark that it is also possible to make
a theoretical prediction for $e_{0}(l)$, which allows to extract
$c_{IR}$ from the TCSA data for $e_{0}(l)$ in principle \cite{BPTW}. However,
experience (\cite{BPTW}) shows that the accuracy of the TCSA data is not sufficient to determine 
$c_{IR}$ precisely in
this way, so we did not attempt to extract $c_{IR}$ directly from the TCSA data.

In the classical case a first order phase transition occurs when the
absolute minimum of the potential becomes a relative minimum and a
previously relative minimum becomes absolute. In the quantum case
this phase transition is characterized by the presence of `runaway
energy levels' with asymptotic behaviour $e(l)\sim cl$ for large $l$
in the neighbourhood of the transition point, where $c$ is a constant
that tends to zero as the transition point is approached. The multiplicity
of the ground state also changes as the transition point is passed
if the two phases have different symmetry properties. We remark that
runaway energy levels are present in general whenever a model
has unstable vacua.

\section{\label{sec: hat}The phase diagram of the two-frequency model in
the case $\frac{\alpha}{\beta}=\frac{1}{3}$, $\delta=\frac{\pi}{3}$}
\markright{\thesection.\ \ PHASE DIAGRAM OF THE TWO-FREQUENCY MODEL}

We assume that $\lambda,\mu>0$, and use the parameter \[
\tilde{\eta}=\frac{\lambda^{x_{\beta}}}{\mu^{x_{\alpha}}+\lambda^{x_{\beta}}}\]
 instead of $\eta$ in this case to conform with \cite{BPTW}.

The classical model with $\alpha/\beta=\frac{1}{3}$, $\delta=\frac{\pi}{3}$
exhibits an Ising type phase transition at $\tilde{\eta}=3^{4}/(1+3^{4})$.
\begin{figure}
\begin{center}\includegraphics[%
  width=0.30\linewidth,
  angle=270]{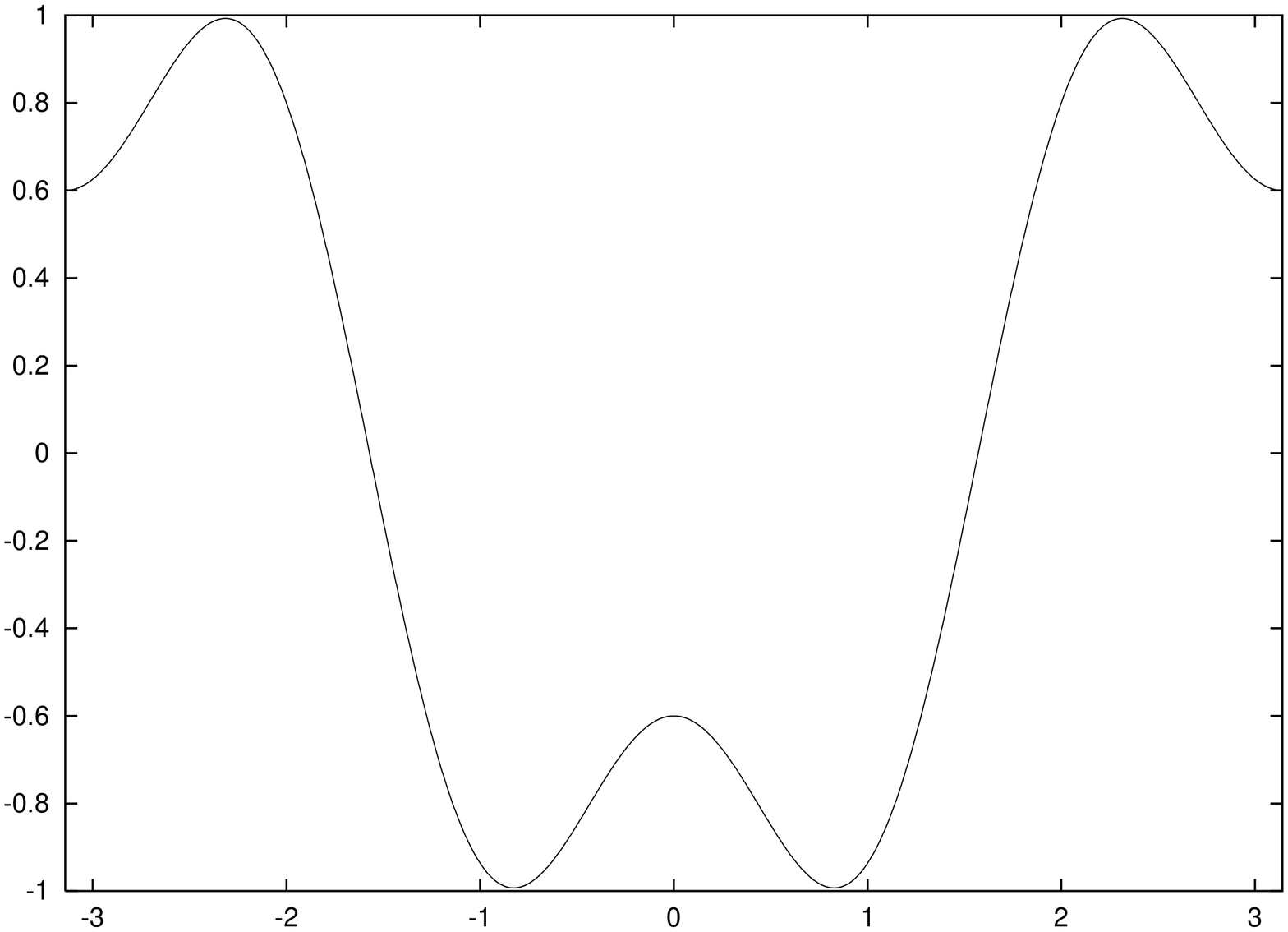}\includegraphics[%
  width=0.30\linewidth,
  angle=270]{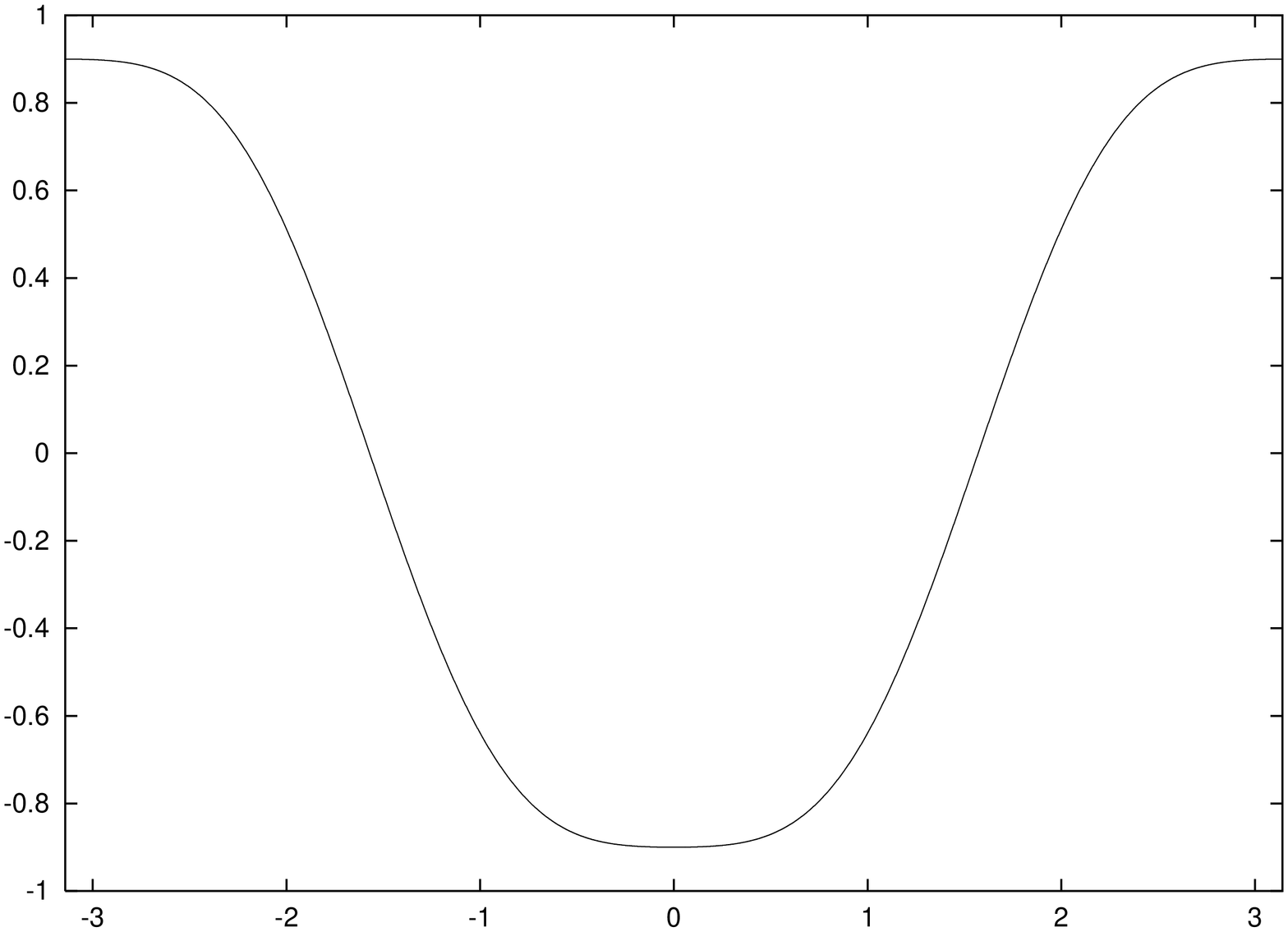}\end{center}

\hspace{4.5cm}a\hspace{7cm}b

\caption{\label{pot12}Typical shape of $V(\Phi)$ in the broken and in the
unbroken symmetry phase}
\end{figure}

Considering the quantum case we proceed along the lines of \cite{BPTW}
in this section. The numerical nature of the TCSA makes it necessary
to choose a finite number of values for $\beta$ and $l$ at which
calculations are done. One should choose as large values for $l$
as possible, the $l\rightarrow\infty$ limit being of interest. However,
the accuracy of TCSA decreases as $l$ grows. The accuracy can be
improved by taking higher $e_{cut}$, but this increases the size
of the TCSA Hamiltonian matrix and the time needed for diagonalization. Experience
shows that accuracy decreases for values of $\beta$ near to $\sqrt{4\pi}$
(this is the value where UV divergences appear in conformal perturbation
theory) although the speed of convergence of the spectrum to the $l\rightarrow\infty$
asymptotic values increases, and the speed of convergence becomes
very low for values of $\beta$ near to $0$. Taking these properties
of the TCSA into consideration and following \cite{BPTW} we performed
calculations at the values $\beta=8\sqrt{\pi}/7$, $4\sqrt{\pi}/3$,
$8\sqrt{\pi}/5$. We also note that the accuracy of the TCSA spectra
is severely decreased if $V$ has several (local) minima.

\begin{figure}
\begin{tabular}{cc}\includegraphics[
  width=0.17\paperwidth,
  height=0.37\paperwidth,
  angle=270]{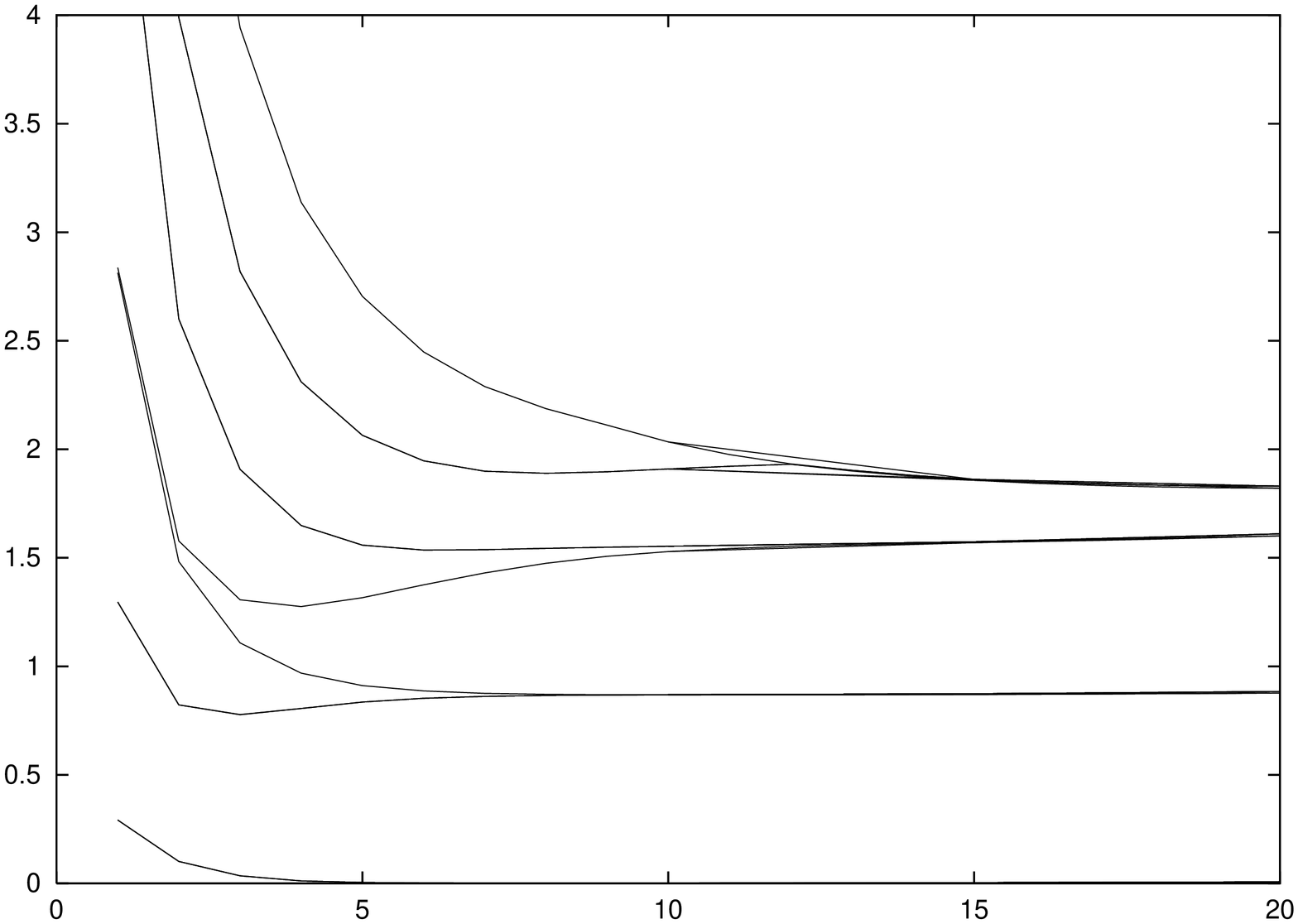}&
\includegraphics[
  width=0.17\paperwidth,
  height=0.37\paperwidth,
  angle=270]{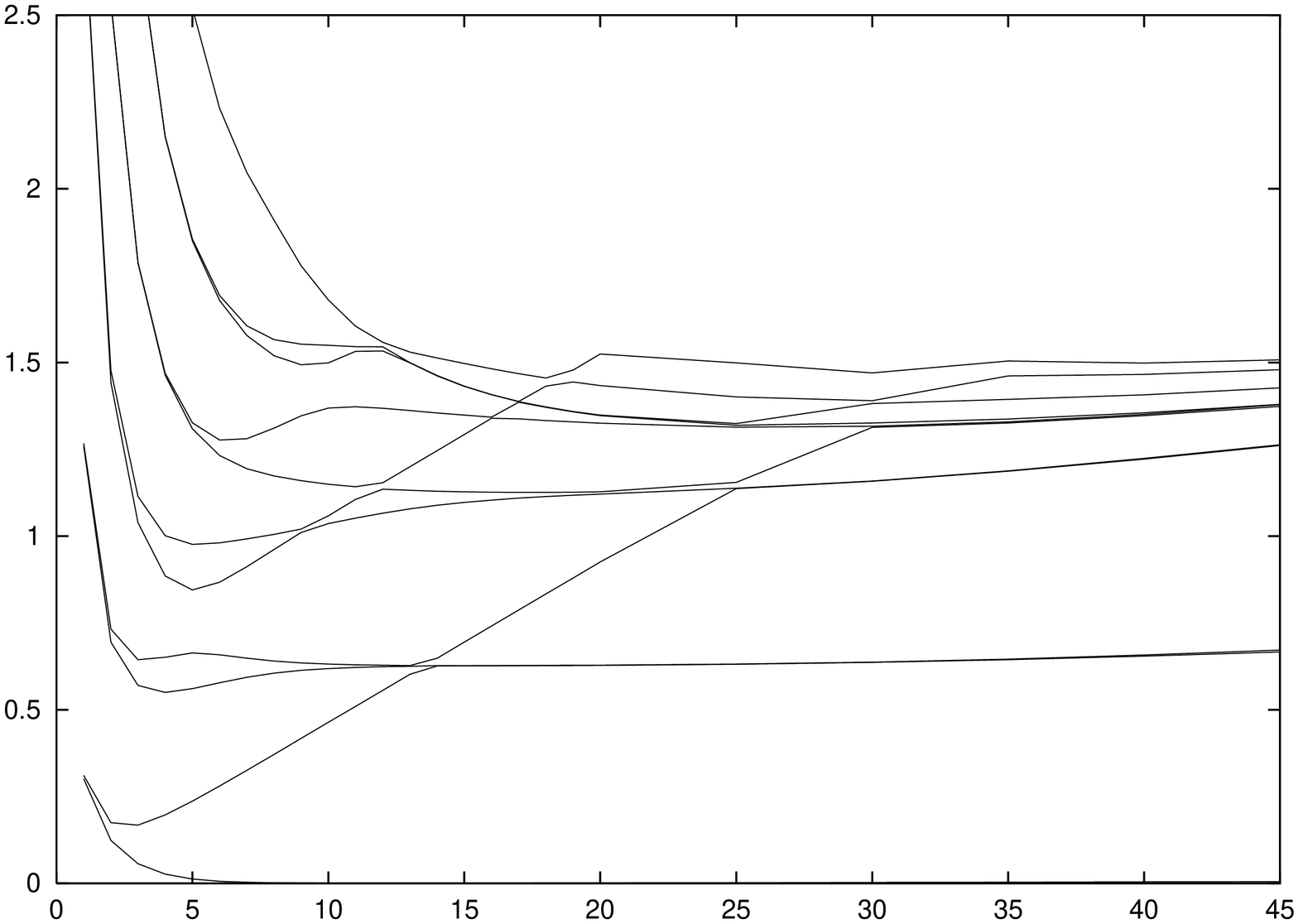}\\
a: $\tilde{\eta}=0$ & b:
$\tilde{\eta}=0.3$ \\
\includegraphics[
  width=0.17\paperwidth,
  height=0.37\paperwidth,
  angle=270]{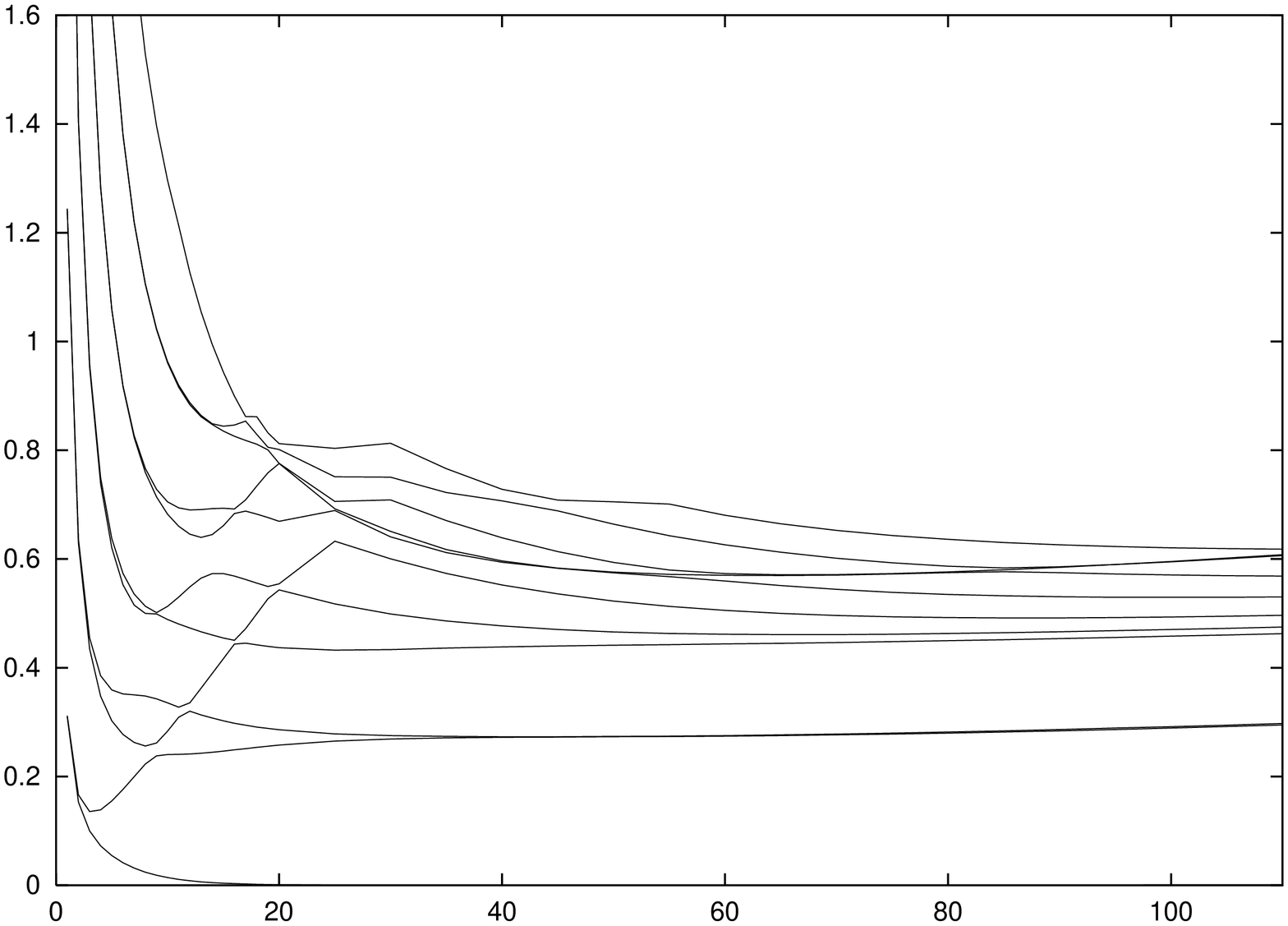}&
\includegraphics[
  width=0.17\paperwidth,
  height=0.37\paperwidth,
  angle=270]{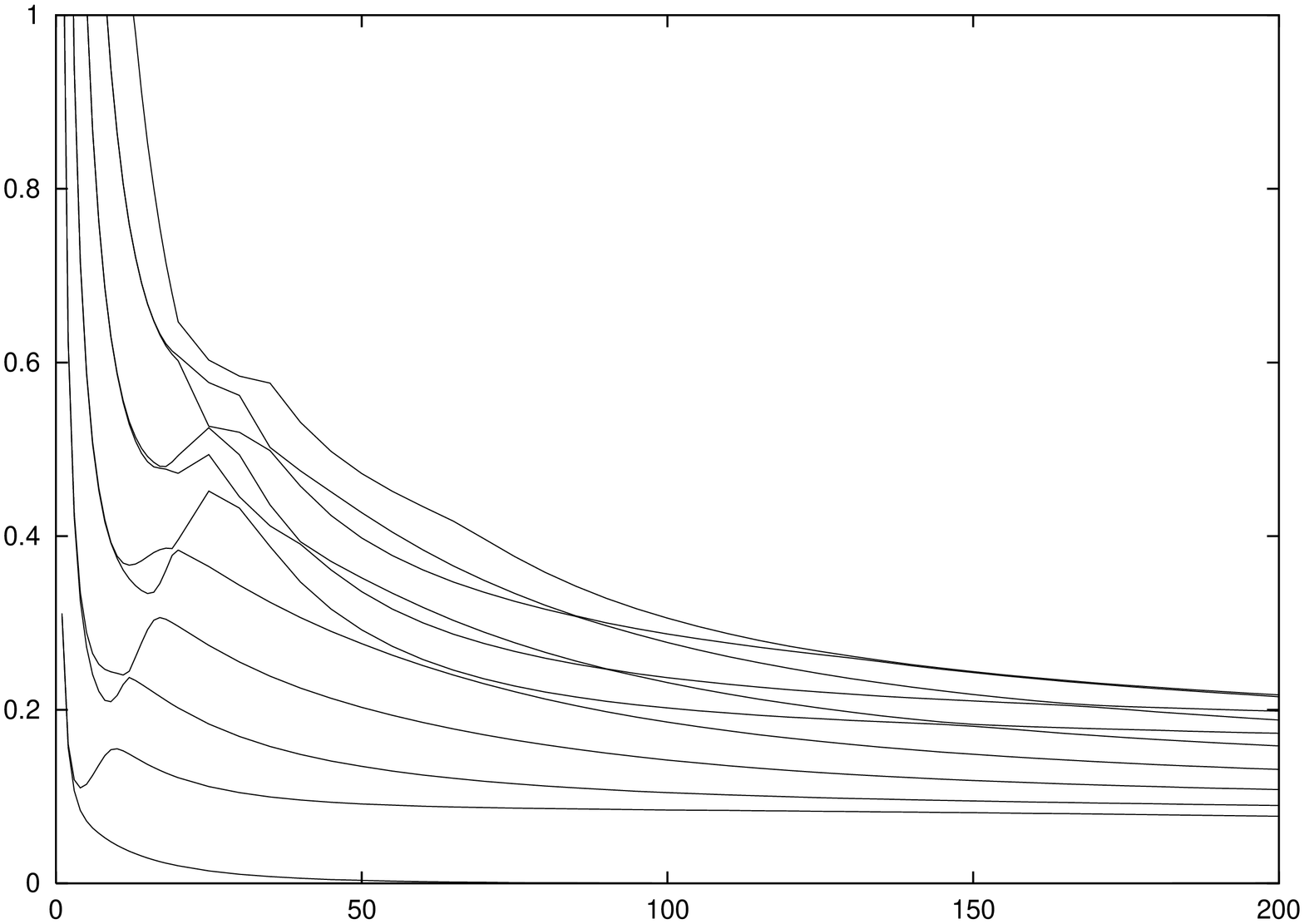}\\
c: $\tilde{\eta}=0.7$ & d:
$\tilde{\eta}=0.92$\\
\includegraphics[
  width=0.17\paperwidth,
  height=0.37\paperwidth,
  angle=270]{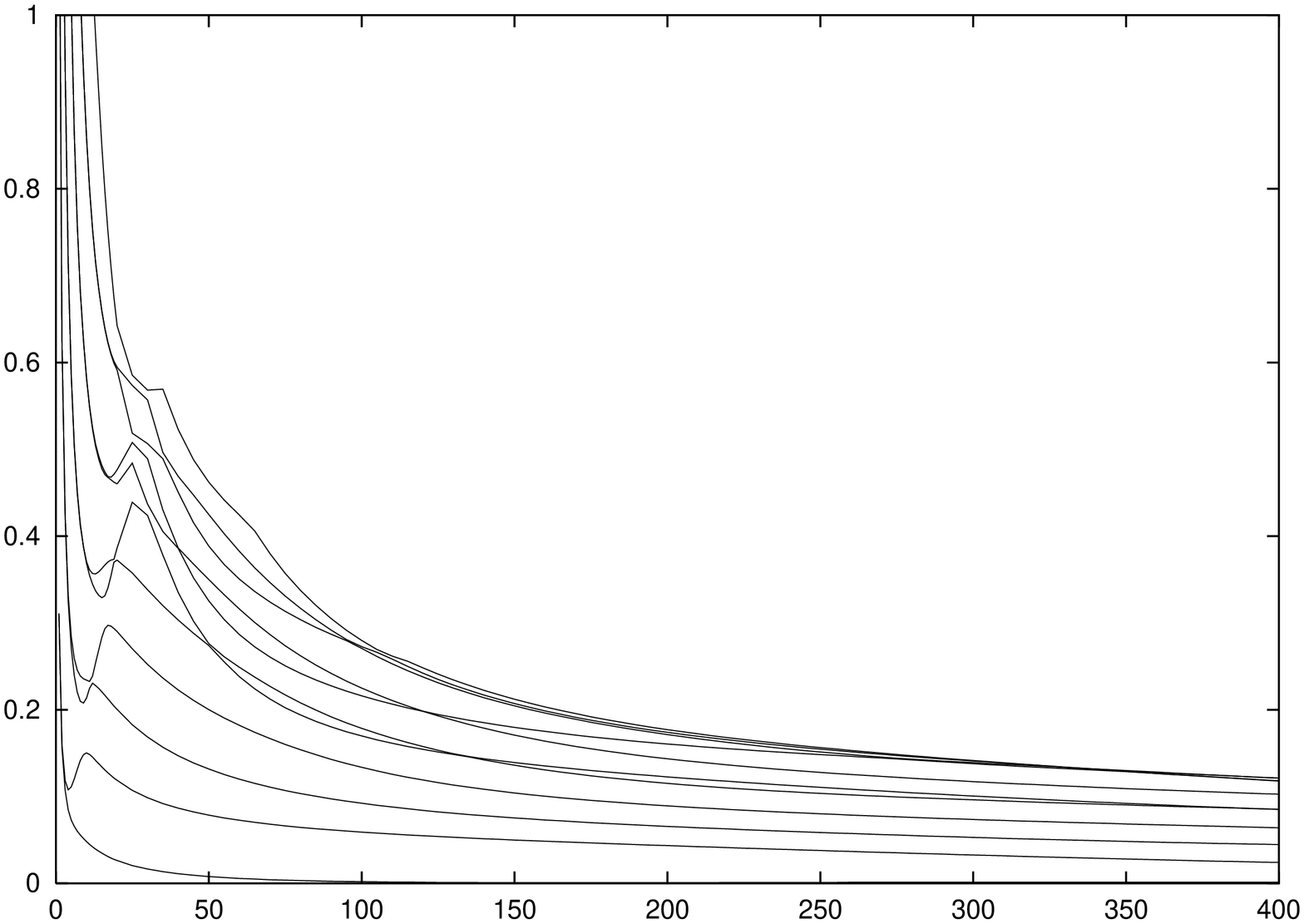}&
\includegraphics[
  width=0.17\paperwidth,
  height=0.37\paperwidth,
  angle=270]{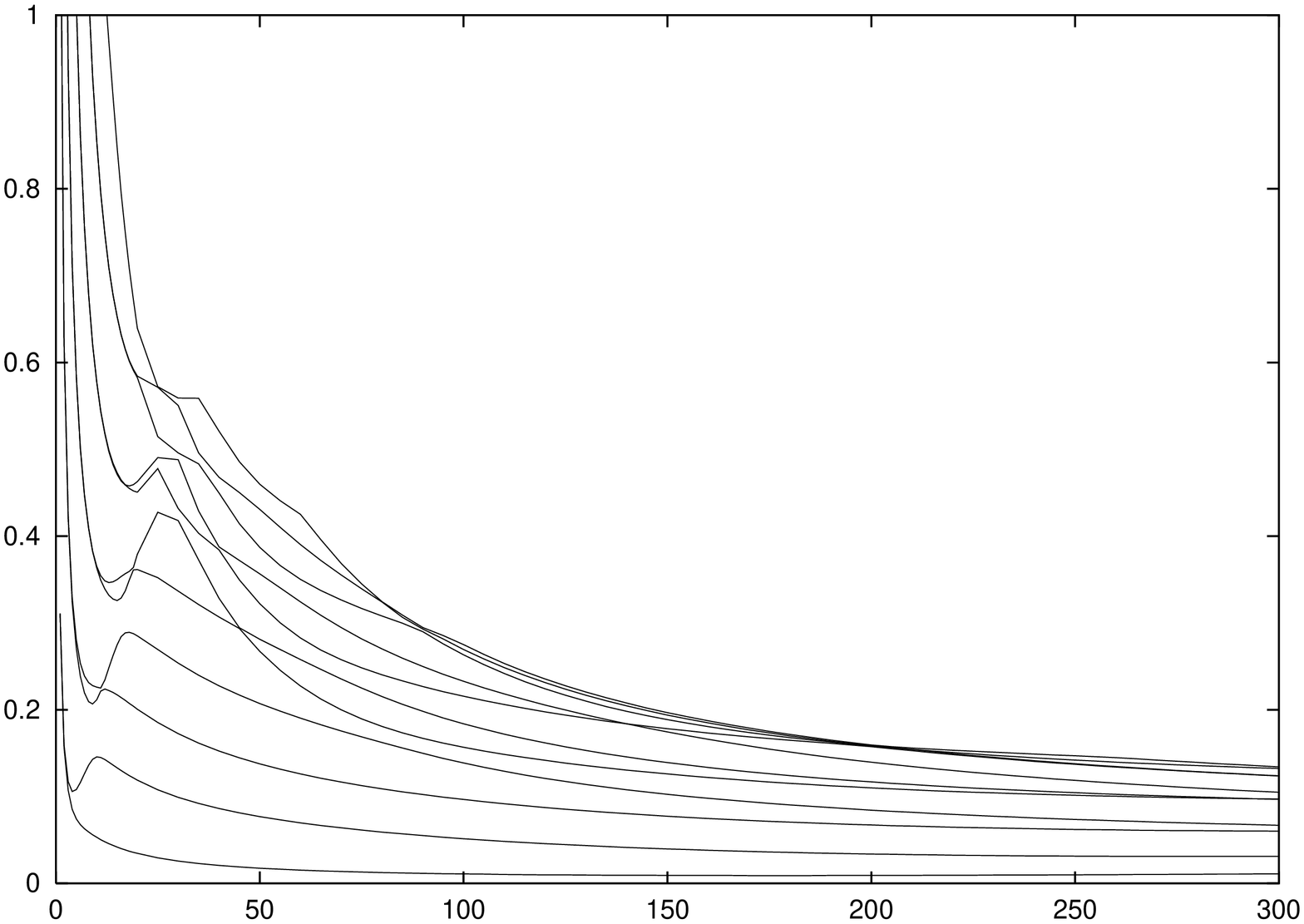}\\
e: $\tilde{\eta}=0.94$ & f:
$\tilde{\eta}=0.96$ 
\end{tabular}
\begin{center}\includegraphics[%
  width=0.17\paperwidth,
  height=0.37\paperwidth,
  angle=270]{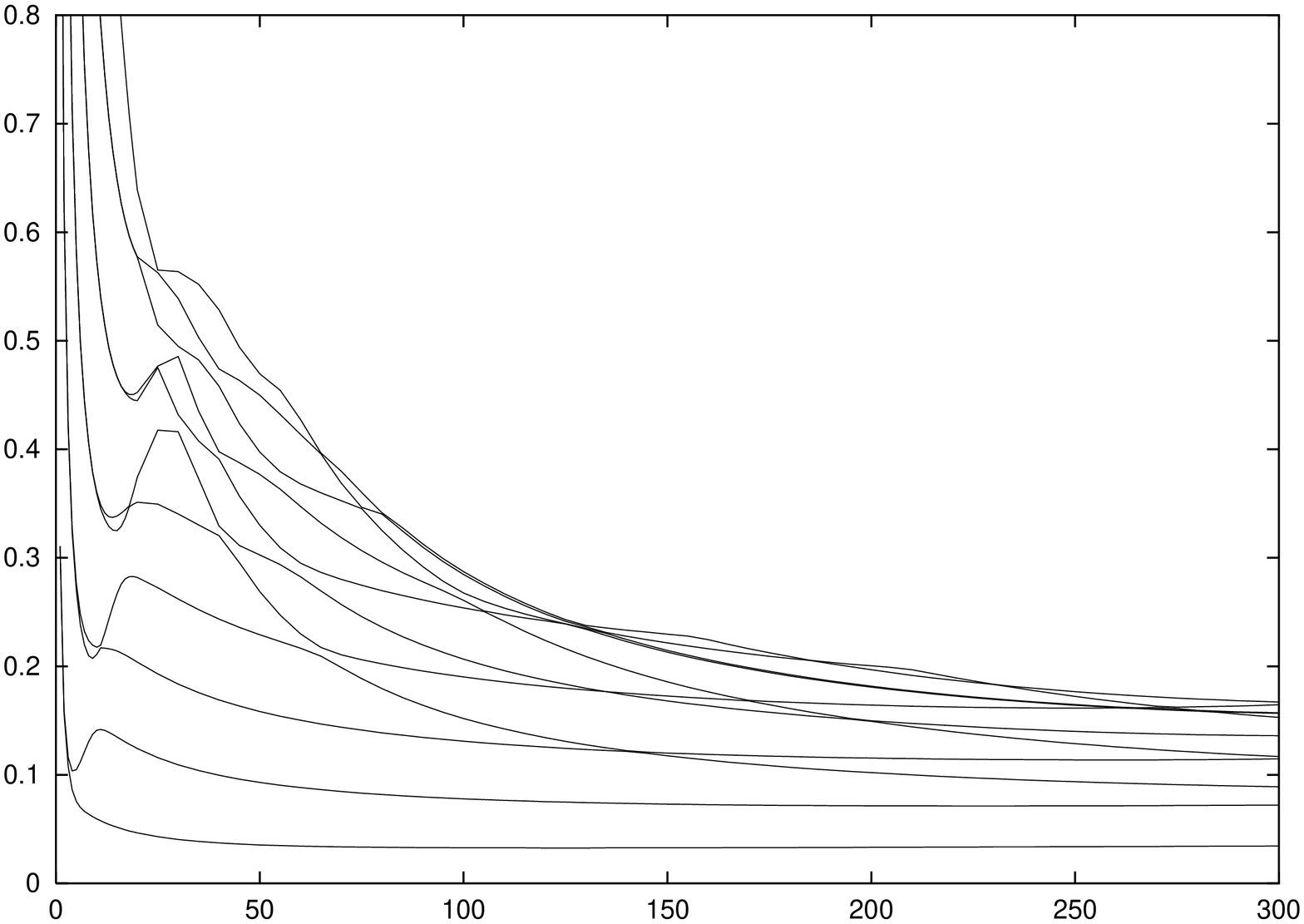}\end{center}

\begin{center}g: $\tilde{\eta}=0.98$\end{center}

\caption{\label{spectra1}Change of the spectrum as $\tilde{\eta}$ varies
from 0 to 1 at $\beta=4\sqrt{\pi}/3$. The first 12 energy levels
(including the ground level) relative to the ground level are shown
as functions of $l$. }
\end{figure}

Figure \ref{pot12}.a and \ref{pot12}.b show the shape of the classical
potential in the phases with broken and unbroken $\ZZ_{2}$-symmetry,
respectively. Figures \ref{spectra1}.a-\ref{spectra1}.g show TCSA
spectra obtained at $\beta=4\sqrt{\pi}/3$ at various values of $\tilde{\eta}$.
The TCSA Hilbert space had dimension $3700$, the first $12$ energy
levels are shown in the figures. The highest values of $l$ are chosen
so that the truncation error still be small (the massgap remain constant).
However, the effect of truncation is perceptible in Figure \ref{spectra1}.b
for instance. It can be seen that in the domain $\tilde{\eta}<0.92$
the ground states and the first massive states are doubly degenerate.
(They are triply degenerate at $\tilde{\eta}=0$.) `Runaway'
energy levels (of constant slope) corresponding to the single local
minimum of the potential can also be seen (especially clearly in Figure
\ref{spectra1}.b). In the domain $\tilde{\eta}>0.98$ the spectra
are massive, but the ground state and the first massive state
are nondegenerate. In the intermediate domain (especially for $\tilde{\eta}\sim0.95$)
the structure of the spectrum changes, no massgap and degeneracy can
clearly be seen. We obtained similar spectra at $\beta=8\sqrt{\pi}/5$
and $\beta=8\sqrt{\pi}/7$ as well. As we did not see `runaway'
energy levels that would have signaled first order phase transition
in the transitional domain of $\tilde{\eta}$ we analyzed the data
by looking for a second order Ising type phase transition at some
critical values $\tilde{\eta}_{c}(\beta)$.

The Ising model contains three primary fields: the identity with weights
$(0,0)$, the $\epsilon$ with weights $(1/2,1/2)$, and $\sigma$
with weights $(1/16,1/16)$. Since the DSG model exhibits the $\ZZ_{2}$-symmetry
for all values of $\tilde{\eta}$, the $\ZZ_{2}$-odd $\sigma$ and
its descendants cannot appear as perturbations in the Hamiltonian operator
(\ref{IR Hamilton}). The only relevant field compatible with the
$\ZZ_{2}$-symmetry is $\epsilon$ (the contribution of the identity
cancels in the relative energy levels). The presence of a relevant
perturbation $\epsilon$ in the Hamiltonian operator leads to a correction
$B_{\Psi}$ or $B_{\Psi}+C_{\Psi}l$ to $e_{\Psi}(l)-e_{0}(l)$. The
term $C_{\Psi}l$ is of second order in conformal perturbation theory. The leading
irrelevant perturbation (compatible with the $\ZZ_{2}$-symmetry)
is the first descendant of $\epsilon$, this gives a correction $A_{\Psi}l^{-2}$
to $e_{\Psi}(l)-e_{0}(l)$ (in first order). Thus we expect that in
a large but finite volume range, near $\tilde{\eta}_{c}(\beta)$,
the volume dependence of the energy levels is described well by the
formula
\begin{equation}
e_{i}(l)-e_{0}(l)=\frac{2\pi}{l}D_{i}+A_{i}l^{-2}+B_{i}+C_{i}l\ .\label{fit1}
\end{equation}
 We fitted this function to the lowest energy levels obtained by TCSA
and determined the `best' $\tilde{\eta}_{c}(\beta)$ value by
tuning $\tilde{\eta}$ in the transition region and looking for whether
$e_{2}(l)-e_{0}(l)$ continues to decrease along the complete $l$
range ($\lim_{l\rightarrow\infty}(e_{2}(l)-e_{0}(l))=0$ only at $\tilde{\eta}_{c}(\beta)$),
and $B_{i}$ and $C_{i}$ are as small as possible. The result is
shown in Table \ref{tabla1}. The fitting was done in the volume ranges
$l=10-105$, $l=55-105$; $l=10-140$, $l=100-190$; $l=20-200$,
$l=150-390$. The errors presented come from the fitting process and
do not contain the truncation errors which are generally larger.

The first two energy levels above the ground state correspond
to the operators $\sigma$ and $\epsilon$ in the Ising model. These
operators have conformal weights $\Delta^{\pm}=1/16$ and $\Delta^{\pm}=1/2$,
so the exact values for $D_{1}$ and $D_{2}$ are \[
D_{1}=0.125\ ,\qquad D_{2}=1\ .\]
 The results of the fits agree quite well with this prediction.

\begin{figure}
\begin{center}\includegraphics[%
  width=0.90\columnwidth,
  height=0.27\textwidth]{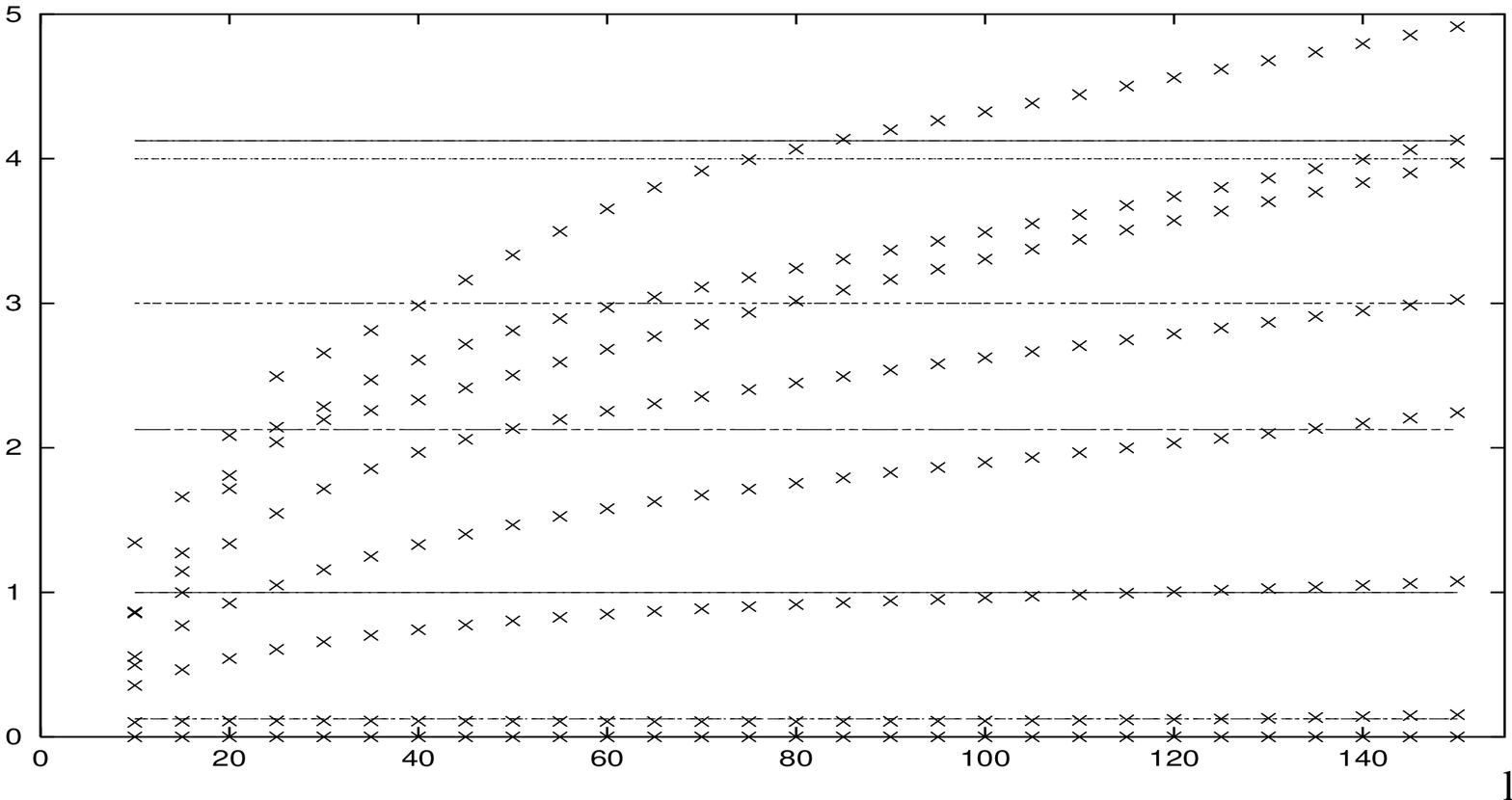}\end{center}

\begin{center}$[e_{i}(l)-e_{0}(l)]\cdot l/2\pi$ at $\beta=8\sqrt{\pi}/5$
and $\tilde{\eta}=0.944$\end{center}

\begin{center}\includegraphics[%
  width=0.90\columnwidth,
  height=0.27\textwidth]{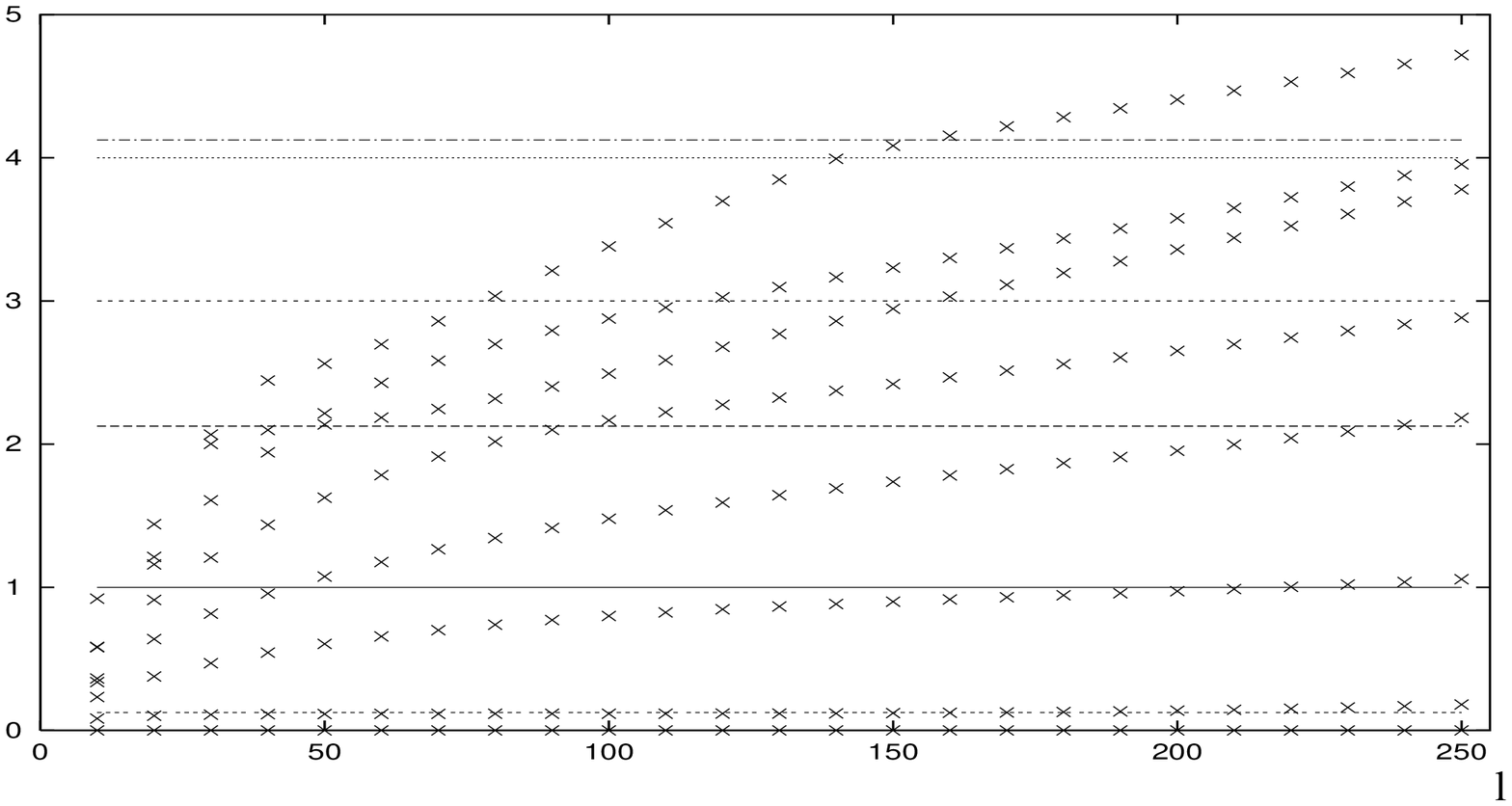}\end{center}

\begin{center}$[e_{i}(l)-e_{0}(l)]\cdot l/2\pi$ at $\beta=4\sqrt{\pi}/3$
and $\tilde{\eta}=0.955$\end{center}

\begin{center}\includegraphics[%
  width=0.90\columnwidth,
  height=0.27\textwidth]{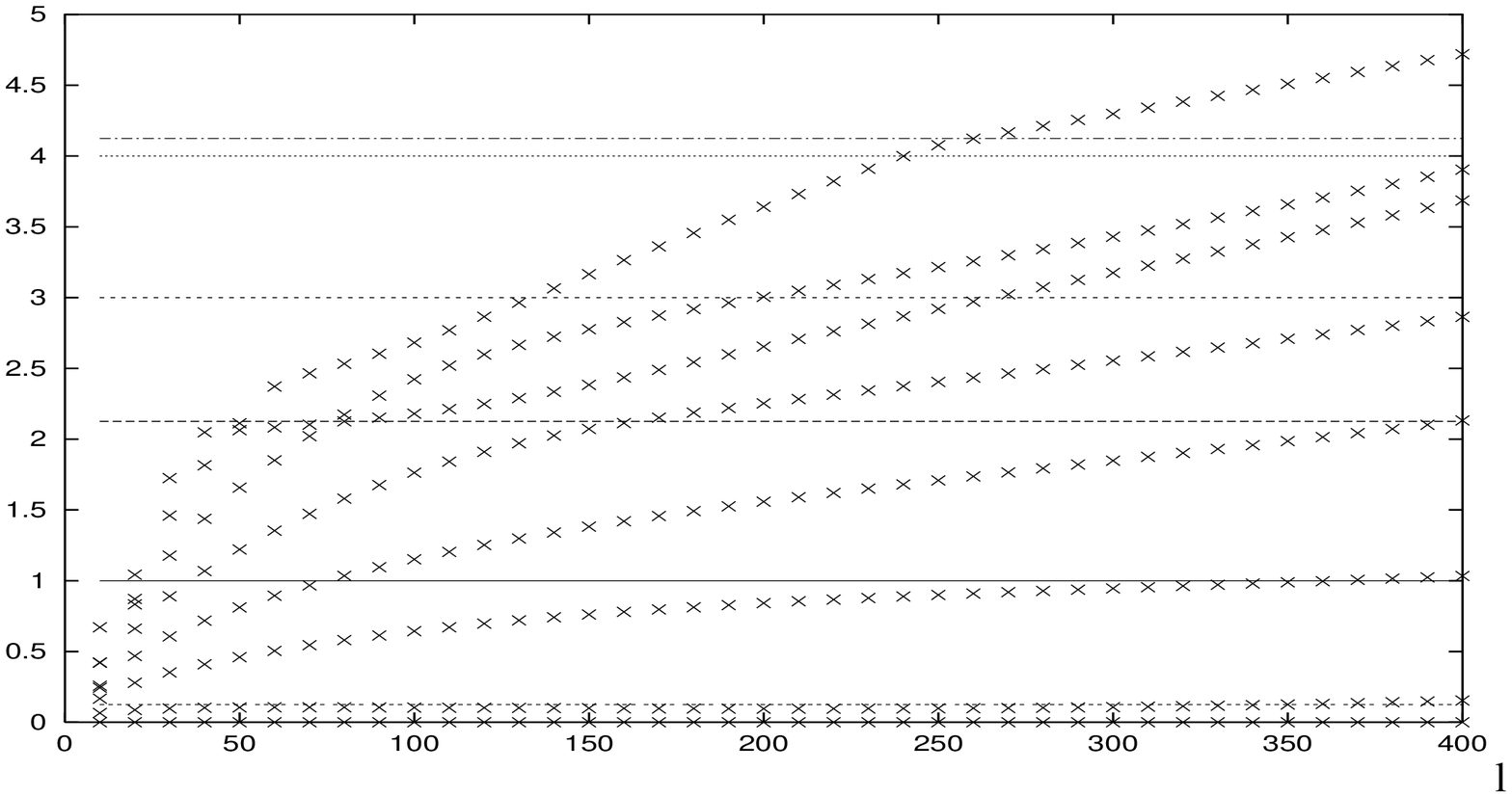}\end{center}

\begin{center}$[e_{i}(l)-e_{0}(l)]\cdot l/2\pi$ at $\beta=8\sqrt{\pi}/7$
and $\tilde{\eta}=0.961$\end{center}

\caption{\label{critspectra}TCSA spectra as functions of $l$ at the estimated
critical values of $\tilde{\eta}$ }
\end{figure}

The TCSA data obtained at the estimated values of $\tilde{\eta}_{c}$
using a truncated space with dimension 4800, 5300 and 5100 are shown
in Figure \ref{critspectra}. These figures show energy levels multiplied
by $l/(2\pi)$ as functions of $l$. The constant lines corresponding
to the Ising model values of $D_{i}$ are also shown in these figures.
\begin{table}
\caption{\label{tabla1}The results of fitting (\ref{fit1}) to the first
two excited levels for various values of $\beta$ at the estimated
critical value of $\tilde{\eta}$ }
\begin{center}$\beta=8\sqrt{\pi}/5$, $\tilde{\eta}=0.944$\end{center}
\begin{center}
\begin{tabular*}{16cm}{l@{\extracolsep{\fill}}cccc}
\hline 
State&
 $D_{i}$&
 $A_{i}$&
 $B_{i}$&
 $C_{i}$\tabularnewline
\hline
$i=1$&
 $0.138\pm0.0005$&
 $-2.0\pm0.02$&
 $-0.0046\pm0.0001$&
 $2.98\cdot10^{-5}\pm9\cdot10^{-7}$\tabularnewline
% \hline
$i=2$&
 $1.00\pm0.01$&
 $-74\pm2$&
 $0.004\pm0.001$&
 $9\cdot10^{-6}\pm4\cdot10^{-6}$ \tabularnewline
\hline
\end{tabular*}
\end{center}

\vspace{0.2cm}
\begin{center}$\beta=4\sqrt{\pi}/3$, $\tilde{\eta}=0.955$\end{center}
\begin{center}
\begin{tabular*}{16cm}{l@{\extracolsep{\fill}}cccc}
\hline 
State&
 $D_{i}$&
 $A_{i}$&
 $B_{i}$&
 $C_{i}$\tabularnewline
\hline
$i=1$&
 $0.125\pm0.001$&
 $-2.63\pm0.025$&
 $-0.0002\pm0.0001$&
 $3\cdot10^{-7}\pm1\cdot10^{-6}$\tabularnewline
% \hline
$i=2$&
 $1.04\pm0.02$&
 $-152\pm6$&
 $0.0014\pm0.0009$&
 $-1.5\cdot10^{-5}\pm2\cdot10^{-6}$ \tabularnewline
\hline
\end{tabular*}
\end{center}

\vspace{0.2cm}
\begin{center}$\beta=8\sqrt{\pi}/7$, $\tilde{\eta}=0.961$\end{center}
\begin{center}
\begin{tabular*}{16cm}{l@{\extracolsep{\fill}}cccc}
\hline 
State&
 $D_{i}$&
 $A_{i}$&
 $B_{i}$&
 $C_{i}$\tabularnewline
\hline
$i=1$&
 $0.125\pm0.001$&
 $-4.4\pm0.1$&
 $-0.0009\pm0.0001$&
 $4\cdot10^{-7}\pm5\cdot10^{-7}$\tabularnewline
%\hline
$i=2$&
 $1.00\pm0.02$&
 $-252\pm9$&
 $0.0002\pm0.0004$&
 $4.3\cdot10^{-6}\pm6\cdot10^{-7}$ \tabularnewline
\hline
\end{tabular*}
\end{center}

% \vspace*{\medskipamount}
\end{table}

To summarize, evaluating the TCSA data we found that the phase transition
is second order and Ising type at the values of $\beta$ chosen. 
We remark that
it is not possible to distinguish a second order transition from a
(very) weakly first order transition by TCSA. 

For values of $\beta$ near $0$ second order phase transition can be expected,
because the model is semi-classical in this region and so the correction
to the classical potential in the effective potential is expected
to be small.

A correction to the classical potential in the effective potential
of frequency $2\beta/3$ is possible in principle. A correction with
this frequency is ---unlike in the case of $\alpha/\beta=1/2$---
relevant for any values of $\beta$, and it can be verified by elementary
calculation that it may change the order of the transition outside
the semiclassical region if its coefficient is sufficiently large
(see also \cite{FGN,BPTW}). However, the accuracy of TCSA did not
allow us to perform calculations at values about $\beta^{2}>8\pi/3$
and to check the nature of phase transition in this domain.

\begin{figure}
\begin{center}\includegraphics[%
  width=0.85\textwidth]{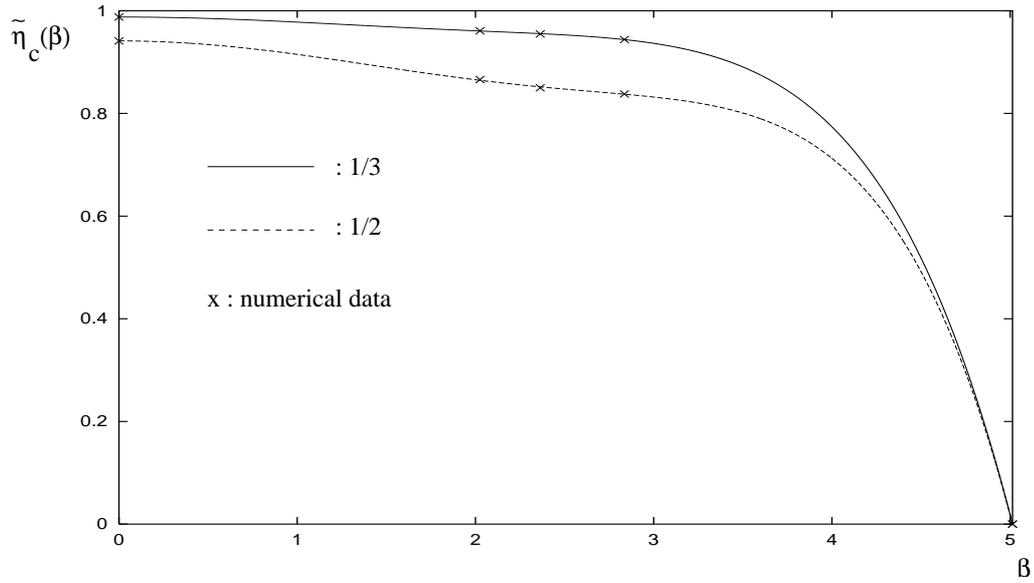}\end{center}

\caption{\label{phdiag1}The phase diagram of the two-frequency model at $\alpha/\beta=1/2$ and $\alpha/\beta=1/3$}
\end{figure}

The phase diagram in the $(\beta,\tilde{\eta})$ plane based on the
data obtained by TCSA can be seen in Figure \ref{phdiag1}. We took
into consideration that $\tilde{\eta}_{c}(0)=3^{4}/(1+3^{4})\approx0.988$
is exactly known $\beta=0$ being the classical limit, and at $\beta=\sqrt{8\pi}$
the term with frequency $\beta$ in the potential becomes irrelevant
and thus for $\beta\rightarrow\sqrt{8\pi}$ the other term and so
the symmetric phase is expected to dominate, so $\lim_{\beta\rightarrow\sqrt{8\pi}}\tilde{\eta}_{c}(\beta)=0$.
The figure shows the three values of $\tilde{\eta}_{c}$ obtained
from the TCSA data and the two values at $\beta=0$ and $\beta=\sqrt{8\pi}$.
The continuous line is obtained by fitting an even polynomial (note
that $\tilde{\eta}_{c}(\beta)=\tilde{\eta}_{c}(-\beta)$) to these
values and is shown in order to get an idea of the phase transition
line. The model is in the symmetric phase above the line and in the
phase with broken symmetry below the line. The data obtained by \cite{BPTW}
for the $\alpha/\beta=1/2$ case are also shown, the dashed line is
fitted to these data.

\section{\label{sec: het}Phase diagram of the three-frequency model}
\markright{\thesection.\ \ PHASE DIAGRAM OF THE THREE-FREQUENCY MODEL}

We take the same values of the parameters of (\ref{eq: 3freq pot})
as in section \ref{sec 4.2}, namely $\beta_{1}=\beta,\beta_{2}=\frac{2}{3}\beta,\beta_{3}=\frac{1}{3}\beta,$
$\delta_{2}=\delta_{3}=0$, $\mu_{1},\mu_{3}<0$, $\mu_{2}>0$, and
investigate the quantum model in this case.

\subsection{The tricritical point}

\begin{figure}
\begin{center}\includegraphics[%
  height=0.70\paperwidth,
  angle=270]{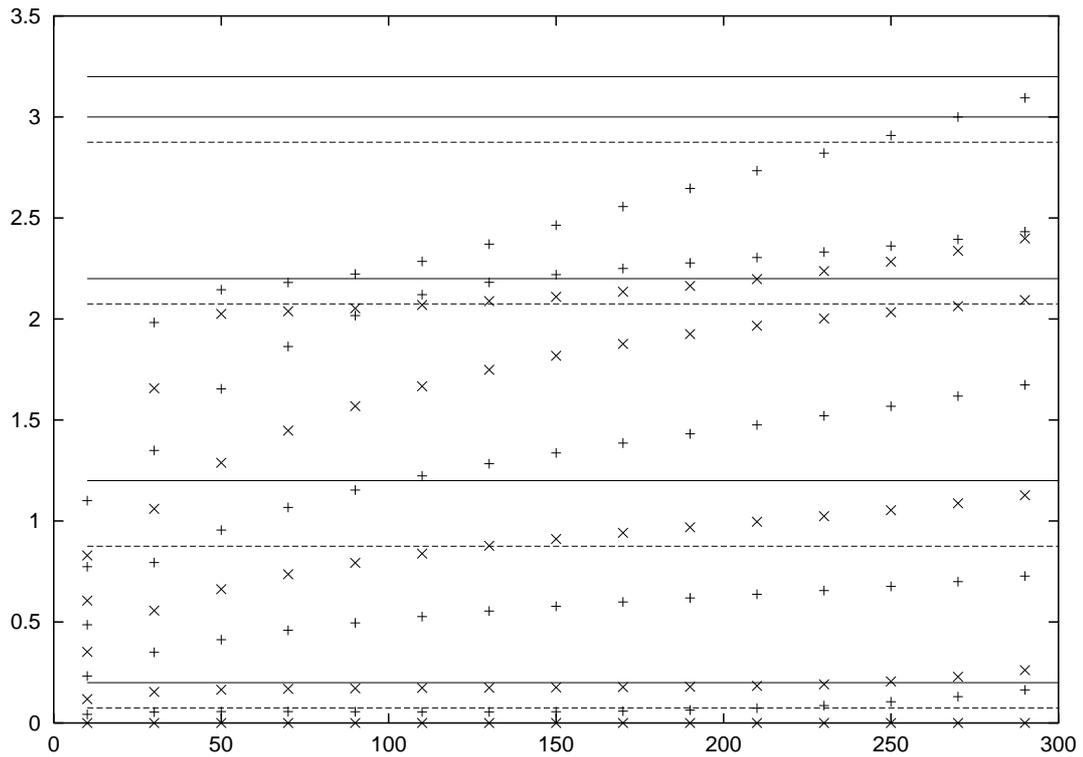}\end{center}

\caption{\label{spectrum3}$[e_{i}(l)-e_{0}(l)]\cdot l/2\pi$ as functions
of $l$ obtained by TCSA at $\beta=8\sqrt{\pi}/7$, $\eta_{1}=0.163$,
$\eta_{2}=0.3518$ }
\end{figure}

The tricritical Ising model contains 6 primary fields with the following
conformal weights:
\begin{equation}
(0,0),\ \left(\frac{1}{10},\frac{1}{10}\right),\
  \left(\frac{3}{5},\frac{3}{5}\right),\
  \left(\frac{3}{2},\frac{3}{2}\right),
\label{eq:tri1}
\end{equation}
 and
\begin{equation}
\left(\frac{3}{80},\frac{3}{80}\right),\
\left(\frac{7}{16},\frac{7}{16}\right)\ .
\label{eq:tri2}
\end{equation}
 The fields corresponding to (\ref{eq:tri1}) are even and those corresponding
to (\ref{eq:tri2}) are odd with respect to parity, so only 
the fields corresponding to (\ref{eq:tri1}) and their descendants
can contribute to the Hamiltonian operator (\ref{IR Hamilton}). Thus the volume
dependence of the energy levels near the tricritical point should
be described well for large $l$ by \[
e_{\Psi}(l)-e_{0}(l)=\frac{2\pi}{l}(\Delta_{IR,\Psi}^{+}+\Delta_{IR,\Psi})+A_{\Psi}l^{-0.2}+B_{\Psi}l^{0.8}+\dots\
,\]
 where only the leading terms are kept. Searching for the tricritical
point we fitted the function 
\begin{equation}
\frac{2\pi}{l}D_{i}+A_{i}l^{-0.2}+B_{i}l^{0.8}
\label{eq:fit2}
\end{equation}
of $l$ to the data obtained by TCSA for $e_{i}(l)-e_{0}(l)$ near the estimated
location of the tricritical point. Best fits are shown in Table \ref{table2}.
(The errors presented come from the fitting process and do not contain
the truncation errors which are generally larger.) The fitting
was done in the volume ranges $l=50-230$, $l=110-230$, the dimension
of the truncated Hilbert space was 13600. The results of the fitting
support the existence of a tricritical point located (approximately)
at $\eta_{1}=0.163$, $\eta_{2}=0.3518$. The exact values of $D_{1}$
and $D_{2}$ in the tricritical Ising model are \[
D_{1}=0.075\ ,\qquad D_{2}=0.2\ .\]
The numerical results agree quite well with this prediction. The
TCSA spectrum obtained at the tricritical point is shown in Figure
\ref{spectrum3}. The values of $D_{i}$ predicted by the tricritical
Ising model are also shown in the figure. The dashed lines and $+$
signs are used for odd parity states, the continuous lines and $\times$
signs are used for even parity states. 
\begin{table}
\caption{\label{table2}The results of fitting (\ref{eq:fit2}) to the first
two excited levels in the estimated tricritical point}
\begin{center}
\begin{tabular}{lccc}
\hline 
State&
 $D_{i}$&
 $A_{i}$&
 $B_{i}$\\
\hline
$i=1$&
 $0.074\pm0.004$&
 $-0.0060\pm0.001$&
 $2.8\cdot10^{-5}\pm5\cdot10^{-6}$\\
% \hline
$i=2$&
 $0.196\pm0.01$&
 $-0.006\pm0.002$&
 $2.4\cdot10^{-5}\pm6\cdot10^{-6}$\\
\hline
\end{tabular}
\end{center}
% \vspace*{\medskipamount}
\begin{center}$\beta=8\sqrt{\pi}/7$, $\eta_{1}=0.163$,
   $\eta_{2}=0.3518$\end{center}
% \vspace*{\medskipamount}
\end{table}

We used a modified version of the TCSA program exploiting the $\ZZ_{2}$-symmetry
of the model by taking even and odd basis vectors and taking the 
Hamiltonian operator on the even and odd subspaces
separately, which reduces the total time needed for diagonalization
and thus allows to take higher $e_{cut}$ values.

The TCSA data for the first two excited states fit quite well to the
prediction of the tricritical Ising model, the energy levels of the
next two excited states also show correspondence with the prediction.
Clear correspondence cannot be seen for higher levels.

\subsection{The critical line}

\begin{table}
\caption{\label{critline}Points of the critical line found by TCSA}
\begin{center}
\begin{tabular}{llllllll}
\hline 
$\eta_{1}$&
 $\eta_{2}$&
 $\eta_{1}$&
 $\eta_{2}$&
 $\eta_{1}$&
 $\eta_{2}$&
 $\eta_{1}$&
 $\eta_{2}$\tabularnewline
\hline
0.01&
 0.354&
 0.06&
 0.355&
 0.11&
 0.359&
 0.16 &
 0.357\tabularnewline
%\hline
0.02&
 0.354&
 0.07&
 0.3555&
 0.12&
 0.36&
 0.163&
 0.3565\tabularnewline
%\hline
0.03&
 0.354&
 0.08&
 0.356&
 0.13&
 0.36&
&
\tabularnewline
%\hline
0.04&
 0.354&
 0.09&
 0.3575&
 0.14&
 0.3595&
&
\tabularnewline
%\hline
0.05&
 0.354&
 0.1&
 0.3585&
 0.15&
 0.359&
&
 \tabularnewline
\hline
\end{tabular}
\end{center}
\end{table}

\begin{figure}[H]
\begin{tabular}{cc}
\includegraphics[
  width=0.17\paperwidth,
  height=0.37\paperwidth,
  angle=270]{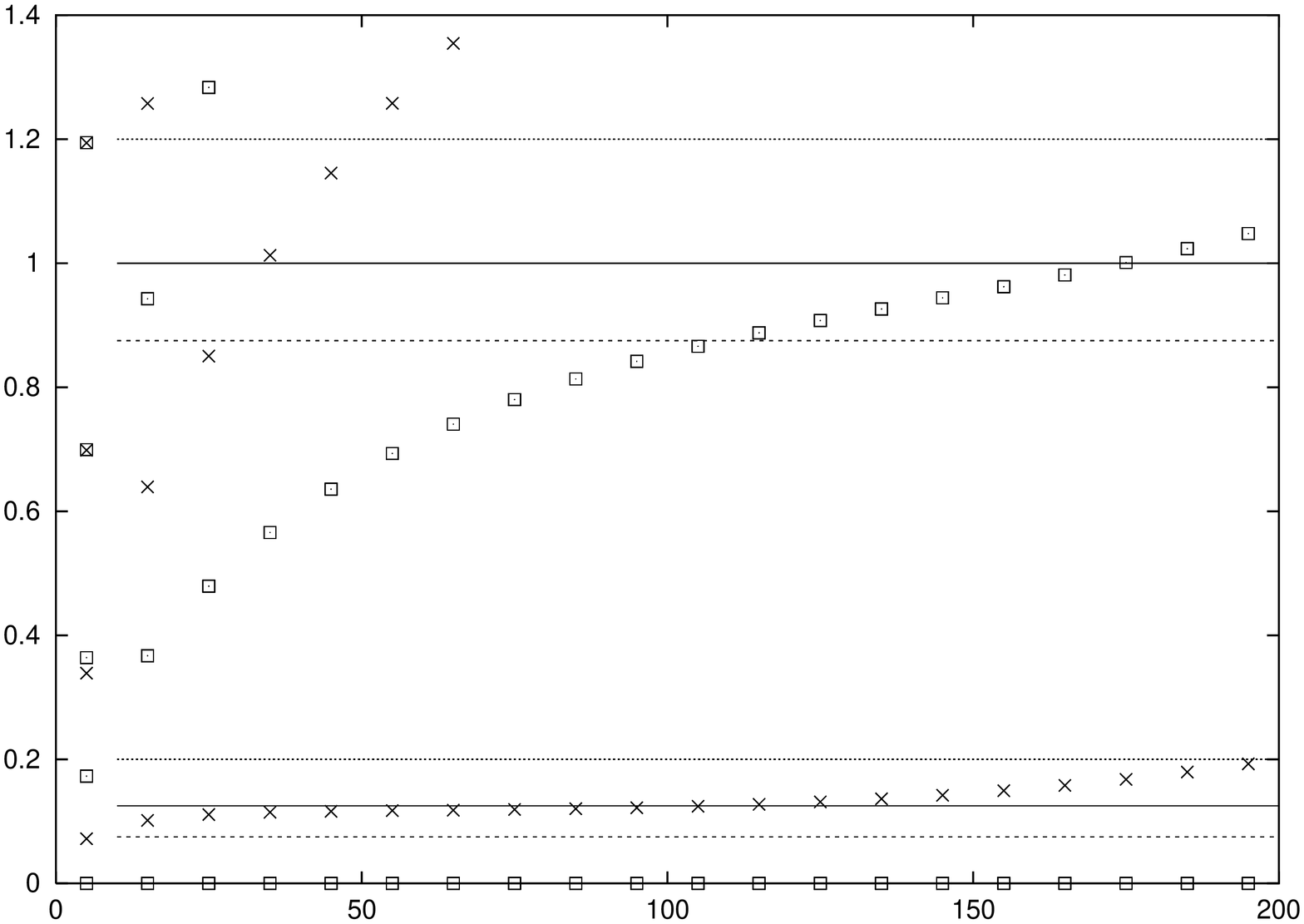}&
\includegraphics[
  width=0.17\paperwidth,
  height=0.37\paperwidth,
  angle=270]{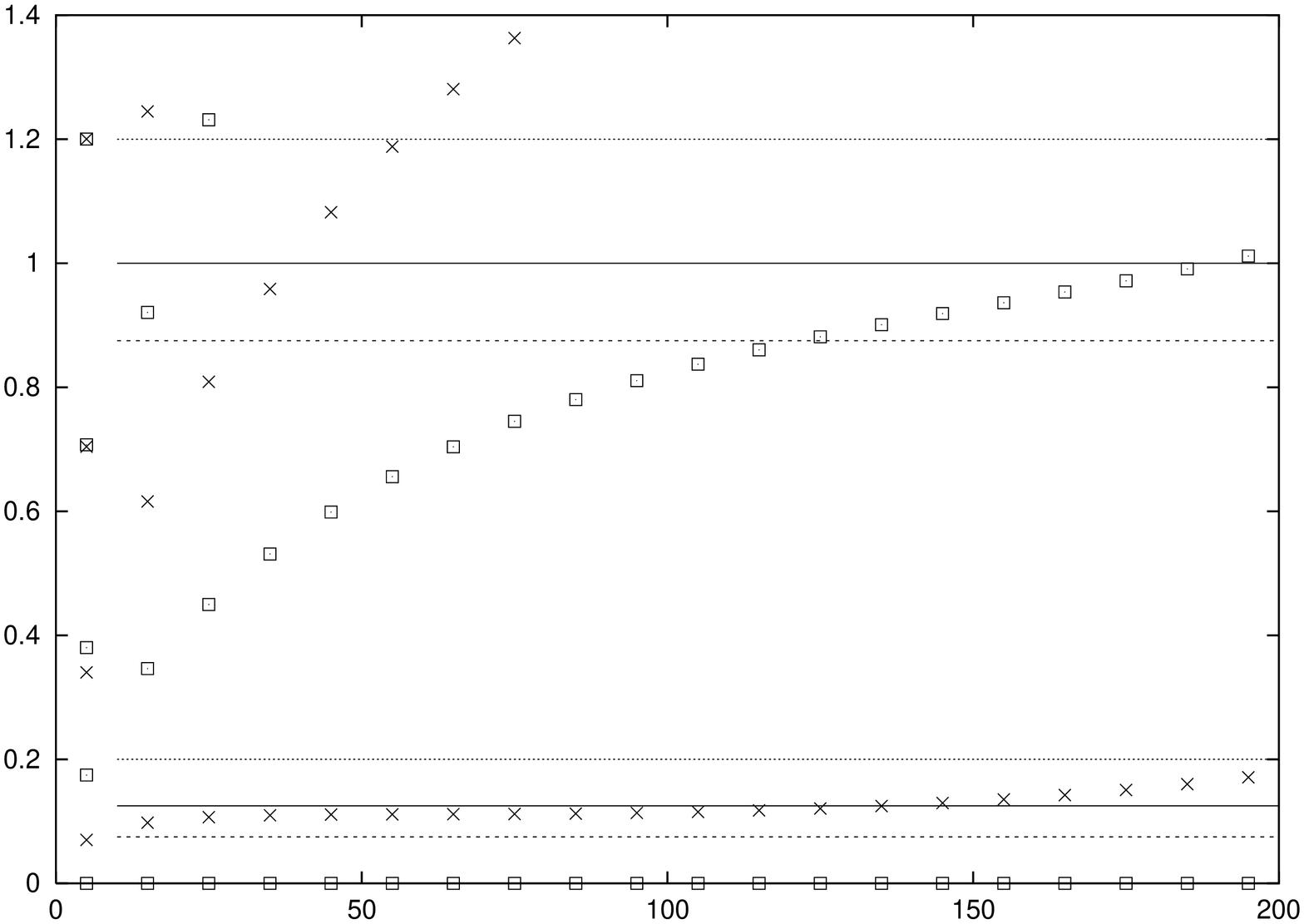}\\
a: $\eta_{1}=0.01$ & b:
$\eta_{1}=0.05$\\
\includegraphics[
  width=0.17\paperwidth,
  height=0.37\paperwidth,
  angle=270]{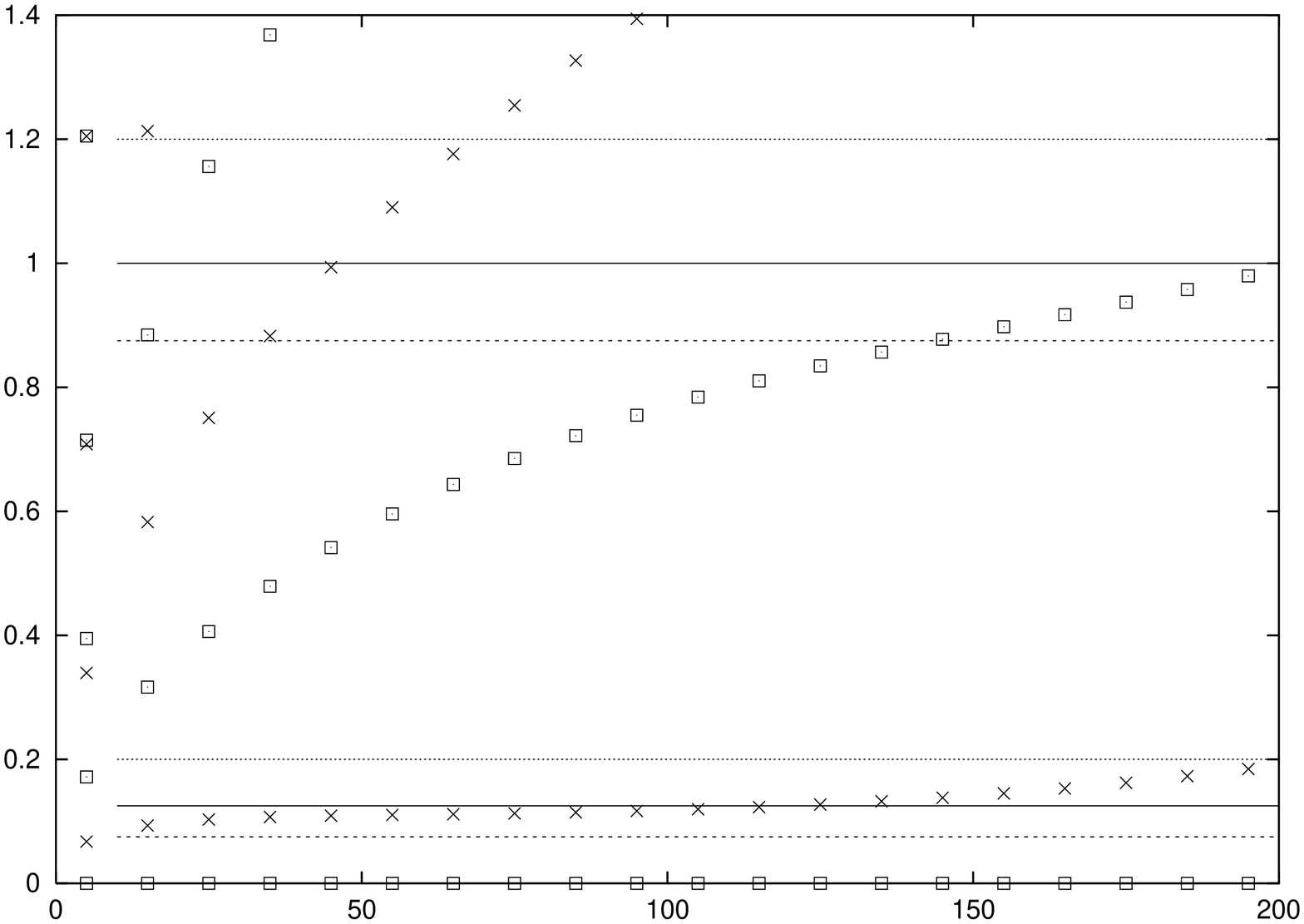}&
\includegraphics[
  width=0.17\paperwidth,
  height=0.37\paperwidth,
  angle=270]{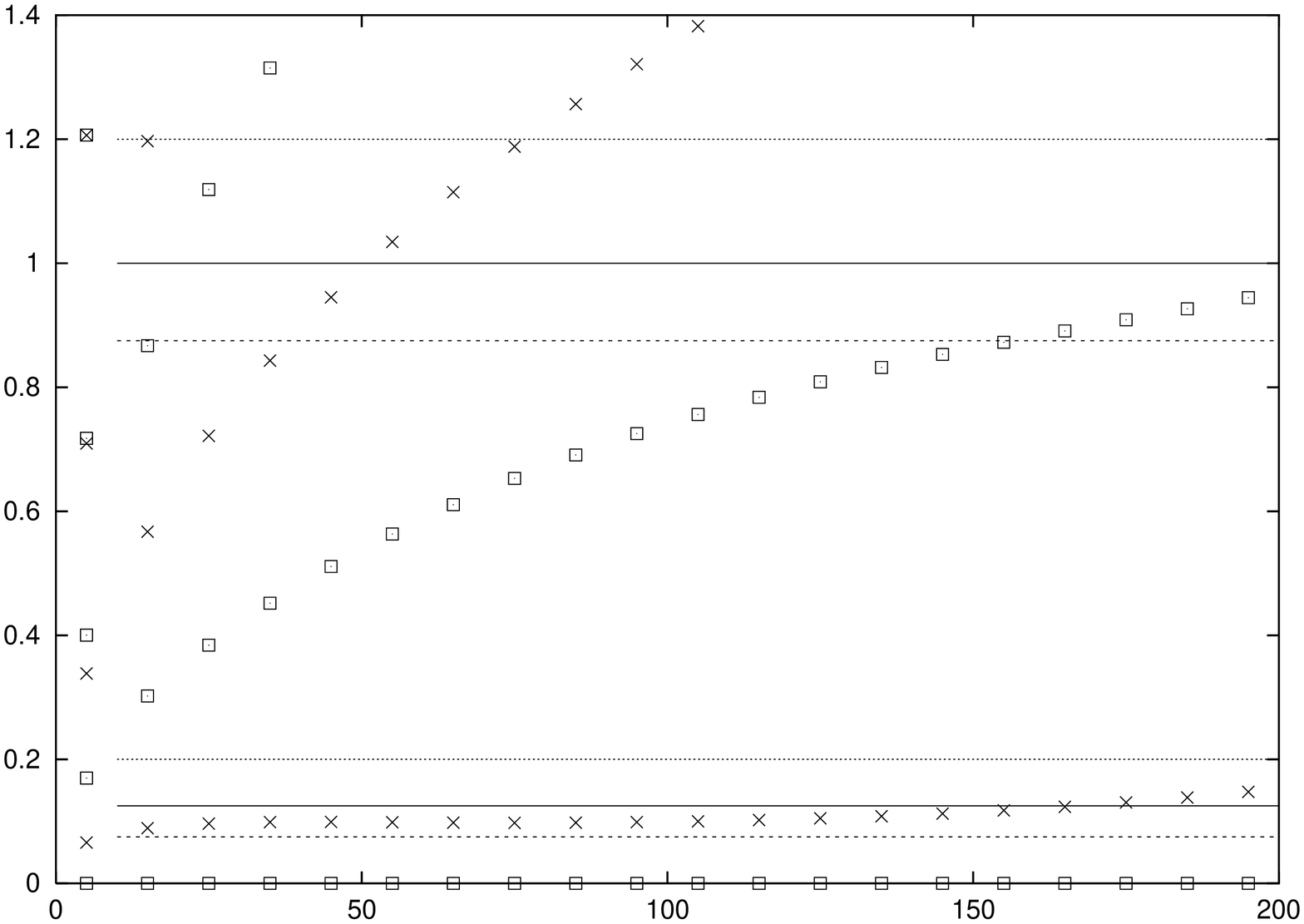}\\
c: $\eta_{1}=0.08$ & d:
$\eta_{1}=0.09$\\
\includegraphics[
  width=0.17\paperwidth,
  height=0.37\paperwidth,
  angle=270]{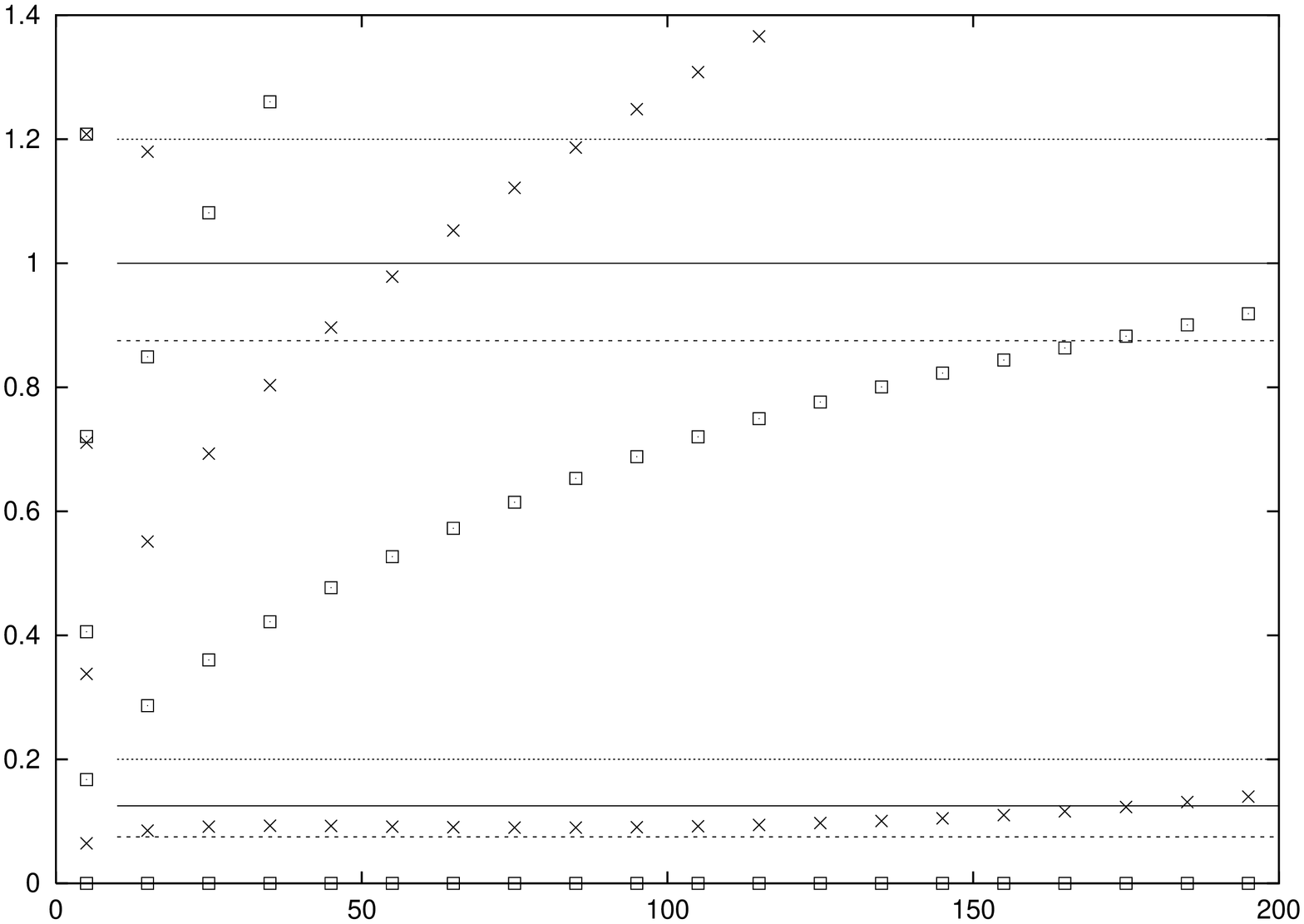}&
\includegraphics[
  width=0.17\paperwidth,
  height=0.37\paperwidth,
  angle=270]{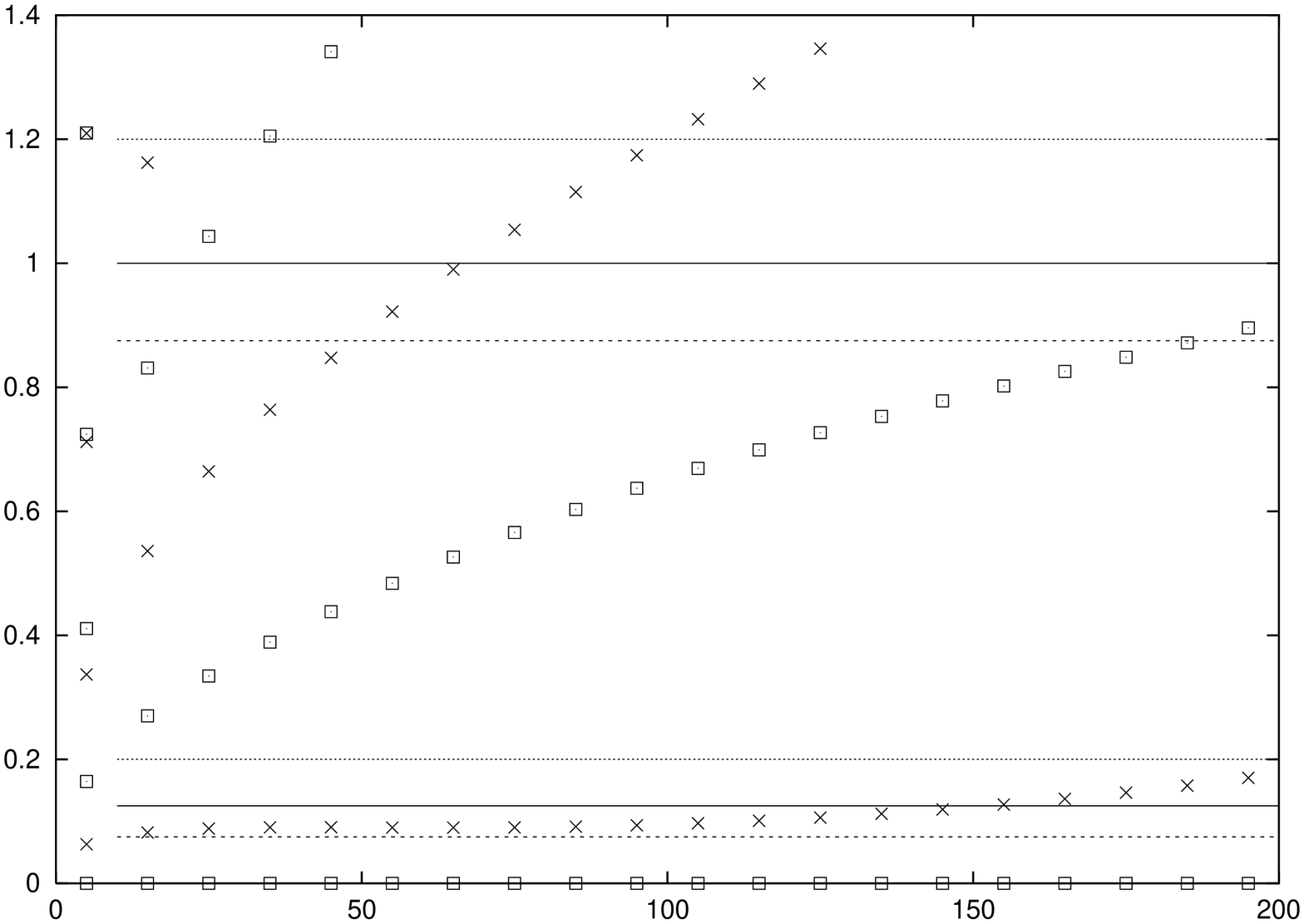}\\
e: $\eta_{1}=0.10$ & f:
$\eta_{1}=0.11$\\
\includegraphics[
  width=0.17\paperwidth,
  height=0.37\paperwidth,
  angle=270]{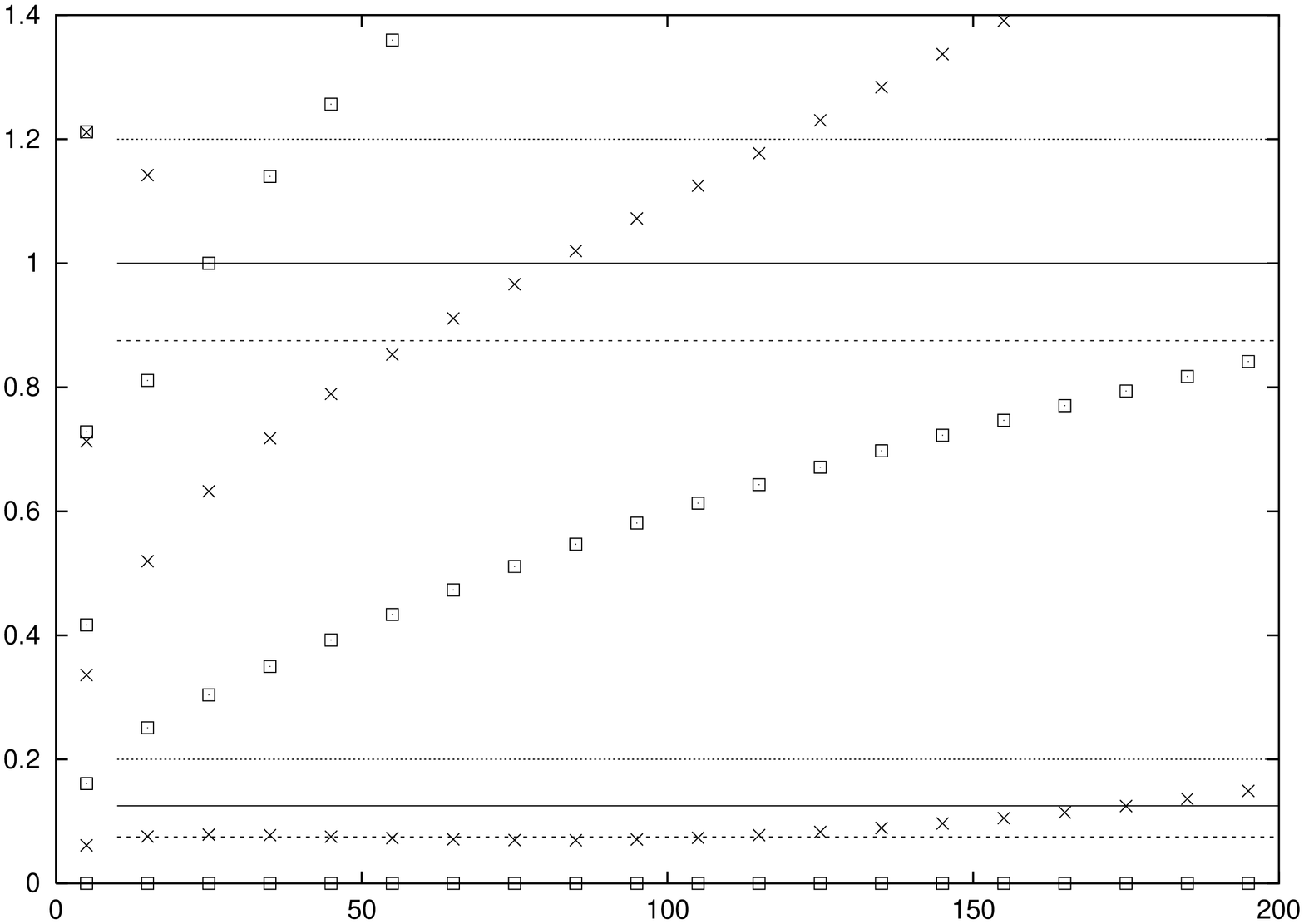}&
\includegraphics[
  width=0.17\paperwidth,
  height=0.37\paperwidth,
  angle=270]{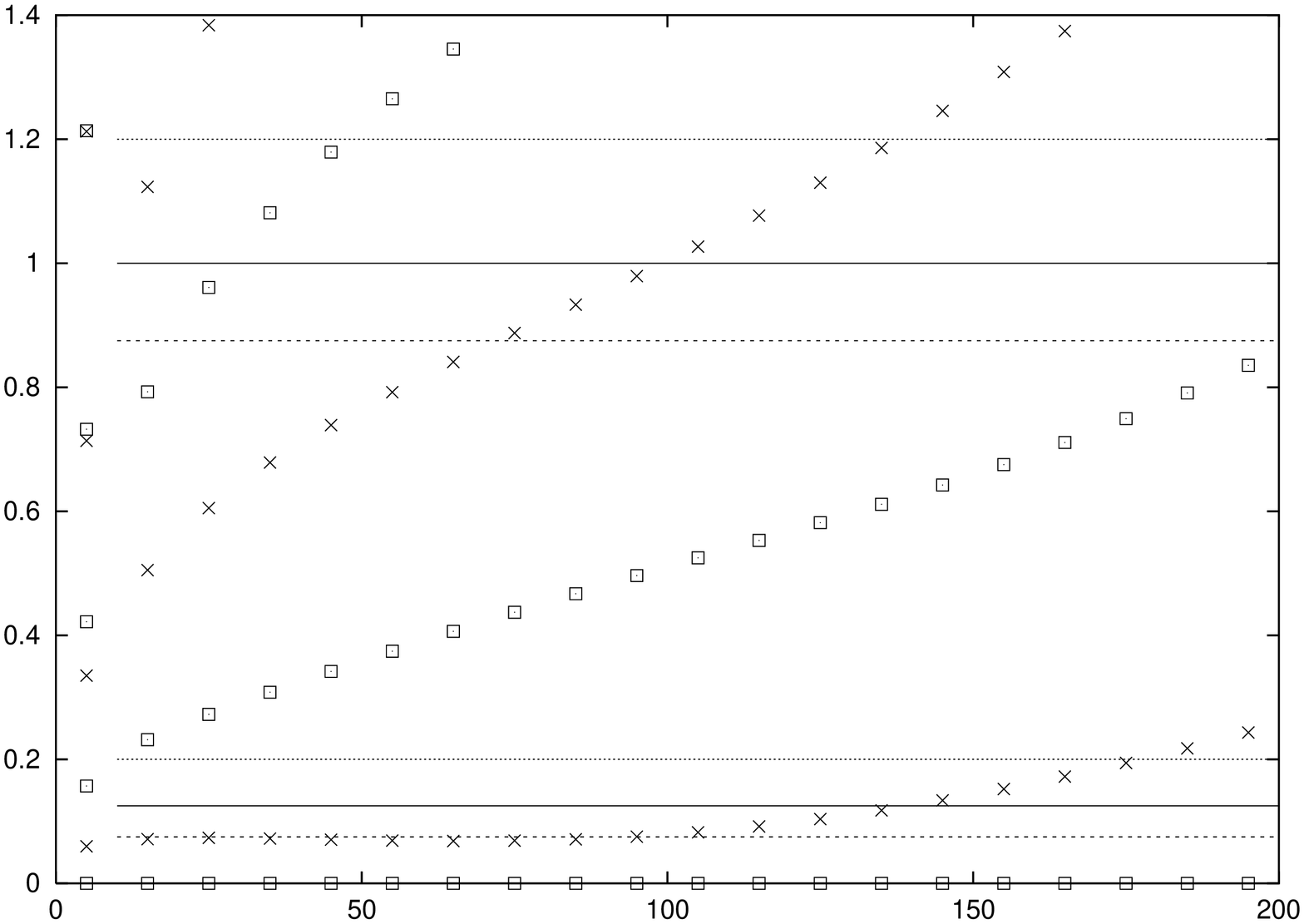}\\
g: $\eta_{1}=0.12$ & h:
$\eta_{1}=0.13$
\end{tabular}
\caption{\label{dikrit vonal}$[e_{i}(l)-e_{0}(l)]\cdot l/2\pi$ as functions
of $l$ obtained by TCSA at $\beta=8\sqrt{\pi}/5$ and at various
points $(\eta_{1},\eta_{2})$ lying on the critical line, the predictions
of the critical Ising model (continuous horizontal lines) for $D_{i}$,
the predictions of the tricritical Ising model (dashed horizontal
lines) for $D_{i}$ }
\end{figure}

The points of the critical line we found using TCSA are listed in
Table \ref{critline}. The value of $\eta_{1}$ was chosen and fixed
in advance, and then $\eta_{2}$ was estimated in the same way as
described in Section \ref{sec: hat}. These points are also marked
in Figure \ref{threefreq phasediag} by crosses. The dimension of
the truncated Hilbert space was $10269$ in these calculations, which
corresponded to $e_{cut}=17$. The value of $\beta$ was $8\sqrt{\pi}/5$.
Figures \ref{dikrit vonal}.a-l show the TCSA spectra (especially
the lowest lying energy levels) obtained in these points as well as
the values of $D_{i}$ corresponding to both the critical and the
tricritical Ising model. Crosses are used for odd parity states, squares 
are used for even parity states.
It can be seen that moving on the critical
line in the phase space towards the tricritical endpoint the finite
volume spectrum changes continuously. In the $\eta_{1}<0.11$ domain
the spectra (especially the first two levels) correspond clearly to
phase transitions in the Ising universality class. At $\eta_{1}=0.11$
the first excited level already appears to correspond to the prediction
of the tricritical Ising model, whereas the second excited level still
has the behaviour predicted by the Ising model. It would be very interesting
to know the large volume behaviour (and infinite volume limit) of
the first excited level, but the precision of TCSA does not allow
to determine it. What we can see is that there is no sign in the TCSA
data that the first excited level follows the predictions of the Ising
model in the large volume limit. In the domain $0.11\leq\eta_{1}\leq0.16$
there is a spectacular rearrangement of the higher energy levels (already
observable in the $\eta_{1}<0.11$ domain), and at $\eta_{1}=0.16$
the second excited level also appears to correspond to the prediction
of the tricritical Ising model. We regard therefore the point $\eta_{1}=0.16$,
$\eta_{2}=0.357$ to be the tricritical endpoint of the critical line,
this point is marked by a square in Figure \ref{threefreq phasediag}.
(We did not aspire to determine the value of $\eta_{1}$ more precisely
for this value of $\beta$.) Each figure shows the spectrum in the
volume interval $l=0 \dots 200$, and for the large values $l\approx200$
the truncation error is always conspicuous.

The value of $l$ where the truncation errors become large
gets smaller and smaller as the tricritical point is approached, which
corresponds to the fact that a tricritical point as renormalization
group fixed point is more repelling than an Ising type one. Increasing
$\eta_{1}$ further the behaviour corresponding to the tricritical
point would rapidly disappear from the finite volume spectrum.
\clearpage

\begin{figure}[H]
\begin{tabular}{cc}
\includegraphics[
  width=0.17\paperwidth,
  height=0.37\paperwidth,
  angle=270]{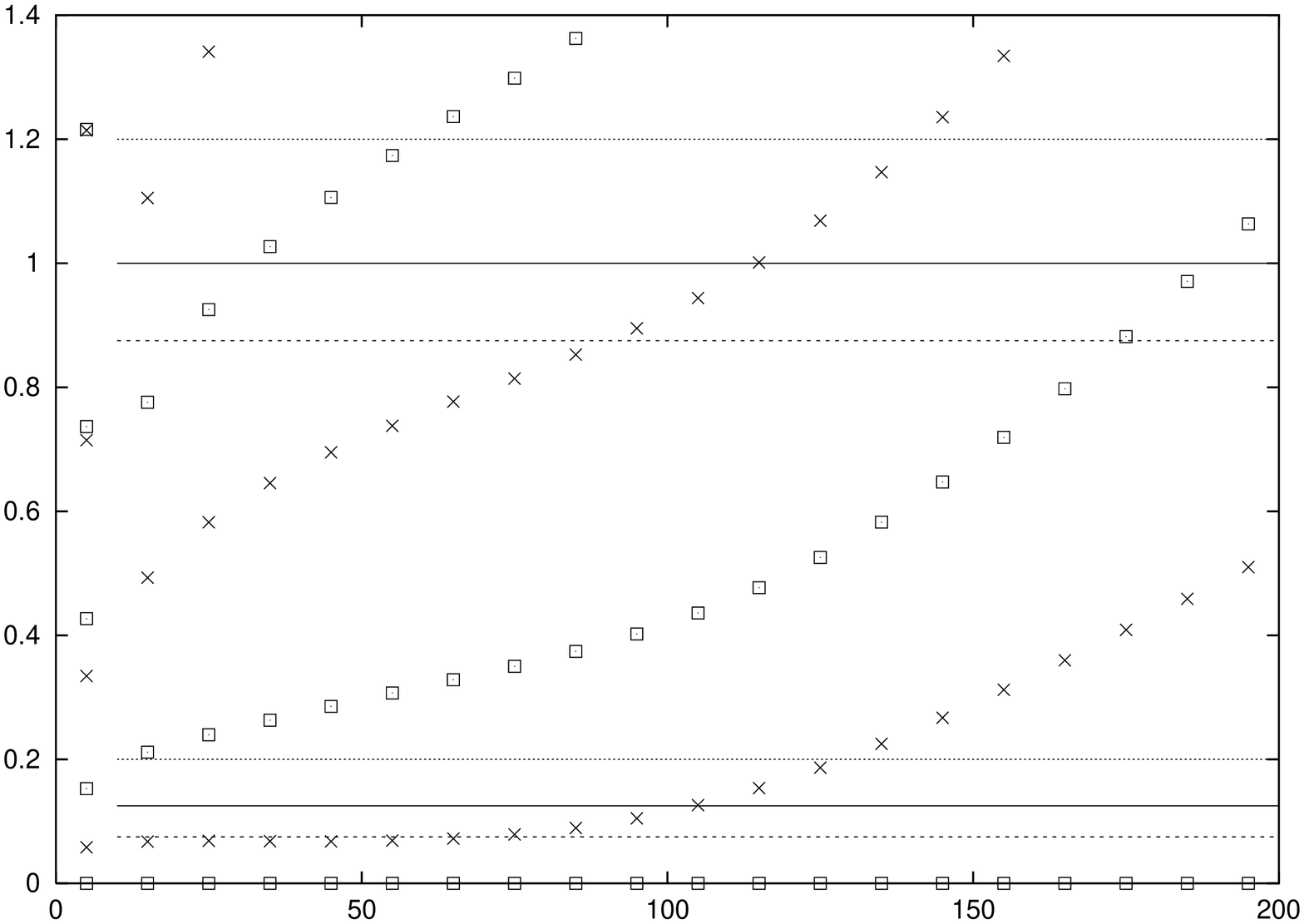}
&\includegraphics[
  width=0.17\paperwidth,
  height=0.37\paperwidth,
  angle=270]{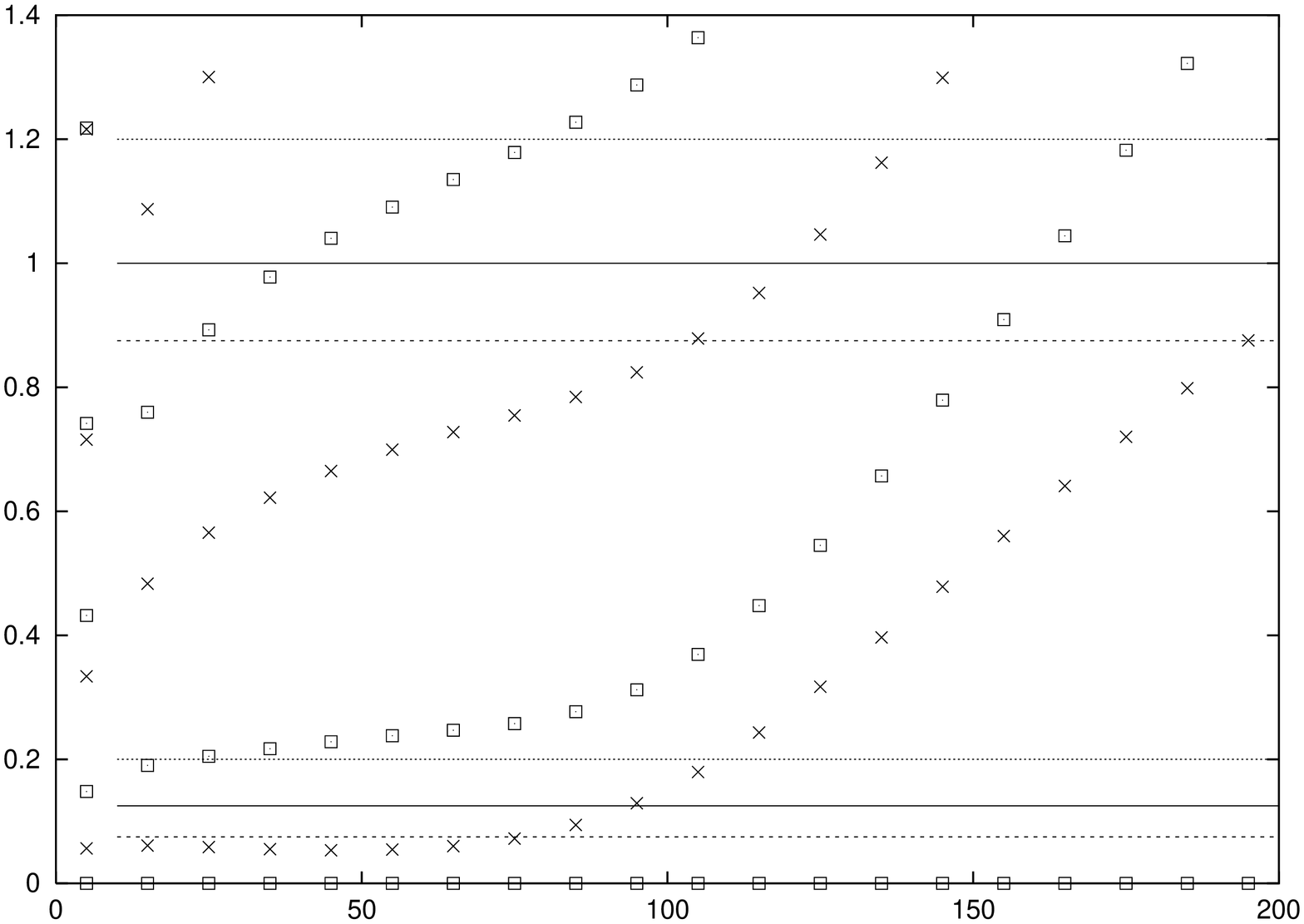}\\
i: $\eta_{1}=0.14$& j:
$\eta_{1}=0.15$ \\
\includegraphics[
  width=0.17\paperwidth,
  height=0.37\paperwidth,
  angle=270]{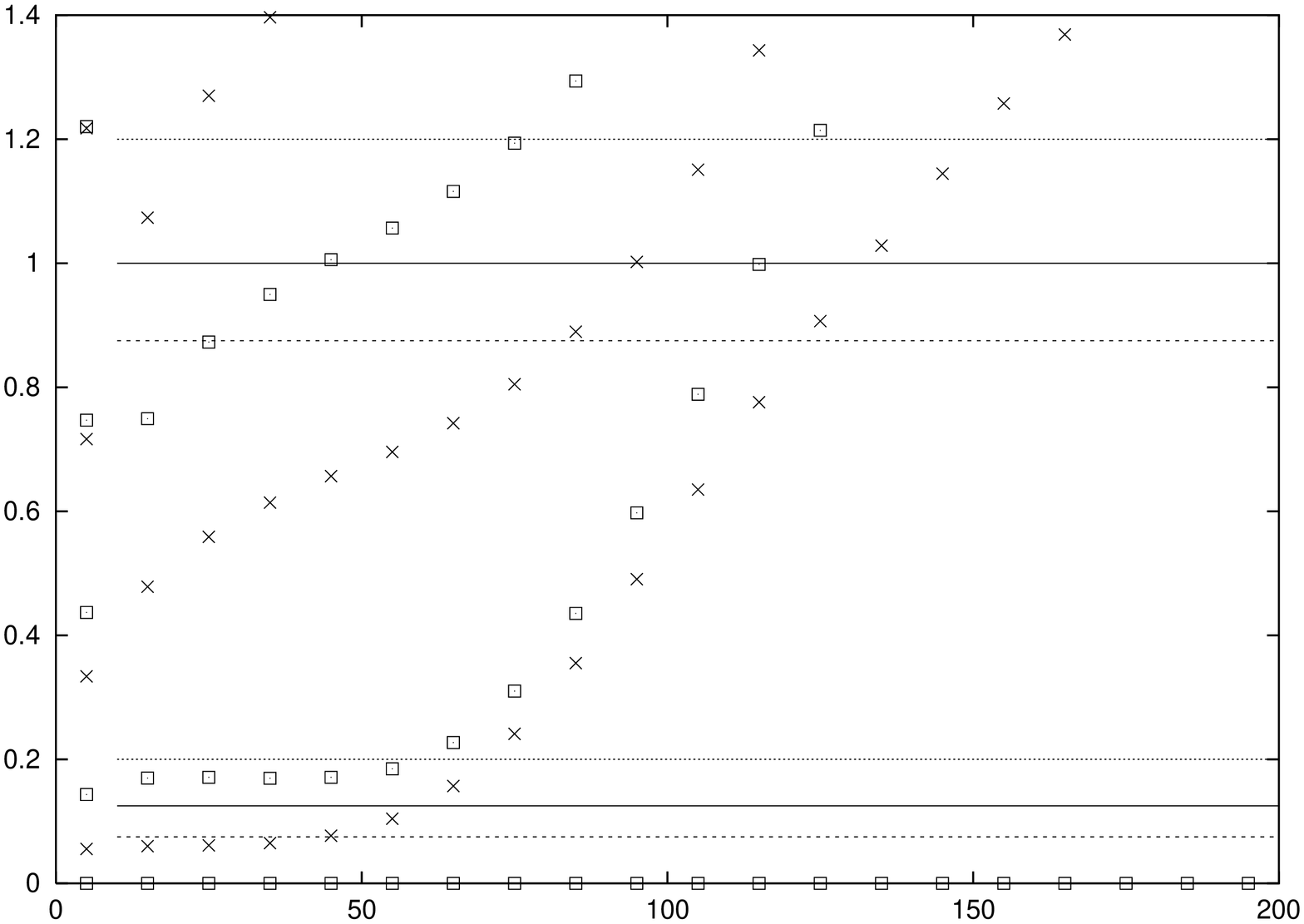}&
\includegraphics[
  width=0.17\paperwidth,
  height=0.37\paperwidth,
  angle=270]{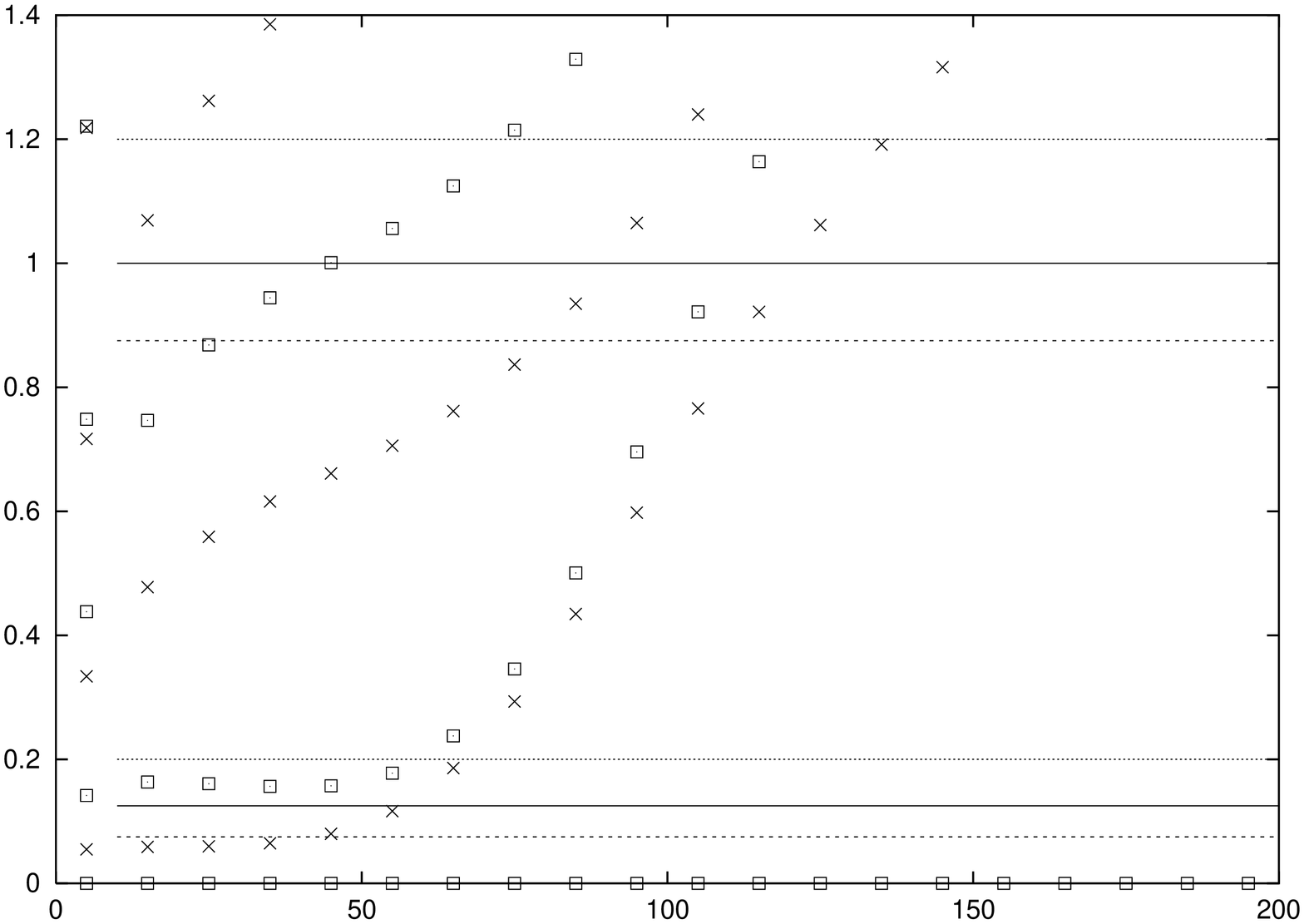}\\
k: $\eta_{1}=0.16$ & l:
$\eta_{1}=0.163$
\end{tabular}
\begin{center}
Figure \ref{dikrit vonal} continued
\end{center}
\end{figure}

\subsection{The line of first order transition}

\begin{figure}[p]
\begin{tabular}{cc}
\includegraphics[
  width=0.15\paperwidth,
  height=0.37\paperwidth,
  angle=270]{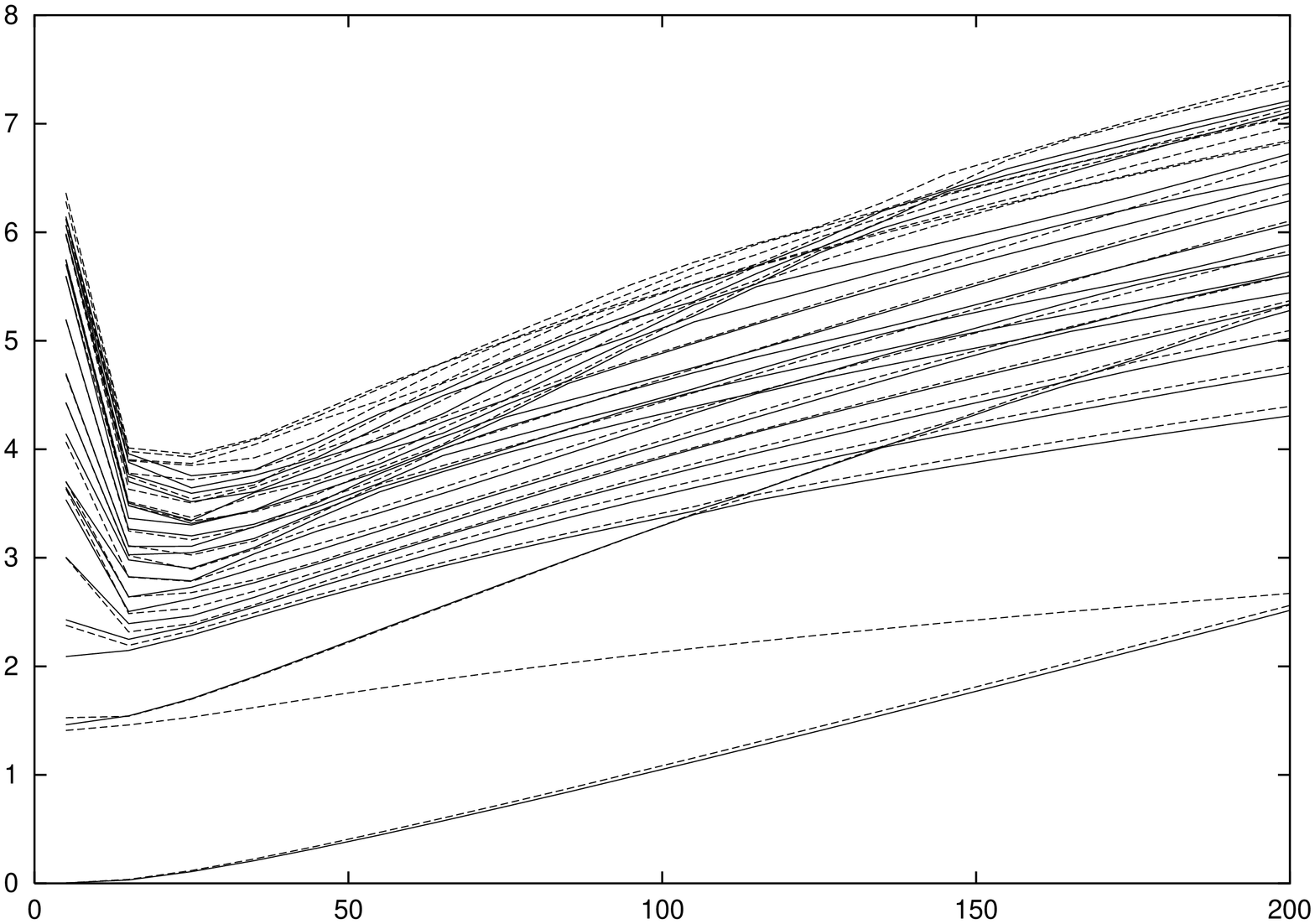} & 
\includegraphics[
  width=0.15\paperwidth,
  height=0.37\paperwidth,
  angle=270]{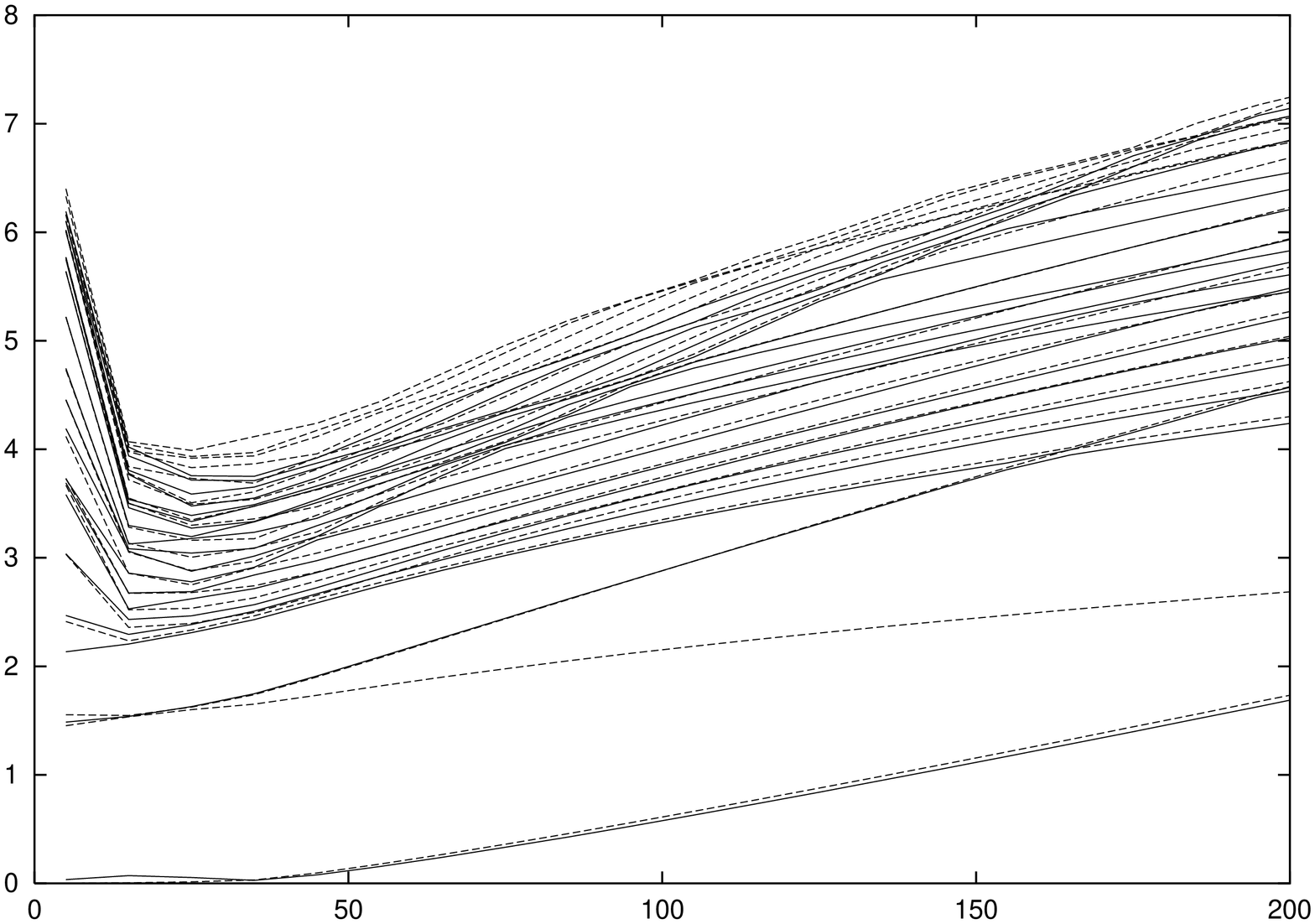}\\
a: $\eta_{2}=0.17$ & b:
$\eta_{2}=0.18$ \\
\includegraphics[
  width=0.15\paperwidth,
  height=0.37\paperwidth,
  angle=270]{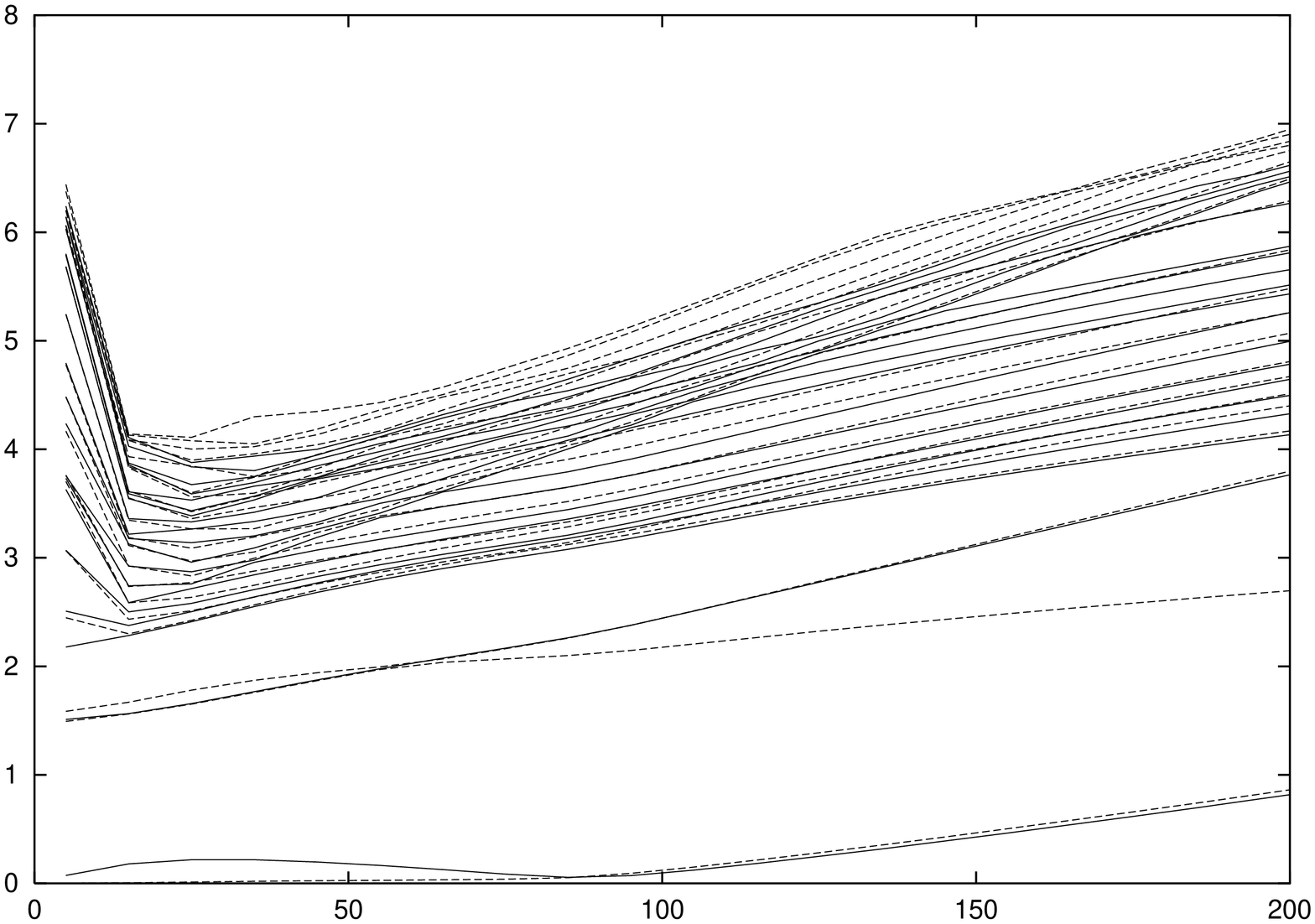}&
\includegraphics[
  width=0.15\paperwidth,
  height=0.37\paperwidth,
  angle=270]{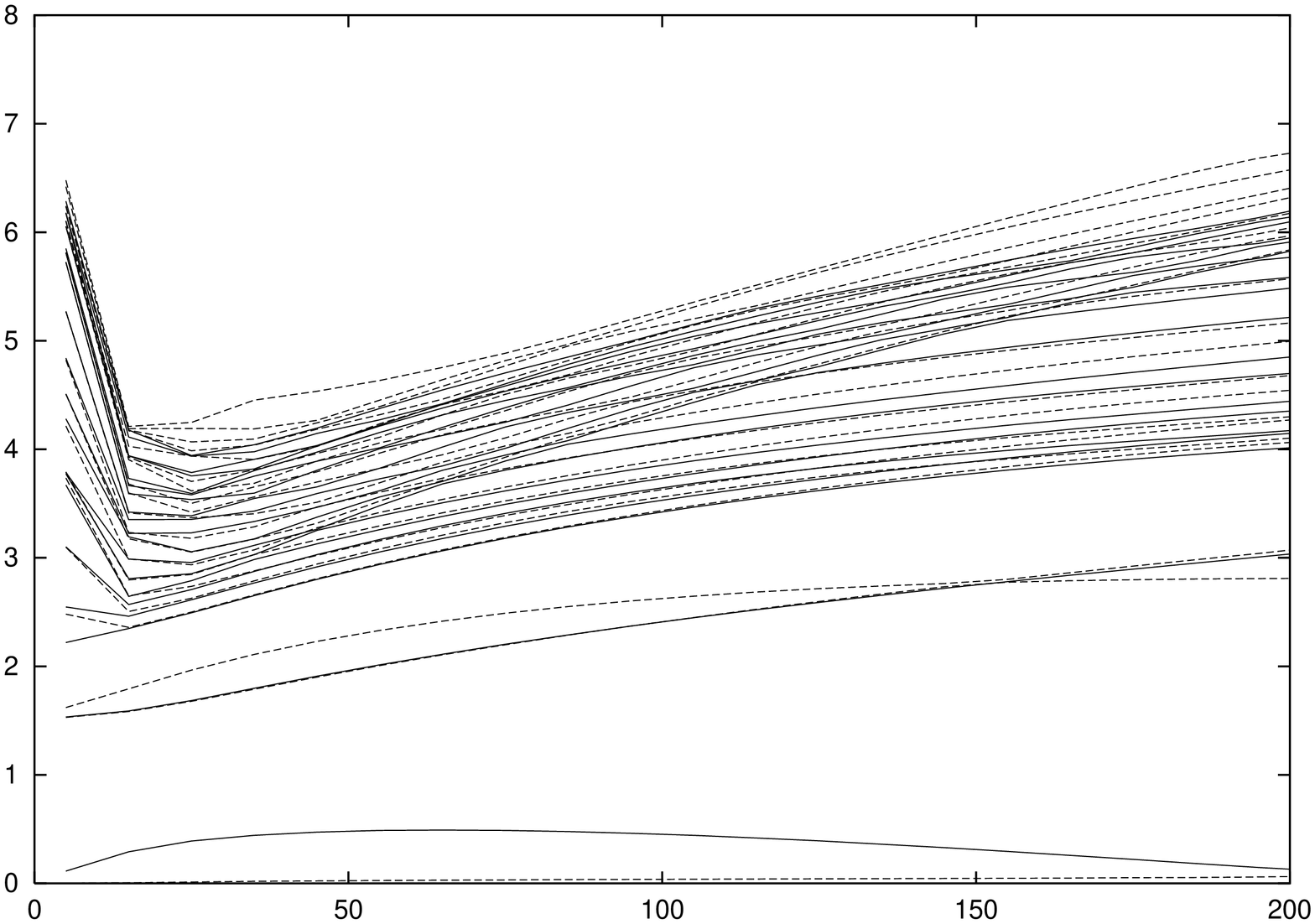}\\
c: $\eta_{2}=0.19$ & d:
$\eta_{2}=0.20$ \\
\includegraphics[
  width=0.15\paperwidth,
  height=0.37\paperwidth,
  angle=270]{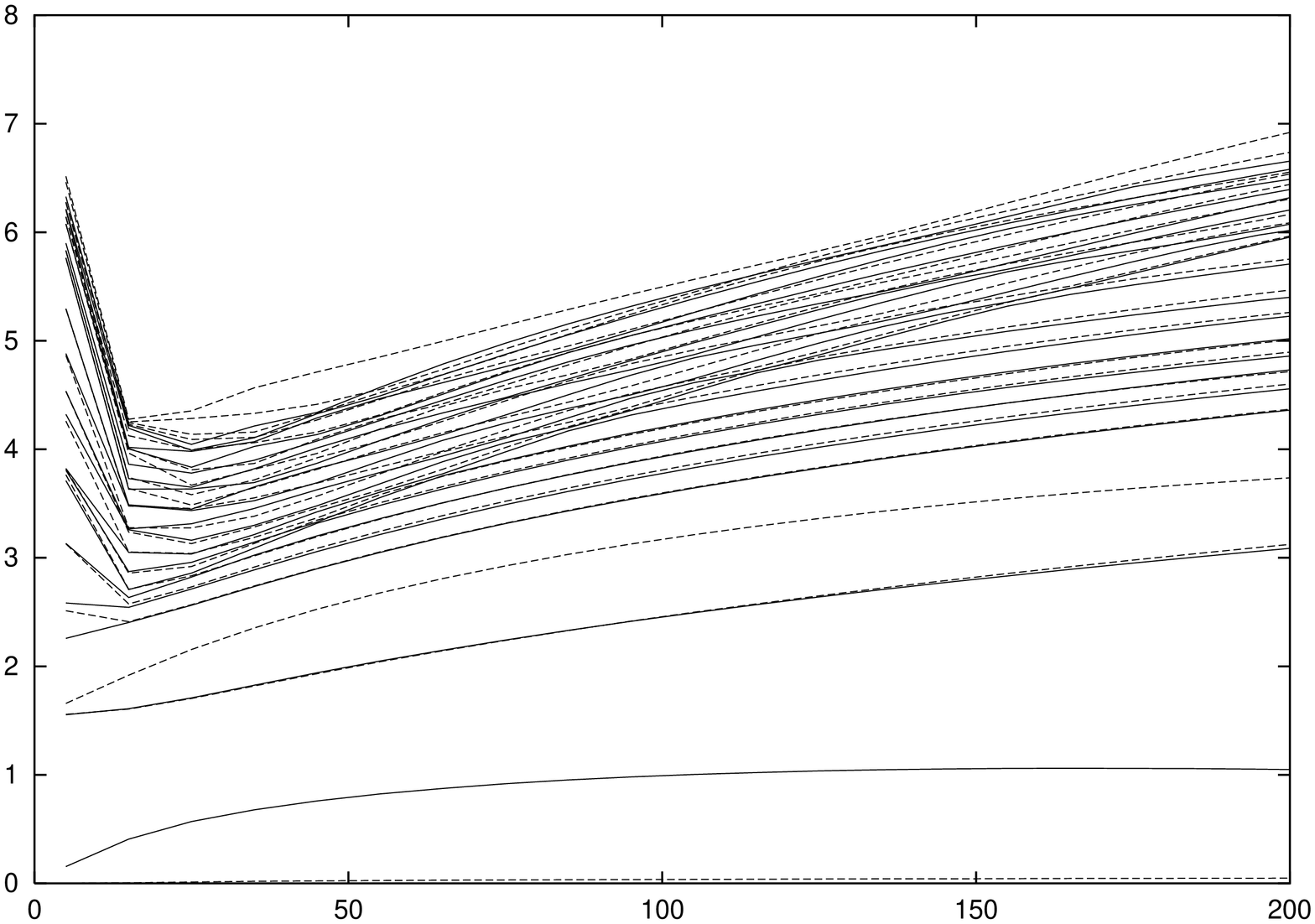}&
\includegraphics[
  width=0.15\paperwidth,
  height=0.37\paperwidth,
  angle=270]{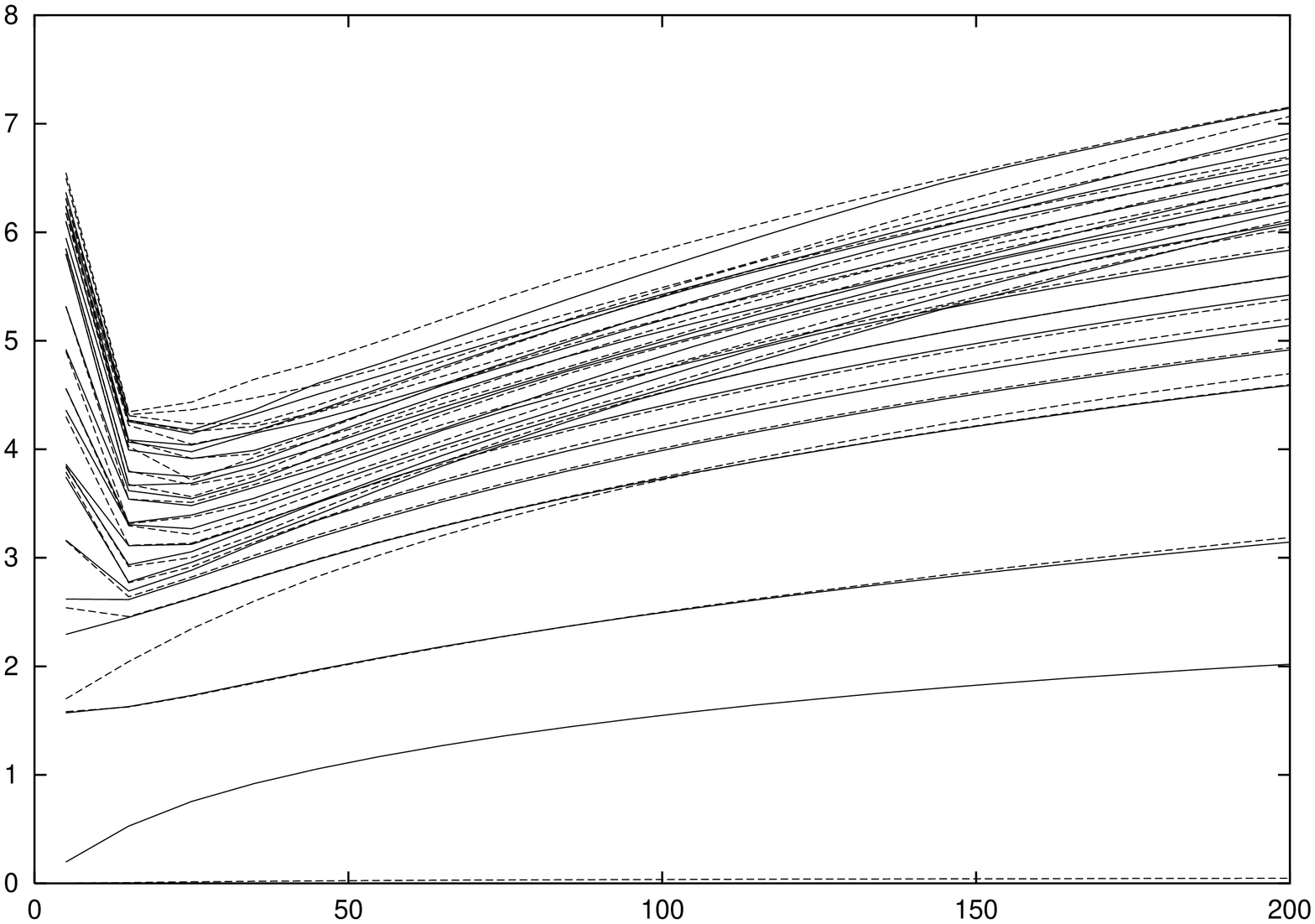}\\
e: $\eta_{2}=0.21$ & f:
$\eta_{2}=0.22$\\
\includegraphics[
  width=0.15\paperwidth,
  height=0.37\paperwidth,
  angle=270]{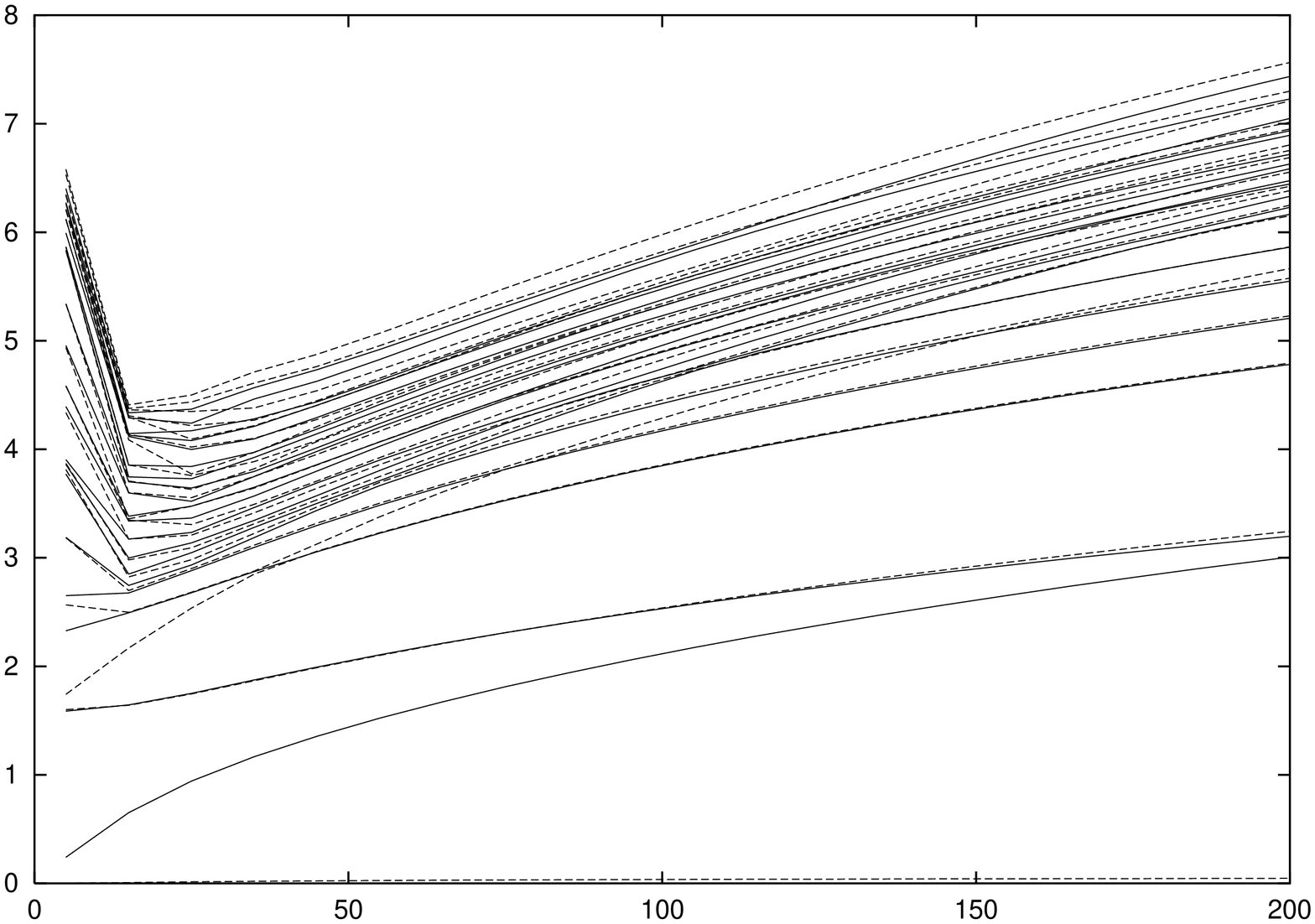}&
\includegraphics[
  width=0.15\paperwidth,
  height=0.37\paperwidth,
  angle=270]{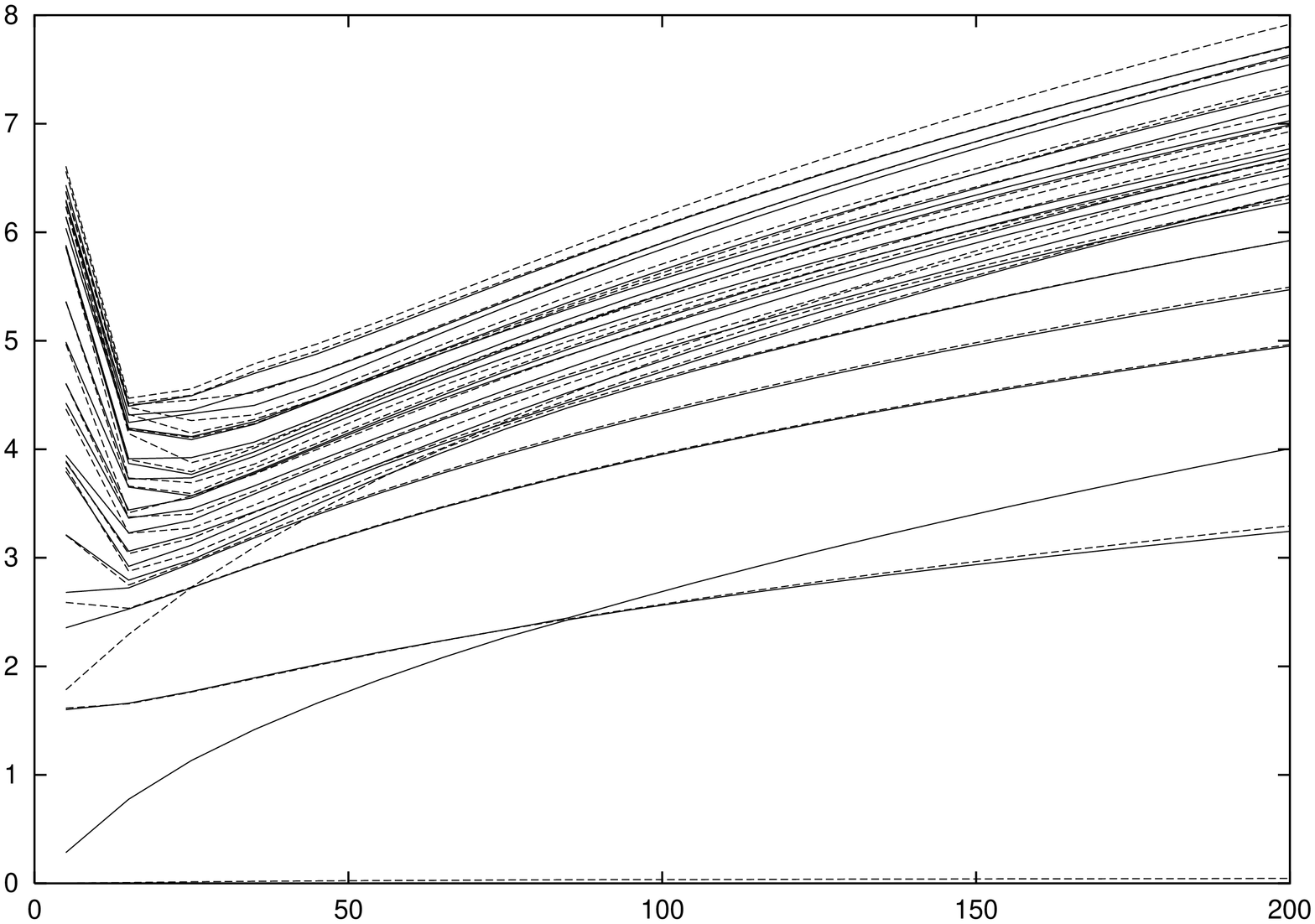}\\
g: $\eta_{2}=0.23$ & h:
$\eta_{2}=0.24$\\
\includegraphics[
  width=0.15\paperwidth,
  height=0.37\paperwidth,
  angle=270]{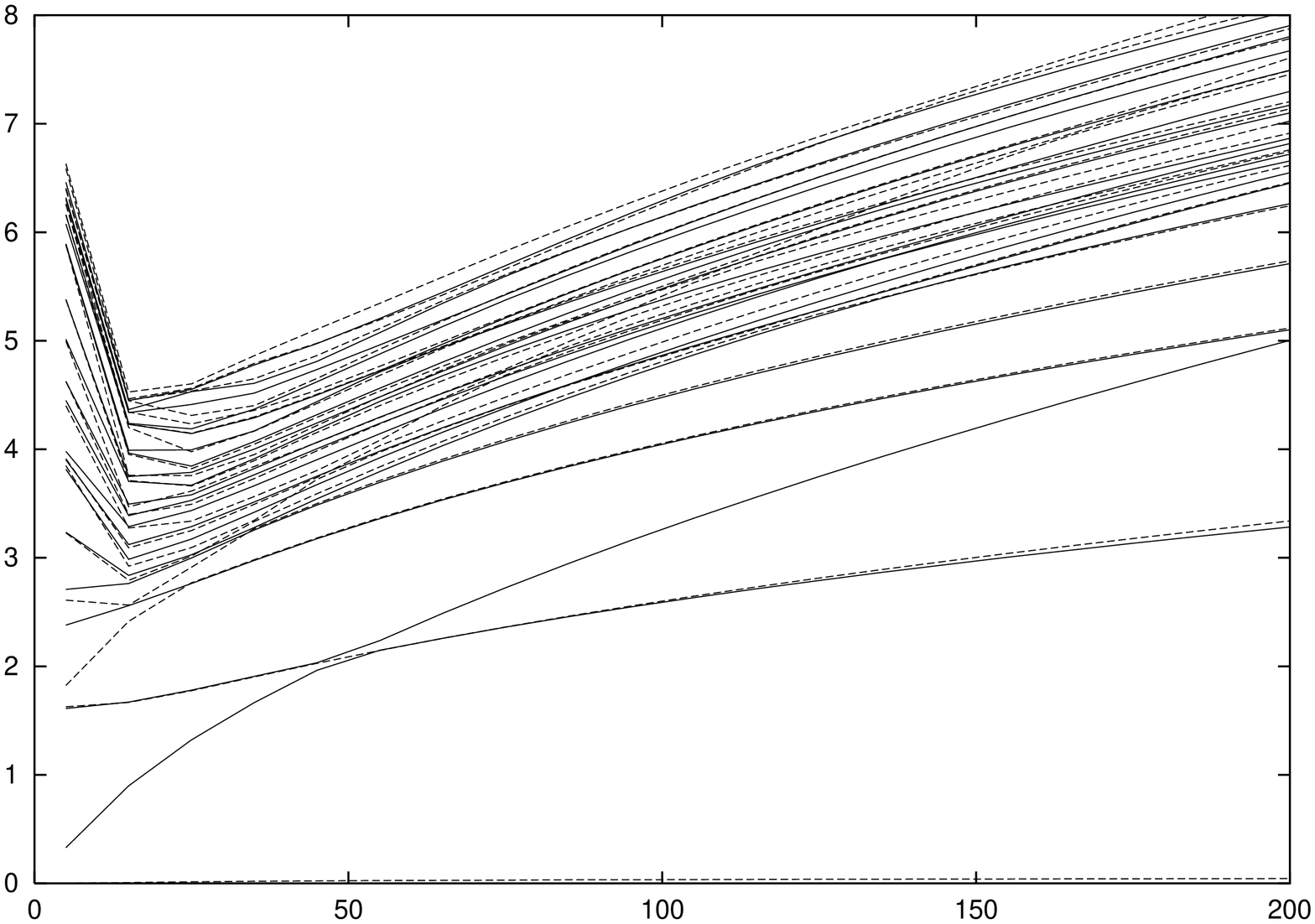}&
\includegraphics[
  width=0.15\paperwidth,
  height=0.37\paperwidth,
  angle=270]{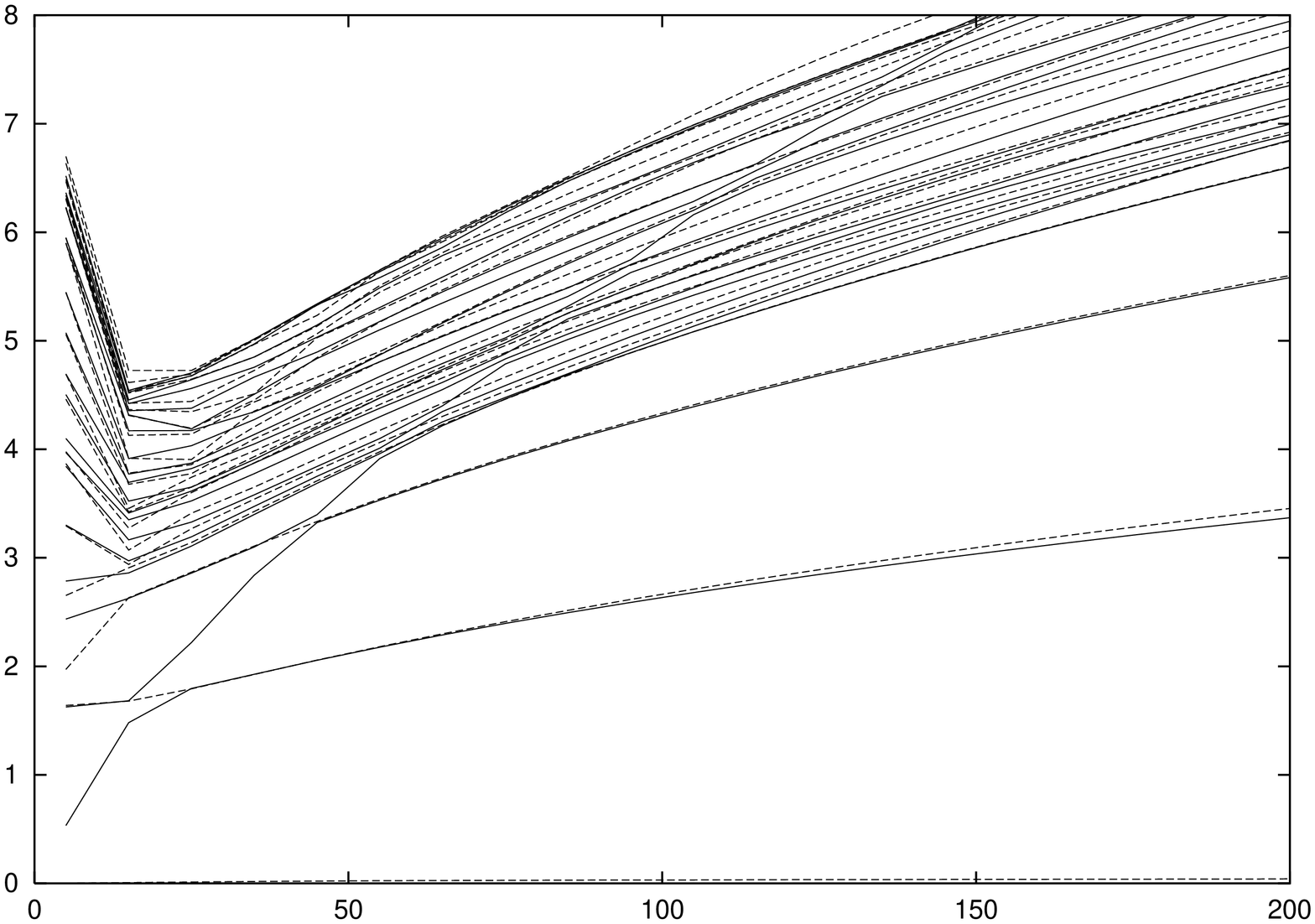}\\
i: $\eta_{2}=0.25$ & j:
$\eta_{2}=0.30$
\end{tabular}
\caption{\label{nemkrit atmenet}$[e_{i}(l)-e_{0}(l)]$, $i=0 \dots 19$ as functions
of $l$ obtained by TCSA at $\beta=8\sqrt{\pi}/5$, $\eta_{1}=0.6$
and at various values of $\eta_{2}$ }
\end{figure}

Figures \ref{nemkrit atmenet}.a-j show TCSA spectra obtained at $\eta_{1}=0.6$
and at various values of $\eta_{2}$ between $0.17$ and $0.30$.
The energy levels are shown as compared to the lowest level. The dimension
of the truncated Hilbert space was $6597$ in these calculations,
which corresponded to $e_{cut}=16$. The value of $\beta$ was $8\sqrt{\pi}/5$.
Dashed lines are used for odd parity levels and continuous lines for
even parity levels.

At $\eta_{2}=0.17$ (and also for $\eta_{2}<0.17$) the ground state
is unique for all values of $l$, whereas doubly degenerate runaway
levels can also be seen. At $\eta_{2}=0.19$, however, the ground
level is doubly degenerate for small values of the volume but becomes
nondegenerate in large volume, 
i.e.\ there is a value of $l$ where the doubly degenerate runaway
level and the single level cross each other. The slope of the runaway
levels become smaller and smaller as $\eta_{2}$ is increased, and
the crossing point also moves towards larger and larger values. At
$\eta_{2}=0.3$ the ground state is already doubly degenerate for
all values of $l$ and nondegenerate runaway levels are present. These
features indicate that a first order phase transition occurs at an
intermediate value of $\eta_{2}$. The behaviour of the finite volume
spectra around the transition point seen in the figures implies that
a precise determination of the transition point from the TCSA data
in a direct way would require precise data for large values of $l$.
For this reason we did not aspire to find many points of the line
of first order phase transition, we did TCSA calculations only at
$\eta_{1}=0.4$ and $\eta_{1}=0.6$, and we roughly estimated the
location of the first order phase transition at these values. Our
estimation based on the TCSA data are $\eta_{2}=0.30$ and $\eta_{2}=0.21$.
These points are also marked in Figure \ref{threefreq phasediag}
by crosses. The Figures \ref{nemkrit atmenet}.a-j show the first
$20$ levels in the domain $l=0 \dots 200$. Unfortunately the truncation
effect is large in most of this domain of $l$, however we expect
that those qualitative features of the spectra which we allude to
are correct.

\clearpage

\section{\label{sec:Conclusions}Discussion}
\markright{\thesection.\ \ DISCUSSION}

We have investigated selected special cases of the multi-frequency
sine-Gordon theory, especially the structure of their phase space, continuing
the work begun in \cite{BPTW}. Concerning the classical limit we
found that the only possible phase transition in the (rational) two-frequency
case is a second order Ising-type transition. The three-frequency
model has some new qualitative features compared to the two-frequency
model: a tricritical point which is at the end of a critical line
can also be found, and first order transition is possible as well.
The phase space of the $n$-frequency model contains $n$-fold critical
points, otherwise we do not expect new qualitative features compared
to the three-frequency model.

The numerical (TCSA) calculations in the quantum case yielded the
following results: we found the same type of phase transitions as
in the classical case, in particular we were able to determine the
second order Ising nature and the location of the phase transition
in the two-frequency model with good precision.  
The accuracy of
the TCSA also allowed us to find the tricritical point in the three-frequency
case, which we regard the main result of this chapter. It demands
more numerical work than the two-frequency model for the following
reasons: the two-frequency model has only one-dimensional phase space,
whereas the phase space is two-dimensional in the three-frequency
model; and the tricritical point as a renormalization group fixed
point is more repelling than the Ising-type critical point, so one
has to take larger truncated space to achieve good precision. Furthermore, in
the three-frequency case we found several
points of the critical line and observed how the TCSA spectrum changes
as the tricritical endpoint is approached. It would be interesting
to investigate this at considerably better precision. Although the
focus was mainly on the tricritical point, we also investigated the
first order phase transition in the three-frequency model and we found
that the first order nature can be established by TCSA, but the location
of the transition can be determined much less accurately than that
of the second-order transition. We expect that multi-critical points
in the universality classes of further elements of the discrete unitary
series could be found in the higher-frequency models, but increasing
numerical accuracy would be needed, because these multi-critical points are more and
more repelling. Finally, we remark that the quantum corrections did
not alter the nature of the phase transitions in any of the cases
we investigated, and the location
of transition points is almost the same  in the classical and quantum theory. The investigation of the
multi-frequency model with irrational frequency ratios is still an
open problem. We also remark that the particle content of the multi-frequency
sine-Gordon model could also be investigated by the method of semiclassical
quantization \cite{key-23}.

\chapter*{Acknowledgements}
\markboth{}{}
\addcontentsline{toc}{chapter}{Acknowledgements}

I would like to express my gratitude to my supervisor L\'aszl\'o Palla for his advice and constant
support during my research.   I am also grateful to G\'abor Tak\'acs and
Zolt\'an Bajnok for  many illuminating discussions. I thank Gustav Delius for
an illuminating discussion on the boundary supersymmetry algebra, and the
Mathematics Department at King's College London, where I spent ten months, for hospitality. 
I also thank G\'erard Watts for acting as a supervisor  while I was at  King's
College London.
Finally, I express my gratitude for the invaluable support 
that was provided to me by the Theoretical Physics Research Group of
the Hungarian Academy of Sciences at the Theoretical Physics Department of E\"otv\"os University.
I acknowledge  support by the Hungarian fund OTKA (T037674), by the Research
Training Network ``EUCLID'' (contract HPRN-CT-2002-00325) of the EU at various times, and by the Marie Curie Training Site (MCFH-2001-00296) ``Strings, Branes
and Boundary Conformal Field Theory'' of the EU at King's
College London.

\clearpage
\
\clearpage
\
\markboth{}{}
\clearpage

\addcontentsline{toc}{chapter}{Abstract}

{\bf \large
\vspace*{-1cm}
\noindent
Summary of the main results\\ }

\noindent
In my thesis I study problems in three areas of $1+1$-dimensional quantum field
theory. These
investigations are described in three largely independent chapters. 

In Chapter \ref{sec.chap2} I deal with the boundary bootstrap for
the scattering theory of supersymmetric 
 massive integrable quantum field theories with a boundary in a special
framework in which the blocks of the full S-matrix and reflection matrix are
assumed to take the form of a product of
 a supersymmetric and a non-supersymmetric factor.
I give a
description of supersymmetry in the presence of a boundary, i.e.\ when the
space is the half-line. I present rules
for the determination of the representations in which higher level boundary bound
states transform, and for the determination of the supersymmetric one-particle
reflection matrix factors for the higher level boundary bound states. These
rules apply under the  assumption that the bulk
particles transform in the kink or in the boson-fermion representation. I also
present examples for the
application of these rules to specific models.

In Chapter \ref{sec.chap3} I investigate the effect of the Hilbert space truncation applied in the
numerical method called  truncated conformal space approach (TCSA) on boundary
flows in the case of the critical Ising model on a strip with magnetic
perturbation on one of the boundaries. The main goal is to show that the effect of truncation on
the spectrum can
be taken into consideration approximately by a change of the coefficients of the terms
in the
Hamiltonian operator. I present the results of numerical and perturbative
calculations,  which support this idea. I also present a comparison with
another truncation method which preserves the solvability of the model. The changing of the
coefficients appears to work for this truncation method as well. The comparison 
reveals that certain qualitative properties of the flows of the  truncated spectra  depend on the
particular truncation method applied. 
The chapter includes an exact quantum field theoretic solution  of the model
under consideration, in particular the calculation of the spectrum and the
matrix elements of the fields. I also propose a
description of the model as a perturbation of its infinite coupling constant 
limit. I present the description of the spectrum by the Bethe-Yang equations,
which gives the exact result in this case.

In Chapter \ref{sec.chap4} I investigate the phase diagrams of the two- and
three-frequency sine-Gordon models by the TCSA method. The focus is mainly on
the finding of a
tricritical point in the case of the three-frequency model.  I give
substantial evidence that this point exists. I also find several points of the critical line in
the phase diagram and present TCSA data showing the change of the finite volume spectrum along
the critical line as the tricritical endpoint is approached. I find a few
points of the line of first order transition as well.

\clearpage

\

\clearpage

{\bf \large
\vspace*{-1cm}
\noindent
A f\H o eredm\'enyek \"osszefoglal\'asa \\ }

\noindent
Doktori \'ertekez\'esemben az $1+1$-dimenzi\'os kvantumt\'erelm\'elet
h\'arom r\'eszter\"ulet\'ehez tartoz\'o probl\'em\'akkal foglalkozom. Ennek
megfelel\H oen az \'ertekez\'es h\'arom l\'enyeg\'eben f\"uggetlen r\'eszre oszlik.

A \ref{sec.chap2}.\ fejezetben a  szuperszimmetrikus 
peremes integr\'alhat\'o kvantumt\'erelm\'eletek sz\'or\'as-elm\'elet\'et
tanulm\'anyozom egy speci\'alis konstrukci\'o
keretei k\"oz\"ott, amelyben a teljes S-m\'atrix \'es reflexi\'os m\'atrix
blokkjai egy szuperszimmetrikus \'es egy nem szuperszimmetrikus r\'esz
szorzatak\'ent \'allnak el\H o. T\'argyalom a szuperszimmetria 
defin\'ici\'oj\'at
perem jelenl\'ete eset\'en, tov\'abb\'a megadok olyan szab\'alyokat, amelyek
seg\'its\'eg\'evel meghat\'arozhat\'ok az egyes magasabb energi\'aj\'u
hat\'ark\"ot\"ott \'allapotokon megval\'osul\'o \'abr\'azol\'asok, \'es az ezen
\'allpotokr\'ol t\"ort\'en\H o r\'eszecskevisszaver\H od\'es m\'atrix\'anak szuperszimmetrikus
r\'eszei. Ezek a szab\'alyok abban az esetben alkalmazhat\'oak, amikor az elm\'elet
r\'eszecsk\'ei a kink vagy a bozon-fermion \'abr\'azol\'as szerint
transzform\'al\'odnak. A szab\'alyok alkalmaz\'as\'at  konkr\'et p\'eld\'akon
is bemutatom.  

A \ref{sec.chap3}.\ fejezetben a lev\'agott konform t\'er k\"ozel\'it\'es
(TCSA) nev\H u numerikus m\'odszer alkalmaz\'asakor v\'egzett
lev\'ag\'asnak a peremes renorm\'al\'asi csoport folyamokra val\'o hat\'as\'at 
 tanulm\'anyozom egy konkr\'et modell, a hat\'aron
m\'agnesesen perturb\'alt, szakaszon \'ertelme-zett kritikus Ising modell
eset\'en.   A kit\H uz\"ott c\'el annak az elk\'epzel\'esnek az igazol\'asa, hogy a lev\'ag\'as
hat\'asa a spektrumra figyelembe vehet\H o a Hamilton oper\'atorban szerepl\H
o tagok egy\"utthat\'oinak megv\'altoztat\'as\'aval. Ismertetem az \'altalam v\'egzett
numerikus \'es perturbat\'iv sz\'amol\'asok eredm\'enyeit, amelyek
al\'at\'amasztj\'ak ezt az elk\'epzel\'est. Elv\'egeztem az eml\'itett
sz\'amol\'asokat egy olyan lev\'ag\'asi elj\'ar\'as eset\'en
is, amelyik az eredeti modell egzakt megoldhat\'os\'ag\'at nem sz\"unteti meg.  
Az egy\"utthat\'ok megv\'altoztat\'asa ebben az esetben is alkalmasnak
l\'atszik a lev\'ag\'as hat\'as\'anak figyelembe v\'etel\'ere, tov\'abb\'a
kider\"ul, hogy (lev\'ag\'as alkalmaz\'asa ut\'an) a folyamok kvalitat\'iv viselked\'ese  f\"ugg az
alkalmazott le-v\'ag\'asi elj\'ar\'ast\'ol. 
Megadom a 
vizsg\'alt modell egy egzakt kvantumt\'erelm\'eleti megold\'as\'at, amely 
mag\'aba foglalja
t\"obbek k\"oz\"ott a spektrum \'es a terek
m\'atrixelemeinek kisz\'am\'it\'as\'at. Javaslom a modellnek egy a v\'egtelen
csatol\'as\'u hat\'areset perturb\'aci\'ojak\'ent val\'o le\'ir\'as\'at. 
Elv\'egzem a spektrum kisz\'am\'it\'as\'at a Bethe-Yang egyenletekkel is.

A \ref{sec.chap4}.\ fejezetben a k\'et- \'es h\'aromfrekvenci\'as sine-Gordon
modell f\'azisszerkezet\'enek  a TCSA m\'odszerrel val\'o
felt\'erk\'epez\'es\'evel foglalkozom. A f\H o eredm\'eny ebben a fejezetben
egy trikritikus pont megtal\'al\'asa. Emellett 
megkeresem annak a   kritikus
vonalnak sz\'amos pontj\'at, amelyiknek a v\'eg\'en  a trikritikus pont tal\'alhat\'o, \'es  bemutatom a  v\'eges
t\'erfogatbeli spektrum v\'altoz\'as\'at  a kritikus vonal ment\'en a trikritikus
pont fel\'e haladva.  Megkeresem a f\'azisdiagramban tal\'alhat\'o els\H
orend\H u f\'azishat\'ar n\'eh\'any pontj\'at is. 

\end{document}